\documentclass[12pt, twoside, english]{article}
\usepackage[utf8]{inputenc}            
\usepackage[T1]{fontenc} 
\usepackage{mathtools}			
\usepackage{amssymb}
\usepackage{a4wide}
\usepackage{graphicx}
\usepackage{fancyhdr}
\usepackage{amsmath, amssymb}
\usepackage{subfigure}
\usepackage{setspace}
\usepackage{verbatim} 
\usepackage[english]{babel}
\usepackage[novbox]{pdfsync} 
\usepackage{caption}
\usepackage{float}
\usepackage{booktabs}

\usepackage{xfrac}

\usepackage{units}

\makeatletter
\usepackage{hyperref}

\usepackage{xcolor}

\hypersetup{
	colorlinks = false,
	linktocpage = true,
	allbordercolors = {0.8 0.8 0.8}
}

\addto\captionsenglish{}
\addto\captionsenglish{}
\addto\extrasenglish{}
\addto\extrasenglish{}
\addto\extrasenglish{}
\addto\extrasenglish{}
\addto\extrasenglish{}
\addto\extrasenglish{}
\addto\extrasenglish{}

\newcommand{\lyxaddress}[1]{
\par {\raggedright #1
\vspace{1.4em}
\noindent\par}
}

\numberwithin{equation}{section}

\newcommand{\vek}[1]{\mathchoice{\displaystyle\boldsymbol#1}
{\textstyle\boldsymbol#1}{\scriptstyle\boldsymbol#1}
{\scriptscriptstyle\boldsymbol#1}}
\newcommand{\mat}[1]{\mathchoice{\displaystyle\mathbf#1}
{\textstyle\mathbf#1}{\scriptstyle\mathbf#1}
{\scriptscriptstyle\mathbf#1}}

\renewcommand{\d}{ \ensuremath{\mathrm{d} }}

\newcommand{\ti}{ \ensuremath{\mathrm{t} }}
\newcommand{\noi}{ \ensuremath{\mathrm{n} }}

\makeatother

\usepackage[backend=biber,style=numeric,natbib=true,hyperref=true,doi=true,sorting=none]{biblatex}
\DeclareFieldFormat[article, inbook]{title}{#1} 
\graphicspath{{Figures/}}

\begin{document}

\title{Simultaneous solution of incompressible Navier--Stokes flows on multiple surfaces}

\author{Michael Wolfgang Kaiser, Thomas-Peter Fries}
\date{February 12, 2025}
\maketitle

\lyxaddress{\begin{center}
Institute of Structural Analysis\\
Graz University of Technology\\
Lessingstr. 25/II, 8010 Graz, Austria\\
\texttt{www.ifb.tugraz.at}\\
\texttt{michael.kaiser@tugraz.at}\\
\texttt{fries@tugraz.at}
\end{center}}

\begin{abstract}
A mechanical model and finite element method for the \emph{simultaneous} solution of Stokes and incompressible Navier--Stokes flows on multiple curved surfaces over a bulk domain are proposed. The two-dimensional surfaces are defined implicitly by all level sets of a scalar function, bounded by the three-dimensional bulk domain. This bulk domain is discretized with hexahedral finite elements which do not necessarily conform with the level sets but with the boundary. The resulting numerical method is a hybrid between conforming and non-conforming finite element methods. Taylor--Hood elements or equal-order element pairs for velocity and pressure, together with stabilization techniques, are applied to fulfil the inf-sup conditions resulting from the mixed-type formulation of the governing equations. Numerical studies confirm good agreement with independently obtained solutions on selected, individual surfaces. Furthermore, higher-order convergence rates are obtained for sufficiently smooth solutions.
\end{abstract}
\tableofcontents{}\newpage{}

\section{Introduction\label{sec:Introduction}}
\begin{figure}
	\centering
	
	\subfigure[]{\includegraphics[width=0.33\textwidth]{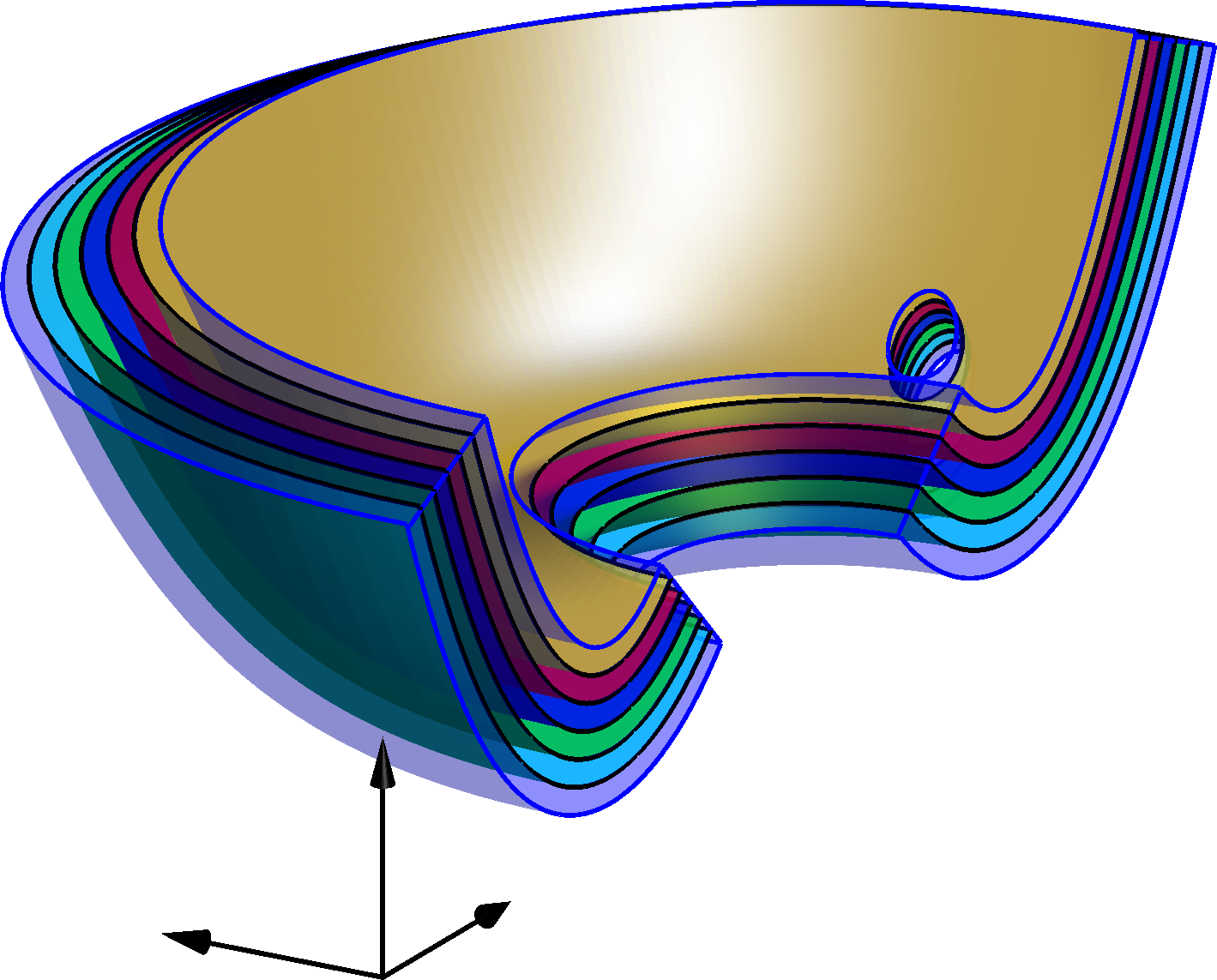}}\hfil
	\subfigure[]{\includegraphics[width=0.33\textwidth]{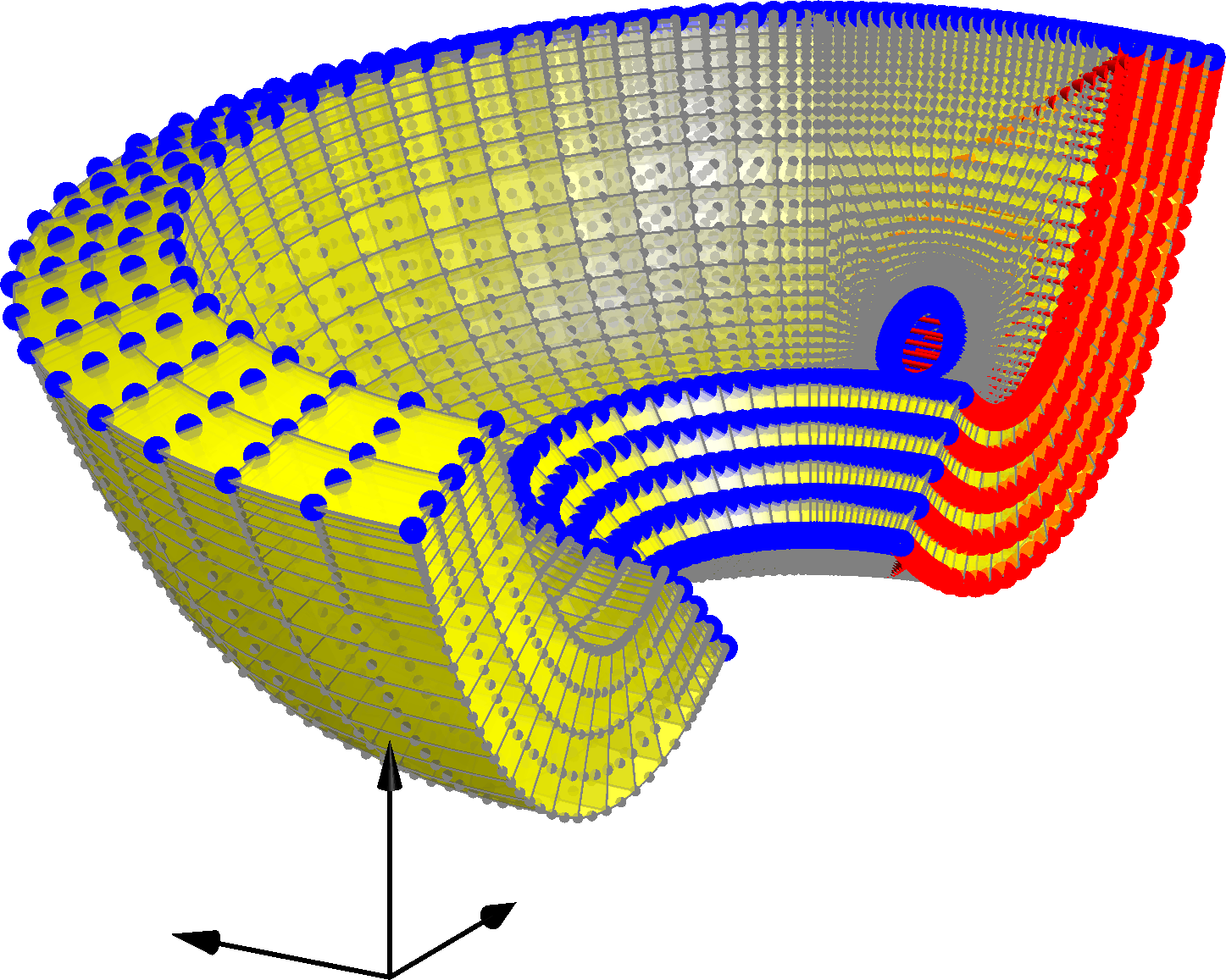}}\hfill
	\subfigure[]{\includegraphics[width=0.33\textwidth]{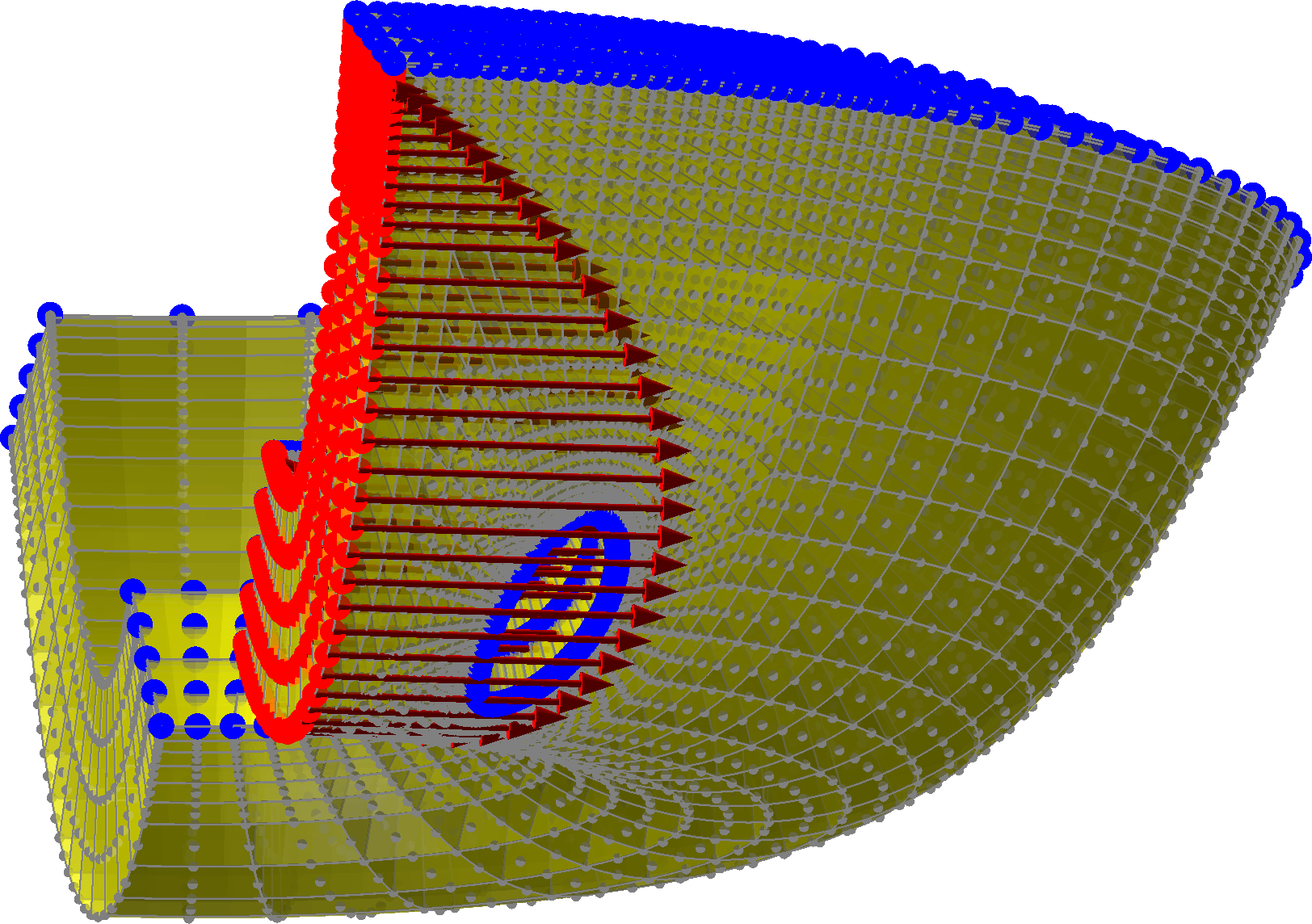}}
	
	\caption{\label{fig:GenExamp} A generic example: (a) The volumetric bulk domain in blue, some level sets shown in different colours. In (b) and (c), some example mesh and nodes with no-slip conditions (blue) and those on the inflow (red) are seen.}
\end{figure}

Flow phenomena are important topics of applied and basic research in several subjects, e.g., physics, chemistry, biology, engineering, and mathematics, e.g., \cite{John_2016a,Zienkiewicz_2014b,Edwards_1991a,Slattery_1990a}. The flow models are usually described by partial differential equations (PDEs) such as the Navier--Stokes equations. For flows on \emph{curved} surfaces, the interaction of the physics, i.e., the flow, which take part on the curved domain, and the geometry of the domain plays a crucial role in the formulation of the mathematical model. Due to more involved definitions of geometric and differential operators in these cases, such models are more advanced than usual models for two- or three-dimensional Euclidean geometries. Flows on curved domains have important applications in nature, e.g., \cite{Napoli_2016a,Gross_2018b}, and engineering, e.g., \cite{Zienkiewicz_2014b,Edwards_1991a,Slattery_1990a}. In recent years, flows on curved manifolds which are embedded in some higher-dimensional background space have gained significant attention, see, e.g., \cite{Jankuhn_2018a,Fries_2018a,Brandner_2022a,Brandner_2022b,Lederer_2020a,Olshanskii_2019a}. Therein, various formulations of the model and approximation methods have been proposed for individual surfaces. In this work, we propose a mechanical model and corresponding finite element method to solve flows on \emph{all} surfaces over a bulk domain \emph{simultaneously}.\\
\\
An overview of different models for surface flows and their derivations is found in \cite{Brandner_2022b}. The equations can be formulated based on curvilinear coordinates \cite{Brandner_2022b,Edwards_1991a} or in a coordinate-free formulation, e.g., based on the tangential differential calculus (TDC) \cite{Delfour_2011a,Delfour_1996a,Gurtin_1975a}. The TDC may be interpreted as the modern perspective on differential geometry to formulate PDEs on manifolds based on surface differential operators rather than based on local curvilinear coordinate systems. A detailed derivation of the incompressible Navier--Stokes equations on an evolving surface from first principles of (continuum) mechanics and formulated in  the TDC-framework is given in \cite{Jankuhn_2018a}. It should be noted that formulations of the Navier--Stokes equations on manifolds which are derived by substituting classical differential operators with their geometric counterparts are not necessarily equivalent to formulations derived from first mechanical principles because there may be (subtle) differences, see \cite{Jankuhn_2018a}, page 364. The numerical solution of flows on manifolds comes with additional challenges compared to the classical $d$-dimensional Euclidean space, e.g., the enforcement of a tangential velocity field.\\
\\
Different methods to solve (incompressible Navier--)Stokes flows on surfaces have been proposed in recent years, e.g., finite difference methods in \cite{Yang_2020a}, (classical) Surface finite element methods (FEM) in \cite{Fries_2018a,Bonito_2020a}, Trace FEM in \cite{Olshanskii_2018a,Olshanskii_2019a,Olshanskii_2022a,Jankuhn_2021a}, a FEM with tangential function spaces in \cite{Demlow_2024a,Lederer_2020a}, and a mesh-free method in \cite{Suchde_2021a}. Furthermore, one may distinguish models formulated for velocity and pressure as primal variables and models in which only scalar quantities are sought, e.g., stream-function formulations, see, e.g., \cite{Brandner_2022a}. In all of these methods (incompressible Navier--)Stokes flows on \emph{one single} surface are considered. In this paper we \emph{simultaneously} solve the (incompressible Navier--)Stokes flows on \emph{all}, i.e., infinitely many, level sets of a scalar function embedded in a three-dimensional bulk domain. We refer to the text books \cite{John_2016a,Zienkiewicz_2014b,Kuzmin_2014a} for the FEM in classical fluid dynamics. For an overview about different approaches of finite element methods for PDEs on \emph{curved} surfaces, the reader is referred to \cite{Dziuk_2013a}.\\
\\
Next, the concept of the simultaneous analysis of PDEs on manifolds in general and for (incompressible Navier--)Stokes flows on manifolds in particular is introduced which is the novelty of this paper. The manifolds are defined implicitly as level sets $\Gamma_{\!c}$ of a level-set function $\phi$ and embedded in a higher-dimensional bulk domain $\Omega$. The $c \in \mathbb{R}$ is some constant value related to an iso-surface. In a usual Trace FEM context, e.g., \cite{Olshanskii_2018a,Olshanskii_2019a,Olshanskii_2022a,Jankuhn_2021a}, a zero-isosurface of a level-set function is considered, i.e., $c=0$. For an introduction to level-set functions and the level-set method see, e.g., \cite{Osher_2006a,Gomes_2009a}. The manifolds are bounded by the boundary of the bulk domain, hence, the boundaries of the manifolds and the bulk domain are conforming. It is important to emphasise that the level sets do not have to be aligned to the mesh which is used to discretize the bulk domain in the applied FEM. Therefore, this method may be seen as a hybrid of conforming methods, e.g., the Surface FEM, and fictitious domain methods, e.g., the Trace FEM, and was labelled \emph{Bulk Trace FEM} by the authors in \cite{Fries_2023a} in the context of structural mechanics. However, the usual challenges of fictitious domain methods such as the Trace FEM, among them the need for stabilization and special quadrature in cut elements, do not apply for the Bulk Trace FEM. Similar approaches have been used in transport problems and diffusion on stationary surfaces in \cite{Dziuk_2008a} and in \cite{Dziuk_2010a} on evolving surfaces. A comparison of the solution of elliptic PDEs on all level sets within some bulk domain between phase-field methods and the simultaneous analysis with the FEM is given in \cite{Burger_2009a}. Historically related to the concept of the simultaneous analysis are narrow-band methods, e.g., \cite{Deckelnick_2010a,Deckelnick_2014a,Bertalmio_2001a, Greer_2006a, Greer_2006b}. One could interpret that the goal of narrow-band methods was to reduce the bulk domain to a minimum around a single surface. This goal was reached with fictitious domain methods where only \emph{one} level set, i.e., usually the zero-isosurface, is considered, see, e.g., \cite{Olshanskii_2014a,Olshanskii_2017a,Chernyshenko_2015a,Grande_2016a,Burman_2015a,Burman_2018a,Cenanovic_2016a,Schoellhammer_2021a,Fries_2020a}. Although only the solution on \emph{one} manifold is often required, it is useful to develop methods for the \emph{simultaneous} solution of \emph{all} manifolds embedded in a prescribed bulk domain. Possible applications are in the design process where variations in the geometry should be studied to find an optimal design. Furthermore, it can be applied for advanced anisotropic material models in structural mechanics and research of biological processes in the context of transport and flow problems. The authors developed mechanical models and applied corresponding Bulk Trace FEMs in the context of structural mechanics for geometrically non-linear membranes in \cite{Fries_2023a,Fries_2023c,Kaiser_2023b}, Reissner--Mindlin shells in \cite{Kaiser_2024a}, and for Timoshenko beams in \cite{Kaiser_2024b}. An overview of the used models and first results for transport and incompressible flow problems are shown in \cite{Fries_2024a} by the authors of this paper.\\
\\
In \cite{Fries_2018a}, a higher-order Surface FEM for (incompressible Navier--)Stokes flows (on single surfaces) is presented by the authors. A crucial aspect for models of surface flows is the enforcement of the tangentiality of the velocities. In \cite{Fries_2018a}, a Lagrange multiplier is used and in this work a (consistent) penalty method is applied, similar to a single surface in \cite{Jankuhn_2018a,Jankuhn_2019a,Olshanskii_2019a}. The governing equations are based on \cite{Jankuhn_2018a}. Herein, we consider stationary Stokes, stationary, and instationary Navier--Stokes flows on spatially fixed, two-dimensional curved surfaces $\Gamma_{\!c}$ embedded in a three dimensional bulk domain $\Omega$. First, the governing equations are formulated in strong form for each surface $\Gamma_{\!c}$. In the derivation of the weak form, the co-area formula is applied to formulate a weak form which is suitable for the \emph{simultaneous} analysis of all embedded level sets in $\Omega$. Fig.~\ref{fig:GenExamp} shows the conceptual idea of the simultaneous analysis: In Fig.~\ref{fig:GenExamp}(a), a bulk domain is depicted in blue and some, arbitrarily selected level sets are depicted in different colours. Figs.~\ref{fig:GenExamp}(b) and (c) show an example mesh, highlighting nodes with prescribed velocities.

In section \ref{sec:DSandDiffOp}, the geometric setup, surface differential operators, and divergence theorems are introduced as preliminaries. For each type of considered flow problems, i.e., stationary Stokes, stationary and instationary Navier--Stokes flows, the strong form, the weak forms, and numerical examples are given in Sections \ref{sec:StatStokes}, \ref{sec:StatNSEQ}, and \ref{sec:InstatNSEQ}, respectively. The paper ends in Section \ref{sec:ConclOutlook} with conclusions and an outlook.

\section{Geometric and mathematical preliminaries}\label{sec:DSandDiffOp}

In this section, the geometrical setup of two-dimensional surfaces embedded in a three-dimensional bulk domain is introduced. The differential operators that are used in the formulation of the mechanical models and related numerical methods are defined. Analogous definitions can be found in previous works by the authors, e.g., \cite{Fries_2023a} for the simultaneous solution of geometrically non-linear ropes and membranes and \cite{Kaiser_2024a} for the simultaneous solution of Reissner--Mindlin shells. 

\subsection{Geometric setup of embedded surfaces}\label{subsec:GeomSetup}

%
%

We consider flow phenomena described by the Stokes and incompressible Navier--Stokes equations on curved two-dimensional surfaces. These surfaces are manifolds with co-dimension 1 embedded in the three-dimensional physical space $\mathbb{R}^3$. A three-dimensional bulk domain $\Omega \subset \mathbb{R}^3$ and a level-set function $\phi\left(\vek{x}\right):\Omega\rightarrow\mathbb{R}$ are given. Within this bulk domain exists a minimal value $\phi^{\min}=\inf\phi\left(\vek{x}\right)$ and a maximum value $\phi^{\max}=\sup\phi\left(\vek{x}\right)$ of the level-set function. The individual manifolds $\Gamma_{\!c}$ defined by level sets of $\phi$ with constant level-set values $c\in\mathbb{R}$, 
\begin{equation}
	\Gamma_{\!c}=\left\{ \vek{x}\in\Omega:\,\phi(\vek{x})=c\in\mathbb{R}\right\},\,\phi^{\min}<c<\phi^{\max} ,\label{eq:LevelSets}
\end{equation}
are curved, two-dimensional manifolds, see, e.g., Fig.~\ref{fig:GenExamp}(a). $\phi^{\min}$ and $\phi^{\max}$ may be defined as the infimum/supremum of the level-set function inside the bulk domain or as user-defined values to restrict some larger bulk domain to a sub-interval of interest. If the surface is bounded, the boundary of some selected manifold $\Gamma_{\!c}$ is denoted as $\partial \Gamma_{\!c}$ and is the intersection curve of the level set $\Gamma_{\!c}$ with the boundary $\partial \Omega$ of the bulk domain. In this work, we only consider stationary surfaces and bulk domains, hence, these are fixed in time. The boundary of the bulk domain $\partial \Omega$ is restricted to the parts of the boundary where $\phi\left(\vek{x}\right)\neq\phi^{\min}$ and $\phi\left(\vek{x}\right)\neq\phi^{\max}$ for the proper definition of vector fields to be used in the mathematical description of the flow. Another requirement for the geometrical setup to state proper (initial) boundary value problems simultaneously on all level sets is that the embedded surfaces vary smoothly without topology changes within the bulk domain, see \cite{Fries_2023a,Kaiser_2024a} for further insights.

\subsection{Vector fields and the tangential projector}\label{subsec:GeomVecs}

\begin{figure}
	\centering
	
	\subfigure[vectors $\vek{n}$, $\vek{m}$, $\vek{t}$, $\vek{q}$]{\includegraphics[width=0.5\textwidth]{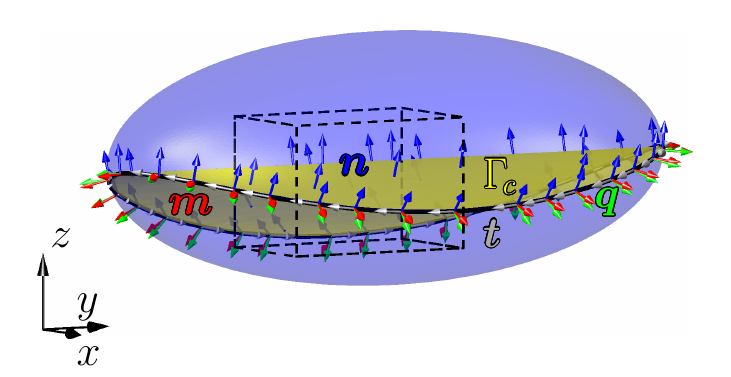}}\qquad\subfigure[zoomed view]{\includegraphics[width=0.4\textwidth]{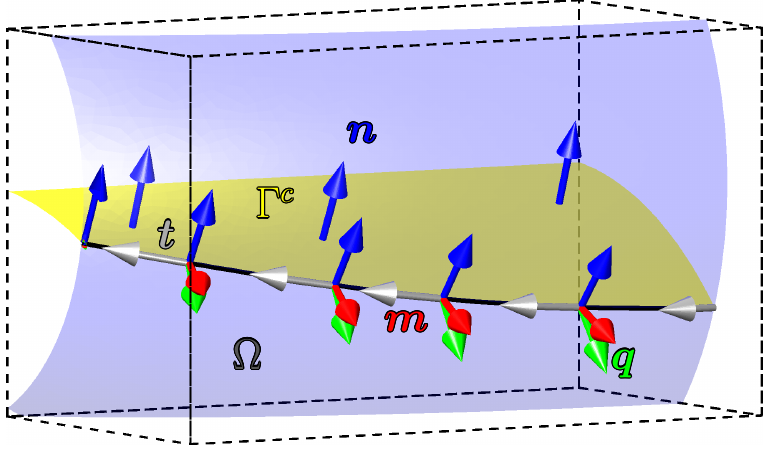}}
	
	\caption{\label{fig:VisVectors}Vector fields in the domain $\Omega$ and
		on the boundary $\partial\Omega$ shown on some level set $\Gamma_{\!c}$ with $c \in \left[\phi_{\min}, \phi_{\max}\right]$. The right figure shows a zoom of the left one. Normal vectors $\vek{n}$ with respect to the level sets $\Gamma_{\!c}$ in $\Omega$ are shown in blue. Normal vectors $\vek{m}$ with respect to $\partial\Omega$ are red, tangential vectors $\vek{t}$ are gray and co-normal vectors $\vek{q}$ are green.}
\end{figure}

We start by introducing some geometrical quantities. The normal vector to the level sets $\Gamma_{\!c}$ can easily be computed via the gradient of the level-set function $\phi$. The unit normal vector (field) $\vek{n}(\vek{x})$ in the whole bulk domain $\Omega$ is obtained by
\begin{equation}
	\vek{n}\left(\vek{x}\right)=\frac{\vek{n}^{\star}}{\left\Vert \vek{n}^{\star}\right\Vert }\quad\textrm{with}\quad\vek{n}^{\star}=\nabla\phi\left(\vek{x}\right),\;\vek{x}\in\Omega,\label{eq:NormalVector_n}
\end{equation}
using the (classical) gradient of the level-set function $\phi$. It is shown in Fig.~\ref{fig:VisVectors}, where for clarity, only one selected level set is plotted. The (tangential) projector field $\mat{P}\left(\vek{x}\right) \in \mathbb{R}^{3\times 3}$ is immediately obtained from the normal vector field as 
\begin{equation}
	\mat{P}\left(\vek{x}\right)=\mat{I}-\vek{n}\left(\vek{x}\right)\otimes\vek{n}\left(\vek{x}\right) \label{eq:tangProjector}
\end{equation} 
where $\mat{I}$ is the identity matrix in $\mathbb{R}^3$. This quantity is crucial to define tangential differential operators in Sec.~\ref{subsec:DiffOp}. Furthermore, it projects quantities onto the tangent space $T_{P}\Gamma$ of a curved surface $\Gamma$ at some point. An arbitrary vector $\vek{v} \in \mathbb{R}^3$ is projected onto the tangent space as $\vek{v}_\mathrm{t} = \mat{P} \cdot \vek{v}$ which will be used frequently in the remainder of this paper. Some important properties of the projector are (i) $\mat{P} = \mat{P}^{\mathrm{T}}$, (ii) $\mat{P} \cdot \mat{P} = \mat{P}$, and (iii) $\mat{P} \cdot \vek{n} = \vek{0}$.\\
\\
Along the boundary $\partial \Omega$ of the bulk domain, a normal vector (field) $\vek{m}(\vek{x}),\,\vek{x}\in\partial\Omega$ is defined. For the computation of the simultaneous flows later on, we assume that the bulk domain is discretized by (higher-order) volumetric (i.e., three-dimensional) finite elements. The computation of $\vek{m}(\vek{x})$ on element boundaries is a standard operation in the FEM.\\
\\
Furthermore, along the boundary $\partial \Omega$ of the bulk domain lives the tangent vector (field) $\vek{t}\left(\vek{x}\right)$ defined as
\begin{equation}
	\vek{t}\left(\vek{x}\right)= \vek{m} \times\vek{n}.\label{eq:TangVec_t}
\end{equation}
With the normal vector of the level sets $\Gamma_{\!c}$ and the tangential vector on $\partial \Omega$, a co-normal vector field $\vek{q}(\vek{x})$ is defined as
\begin{equation}
	\vek{q}\left(\vek{x}\right)=\frac{\vek{q}^{\star}}{\left\Vert \vek{q}^{\star}\right\Vert }\quad\textrm{with}\quad\vek{q}^{\star}=\vek{n} \times \vek{t}.\label{eq:Co-normalVector_q}
\end{equation}
These co-normal vectors play an important role in the formulation of the weak form of the (intitial) boundary value problem as fundamental parts of divergence theorems. Fig.~\ref{fig:VisVectors} shows normal vectors to the level sets as blue arrows, normal vectors to the boundary of the bulk domain as red arrows, tangential vectors as gray arrows, and co-normal vectors as green arrows.

\subsection{Differential operators on manifolds}\label{subsec:DiffOp}
In the governing equations of flows on surfaces, classical differential operators w.r.t.~the embedding three-dimensional space and \emph{tangential} or \emph{surface} differential operators w.r.t.~the curved, embedded, two-dimensional surfaces (level sets) must be distinguished. A subscript $\Gamma$ is used for surface quantities, e.g., $\mathrm{div}_{\Gamma}$ for the surface divergence. The resulting coordinate-free definition does not rely on the introduction of (local) curvilinear coordinates. This approach is sometimes labelled \emph{Tangential Differential Calculus} (TDC), cf.~\cite{Delfour_2011a}. The surface gradient of a scalar function $f\!\left(\vek{x}\right):\Omega\to\mathbb{R}$ is obtained as \cite{Delfour_2011a,Dziuk_2010a,Jankuhn_2018a,Fries_2018a}
\begin{equation}
	\nabla_{\Gamma}f=\mat{P}\cdot\nabla f,\label{eq:SurfGradScalarImplicit}
\end{equation}
where $\nabla f$ is the classical gradient in the three-dimensional
space $\mathbb{R}^3$ and $\mat{P}$ the projector defined in Eq.~(\ref{eq:tangProjector}).

For a vector function $\vek{v}\left(\vek{x}\right):\Omega\to\mathbb{R}^{3}$, the directional surface gradient is defined as
\begin{equation}
	\nabla_{\Gamma}^{\mathrm{dir}}\vek{v} = \nabla\vek{v}\cdot\mat{P}. \label{eq:SurfGradVectorImplicit}
\end{equation}
It is important to distinguish \emph{directional} and \emph{covariant} surface gradients for vector-valued functions, the latter defined as
\begin{equation}
	\nabla_{\Gamma}^{\mathrm{cov}}\vek{v}\;=\;\mat{P}\cdot\nabla_{\Gamma}^{\mathrm{dir}}\vek{v}\;=\;\mat{P}\cdot\nabla\vek{v}\cdot\mat{P}.\label{eq:CovariantSurfaceGradient}
\end{equation}
Note that the covariant gradient is an in-plane quantity, i.e., $ \nabla_{\Gamma}^{\mathrm{cov}}\vek{v}\in T_{P}\Gamma_{\!c}$, while the directional gradient is generally not in the tangent space of $\Gamma_{\!c}$, i.e., $ \nabla_{\Gamma}^{\mathrm{dir}}\vek{v}\notin T_{P}\Gamma_{\!c}$.\\
\\
The \emph{surface divergence} of vector-valued functions $\vek{v}\left(\vek{x}\right) : \Omega \to \mathbb{R}^3$ and second-order tensor-valued functions $\mat{T}\left(\vek{x}\right):\Omega\to\mathbb{R}^{3\times 3}$ are defined as
\begin{eqnarray}
	\mathrm{div}_{\Gamma}\,\vek{v} & = & \mathrm{tr}\left(\nabla_{\Gamma}^{\mathrm{dir}}\vek{v}\right)=\mathrm{tr}\left(\nabla_{\Gamma}^{\mathrm{cov}}\vek{v}\right)\eqqcolon\nabla_{\Gamma}\cdot\vek{v},\label{eq:DivergenceVector}\\
	\mathrm{div}_{\Gamma}\,\mat{T} & = & \left[\begin{array}{c}
		\mathrm{div}_{\Gamma}\left(T_{11},T_{12},T_{13}\right)\\
		\mathrm{div}_{\Gamma}\left(T_{21},T_{22},T_{23}\right)\\
		\mathrm{div}_{\Gamma}\left(T_{31},T_{32},T_{33}\right)
	\end{array}\right]\eqqcolon\nabla_{\Gamma}\cdot\mat{T},\label{eq:DivergenceTensor}
\end{eqnarray}
respectively.

\subsection{Weingarten map and curvature} \label{subsec:Curvature}

For the formulation of (initial) boundary value problems on curved surfaces the curvature of these domains is an important quantity. We use the Weingarten map \cite{Delfour_2011a,Jankuhn_2018a} to quantify the curvature. It is a symmetric, in-plane tensor defined as
\begin{equation}
	\mat{H} = \nabla_{\Gamma}^{\mathrm{dir}}\vek{n} = \nabla_{\Gamma}^{\mathrm{cov}}\vek{n}. \label{eq:Weingarten}
\end{equation}
The two non-zero eigenvalues are the principal curvatures, $\kappa_{1,2} = -\mathrm{eig}(\mat{H})$. The Gauß curvature is obtained as $K=\kappa_1 \cdot \kappa_2$ and the mean curvature as $\varkappa = \mathrm{tr}(\mat{H}) = \mathrm{div}\vek{n} = \mathrm{div}_{\Gamma} \vek{n}$ \cite{Schoellhammer_2019a, Delfour_2011a}.

\subsection{Integral theorems}\label{sec: IntThBTF}
Integral theorems are required to formulate the weak form of (partial) differential equations, needed for the resulting FEM formulation. For the simultaneous analysis of the flow fields on all level sets as proposed in this work, the relation between the integration over all level sets $\Gamma_{\!c}$ and the integration over the bulk domain $\Omega$ is given by the \emph{co-area formula} \cite{Dziuk_2008a,Federer_1969a,Morgan_1988a,Burger_2009a,Delfour_1995a}. The co-area formula for the integration of an arbitrary scalar function $f\left(\vek{x}\right)$ over all embedded surfaces $\Gamma_{\!c}$ in the level-set interval $\left[\phi^{\min},\;\phi^{\max}\right]$ is defined as
\begin{equation}
	\int_{\phi^{\min}}^{\phi^{\max}}\int_{\Gamma_{\!c}}f\left(\vek{x}\right)\;\ensuremath{\mathrm{d\ensuremath{\Gamma}}}\;\ensuremath{\mathrm{d}c}=\int_{\Omega}f\left(\vek{x}\right)\cdot\left\Vert \nabla\phi\right\Vert\;\ensuremath{\mathrm{d\ensuremath{\Omega}}}.\label{eq:CoareaFormulaDomain}
\end{equation}
Analogously, the co-area formula for the integration over the boundary $\partial\Gamma_{\!c}$  is defined as
\begin{equation}
		\int_{\phi^{\min}}^{\phi^{\max}}\int_{\partial\Gamma_{\!c}}f\left(\vek{x}\right)\cdot\vek{q}\;\ensuremath{\mathrm{d\ensuremath{\partial\Gamma}}}\;\ensuremath{\mathrm{d}c}=\int_{\partial\Omega}f\left(\vek{x}\right)\cdot\vek{q}\cdot\left(\vek{q}\cdot\vek{m}\right)\cdot\left\Vert \nabla\phi\right\Vert \;\ensuremath{\mathrm{d\ensuremath{\partial\Omega}}}.\label{eq:CorareaFormulaBoundary}
\end{equation}
In the co-area formulas, the norm of the \emph{classical gradient} of the level-set function $\phi$ is considered at the right hand side in the integration over the bulk domain $\Omega$ and its boundary $\partial\Omega$, respectively. It is seen that the co-normal vector $\vek{q}$ as defined in Eq.~(\ref{eq:Co-normalVector_q}) and the normal vector $\vek{m}$ on $\partial \Omega$ occur in the co-area formula, Eq.~(\ref{eq:CorareaFormulaBoundary}), see \cite{Fries_2023a,Dziuk_2013a} for further details.\\
\\
For the derivation of the weak form, divergence theorems are needed. The divergence theorem for a vector-valued function $\vek{v}(\vek{x}) : \Gamma_{\!c} \to \mathbb{R}^3$ and for a tensor-valued function $\mat{T}\!\left(\vek{x}\right)  : \Gamma_{\!c} \to\mathbb{R}^{3 \times 3}$ on a \emph{single} surface $\Gamma_{\!c}$ is defined as \cite{Delfour_1996a,Delfour_2011a,Fries_2018a}
\begin{equation}
	\int_{\Gamma_{\!c}}\vek{v}\cdot\mathrm{div}_{\Gamma}\,\mat{T}\,\mathrm{d}\Gamma=-\int_{\Gamma_{\!c}}\nabla_{\Gamma}^{\mathrm{dir}} \vek{v} : \mat{T}\,\mathrm{d}\Gamma+\int_{\Gamma_{\!c}}\varkappa\cdot \vek{v}\cdot\left(\mat{T}\cdot\vek{n}\right)\,\mathrm{d}\Gamma+\int_{\partial\Gamma_{\!c}}\vek{v}\cdot\left(\mat{T}\cdot\vek{q}\right)\,\mathrm{d}\partial\Gamma,\label{eq:DivTheoremVector}
\end{equation}
where $\nabla_{\Gamma}^{\mathrm{dir}} \vek{v} : \mat{T} = \mathrm{tr}\left(\nabla_{\Gamma}^{\mathrm{dir}} \vek{v} \cdot \mat{T}^{\mathrm{T}}\right)$. Note that the mean curvature $\varkappa$, the normal vector $\vek{n}$, and the co-normal vector $\vek{q}$ are involved. The term which includes the mean curvature $\varkappa$ on the right hand side vanishes for an in-plane tensor-valued function $\mat{T}(\vek{x})$, i.e., $\mat{T} = \mat{T}_{\mathrm{t}} = \mat{P} \cdot \mat{T} \cdot \mat{P} \in T_{P}\Gamma_{\!c}$ because $\mat{T}_{\mathrm{t}} \cdot \vek{n} = 0$. The combination of this divergence theorem for \emph{one} surface with the co-area formulas,  Eqs.~(\ref{eq:CoareaFormulaDomain}) and (\ref{eq:CorareaFormulaBoundary}), results in a divergence theorem for \emph{all} level sets in the bulk domain as \cite{Fries_2023a}
\begin{align}
	\begin{split}
		\int_{\Omega}\vek{v}\cdot\mathrm{div}_{\Gamma}\,\mat{T}\cdot\left\Vert 	\nabla\phi\right\Vert \,\mathrm{d}\Omega= & -\int_{\Omega}\left(\nabla_{\Gamma}^{\mathrm{dir}}\vek{v}:\mat{T}\right)\cdot\left\Vert \nabla\phi\right\Vert \,\mathrm{d}\Omega+\int_{\Omega}\varkappa\cdot\vek{v}\cdot\left(\mat{T}\cdot\vek{n}\right)\cdot\left\Vert \nabla\phi\right\Vert \,\mathrm{d}\Omega\\
		& +\int_{\partial\Omega}\vek{v}\cdot\left(\mat{T}\cdot\vek{q}\right)\cdot\left(\vek{q}\cdot\vek{m}\right)\cdot\left\Vert \nabla\phi\right\Vert \,\mathrm{d}\partial\Omega,\label{eq:DivTheoremTensorBTF}
	\end{split}
\end{align}
where again, the curvature term vanishes for in-plane tensors $\mat{T}_\ti$.

\section{Mechanical preliminaries}\label{sec:MechPrem}
The governing equations for flows on manifolds given in the following sections are derived from first principles of continuum mechanics. For a detailed derivation of the instationary Navier--Stokes equations on a moving domain we refer to, e.g., \cite{Jankuhn_2018a}. The flow models (for \emph{one} surface) considered in this work are special cases of those derived in \cite{Jankuhn_2018a} and can also be found in other works, e.g., \cite{Fries_2018a,Bothe_2010a,Jankuhn_2019a,Koba_2017a}. A \emph{stationary} manifold is considered for all flow problems in this work, i.e., the surfaces are fixed in time. In this section, the quantities which are used in the formulation of the $[$initial$]$ boundary value problems ($[$I$]$BVP) given in Sec.~\ref{sec:StatStokes} to Sec.~\ref{sec:InstatNSEQ} are introduced.\\
\\
The velocity field lives in the tangent space of $\Gamma$, i.e., $\vek{u}_\ti = \mat{P} \cdot \vek{u}$ with some three-dimensional velocity field  $\vek{u}\left(\vek{x}\right)$ on the surface. In this work, we use the tangential velocity field $\vek{u}_\ti$ in the formulation of the considered models for (Navier--)Stokes flows on manifolds. Therefore, in the formulation of the weak form later on, it is necessary to multiply the test function $\vek{w}_{\vek{u}}$ with the projector $\mat{P}$. An alternative is to use a general (arbitrary) velocity field $\vek{u}$ which is then constrained to live in the tangent space of $\Gamma_{\!c}$ using additional Lagrange multipliers, see \cite{Jankuhn_2018a,Fries_2018a}. The velocity field may be split in a tangential and a normal part as \cite{Jankuhn_2018a}
\begin{equation}
	\vek{u} = \mat{P}\cdot \vek{u} + (\vek{u} \cdot \vek{n})\cdot\vek{n} = \vek{u}_\ti + u_\noi \cdot \vek{n}.\label{eq:VelSplit}
\end{equation}
Furthermore, there is a pressure field  $p\left(\vek{x}\right)$ and a tangential body force $\vek{f}_\ti\left(\vek{x}\right)$ which is often expressed as $\vek{f}_\ti\left(\vek{x}\right) = \rho \cdot \vek{g}_\ti$ with the density $\rho \in \mathbb{R}^+$ of the fluid and $\vek{g}_\ti = \mat{P} \cdot [0,0,-9.81]^{\mathrm{T}}$ when gravity is considered \cite{Fries_2018a}. Note that the subscript t, i.e., $\square_\ti$ indicates tangential quantities, while $t$ stands for time below.

\paragraph{Stress and strain tensors.}

A \emph{directional} and a \emph{covariant} strain tensor are introduced as \cite{Fries_2018a}
\begin{eqnarray}
	\vek\varepsilon^{\mathrm{dir}}\left(\vek{u}_\ti\right) & = & \frac{1}{2}\cdot\left(\nabla_{\Gamma}^{\mathrm{dir}}\vek{u}_\ti+\left(\nabla_{\Gamma}^{\mathrm{dir}}\vek{u}_\ti\right)^{\mathrm{T}}\right),\label{eq:StrainTensorDir}\\
	\vek\varepsilon^{\mathrm{cov}}\left(\vek{u}_\ti\right) & = & \frac{1}{2}\cdot\left(\nabla_{\Gamma}^{\mathrm{cov}}\vek{u}_\ti+\left(\nabla_{\Gamma}^{\mathrm{cov}}\vek{u}_\ti\right)^{\mathrm{T}}\right),\label{eq:StrainTensorCov}
\end{eqnarray}
respectively and related by $\vek\varepsilon^{\mathrm{cov}}\left(\vek{u}_\ti\right)=\mat{P}\cdot\vek\varepsilon^{\mathrm{dir}}\left(\vek{u}_\ti\right)\cdot\mat{P}$. Using Eq.~(\ref{eq:VelSplit}), the relation $ \vek\varepsilon^{\mathrm{cov}}\left(\vek{u}\right) = \vek\varepsilon^{\mathrm{cov}}\left(\vek{u}_\ti\right) + u_\noi \cdot \mat{H}$, c.f., \cite{Jankuhn_2018a}, is useful for the implementation later on.\\
\\The stress tensor is defined as
\begin{equation}
\vek{\sigma}_\ti = \vek{\sigma}\left(\vek{u}_\ti,p\right)=-p\cdot\mat{P}+2\mu\cdot\vek\varepsilon^{\mathrm{cov}}\left(\vek{u}_\ti\right)\label{eq:StressTens}
\end{equation}
which is the Boussinesq--Scriven surface stress tensor for stationary surfaces \cite{Bothe_2010a,Gurtin_1975a,Scriven_1960a,Jankuhn_2018a} and $\mu\in\mathbb{R}^{+}$ is the (constant) dynamic viscosity.

\paragraph{Boundary conditions.}

The boundary $\partial\Gamma_{\!c}$ of a manifold is decomposed into two non-overlapping parts, the Dirichlet boundary $\partial\Gamma_{\!c,\mathrm{D}}$ and the Neumann boundary $\partial\Gamma_{\!c,\mathrm{N}}$.
The boundary conditions are given as 
\begin{equation}
	\vek{u}_\ti\left(\vek{x}\right)  =  \hat{\vek{u}}_\ti\left(\vek{x}\right) \quad  \text{on }\partial\Gamma_{\!c,\mathrm{D}},\label{eq:DirBoundaryConditions}
\end{equation}
with prescribed velocities $\hat{\vek{u}}_\ti$ along the Dirichlet boundary and
\begin{equation}
	\vek{\sigma}_\ti\left(\vek{x}\right)\cdot\vek{q} \left(\vek{x}\right) =  \hat{\vek{t}}_\ti\left(\vek{x}\right) \quad \text{on }\partial\Gamma_{\!c,\mathrm{N}},\label{eq:NeumBoundaryConditions}
\end{equation}
with given tractions $\hat{\vek{t}}_\ti$ along the Neumann boundary. Note that $\hat{\vek{u}}_\ti$ and $\hat{\vek{t}}_\ti$ are in the tangent space of $\Gamma_{\!c}$, i.e., $\hat{\vek{u}}_\ti\cdot\vek{n}=\hat{\vek{t}}_\ti\cdot\vek{n}=0.$\\
\\
There are usually no explicit boundary conditions needed for the pressure $p$. However, when no Neumann boundary is present, i.e., $\partial\Gamma_{c,\mathrm{N}}=\emptyset$ and $\partial\Gamma_{c,\mathrm{D}}=\partial\Gamma$ or in the case of compact manifolds where $\partial\Gamma=\emptyset$, the pressure is defined up to a constant \cite{Donea_2003a,Gresho_2000a,Fries_2018a}. For such cases, the pressure is prescribed at a given point on $\Gamma_{\!c}$ or imposed by the constraint $\int_{\Omega}p \cdot \lVert \nabla \phi \rVert\;\mathrm{d}\Omega=0$ in the weak form for all level sets within a bulk domain.

\paragraph{Vorticity on manifolds.}

A physical quantity which is often computed in the context of flow phenomena is the vorticity $\vek\omega$. For flows on manifolds it is defined as \cite{Fries_2018a}
\begin{equation}
	\vek\omega=\nabla_{\Gamma}^{\mathrm{cov}}\times\vek{u}_\ti.\label{eq:RegularVorticity}
\end{equation}
The vorticity $\vek\omega$ is co-linear to the normal vector $\vek{n}$ which leads to a zero-vector when it is projected onto the tangential space because $\mat{P}\cdot\vek\omega=\vek0$. For this reason, a scalar quantity $\omega^{\star}$ is
determined which is the signed magnitude of $\vek\omega$ defined as
\begin{equation}
	\omega^{\star}\left(\vek{x}\right)=\vek\omega\cdot\vek{n}=\left(\nabla_{\Gamma}^{\mathrm{dir}}\times\vek{u}_\ti\right)\cdot\vek{n}=\pm\left\Vert \vek\omega\right\Vert \qquad\forall\vek{x}\in\Omega.\label{eq:SpecialVorticity}
\end{equation}

\section{Stationary Stokes flow}\label{sec:StatStokes}
\subsection{Strong form for one level set}\label{subsec:StStok-sfEq1LS}
Stationary Stokes flow on a manifold in stress-divergence form \cite{Donea_2003a,Fries_2018a,Jankuhn_2018a} is formulated in the governing field equations to be fulfilled $\forall\vek{x}\in\Gamma_{\!c}$ as
\begin{eqnarray}
	-\mat{P}\cdot\mathrm{div}_{\Gamma}\,\vek{\sigma}\left(\vek{u}_\ti,p\right) & = & \vek{f}_\ti,\label{eq:MomEqtStatStokes}\\
	\mathrm{div}_{\Gamma}\,\vek{u}_\ti & = & 0.\label{eq:ContinuityConstraint}
\end{eqnarray}
Three momentum equations are expanded from Eq.~(\ref{eq:MomEqtStatStokes}), and Eq.~(\ref{eq:ContinuityConstraint}) is the incompressibility constraint. It is easily shown that
\begin{equation}
-\mat{P}\cdot\mathrm{div}_{\Gamma}\,\vek{\sigma}\left(\vek{u}_\ti,p\right)=\nabla_{\Gamma}p-2\mu\mat{P}\cdot\mathrm{div}_{\Gamma}\,\vek\varepsilon^{\mathrm{cov}}\left(\vek{u}_\ti \right).\label{eq:DivSigProp}
\end{equation}

\subsection{Weak form for one level set}\label{subsec:StStok-wfEq1LS}
Function spaces are defined to formulate the weak form of the governing equations. For the weak form on \emph{one single} manifold, the following function spaces are introduced \cite{Fries_2018a}:
\begin{eqnarray}
	\mathcal{S}_{\vek{u}}^{\Gamma} & = & \left\{\vek{u}\in\left[\mathcal{H}^{1}\left(\Gamma_{\!c}\right)\right]^{3},\:\vek{u}=\hat{\vek{u}}\:\textrm{on}\:\partial\Gamma_{\!c,\mathrm{D}}\right\} ,\label{eq:TrialU}\\
	\mathcal{V}_{\vek{u}}^{\Gamma} & = & \left\{ \vek{w}_{\vek{u}}\in\left[\mathcal{H}^{1}\left(\Gamma_{\!c}\right)\right]^{3},\:\vek{w}_{\vek{u}}=\vek{0}\:\textrm{on}\:\partial\Gamma_{\!c,\mathrm{D}}\right\} ,\label{eq:TestU}\\
	\mathcal{S}_{p}^{\Gamma}=\mathcal{V}_{p}^{\Gamma} & = & \mathcal{L}_{2}\left(\Gamma_{\!c}\right),\label{eq:TrialAndTestP}
\end{eqnarray}
where $\mathcal{H}^{1}$ is the Sobolev space of functions with square integrable first derivatives and $\mathcal{L}_{2}$ is the Lebesque space. The function space for the pressure $\mathcal{S}_{p}^{\Gamma}$ may be replaced by 
\begin{equation}
\mathcal{S}_{p}^{\Gamma,0}=\bigg\{ p\in\mathcal{L}_{2}\left(\Gamma_{\!c}\right),\:\int_{\Gamma}p\;\mathrm{d}A=0\bigg\},
\end{equation}
in the case where no Neumann boundary exists, as described above.\\
\\
Using the introduced function spaces, the weak form of the stationary Stokes flow on \emph{one} surface is obtained as usual, that is, by multiplication of the strong form of the governing equations with suitable test functions, i.e., $\vek{w}_{\vek{u},\ti}$ and $w_p$, and integration over the domain, including the application of the divergence theorem, given in Eq.~(\ref{eq:DivTheoremVector}). The resulting continuous weak form is stated as: Given a (constant) shear viscosity $\mu \in \mathbb{R}^+$, penalty parameter $\alpha$, body forces $\vek{f}\left(\vek{x}\right)$ on $\Gamma_{\!c}$, and boundary tractions $\hat{\vek{t}}\left(\vek{x}\right)$ on $\partial \Gamma_{\!c,\mathrm{N}}$, find the velocity field $\vek{u}\left(\vek{x}\right)\in\mathcal{S}_{\vek{u}}^{\Gamma}$ and the pressure field $p\left(\vek{x}\right)\in\mathcal{S}_{p}^{\Gamma}$ such that for all test functions $\left(\vek{w}_{\vek{u}},w_{p}\right)\in\mathcal{V}_{\vek{u}}^{\Gamma}\times\mathcal{V}_{p}^{\Gamma}$, there holds in $\Gamma_{\!c}$, see, e.g., \cite{Jankuhn_2021a,Olshanskii_2019a,Jankuhn_2018a}, 
\begin{align}
	\begin{split}
		\int_{\Gamma_{\!c}}\nabla_{\Gamma}^{\mathrm{dir}}\vek{w}_{\vek{u},\ti}:\vek{\sigma}\left(\vek{u}_\ti,p\right)\mathrm{d}\Gamma\,+\,\alpha\,\cdot & \int_{\Gamma_{\!c}} \left(\vek{u} \cdot \vek{n} \right) \cdot \left(\vek{w}_{\boldsymbol{u}} \cdot \vek{n} \right) \mathrm{d}\Gamma \\ =  \int_{\Gamma_{\!c}}\vek{w}_{\vek{u},\ti}\cdot\vek{f}_t\,\mathrm{d}\Gamma\,+\, & \int_{\partial\Gamma_{\!c,\mathrm{N}}}\!\!\!\vek{w}_{\vek{u},\ti}\cdot\hat{\vek{t}}_\ti\,\mathrm{d}\partial\Gamma,\label{eq:wf1srfStatStokes-mom}
	\end{split}\\
	\int_{\Gamma_{\!c}}w_{p}\cdot&\mathrm{div}_{\Gamma}\,\vek{u}_\ti\:\mathrm{d}\Gamma = 0.\label{eq:wf1srfStatStokes-cont}
\end{align}

Note that the test and trial functions $\vek{u}_\ti = \mat{P} \cdot \vek{u}$ and $\vek{w}_{\vek{u},\ti} = \mat{P} \cdot \vek{w}_{\vek{u}}$, which live in the tangent space of the manifold, are used in the weak form except in the penalty term where the (arbitrary) three-dimensional $\vek{u}$ and $\vek{w}_{\vek{u}}$ are used. With the definition of the stress tensor, i.e., Eq.~(\ref{eq:StressTens}), the first term on the left hand side may be written as
\begin{equation}
\int_{\Gamma_{\!c}}\nabla_{\Gamma}^{\mathrm{dir}}\vek{w}_{\vek{u},\ti}:\vek{\sigma}\left(\vek{u}_\ti,p\right)\mathrm{d}\Gamma=-\int_{\Gamma_{\!c}}\nabla_{\Gamma}^{\mathrm{dir}}\vek{w}_{\vek{u},\ti}:\left(p\cdot\mat{P}\right)\mathrm{d}\Gamma+2\mu\cdot\int_{\Gamma_{\!c}}\nabla_{\Gamma}^{\mathrm{dir}}\vek{w}_{\vek{u},\ti}:\vek\varepsilon^{\mathrm{cov}}\left(\vek{u}_\ti\right)\mathrm{d}\Gamma \label{eq:FrobNormStressSRF}
\end{equation}
with the following relations \cite{Fries_2018a}
\begin{eqnarray}
	\nabla_{\Gamma}^{\mathrm{dir}}\vek{w}_{\vek{u},\ti}:\left(p\cdot\mat{P}\right) & = & p\cdot\mathrm{tr}\left(\nabla_{\Gamma}^{\mathrm{dir}}\vek{w}_{\vek{u},\ti}\cdot\mat{P}\right)\nonumber \\
	& = & p\cdot\mathrm{div}_{\Gamma}\,\vek{w}_{\vek{u},\ti},\nonumber \\
	\nabla_{\Gamma}^{\mathrm{dir}}\vek{w}_{\vek{u},\ti}:\vek\varepsilon^{\mathrm{cov}}\left(\vek{u}_\ti\right) & = & \mathrm{tr}\left(\nabla_{\Gamma}^{\mathrm{dir}}\vek{w}_{\vek{u},\ti}\cdot\vek{\varepsilon}^{\mathrm{cov}}\left(\vek{u}_\ti\right)\right)\nonumber \\
	& = & \mathrm{tr}\left(\vek\varepsilon^{\mathrm{cov}}\left(\vek{w}_{\vek{u},\ti}\right)\cdot\vek\varepsilon^{\mathrm{cov}}\left(\vek{u}_\ti\right)\right)\nonumber \\
	& = & \mathrm{tr}\left(\mat{P}\cdot\nabla_{\Gamma}^{\mathrm{dir}}\vek{w}_{\vek{u},\ti}\cdot\vek{\varepsilon}^{\mathrm{dir}}\left(\vek{u}_\ti\right)\cdot\mat{P}\right).\nonumber
\end{eqnarray}
In \cite{Fries_2018a} a similar weak form is stated where the tangentiality of the velocity field is enforced by a Lagrange multiplier instead of the penalty method as used herein. Further discussions about weak forms with different strategies to enforce the tangentialty of the velocities with a focus on mathematical details including the definition of the applied function spaces are found in \cite{Jankuhn_2018a}. The weak form given here in Eqs.~(\ref{eq:wf1srfStatStokes-mom}) and (\ref{eq:wf1srfStatStokes-cont}) is used to obtain the FE solution for \emph{one single} surface.

\subsection{Weak form for all level sets in a bulk domain}\label{subsec:StStok-wfEqAllLS}

\paragraph{Continuous weak form.} To obtain the weak form which is required for the \emph{simultaneous} solution of \emph{all} level sets $\Gamma_{\!c}$ embedded in a bulk domain $\Omega$, Eqs.~(\ref{eq:wf1srfStatStokes-mom}) and (\ref{eq:wf1srfStatStokes-cont}) are integrated over the level-set interval from $\phi^{\min}$ to $\phi^{\max}$. There follows for the weak form of stationary Stokes flow, analogously to Eqs.~(\ref{eq:wf1srfStatStokes-mom}) and (\ref{eq:wf1srfStatStokes-cont}),
\begin{align}
	\begin{split}
		\int_{\phi^{\min}}^{\phi^{\max}}\int_{\Gamma_{\!c}}\nabla_{\Gamma}^{\mathrm{dir}}\vek{w}_{\vek{u},\ti}:\vek{\sigma}\left(\vek{u}_\ti,p\right)\mathrm{d}\Gamma\;\mathrm{d}c\,+\,\alpha\,\cdot & \int_{\phi^{\min}}^{\phi^{\max}}\int_{\Gamma_{\!c}} \left(\vek{u} \cdot \vek{n} \right) \cdot \left(\vek{w}_{\boldsymbol{u}} \cdot \vek{n} \right) \mathrm{d}\Gamma\;\mathrm{d}c \\ = \int_{\phi^{\min}}^{\phi^{\max}}\int_{\Gamma_{\!c}}\vek{w}_{\vek{u},\ti}\cdot\vek{f}_t\,\mathrm{d}\Gamma\;\mathrm{d}c\,+\, &\int_{\phi^{\min}}^{\phi^{\max}}\int_{\partial\Gamma_{\!c,\mathrm{N}}}\!\!\!\vek{w}_{\vek{u},\ti}\cdot\hat{\vek{t}}_\ti\,\mathrm{d}\partial\Gamma\;\mathrm{d}c,\label{eq:wfBTF2IntStatStokes-mom}
	\end{split}\\
	\int_{\phi^{\min}}^{\phi^{\max}}\int_{\Gamma_{\!c}}w_{p} & \cdot\mathrm{div}_{\Gamma}\,\vek{u}_\ti\:\mathrm{d}\Gamma\;\mathrm{d}c = 0.\label{eq:wfBTF2IntStatStokes-cont}
\end{align}
Applying the co-area formula over the domain, i.e., Eq.~(\ref{eq:CoareaFormulaDomain}), the double integrals $\int_{\phi^{\min}}^{\phi^{\max}}\int_{\Gamma_{\!c}} \bullet \,\mathrm{d}\Gamma\;\mathrm{d}c$ over the interval of level sets and the domain of one level set can be converted to an integral over the bulk domain $\int_{\Omega} \bullet \cdot \lVert \nabla \phi \rVert\,\mathrm{d}\Omega$. With Eq.~(\ref{eq:CorareaFormulaBoundary}), the procedure is analogous for the boundary terms. Additionally, the following function spaces are introduced:
\begin{eqnarray}
	\mathcal{S}_{\vek{u}}^{\Omega} & = & \left\{\vek{u}\in\left[\mathcal{H}^{1}\left(\Omega\right)\right]^{3},\:\vek{u}=\hat{\vek{u}}\:\textrm{on}\:\partial\Omega_{\mathrm{D}}\right\} ,\label{eq:TrialUBTF}\\
	\mathcal{V}_{\vek{u}}^{\Omega} & = & \left\{ \vek{w}_{\vek{u}}\in\left[\mathcal{H}^{1}\left(\Omega\right)\right]^{3},\:\vek{w}_{\vek{u}}=\vek{0}\:\textrm{on}\:\partial\Omega_{\mathrm{D}}\right\} ,\label{eq:TestUBTF}\\
	\mathcal{S}_{p}^{\Omega}=\mathcal{V}_{p}^{\Omega} & = & \mathcal{L}_{2}\left(\Omega\right).\label{eq:TrialAndTestPBTF}
\end{eqnarray}
These function spaces may be seen as the bulk-equivalent over $\Omega$ compared to those given in Eqs. (\ref{eq:TrialU}) to (\ref{eq:TrialAndTestP}) w.r.t.~individual surfaces $\Gamma_c$. Now, the weak form of the stationary Stokes flow on \emph{all} manifolds $\Gamma_{\!c}$ over the bulk domain $\Omega$ can be formulated similar to the weak form on \emph{one} manifold, see Sec.~\ref{subsec:StNSEQ-sfEq1LS}. Using the introduced function spaces, the trial and test functions $\vek{u}_\ti = \mat{P} \cdot \vek{u}$ and $\vek{w}_{\vek{u},\ti} = \mat{P} \cdot \vek{w}_{\vek{u}}$, respectively, the weak form of the stationary Stokes flow on \emph{all} surfaces embedded in a bulk domain is obtained as: Given a (constant) shear viscosity $\mu \in \mathbb{R}^+$, penalty parameter $\alpha$, body forces $\vek{f}\left(\vek{x}\right)$, and boundary tractions $\hat{\vek{t}}\left(\vek{x}\right)$ on $\partial \Omega_{\mathrm{N}}$, find the velocity field $\vek{u}\left(\vek{x}\right)\in\mathcal{S}_{\vek{u}}^{\Omega}$ and the pressure field $p\left(\vek{x}\right)\in\mathcal{S}_{p}^{\Omega}$ such that for all test functions $\left(\vek{w}_{\vek{u}},w_{p}\right)\in\mathcal{V}_{\vek{u}}^{\Omega}\times\mathcal{V}_{p}^{\Omega}$, there holds in $\Omega$ 
\begin{align}
	\begin{split}
		\int_{\Omega} \big(\nabla_{\Gamma}^{\mathrm{dir}}\vek{w}_{\vek{u},\ti}:\vek{\sigma}\left(\vek{u}_\ti,p\right)\big) & \cdot \lVert \nabla \phi \rVert\;\mathrm{d}\Omega\,+\,\alpha\,\cdot \int_{\Omega} \left(\vek{u} \cdot \vek{n} \right) \cdot \left(\vek{w}_{\boldsymbol{u}} \cdot \vek{n} \right) \cdot \lVert \nabla \phi \rVert\;\mathrm{d}\Omega \\ = \int_{\Omega}\vek{w}_{\vek{u},\ti}\cdot\vek{f}_\ti & \cdot \lVert \nabla \phi \rVert\;\mathrm{d}\Omega \,+ \int_{\partial\Omega_{\mathrm{N}}}\!\!\!\vek{w}_{\vek{u}}\cdot\hat{\vek{t}}_\ti \cdot \left(\vek{q} \cdot \vek{m}\right)\cdot \lVert \nabla \phi \rVert\;\mathrm{d}\partial\Omega,
	\end{split}\\
	& \int_{\Omega}w_{p} \cdot\mathrm{div}_{\Gamma}\,\vek{u}_\ti \: \cdot \lVert \nabla \phi \rVert\;\mathrm{d}\Omega = 0.
\end{align}
For the case of the consideration of all level sets embedded in the bulk domain follows for the term with the stress tensor analogously to Eq.~(\ref{eq:FrobNormStressSRF})
\begin{align}
	\begin{split}
		\int_{\Omega}\big(\nabla_{\Gamma}^{\mathrm{dir}}\vek{w}_{\vek{u},\ti}:\vek{\sigma}\left(\vek{u}_\ti,p\right)\big)\cdot \lVert \nabla \phi \rVert\;\mathrm{d}\Omega=-\int_{\Omega} \big(\nabla_{\Gamma}^{\mathrm{dir}}\vek{w}_{\vek{u},\ti}:\left(p\cdot\mat{P}\right) \big) \cdot \lVert \nabla \phi \rVert\; & \mathrm{d}\Omega \\ +\, 2\mu\cdot\int_{\Omega}\big(\nabla_{\Gamma}^{\mathrm{dir}}\vek{w}_{\vek{u},\ti}:\vek\varepsilon^{\mathrm{cov}}\left(\vek{u}_\ti\right)
	\big)\cdot \lVert \nabla \phi \rVert\; & \mathrm{d}\Omega. \label{eq:FrobNormStressBTF}
	\end{split}
\end{align}

\paragraph{Discretization of the weak form.} The bulk domain $\Omega$ is discretized by a conforming, three-dimensional mesh of higher-order  tetrahedral or hexahedral Lagrange elements of order $q_k$. The resulting mesh is an approximation $\Omega_{q_k}^{h}$ of $\Omega$ with the nodal coordinates denoted as $\vek{x}_{i}$ where $i=1,\dots,n_{q_k}$ and $n_{q_k}$ being the number of nodes in the mesh. Note that the mesh is conforming to the boundary of the bulk domain $\partial \Omega$, however, it does not have to be aligned with the level sets $\Gamma_{\!c}$. This is why the resulting FEM can be seen as a hybrid between the conforming Surface FEM (i.e., classical FEM) and non conforming fictitious domain methods, e.g., the Trace FEM.\\
\\
Finite element spaces of different orders are involved as usual when the FEM is applied to (Navier--)Stokes flow formulated in stress-divergence form (e.g., Taylor-Hood elements \cite{Taylor_1973a}). Global $C^{0}$-continuous basis functions $B_{i}^{q_k}\!\left(\vek{x}\right)$ are introduced with $i = 1,\ldots,{n_{q_k}}$. These basis functions span a $C^{0}$-continuous finite element space of order $q_k$ defined as
\begin{align}
	\mathcal{Q}_{q_k}^{h}:=\left\lbrace v_{h}\in C^{0}(\Omega_{q_{\mathrm{geom}}}^{h}):\ v_{h}=\sum_{i=1}^{n_{q_k}}B_{i}^{q_k}(\vek{x})\cdot\hat{v}_{i}\text{ with }\hat{v}_{i}\in\mathbb{R}\right\rbrace \subset\mathcal{H}^{1}(\Omega_{q_{\mathrm{geom}}}^{h}).\label{eq:BTFFctSpace}
\end{align} 
Note that only the coordinates of the geometry mesh are needed to generate the basis $\{B_{i}^{q_k}\!\left(\vek{x}\!\left(\vek{r}\right)\right)\}$. Different degrees of polynomial orders are introduced: (i) $q_{\mathrm{geom}}$ for the mesh which describes the geometry of the bulk domain approximately, (ii) $q_{\vek{u}}$ for the velocity field, and (iii) $q_{p}$ for the pressure field, i.e., in Eq.~(\ref{eq:BTFFctSpace}) holds $k\in \left\{\mathrm{geom},\vek{u},p\right\}$. Note that this is only an isoparametric map if $q_k = q_{\mathrm{geom}}$. Analogously to a typical setup in classical FEM for (Navier--)Stokes equations, Taylor--Hood elements where $q_{p}$ = $q_{\vek{u}}-1$, c.f., \cite{Taylor_1973a}, are used. For the geometry, $q_{\mathrm{geom}}$ = $q_{\vek{u}}+1$ is chosen, c.f., \cite{Fries_2018a}. It is also possible to apply stabilization techniques to obtain a stable FEM with respect to the well-known Ladyzhenskaya--Babu\v ska--Brezzi (LBB) condition \cite{Babuska_1971a,Brezzi_1974,Ladyzhenskaya_1969a,Franca_1988a} in combination with equal element orders for the velocity and pressure. This is discussed in more detail in the context of Navier--Stokes equations below and in a forthcoming publication. Furthermore, the level-set function $\phi$ is replaced by its interpolation $\phi^{h}\left(\vek{x}\right)\in\mathcal{Q}_{q_{\mathrm{geom}}}^{h}$
with prescribed nodal values $\hat{\phi}_{i}=\phi\left(\vek{x}_{i}\right)$. Based on Eq.~(\ref{eq:BTFFctSpace}), the following discrete
test and trial function spaces are introduced 
\begin{eqnarray}
	\mathcal{S}_{\vek{u}}^{\Omega,h} & = & \left\{ \vek{u}^{h}\in\left[\mathcal{Q}_{q_{\vek{u}}}^{h}\right]^{3},\:\vek{u}^{h}=\hat{\vek{u}}^{h}\:\textrm{on}\:\partial\Omega_{\mathrm{D}}^{h}\right\} ,\label{eq:TrialU-1}\\
	\mathcal{V}_{\vek{u}}^{\Omega,h} & = & \left\{ \vek{w}_{\vek{u}}^{h}\in\left[\mathcal{Q}_{q_{\vek{u}}}^{h}\right]^{3},\:\vek{w}_{\vek{u}}^{h}=\vek{0}\:\textrm{on}\:\partial\Omega_{\mathrm{D}}^{h}\right\} ,\label{eq:TestU-1}\\
	\mathcal{S}_{p}^{\Omega,h}=\mathcal{V}_{p}^{\Omega,h} & = & \mathcal{Q}_{q_{p}}^{h}.\label{eq:TrialAndTestP-1}
\end{eqnarray}

Discretizing the continuous weak form of the stationary Stokes flow given above, i.e., Eqs.~(\ref{eq:wfBTF2IntStatStokes-mom}) and (\ref{eq:wfBTF2IntStatStokes-cont}), leads to the discrete weak form which is stated as: Given a shear viscosity $\mu \in \mathbb{R}^+$, penalty parameter $\alpha$, body forces $\vek{f}^h\left(\vek{x}\right)$, and boundary tractions $\hat{\vek{t}}^h\left(\vek{x}\right)$ on $\partial \Omega^h_\mathrm{N}$, find the velocity field $\vek{u}^h\left(\vek{x}\right)\in\mathcal{S}_{\vek{u}}^{\Omega,h}$ and the pressure field $p^h\left(\vek{x}\right)\in\mathcal{S}_{p}^{\Omega,h}$ such that for all test functions $\left(\vek{w}^h_{\vek{u}},w_{p}^h\right)\in\mathcal{V}_{\vek{u}}^{\Omega,h}\times\mathcal{V}_{p}^{\Omega,h}$, there holds in $\Omega^h$ 
\begin{align}
	\begin{split}
		\int_{\Omega^h} \big(\nabla_{\Gamma}^{\mathrm{dir}}\vek{w}^h_{\vek{u},\ti}:\vek{\sigma}\left(\vek{u}^h_\ti,p\right)\big) & \cdot \lVert \nabla \phi \rVert\;\mathrm{d}\Omega\,+\,\alpha\,\cdot \int_{\Omega^h} \left(\vek{u}^h \cdot \vek{n} \right) \cdot \left(\vek{w}^h_{\boldsymbol{u}} \cdot \vek{n} \right) \cdot \lVert \nabla \phi \rVert\;\mathrm{d}\Omega \\ = \int_{\Omega^h}\vek{w}^h_{\vek{u},\ti}\cdot\vek{f}^h_\ti & \cdot \lVert \nabla \phi \rVert\;\mathrm{d}\Omega \,+ \int_{\partial\Omega_{\mathrm{N}}^h}\!\!\!\vek{w}^h_{\vek{u}}\cdot\hat{\vek{t}}^h_\ti \cdot \left(\vek{q} \cdot \vek{m}\right)\cdot \lVert \nabla \phi \rVert\;\mathrm{d}\partial\Omega,
	\end{split}\\
	& \int_{\Omega^h}w^h_{p} \cdot\mathrm{div}_{\Gamma}\,\vek{u}^h_\ti \: \cdot \lVert \nabla \phi \rVert\;\mathrm{d}\Omega = 0.
\end{align}
Note that we omit the superscript $h$ on geometric quantities and differential operators, e.g., the normal vector $\vek{n}^h \rightarrow \vek{n}$, the level-set function $\phi^h \rightarrow \phi$, and $\nabla_{\Gamma}^{\mathrm{dir},h} \rightarrow \nabla_{\Gamma}^{\mathrm{dir}}$, for brevity.\\
\\
This leads to a system of equations with a saddle point structure. The numerical results shown next confirm higher-order convergence rates for the application of the Taylor--Hood elements introduced above.

\subsection{Numerical results for stationary Stokes flow}\label{subsec:StStok-NumRes}

\begin{figure}
	\centering
	
	\subfigure[]{\includegraphics[width=0.3\textwidth]{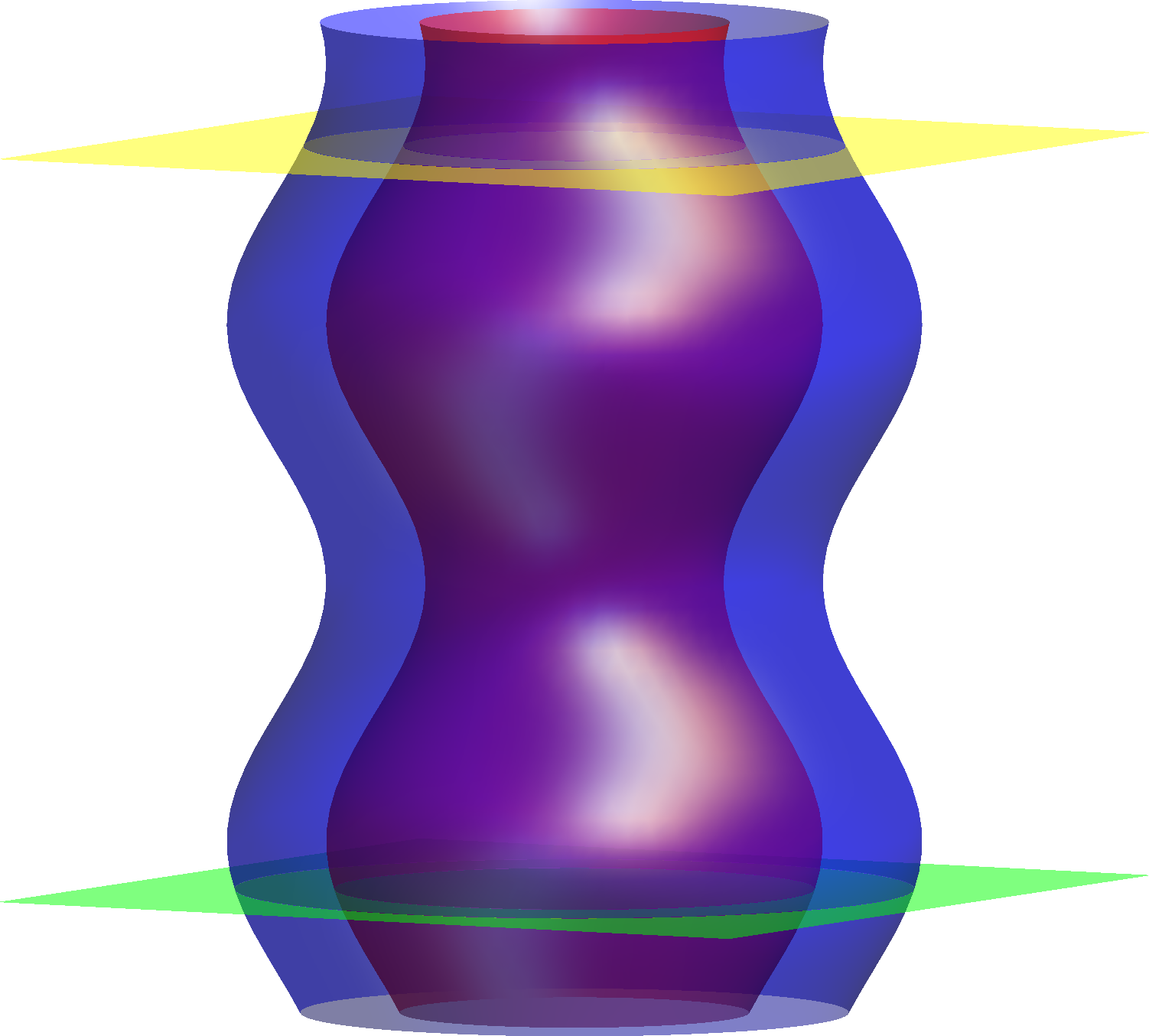}}
	\hspace{1.5cm}
	\subfigure[]{\includegraphics[width=0.2\textwidth]{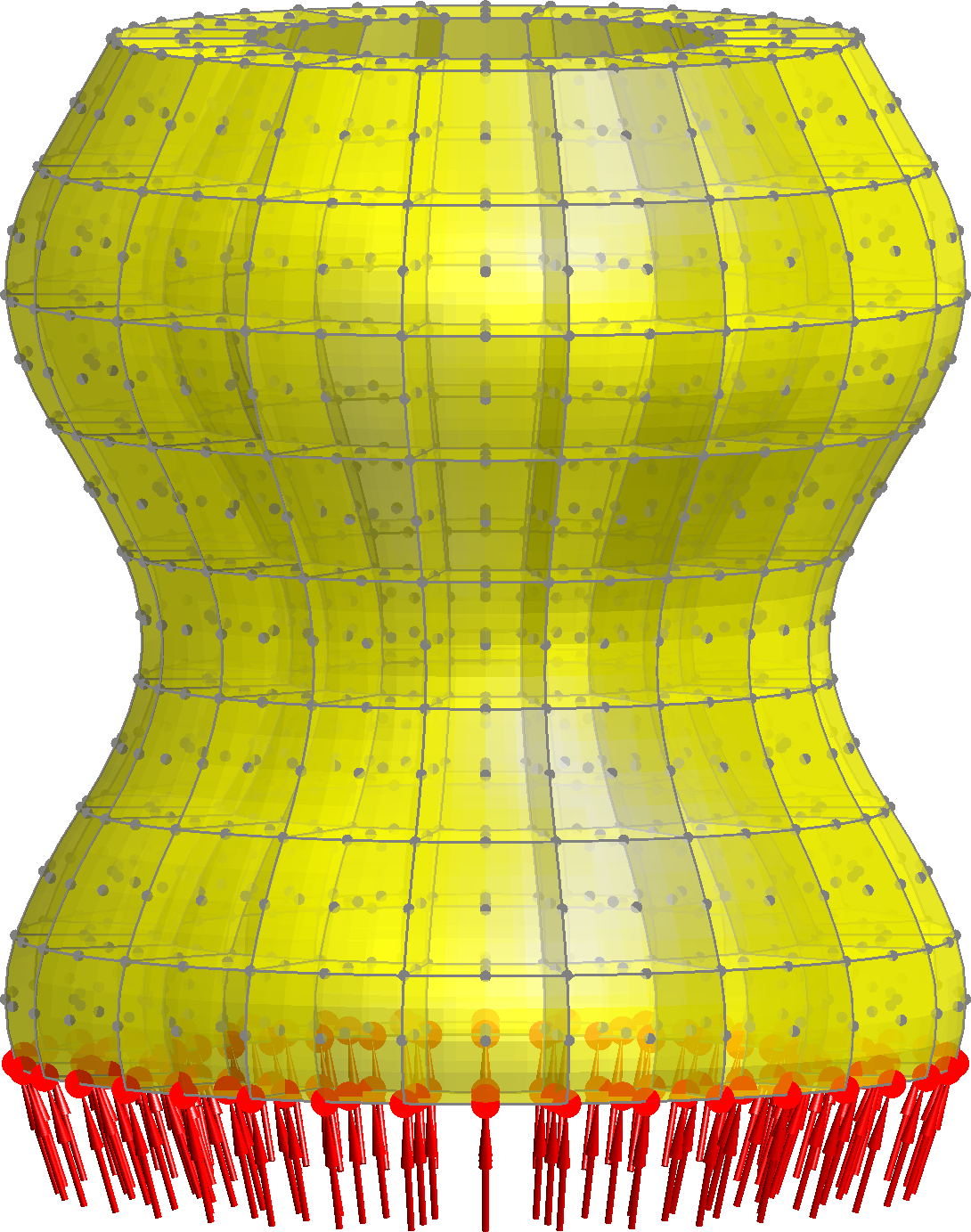}}
	\hspace{2cm}
	\subfigure[]{\includegraphics[width=0.2\textwidth]{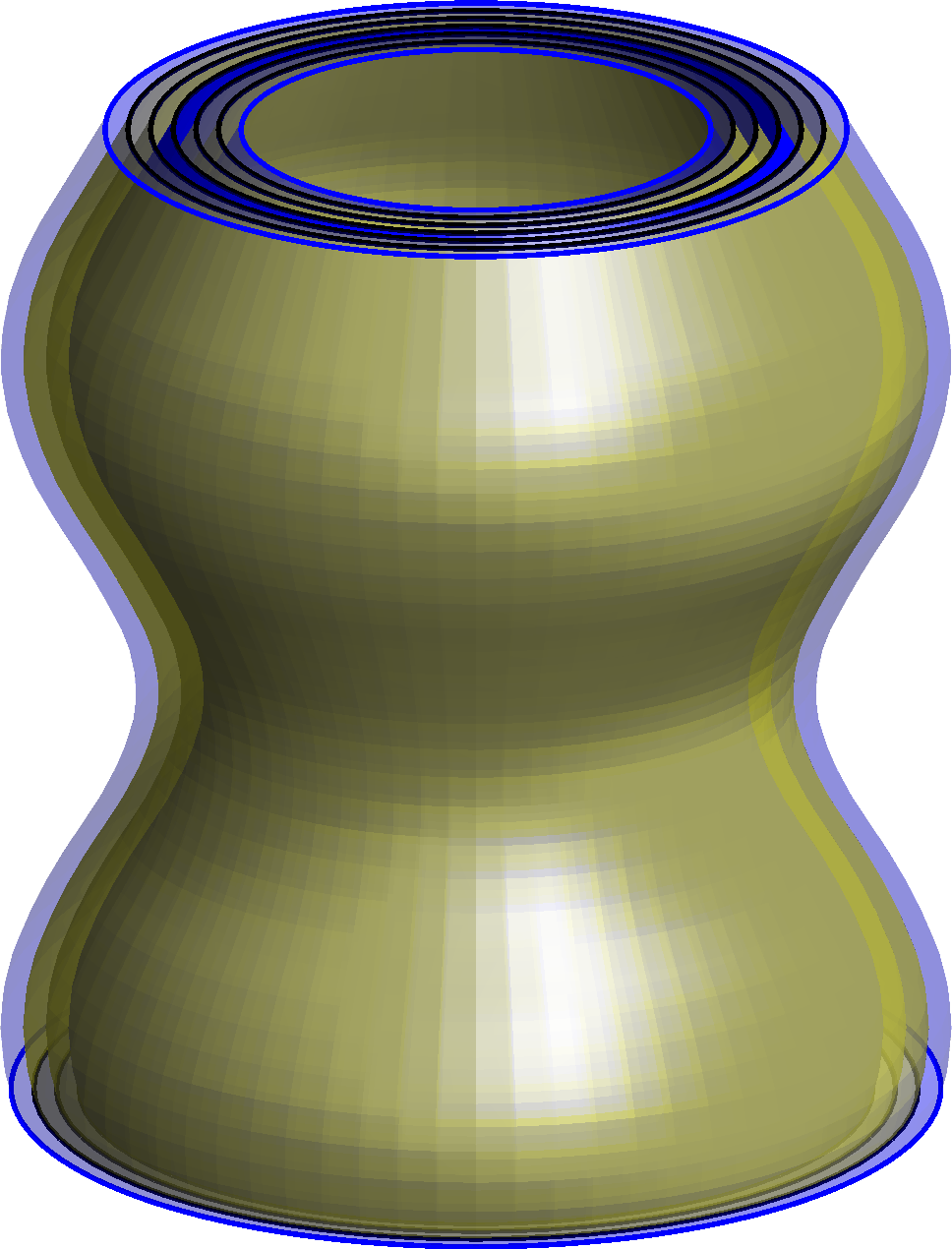}}
	
	\caption{\label{fig:TC1-setup} Setup for the first numerical example: (a) The level sets which define the bulk domain $\Omega$, (b) some arbitrary mesh and Dirichlet boundary conditions at the inflow, and (c) the bulk domain in gray and some selected level sets in yellow and the level set with $r_0=1.0$ in blue.}
\end{figure}

\begin{figure}
	\centering
	
	\subfigure[Bulk Trace FEM - $\lVert\vek{u}\rVert$]{\includegraphics[width=0.35\textwidth]{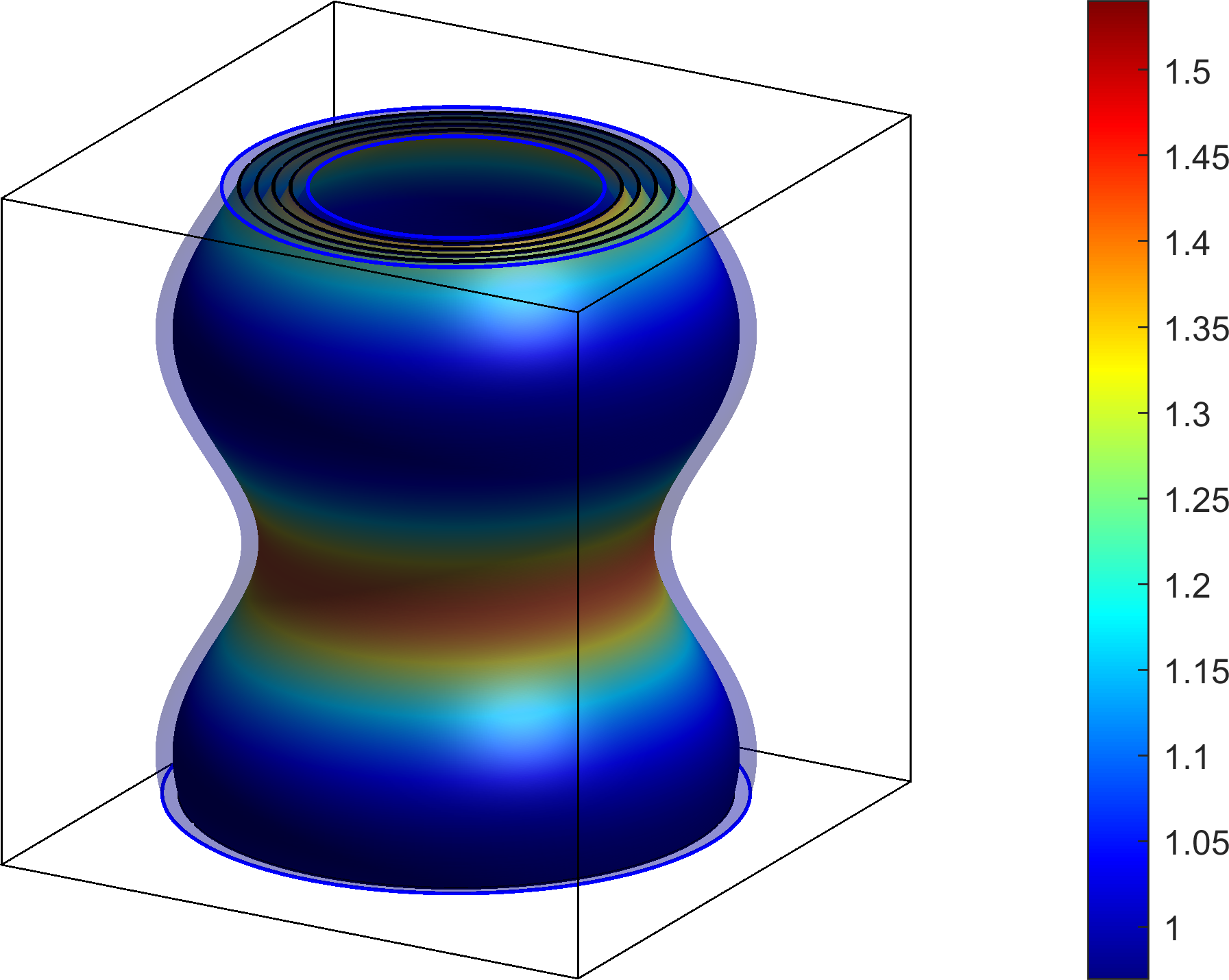}}\hspace{0.1\textwidth}
	\subfigure[Bulk Trace FEM - $p$]{\includegraphics[width=0.35\textwidth]{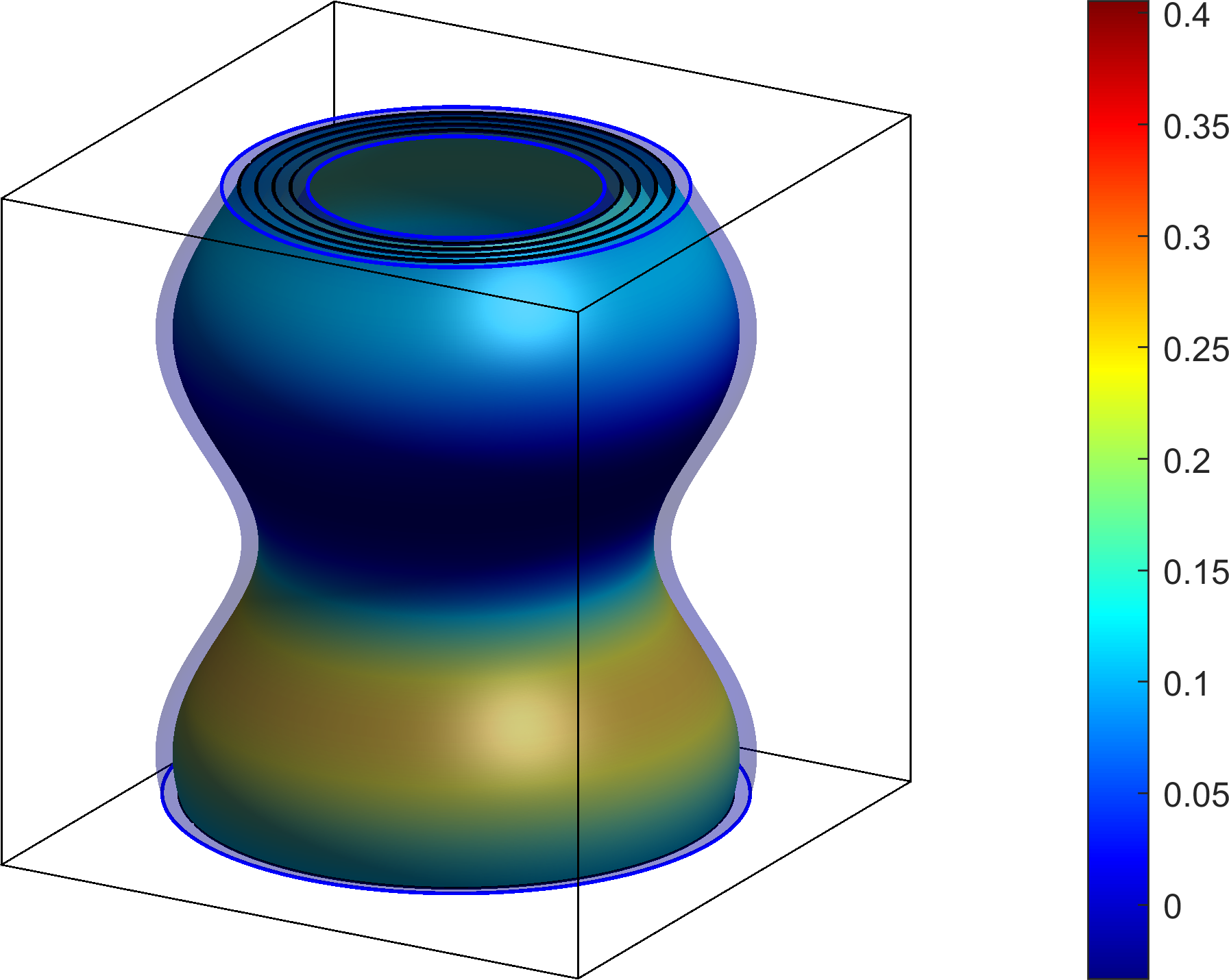}}
	\subfigure[Bulk Trace FEM - $\lVert\vek{u}\rVert$]{\includegraphics[width=0.35\textwidth]{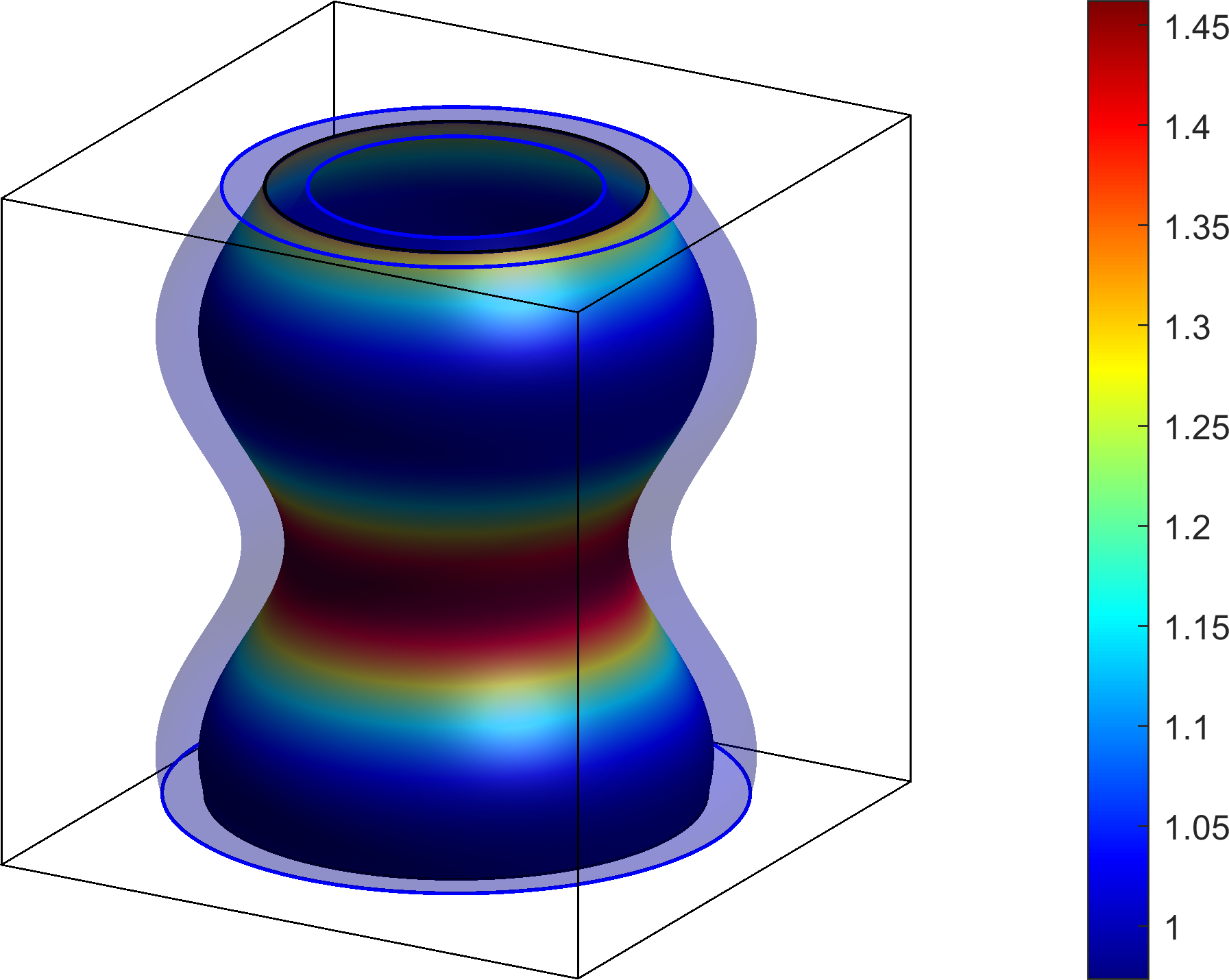}}\hspace{0.1\textwidth}
	\subfigure[Bulk Trace FEM - $p$]{\includegraphics[width=0.35\textwidth]{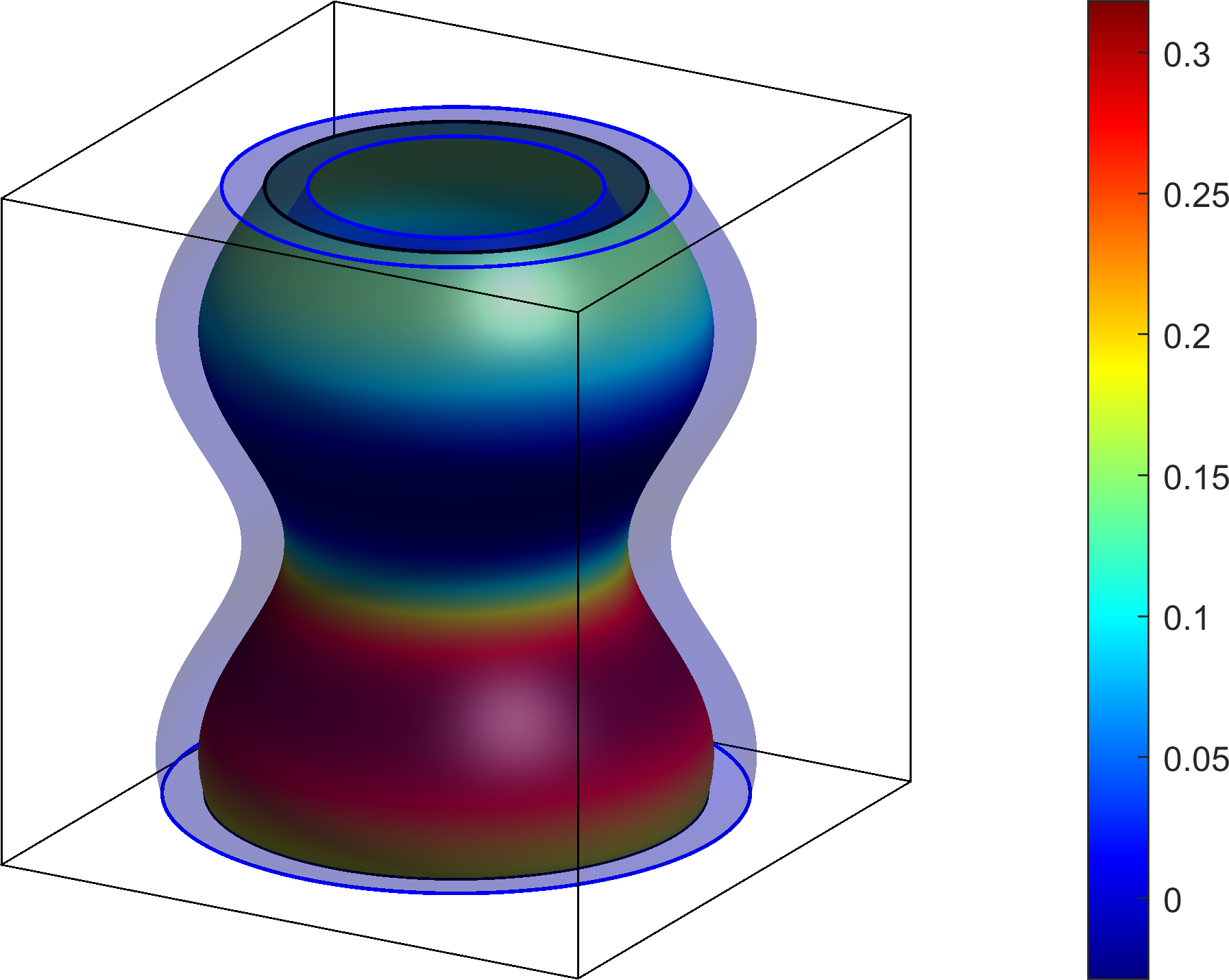}}
	\subfigure[Surface FEM - $\lVert\vek{u}\rVert$]{\includegraphics[width=0.35\textwidth]{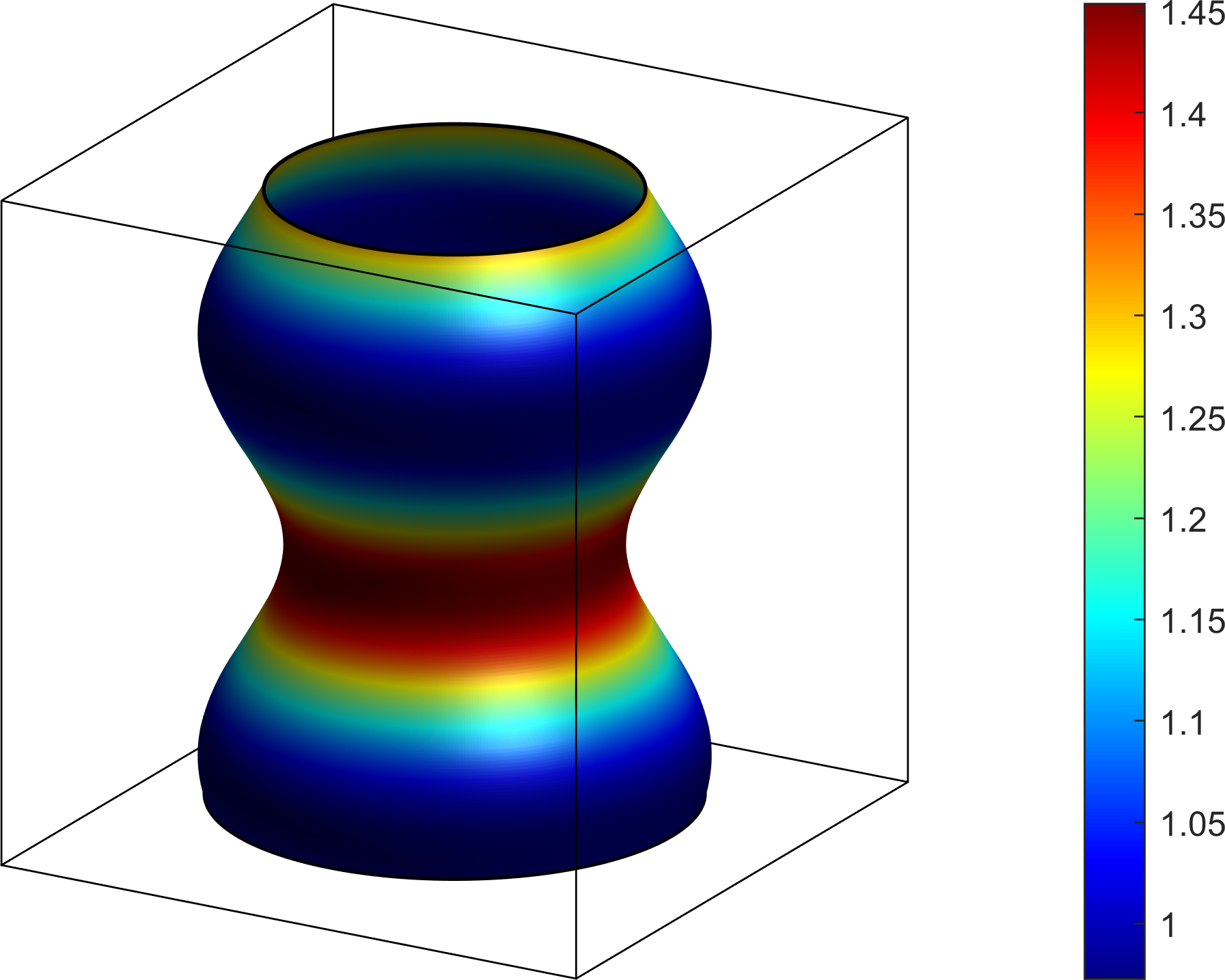}}\hspace{0.1\textwidth}
	\subfigure[Surface FEM - $p$]{\includegraphics[width=0.35\textwidth]{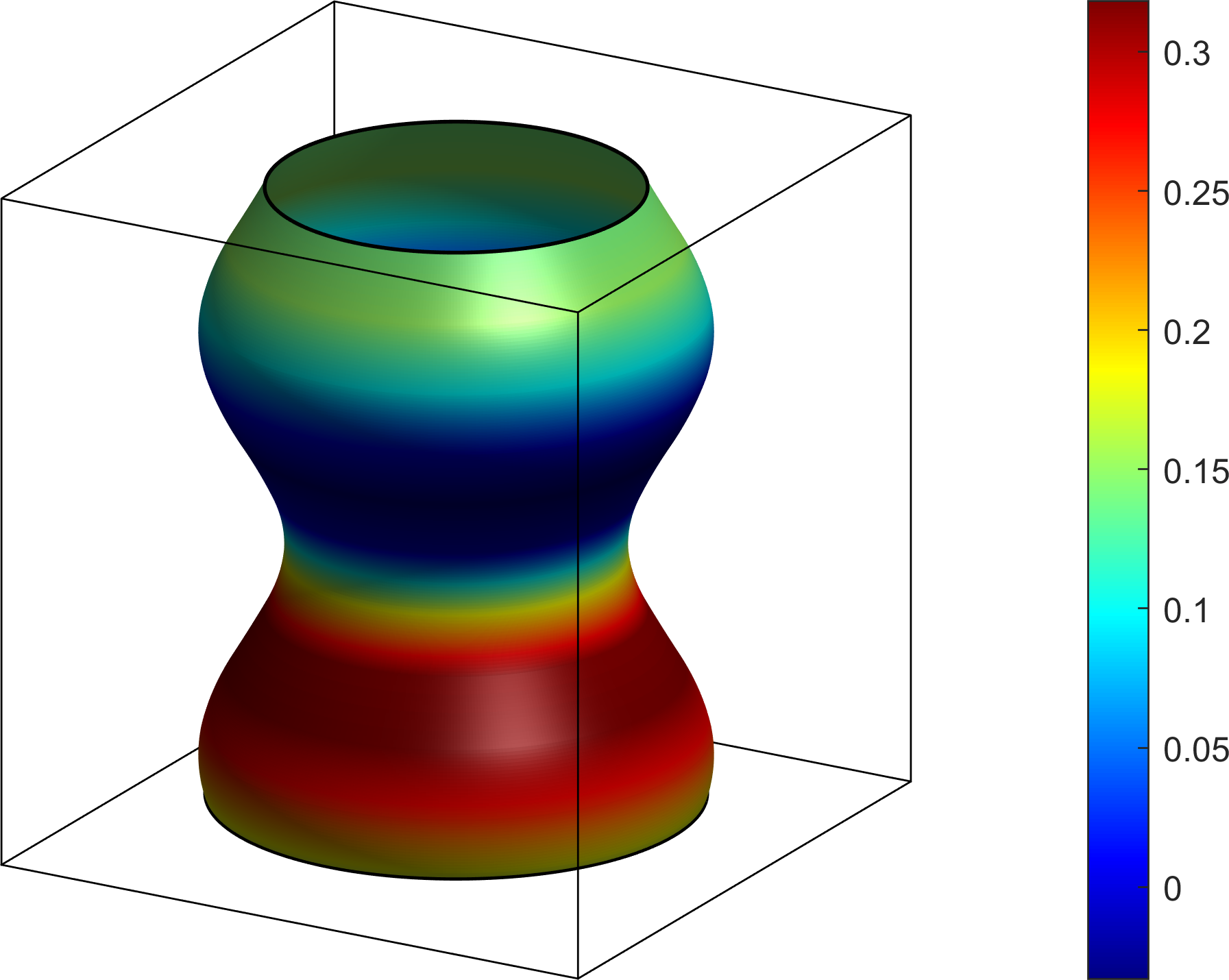}}
	
	\caption{\label{fig:TC1-res} Velocity magnitudes and pressure of the first numerical example are shown. (a) to (d) show results obtained with the Bulk Trace FEM, while in (a) and (b) some arbitrarily selected level sets are shown, (c) and (d) show the surface with $r_0 = 1.0$. (e) and (f) show results obtained with the Surface FEM where the \emph{one} considered surface has a radius of $1.0$ at $z=0$ and, therefore, (c) to (f) can be used to compare a Bulk Trace FEM solution with a Surface FEM solution.}
\end{figure}

\begin{figure}
	\centering
	
	\subfigure[$\varepsilon_{\vek{u}}$]{\includegraphics[width=0.45\textwidth]{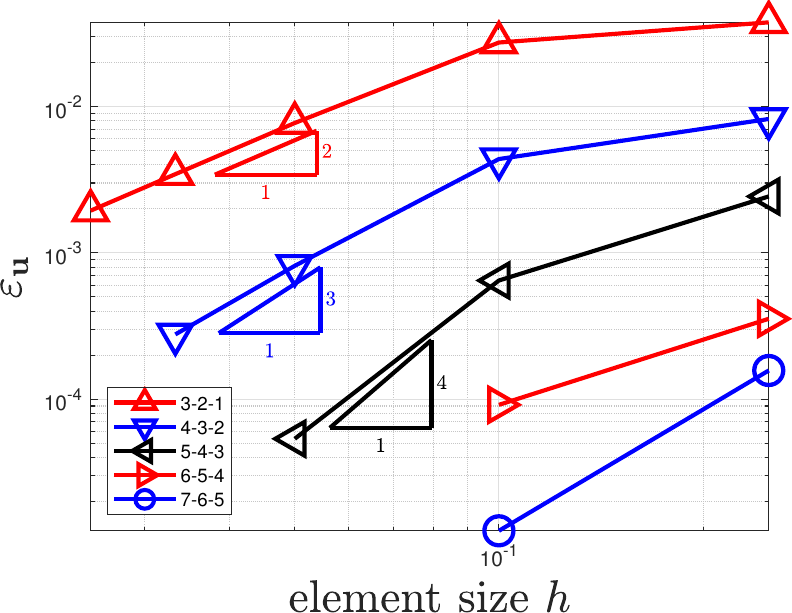}}\qquad\subfigure[$\varepsilon_{e}$]{\includegraphics[width=0.45\textwidth]{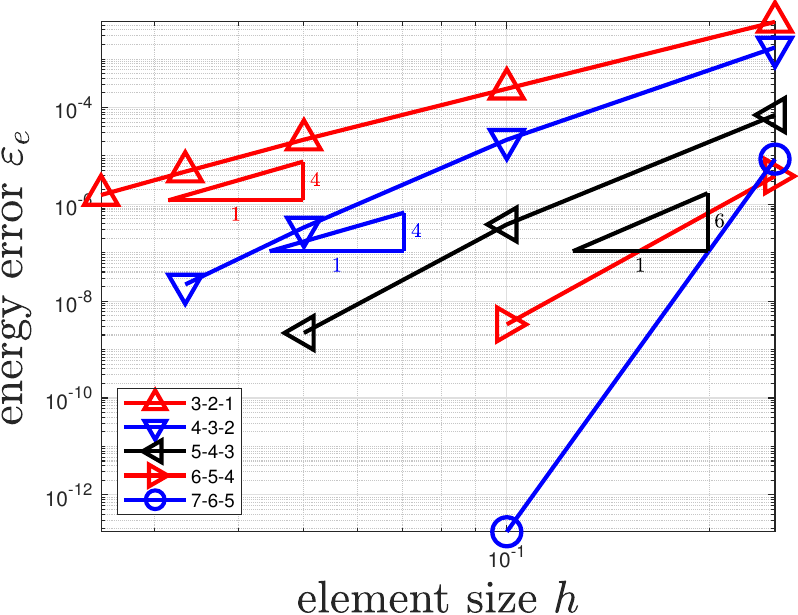}}
	
	\subfigure[$\varepsilon_{\mathrm{mom}}$]{\includegraphics[width=0.45\textwidth]{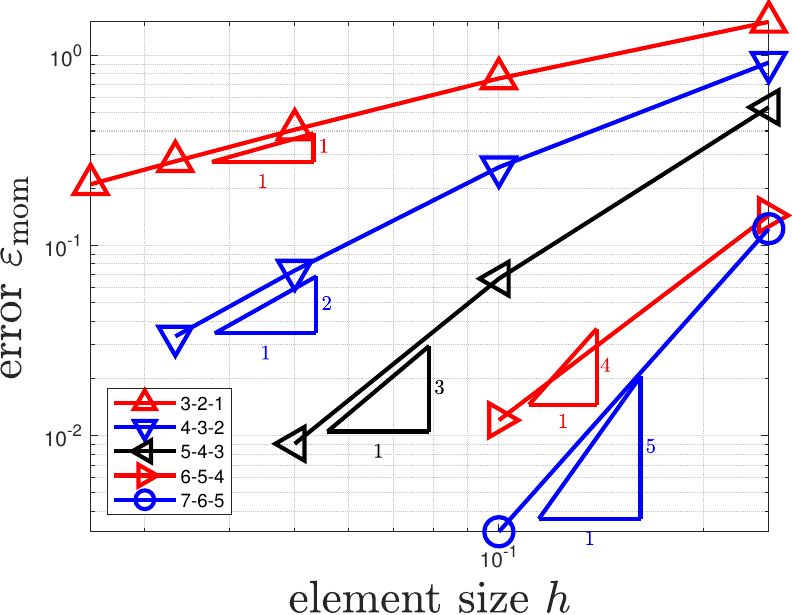}}\qquad\subfigure[$\varepsilon_{\mathrm{cont}}$]{\includegraphics[width=0.45\textwidth]{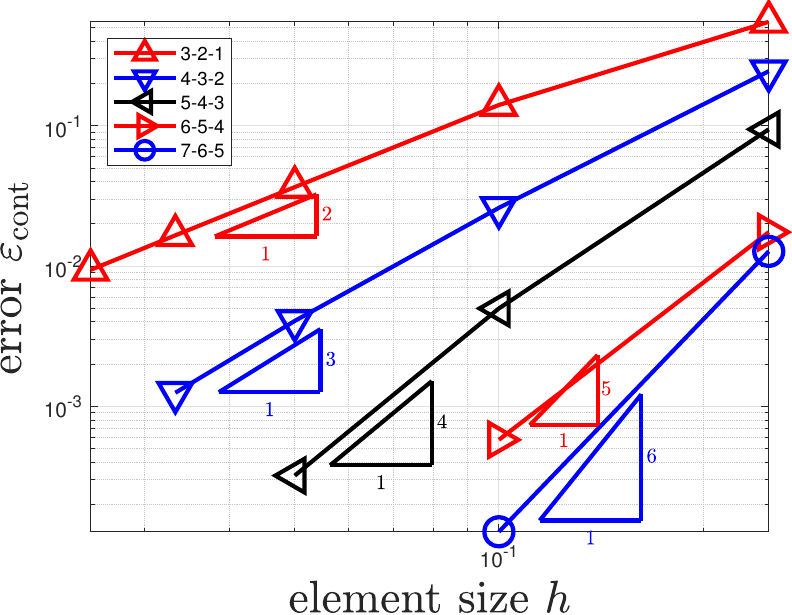}}
	
	\caption{\label{fig:TC1-convStudies} Convergence results for the simultaneous solution of the axisymmetric test case, (a) $L_2$-error of the velocities, (b) (pseudo-)energy error, (c) residual error in the momentum equations, and (d) residual error in the continuity equation. The numbers in the legends are the polynomial orders $\{q_{\mathrm{geom}},\, q_{\vek{u}},\,q_p\}$ of the FE function spaces.}
\end{figure}

In this section, we show higher-order convergence studies to verify the simultaneous solution method. With a known exact (analytic) velocity field $\vek{u}^{\mathrm{ex}}$, the error $\varepsilon_{\vek{u}}$ in the velocity components is defined as
\begin{equation}
	\varepsilon_{\vek{u}} = \sum_{i=1}^{3}\sqrt{\int_{\Omega}\left(u^h_i - u^{\mathrm{ex}}_i\right) \cdot \lVert \nabla \phi \rVert\,\mathrm{d}\Omega}.\label{eq:VelErrorNorm}
\end{equation}
An error measure based on (pseudo-)energy \cite{Auricchio_2017a} is introduced as 
\begin{equation}
	\varepsilon_e = \lvert \mathfrak{e}\left(\vek{u}^h\right) - \mathfrak{e}\left(\vek{u}^{\mathrm{ex}}\right)\rvert, \qquad \mathfrak{e}\left(\vek{u}\right) = \frac{1}{2} \mu \cdot \int_{\Omega} \left(\nabla_{\Gamma}^{\mathrm{cov}} \vek{u} : \nabla_{\Gamma}^{\mathrm{cov}} \vek{u}\right) \cdot \lVert \nabla \phi \rVert \,\mathrm{d}\Omega.\label{eq:EnergyError}
\end{equation} 
Furthermore, the integrated residual errors w.r.t.~the momentum equation (\ref{eq:MomEqtStatStokes}), $\varepsilon_{\mathrm{mom}}$, and the continuity equation (\ref{eq:ContinuityConstraint}), $\varepsilon_{\mathrm{cont}}$, are evaluated. These error measures do not rely on an analytical solution, hence, these can be used to verify also more advanced test cases where no analytical solutions are known. Similar error measures for (incompressible Navier--) Stokes flow on \emph{one} surface have been used in \cite{Fries_2018a} and for the \emph{simultaneous} solution in the context of structural  mechanics, e.g., in \cite{Fries_2023a,Kaiser_2024a}. For stationary Stokes flow on all embedded manifolds, these error measures are defined as
\begin{align}
	\varepsilon_{\mathrm{mom}} &= \sqrt{\sum_{i=1}^{n_{\mathrm{el}}} \int_{\Omega^{\mathrm{el},i}}\left(\mat{P} \cdot \mathrm{div}_{\Gamma}\,\vek{\sigma}\left(\vek{u}_\ti^h,p^h\right) +  f_\ti^h\right)^2\cdot \lVert \nabla \phi \rVert\,\mathrm{d}\Omega},\label{eq:ResErrorMom}\\
	\varepsilon_{\mathrm{cont}} &= \sqrt{\sum_{i=1}^{n_{\mathrm{el}}} \int_{\Omega^{\mathrm{el},i}}\left(\mathrm{div}_{\Gamma}\,\vek{u}_\ti^h\right)^2\cdot \lVert \nabla \phi \rVert\,\mathrm{d}\Omega}.\label{eq:ResErrorCont}
\end{align}
Note the summation of contributions over element interiors $\Omega^{\mathrm{el},i}$.\\
\\
The following numerical example for stationary Stokes flow is inspired by \cite{Fries_2018a} and a summary of the results presented herein is given in \cite{Fries_2024a} by the authors. Axisymmetric surfaces are considered featuring a height of $L = 3$ and a variable radius of $r(z,r_0) = r_0 + \nicefrac{1}{5} \cdot \sin(1+3\cdot z)$ with $z \in \left[0,L\right]$ and the radius $r_0 \in \left[0.8,1.2\right]$ at the bottom, i.e., at $z=0$. The outer surface of the bulk domain $\Omega$ coincides with the axisymmetric surface with radius $r(z,r_0 = 0.8)$ and $r(z,r_0 = 1.2)$. At the bottom, i.e., at $z=0$, and the top, i.e., at $z=3$, the bulk domain is bounded by horizontal planes, see Fig.~\ref{fig:TC1-setup}(a). The fluid's viscosity is $\mu = 0.1$ and the density is $\rho = 1$. As in \cite{Fries_2018a} for \emph{one} surface, the lower boundary at $z=0$ is the Dirichlet boundary $\partial\Omega_{\mathrm{D}}$ where the inflow in co-normal direction to the surfaces $\Gamma_{\!c}$ is prescribed and the upper boundary $\partial\Omega_{\mathrm{N}}$ at $z=3$ is the outflow boundary where zero tractions are applied, see Fig.~\ref{fig:TC1-setup}(b). The bulk domain $\Omega$ and some selected surfaces $\Gamma_{\!c}$ are shown in Fig.~\ref{fig:TC1-setup}(c).
The exact velocity which is required for the convergence study in $\varepsilon_{\vek{u}}$ is computed analogously to \cite{Fries_2024a} as
\[
\left[\begin{array}{l}
	u_{\mathrm{ex}}(\vek{x})\\
	v_{\mathrm{ex}}(\vek{x})\\
	w_{\mathrm{ex}}(\vek{x})
\end{array}\right]=\frac{\hat{r}_{0}}{r\cdot\sqrt{1+\left(\frac{\mathrm{d}r}{\mathrm{d}z}\right)^{2}}}\cdot\left[\begin{array}{l}
	\frac{\mathrm{d}r}{\mathrm{d}z}\cdot\nicefrac{x}{r}\\
	\frac{\mathrm{d}r}{\mathrm{d}z}\cdot\nicefrac{y}{r}\\
	1
\end{array}\right]
\]
with $\hat{r}_0 = r(0,r_0)$ and $r = r(z,r_0)$. Fig.~\ref{fig:TC1-res} shows the results of this test case for the velocities and pressure on selected level sets $\Gamma_{\!c}$ and in comparison to the Surface FEM evaluated for the surface with $r_0 = 1.0$.
Fig.~\ref{fig:TC1-convStudies}(a) shows the results of the error in the velocities $\varepsilon_{\vek{u}}$ which converges with the expected optimal rate of $q_{\vek{u}}+1$. The (pseudo-)energy error $\varepsilon_e$ converges with $q_{\vek{u}}+2$ for even element orders and with $q_{\vek{u}}+1$ for odd element orders as seen in Fig.~\ref{fig:TC1-convStudies}(b). More details about the convergence properties of this error measure can be found in \cite{Fries_2023a} in the context of the stored elastic energy of membranes. Furthermore, the residual errors converge with the expected rates $q_{\vek{u}}-1$ for $\varepsilon_{\mathrm{mom}}$ where second-order derivatives are involved and with $q_{\vek{u}}$ for $\varepsilon_{\mathrm{cont}}$ where only first-order derivatives occur. The convergence plots for $\varepsilon_{\mathrm{mom}}$ and $\varepsilon_{\mathrm{cont}}$ are shown in Fig.~\ref{fig:TC1-convStudies}(c) and (d), respectively.

\section{Stationary Navier--Stokes flow}\label{sec:StatNSEQ}

\subsection{Strong form for one level set}\label{subsec:StNSEQ-sfEq1LS}

Stationary Navier--Stokes flow includes a non-linear advection term in the momentum equation. Hence, the strong form of stationary Stokes flow, given in Eqs.~(\ref{eq:MomEqtStatStokes}) and (\ref{eq:ContinuityConstraint}), is similar to the strong form of stationary Navier--Stokes flow. While the incompressibility constraint, i.e., Eq.~(\ref{eq:ContinuityConstraint}) and the boundary conditions remain unchanged, the advection term is added to Eq.~(\ref{eq:MomEqtStatStokes}) leading to \cite{Fries_2018a}
\begin{equation}
	\varrho\cdot\left(\vek{u}_\ti\cdot\nabla_{\Gamma}^{\mathrm{cov}}\right)\vek{u}_\ti-\mat{P}\cdot\mathrm{div}_{\Gamma}\,\vek{\sigma}\left(\vek{x}\right)=\vek{f}_\ti\left(\vek{x}\right),\label{eq:MomentumEqtNSstat}
\end{equation}
where $\varrho \in \mathbb{R}^+$ is the (constant) fluid density. 

\subsection{Weak form for one level set}\label{subsec:stNSEQ-wfEq1LS}

The weak form of the stationary Navier--Stokes equations on \emph{one} surface is obtained analogously to the weak form of stationary Stokes flow, described in Sec.~\ref{subsec:StStok-wfEq1LS}. The function spaces are the same as given in Eqs.~(\ref{eq:TrialU}) to (\ref{eq:TrialAndTestP}). Furthermore, the weak form of the incompressiblilty constraint equals Eq.~(\ref{eq:wf1srfStatStokes-cont}). Hence, the only equation which is different in the weak form of stationary Navier--Stokes flow compared to stationary Stokes flow is the momentum equation \cite{Fries_2018a}, calculated as
\begin{align}
	\begin{split}
		\varrho\,\cdot&\int_{\Gamma_{\!c}}\vek{w}_{\vek{u},\ti}\cdot\left(\vek{u}_\ti\cdot\nabla_{\Gamma}^{\mathrm{cov}}\right)\vek{u}_\ti\,\mathrm{d}\Gamma+\int_{\Gamma_{\!c}}\nabla_{\Gamma}^{\mathrm{dir}}\vek{w}_{\vek{u},\ti}:\vek{\sigma}_\ti\left(\vek{u}_\ti,p\right)\mathrm{d}\Gamma \\ + \,\alpha\, \cdot &\int_{\Gamma_{\!c}} \left(\vek{u} \cdot \vek{n} \right) \cdot \left(\vek{w}_{\boldsymbol{u}} \cdot \vek{n} \right) \mathrm{d}\Gamma
		=\int_{\Gamma_{\!c}}\vek{w}_{\vek{u},\ti}\cdot\vek{f}_\ti\,\mathrm{d}\Gamma+\int_{\partial\Gamma_{\!c,\mathrm{N}}}\!\!\!\vek{w}_{\vek{u},\ti}\cdot\hat{\vek{t}}_\ti\,\mathrm{d}\partial\Gamma.
	\end{split}
\end{align}

\subsection{Weak form for all level sets in a bulk domain}\label{subsec:StNSEQ-wfEqAllLS}

Again the only change is in the momentum equation while the other equations and function spaces remain unchanged from Sec.~\ref{subsec:StStok-wfEq1LS}. For the momentum equation follows
\begin{align}
	\begin{split}
		\varrho\,\cdot &\int_{\Omega}\vek{w}_{\vek{u},\ti}\cdot\left(\vek{u}_\ti\cdot\nabla_{\Gamma}^{\mathrm{cov}}\right)\vek{u}_\ti \cdot \lVert \nabla \phi \rVert\;\mathrm{d}\Omega + \int_{\Omega} \big(\nabla_{\Gamma}^{\mathrm{dir}} \vek{w}_{\vek{u},\ti}:\vek{\sigma}_\ti\left(\vek{u}_\ti,p\right)\big) \cdot \lVert \nabla \phi \rVert\;\mathrm{d}\Omega \\ & + \, \alpha \,\cdot \int_{\Omega} \left(\vek{u} \cdot \vek{n} \right) \cdot \left(\vek{w}_{\boldsymbol{u}} \cdot \vek{n} \right) \cdot \lVert \nabla \phi \rVert\;\mathrm{d}\Omega =  \int_{\Omega}\vek{w}_{\vek{u},\ti}\cdot\vek{f}_\ti \cdot \lVert \nabla \phi \rVert\;\mathrm{d}\Omega \\ & + \, \int_{\partial\Omega_{\mathrm{N}}}\!\!\!\vek{w}_{\vek{u},\ti}\cdot\hat{\vek{t}}_\ti\cdot\left(\vek{q}\cdot\vek{m}\right)\cdot \lVert \nabla \phi \rVert\;\mathrm{d}\partial\Omega.
	\end{split}
\end{align}
The resulting continuous weak form is discretized analogously to the case of the stationary Stokes flow as described in Sec.~\ref{subsec:StStok-wfEqAllLS}. This leads to the discretized weak form of the stationary Navier--Stokes flow for the \emph{simultaneous} analysis on \emph{all} level sets which are embedded in a bulk domain: Given a shear viscosity $\mu \in \mathbb{R}^+$, fluid density $\varrho \in \mathbb{R}^+$, penalty parameter $\alpha$, body forces $\vek{f}^h\left(\vek{x}\right)$, and boundary tractions $\hat{\vek{t}}^h\left(\vek{x}\right)$ on $\partial \Omega^h_\mathrm{N}$, find the velocity field $\vek{u}^h\left(\vek{x}\right)\in\mathcal{S}_{\vek{u}}^{\Omega,h}$ and the pressure field $p^h\left(\vek{x}\right)\in\mathcal{S}_{p}^{\Omega,h}$ such that for all test functions $\left(\vek{w}^h_{\vek{u}},w_{p}^h\right)\in\mathcal{V}_{\vek{u}}^{\Omega,h}\times\mathcal{V}_{p}^{\Omega,h}$, there holds in $\Omega^h$ 
\begin{align}
	\begin{split}
		\varrho\,\cdot & \int_{\Omega^h}\vek{w}^h_{\vek{u},\ti}\cdot\left(\vek{u}^h_\ti\cdot\nabla_{\Gamma}^{\mathrm{cov}}\right)\vek{u}^h_\ti \cdot \lVert \nabla \phi \rVert\;\mathrm{d}\Omega +  \int_{\Omega^h} \big(\nabla_{\Gamma}^{\mathrm{dir}} \vek{w}^h_{\vek{u},\ti}:\vek{\sigma}_\ti\left(\vek{u}^h_\ti,p^h\right)\big) \cdot \lVert \nabla \phi \rVert\;\mathrm{d}\Omega \\ & + \, \alpha \,\cdot \int_{\Omega^h} \left(\vek{u}^h \cdot \vek{n} \right) \cdot \left(\vek{w}^h_{\boldsymbol{u}} \cdot \vek{n} \right) \cdot \lVert \nabla \phi \rVert\;\mathrm{d}\Omega = \int_{\Omega^h}\vek{w}^h_{\vek{u},\ti}\cdot\vek{f}^h_\ti \cdot \lVert \nabla \phi \rVert\;\mathrm{d}\Omega \\ & + \, \int_{\partial\Omega_{\mathrm{N}}^h}\!\!\!\vek{w}^h_{\vek{u},\ti}\cdot\hat{\vek{t}}^h_\ti\cdot\left(\vek{q}\cdot\vek{m}\right)\cdot \lVert \nabla \phi \rVert\;\mathrm{d}\partial\Omega,\label{eq:DisWFstatNSEQ} 
	\end{split}\\
	&\qquad\qquad\qquad\qquad\qquad\int_{\Omega^h}w^h_{p} \,\cdot\, \mathrm{div}_{\Gamma}\,\vek{u}^h_\ti \: \cdot \lVert \nabla \phi \rVert\;\mathrm{d}\Omega = 0.
\end{align}
Picard's iteration, a fixed-point iteration scheme, is applied to solve the nonlinear system of equations which results due to the advection term in the (stationary) Navier--Stokes equations. For further details on this iteration scheme, see, e.g., Sec.~6.3 in \cite{John_2016a}. As an alternative to Picard's iteration, Newton's method could be applied.\\
\\
For the discretization, Taylor--Hood elements, i.e., $q_{p}$ = $q_{\vek{u}}-1$, can be applied as in Sec.~\ref{sec:StatStokes}. In addition, element pairs of equal-order for the velocity and pressure discretization, i.e., $q_{p}$ = $q_{\vek{u}}$ can be used within the PSPG stabilization framework, c.f., \cite{Hughes_1986e,John_2016a,Tezduyar_2003a,Pacheco_2021a}. Then, we add  the PSPG stabilization term to the left hand side of Eq.~(\ref{eq:DisWFstatNSEQ}), defined as
\begin{equation}
	+ \sum_{i=1}^{n_{\mathrm{el}}} \int_{\Omega^{\mathrm{el},i}} \tau_{\mathrm{PSPG}} \cdot \frac{1}{\varrho} \cdot \left(\nabla_{\Gamma} w_p^h \right) \left[ \varrho\cdot\left(\vek{u}_\ti^h\cdot\nabla_{\Gamma}^{\mathrm{cov}}\right)\vek{u}_\ti^h-\mat{P}\cdot\mathrm{div}_{\Gamma}\,\vek{\sigma}_\ti\left(\vek{u}^h_\ti,p^h\right)-\vek{f}_\ti^h\right] \cdot \lVert \nabla \phi \rVert\,\mathrm{d}\Omega. \label{eq:PSPGstatNSEQ}
\end{equation} 
For the geometry discretization we use elements of order $q_{\mathrm{geom}}$ = $q_{\vek{u}}+1$. The stabilization parameter $\tau_{\mathrm{PSPG}}$ is defined as a function of the element size $h$.\\
\\
For large Reynolds numbers, the solution, either obtained with Taylor--Hood elements or the PSPG stabilization, may become unstable. In such cases, the SUPG method could be applied \cite{John_2016a,Tezduyar_2003a,Hughes_1979a,Brooks_1982a}. However, in the context of the simultaneous solution of flow problems on all level sets within a bulk domain, this is beyond the scope of this paper.

\subsection{Numerical results for stationary Navier--Stokes flow}\label{subsec:StNSEQ-NumRes}

\subsubsection{Flow around an obstacle}\label{subsec:CylFlowStat}
\begin{figure}
	\centering
	
	\includegraphics[width=1\textwidth]{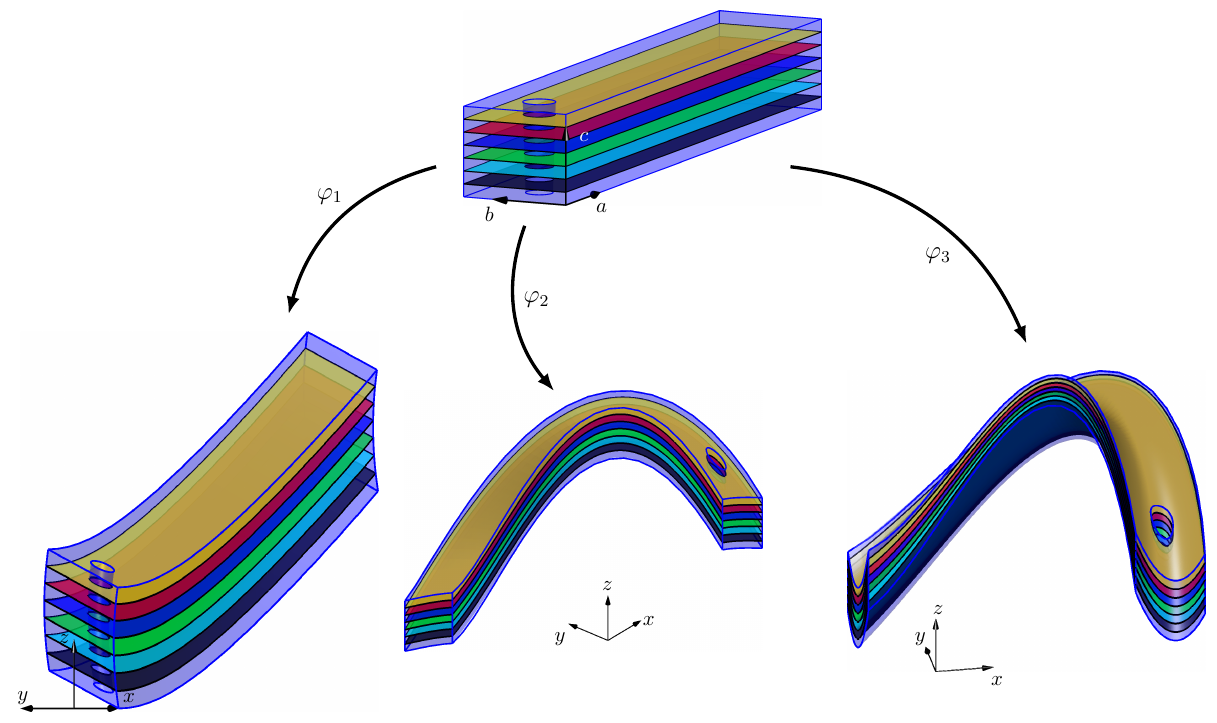}
	
	\caption{\label{fig:TurekMaps} The three different mappings $\varphi_i$ with $i \in [1,2,3]$ as defined in Eq.~(\ref{eq:TurekMaps}) for the geometry definition of the cylinder flow test case. The bulk domain is shown in light blue and some arbitrarily selected level sets are shown in different colours. Note that the view point is different for each geometry for a better visualization.}
\end{figure}

These test cases are inspired by the two-dimensional bench mark in \cite{Schaefer_1996a}. First, two-dimensional geometries are embedded in a flat three-dimensional bulk domain. The bulk domain including the embedded surfaces is then mapped to obtain a curved bulk domain with curved embedded surfaces on which the stationary Navier--Stokes flows are simulated. Three different mappings are considered, see Fig.~\ref{fig:TurekMaps} for the resulting geometries, given as
\begin{align}
	\varphi_1 &= \begin{Bmatrix*}[l]
		x &=& a + 0.1b + 0.2c + 0.1 \cdot \sin(a) + 0.2 \cdot  \sin(b) + 0.3 \cdot \sin(2c)\\
		y &=& 0.1a + b - 0.2c + 0.3 \cdot \sin(a) + 0.2 \cdot \sin(b) + 0.1 \cdot \sin(2c) - (c-0.5)^2\\
		z &=&-0.2a + 0.3b + c + 0.2 \cdot \sin(a) + 0.1 \cdot \sin(b) + 0.3 \cdot \sin(2c) + 0.25a^2
	\end{Bmatrix*},\nonumber\\
	\varphi_2 &= \begin{Bmatrix*}[l]
		x &=& \cos(0.25\cdot \pi \cdot a) \cdot (\bar{s}+1.2)\\
		y &=& \sin(0.25\cdot \pi \cdot a) \cdot (\bar{s}+1.2)\\
		z &=& c + 0.5 \cdot \sin(2a) + 0.2\cdot \sin(2b)
	\end{Bmatrix*},\label{eq:TurekMaps}\\
	\varphi_3 &= \begin{Bmatrix*}[l]
		x &=& \cos(0.25\cdot \pi \cdot a) \cdot  (\tilde{s}+1.2)\\
		y &=& \sin(0.25\cdot \pi \cdot a)  \cdot (\tilde{s}+1.2)\\
		z &=& \frac{6a}{11} + c - \frac{10a\cdot c}{11} + \frac{50a^2\cdot c}{121} - \frac{30a^2}{121} + 5(b-0.205)^2 + 0.5\cdot \sin(2a) + 0.2 \cdot \sin(2b)
	\end{Bmatrix*}\nonumber,		
	\end{align}
with $\bar{s} = -(1+\bar{q}) \cdot (b-0.205) \cdot \cos(\pi/6\cdot(1-a))$, $\bar{q} = -0.1a^2 + 0.2a$, and $ \tilde{s} = -(1-0.1a^2 + 0.2a) \cdot (b-0.205) \cdot \cos(\pi/6\cdot(1-a))$. $a$, $b$, and $c$ are the Cartesian coordinates in the undeformed, i.e., flat domain and $x$, $y$, and $z$ are the Cartesian coordinates in which the deformed, i.e., curved domain, is described. The flat bulk domain is defined as the channel $[0,2.20] \times [0,0.41] \times [0,1/3]$ and the obstacle's radius (the cylinder in the flat $2$-dimensional channel) is $r_b = 0.05$ at the bottom $(c=0)$, $r_t = 0.06$ at the top $(c=1/3)$, and varies linearly in between. The obstacle is placed slightly unsymmetrically in the channel in $y$-direction. The level-set function is defined as $\phi\left(\vek{a}\right)=c$. Fig.~\ref{fig:TurekMaps} visualizes the geometries which are defined by these mappings.

\begin{figure}
	\centering
	
	\subfigure[]{\includegraphics[width=0.2\textwidth]{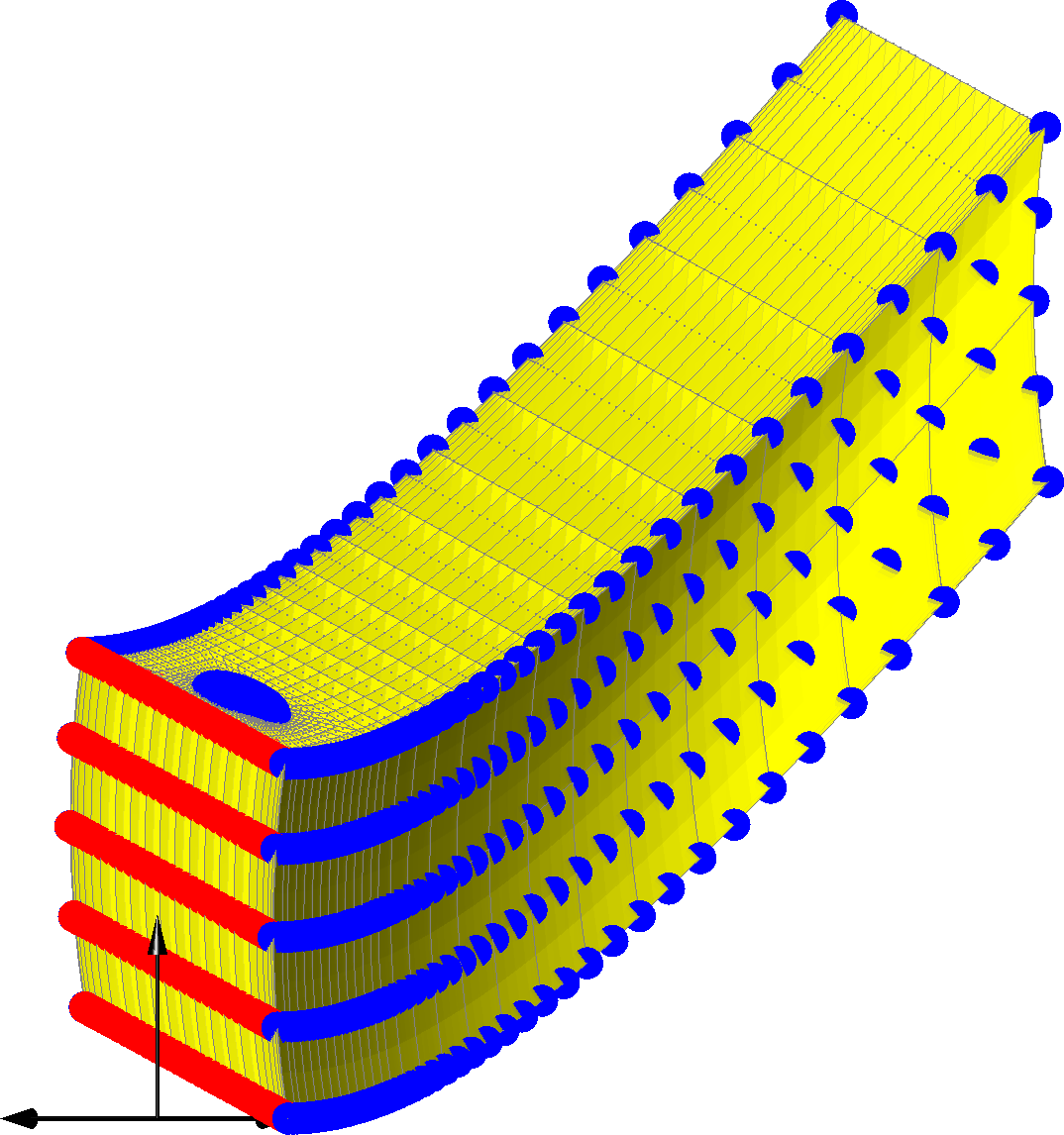}}
	\hspace{0.5cm}
	\subfigure[]{\includegraphics[width=0.25\textwidth]{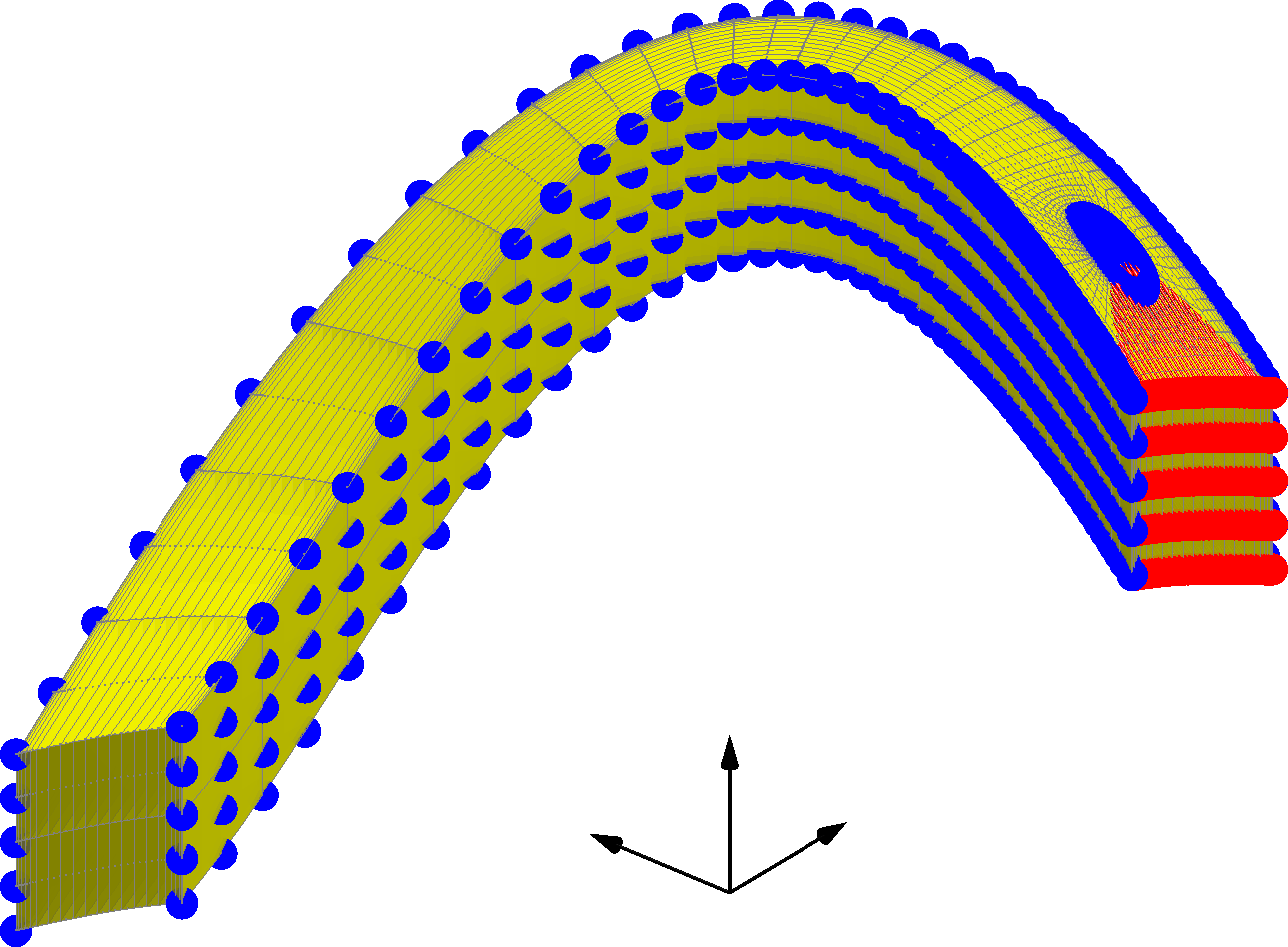}}
	\hspace{0.8cm}
	\subfigure[]{\includegraphics[width=0.25\textwidth]{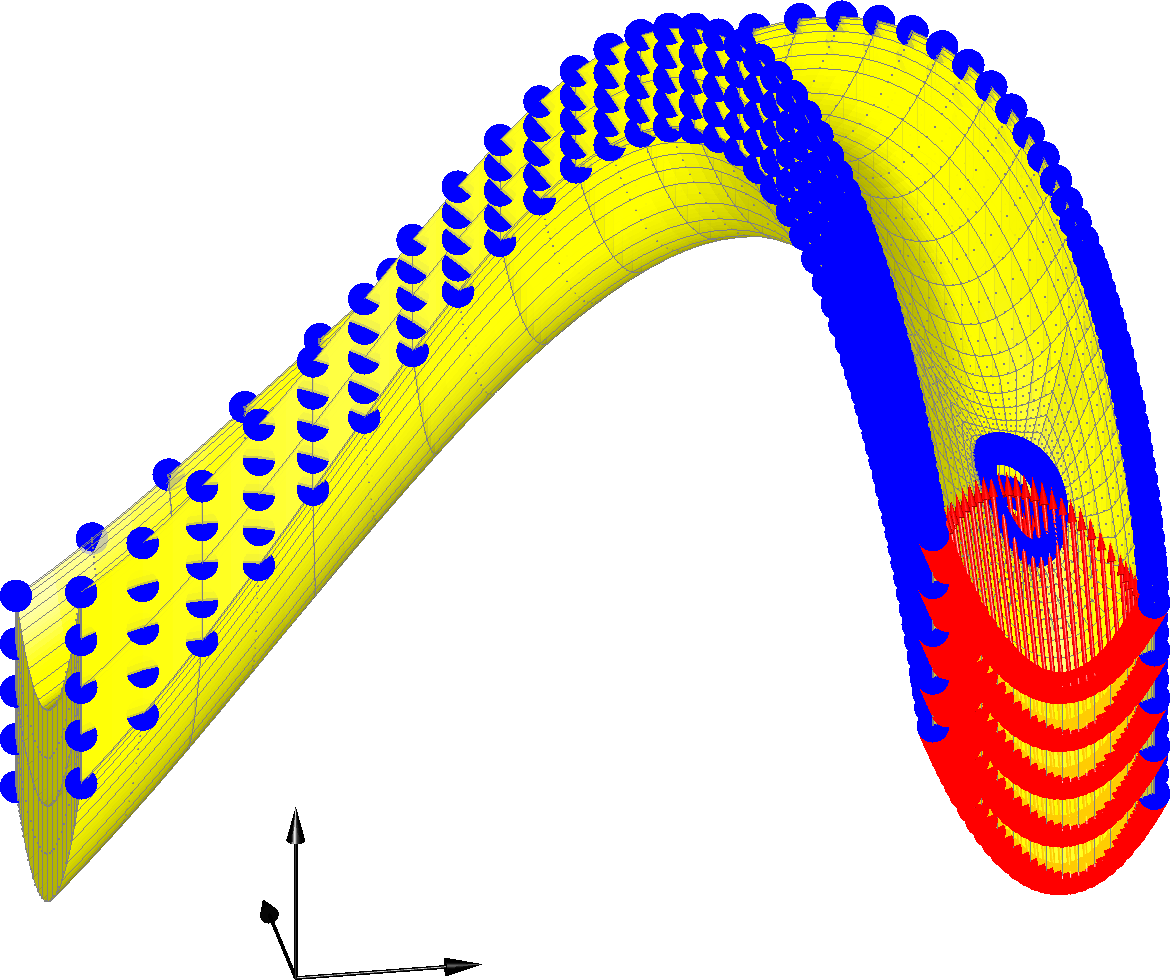}}
	
	\caption{\label{fig:TurekMeshBCs} Some mesh with highlighted nodes where velocities are prescribed. Mapping function $\varphi_1$ in (a), $\varphi_2$ in (b), and $\varphi_3$ in (c), respectively. Red arrows are shown for the prescribed velocities at the inflow boundary, blue dots indicate the no-slip condition along the channel walls and the obstacle.}
\end{figure}

Fig.~\ref{fig:TurekMeshBCs} shows the discretization of the three geometries with some example mesh (coarser than the mesh used for the computation) including the velocity profiles prescribed at the inflow. The red arrows indicate the inflow boundary condition and the blue dots are nodes on the no-slip boundary. Note that the geometry and the inflow velocities (length of the red arrows) are scaled differently for each map in this visualization. The inflow boundary condition is the quadratic velocity profile with $u_{\mathrm{max}} = 1.5$ and $v=0$ in the flat case and is mapped by the Jacobians of the respective mappings to ensure tangentiality of the velocities at the inflow boundary. At the outflow, traction-free boundary conditions are prescribed. The density is defined as $\varrho = 1$ and the viscosity is $\mu = 0.01$. For the computation, the bulk domain $\Omega$ is discretized by a mesh with 3760 elements. The orders for the geometry, the velocity, and the pressure meshes are $q_{\mathrm{geom}} = 3$, $q_{\vek{u}} = 2$, and $q_{p} = 1$, respectively.

\begin{figure}
	\centering
	
	\subfigure[Bulk Trace FEM - $\lVert\vek{u}\rVert$]{\includegraphics[width=0.35\textwidth]{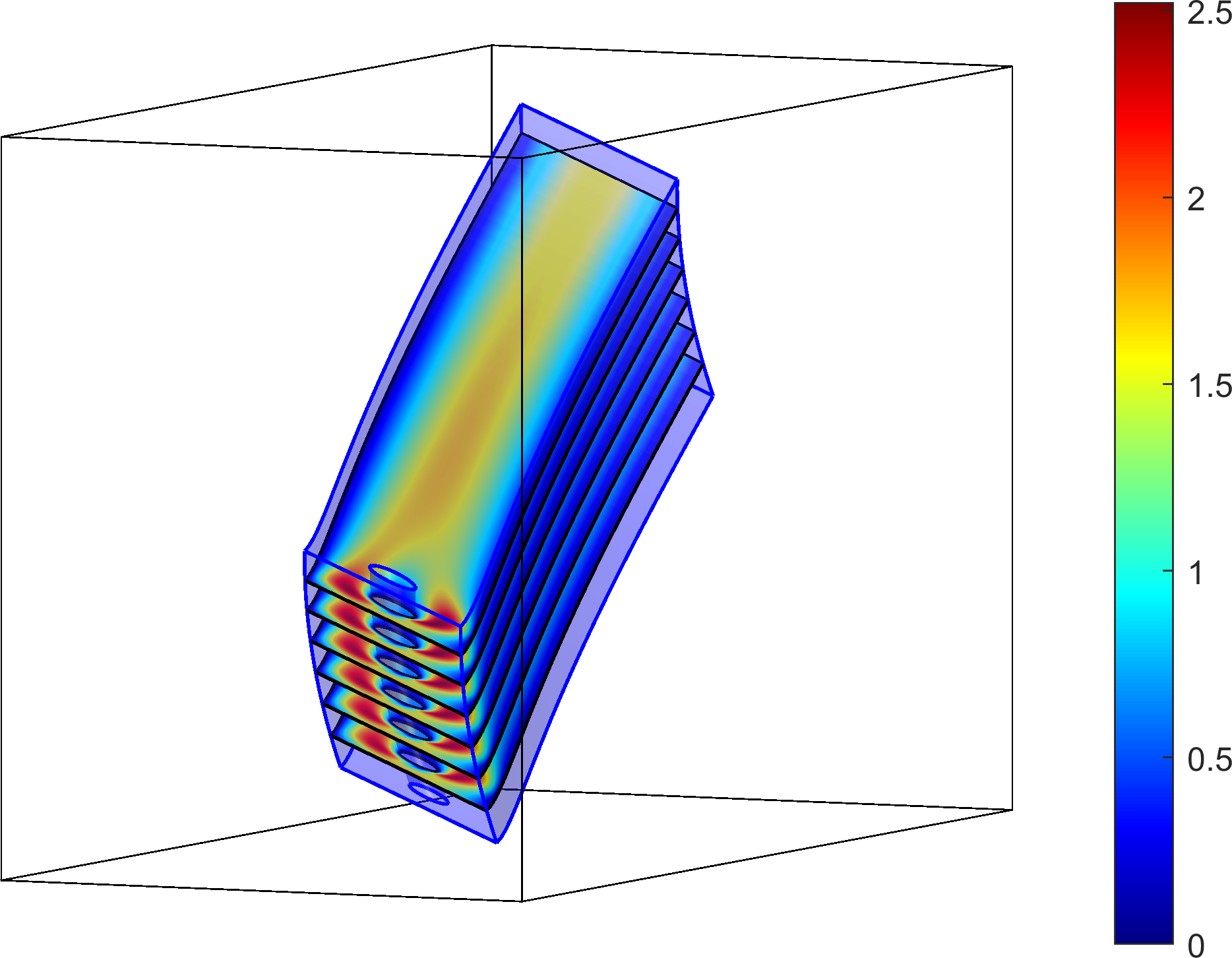}}\hspace{0.1\textwidth}
	\subfigure[$\Delta p$ over selected level sets $\phi$]{\includegraphics[width=0.35\textwidth]{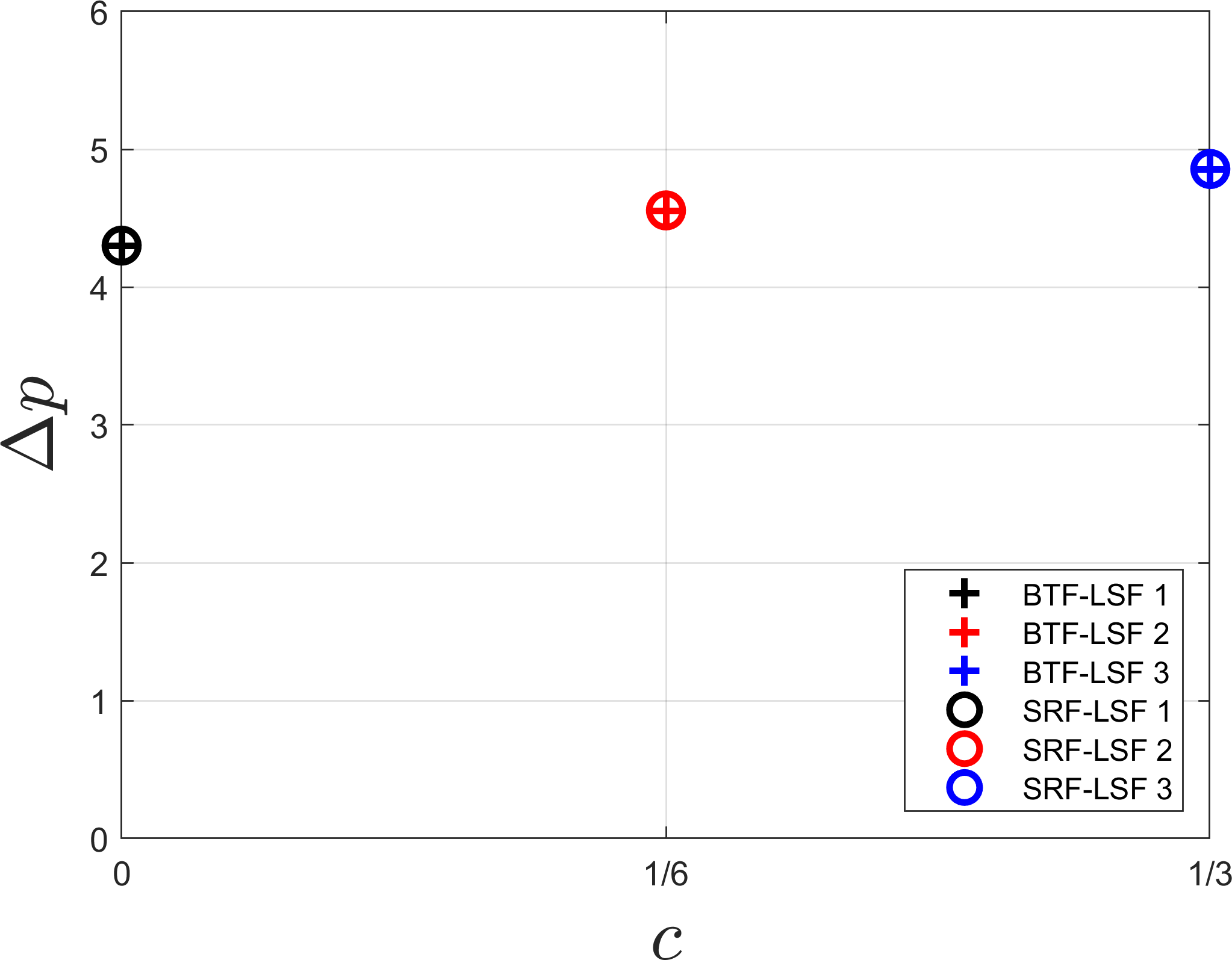}}
	\subfigure[Bulk Trace FEM - $\lVert\vek{u}\rVert$]{\includegraphics[width=0.35\textwidth]{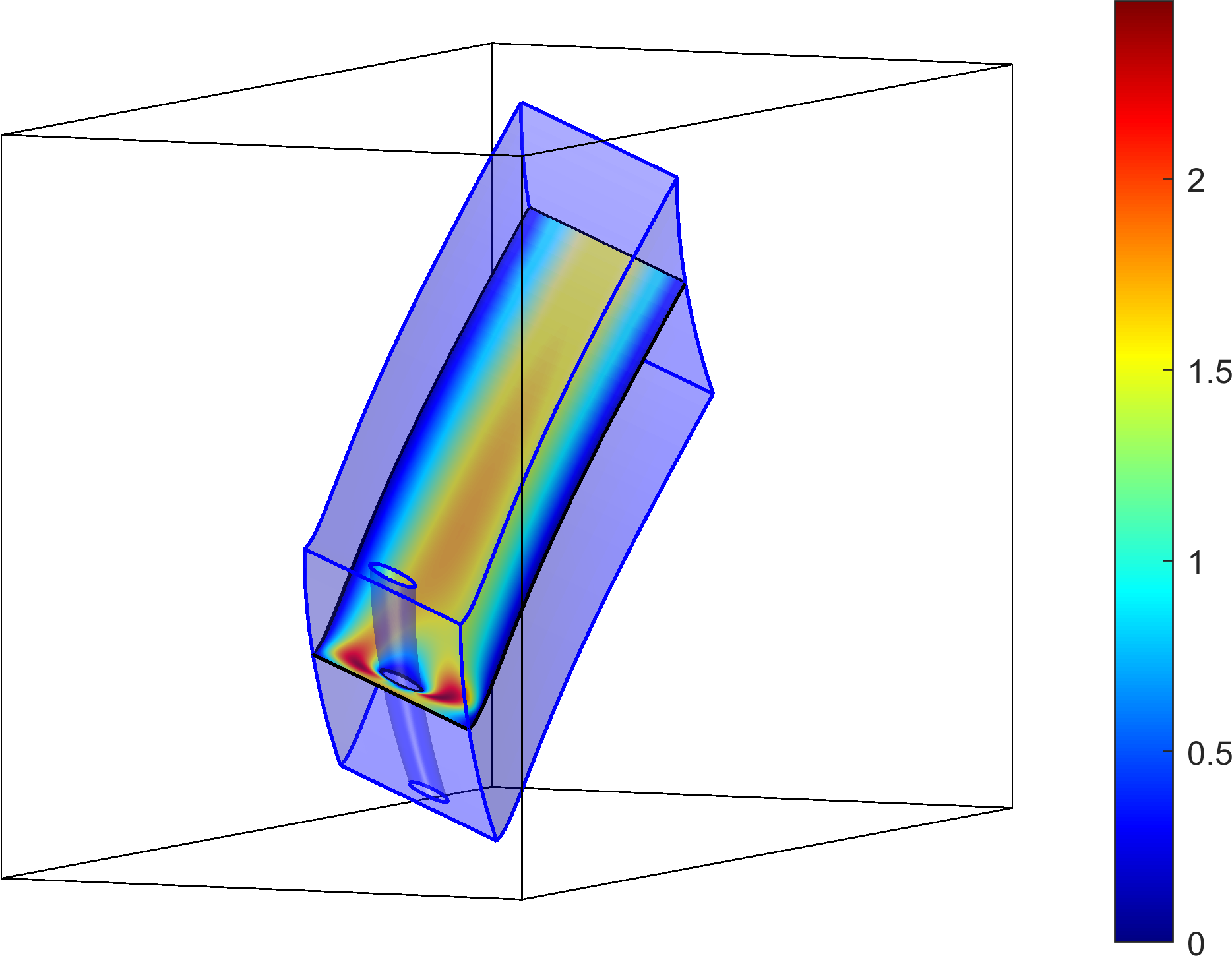}}\hspace{0.1\textwidth}
	\subfigure[Surface FEM - $\lVert\vek{u}\rVert$]{\includegraphics[width=0.35\textwidth]{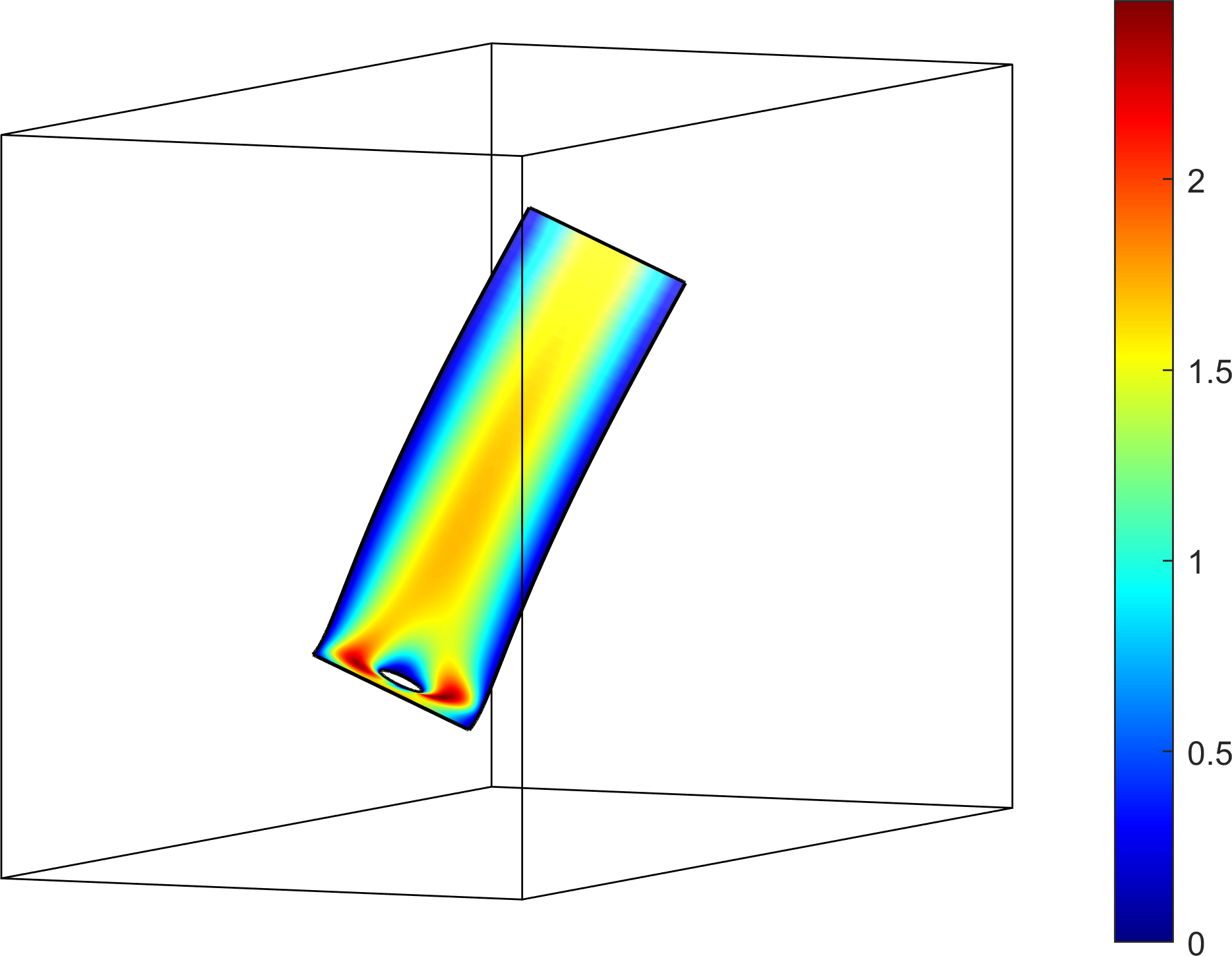}}
	\subfigure[Bulk Trace FEM - $p$]{\includegraphics[width=0.35\textwidth]{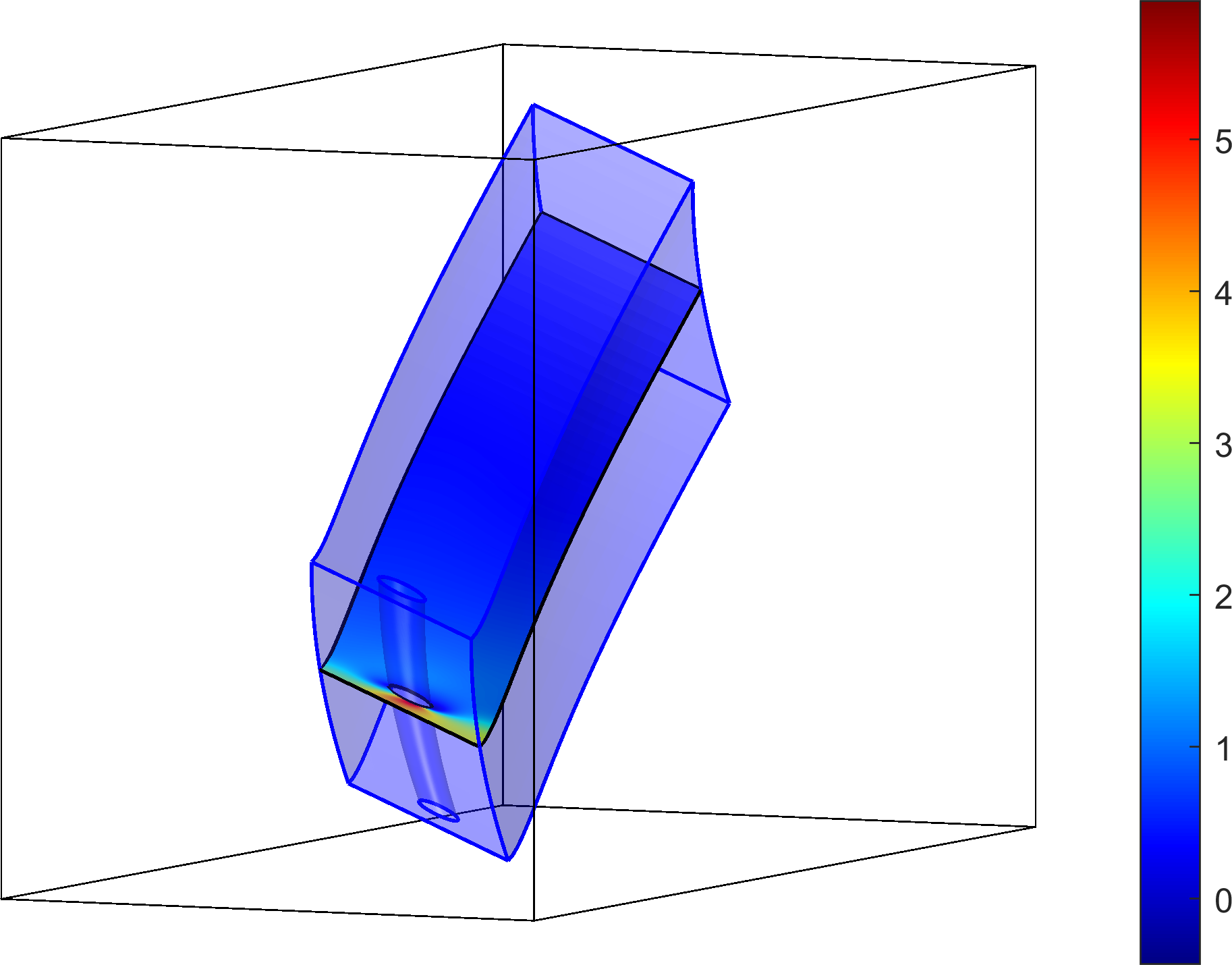}}\hspace{0.1\textwidth}
	\subfigure[Surface FEM - $p$]{\includegraphics[width=0.35\textwidth]{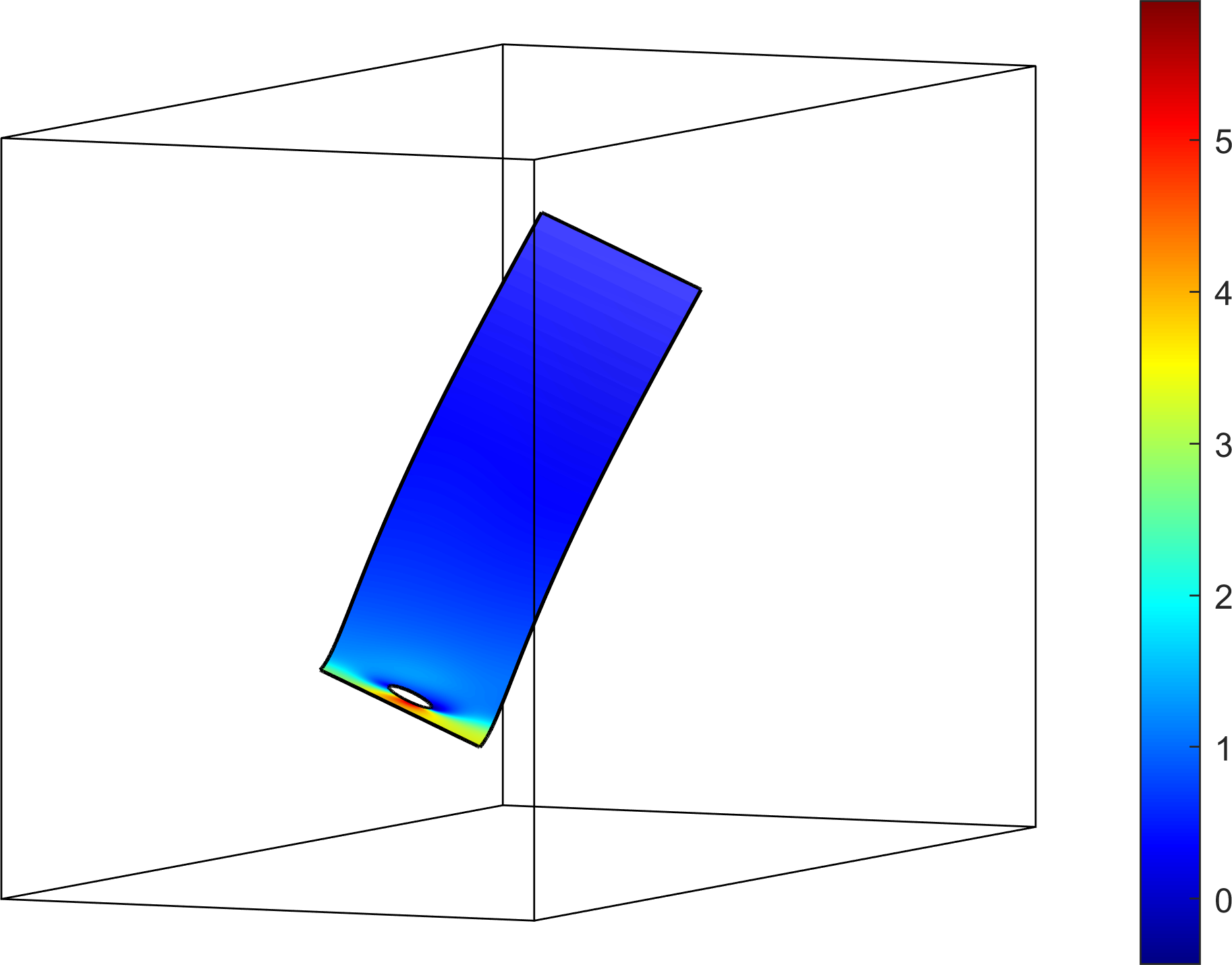}}
	
	\caption{\label{fig:TurekStatNSEQ-Res1} Results for mapping $\varphi_1$. The velocity magnitudes obtained with the Bulk Trace FEM on selected level sets are shown in (a) and (c), while in (d), a Surface FEM solution is shown on one level set. The pressure difference at two nodes on the cylindrical obstacle, obtained with the Bulk Trace FEM is shown in (b), the pressure field obtained with the Bulk Trace FEM in (e) and with the Surface FEM in (f).}
\end{figure}

\begin{figure}
	\centering
	
	\subfigure[Bulk Trace FEM - $\lVert\vek{u}\rVert$]{\includegraphics[width=0.35\textwidth]{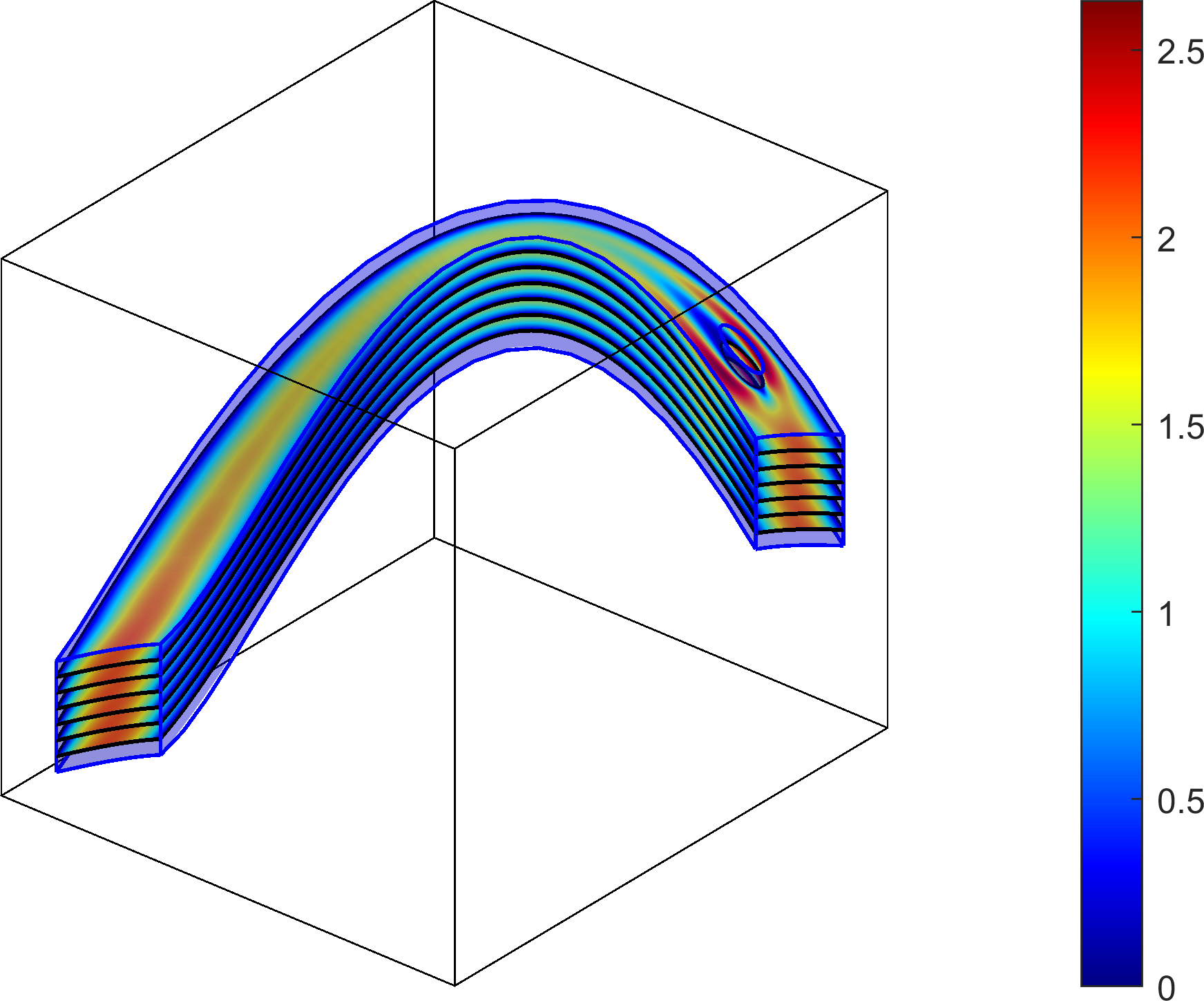}}\hspace{0.1\textwidth}
	\subfigure[$\Delta p$ over selected level sets $\phi$]{\includegraphics[width=0.35\textwidth,height=0.3\textwidth]{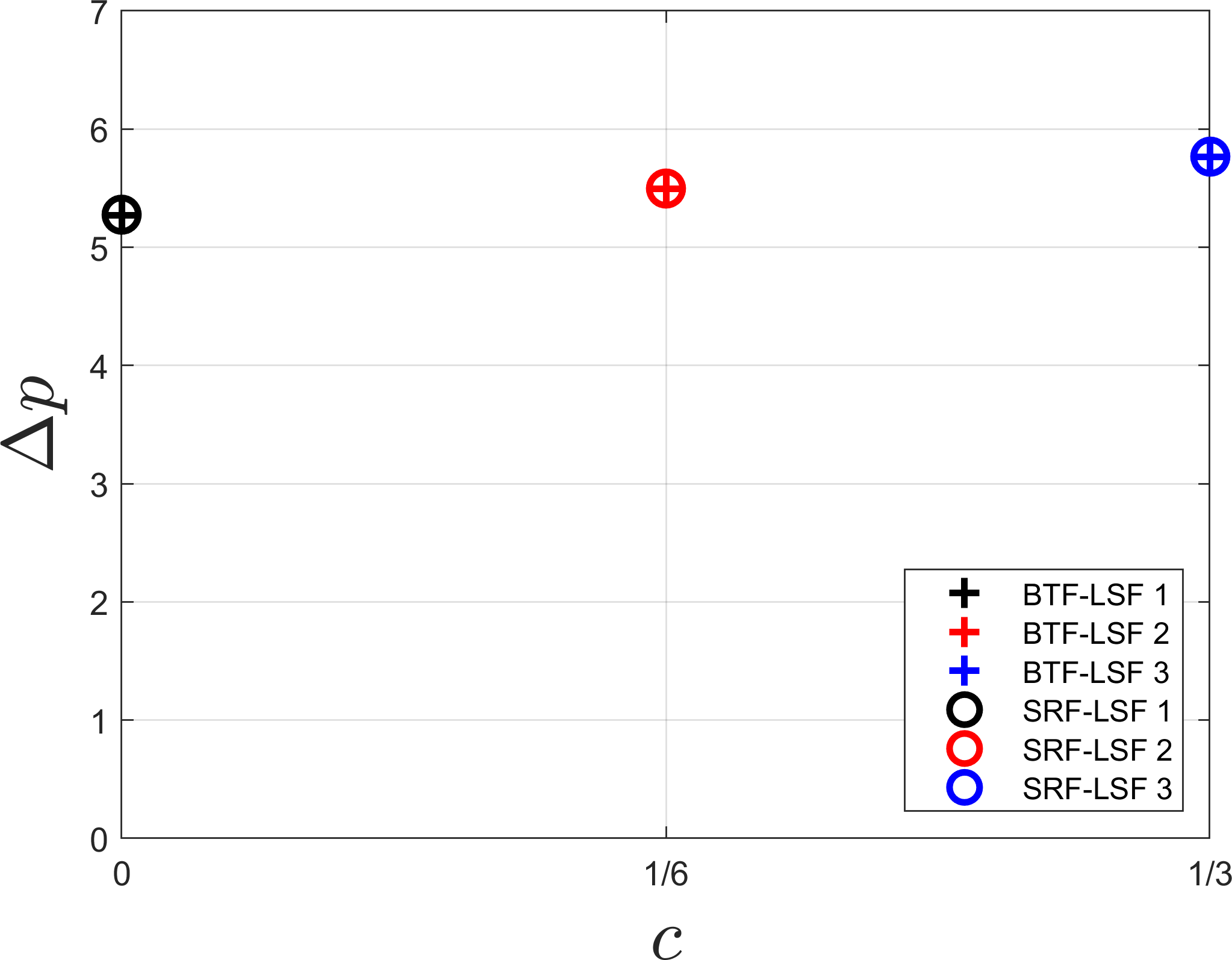}}
	\subfigure[Bulk Trace FEM - $\lVert\vek{u}\rVert$]{\includegraphics[width=0.35\textwidth]{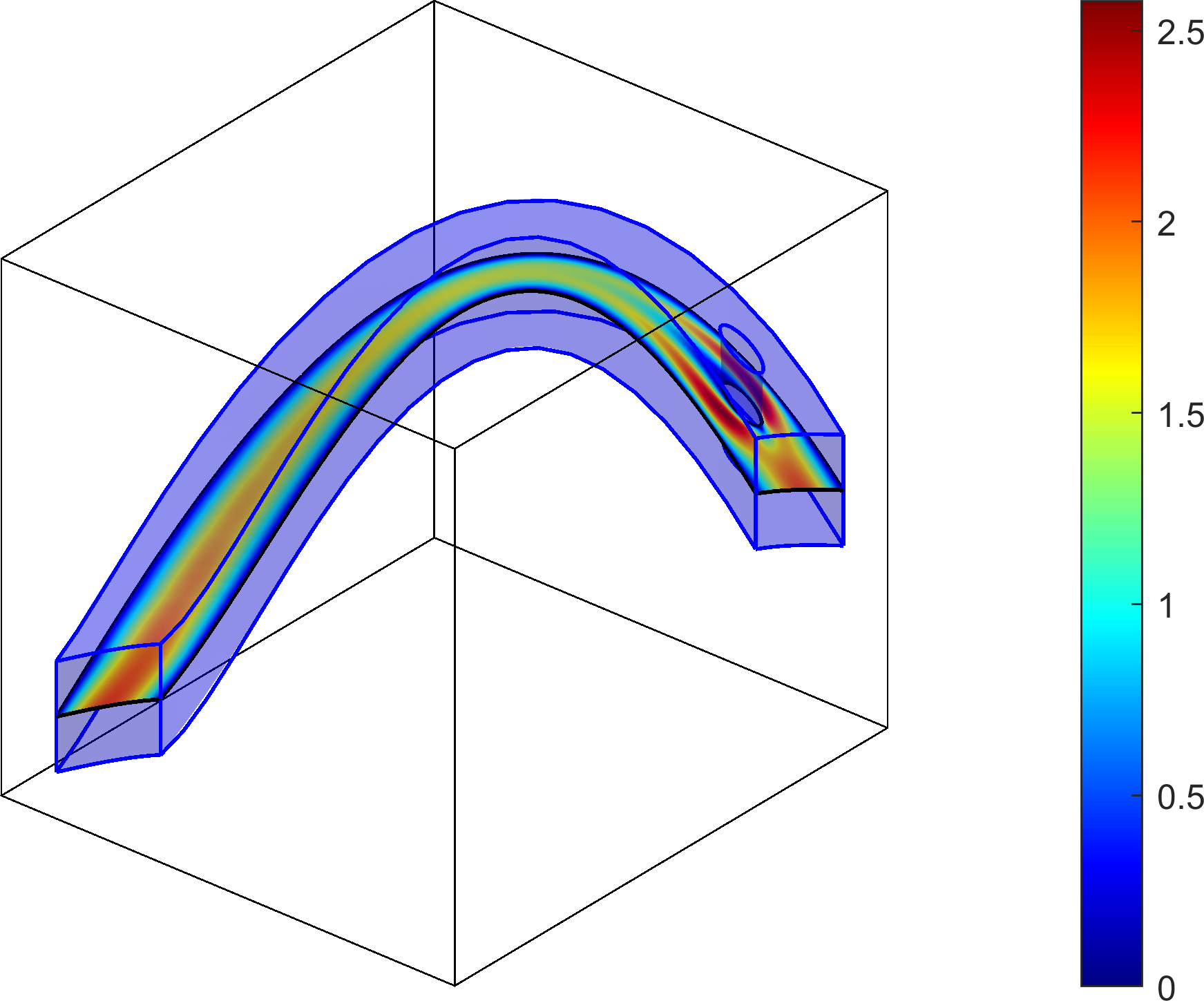}}\hspace{0.1\textwidth}
	\subfigure[Surface FEM - $\lVert\vek{u}\rVert$]{\includegraphics[width=0.35\textwidth]{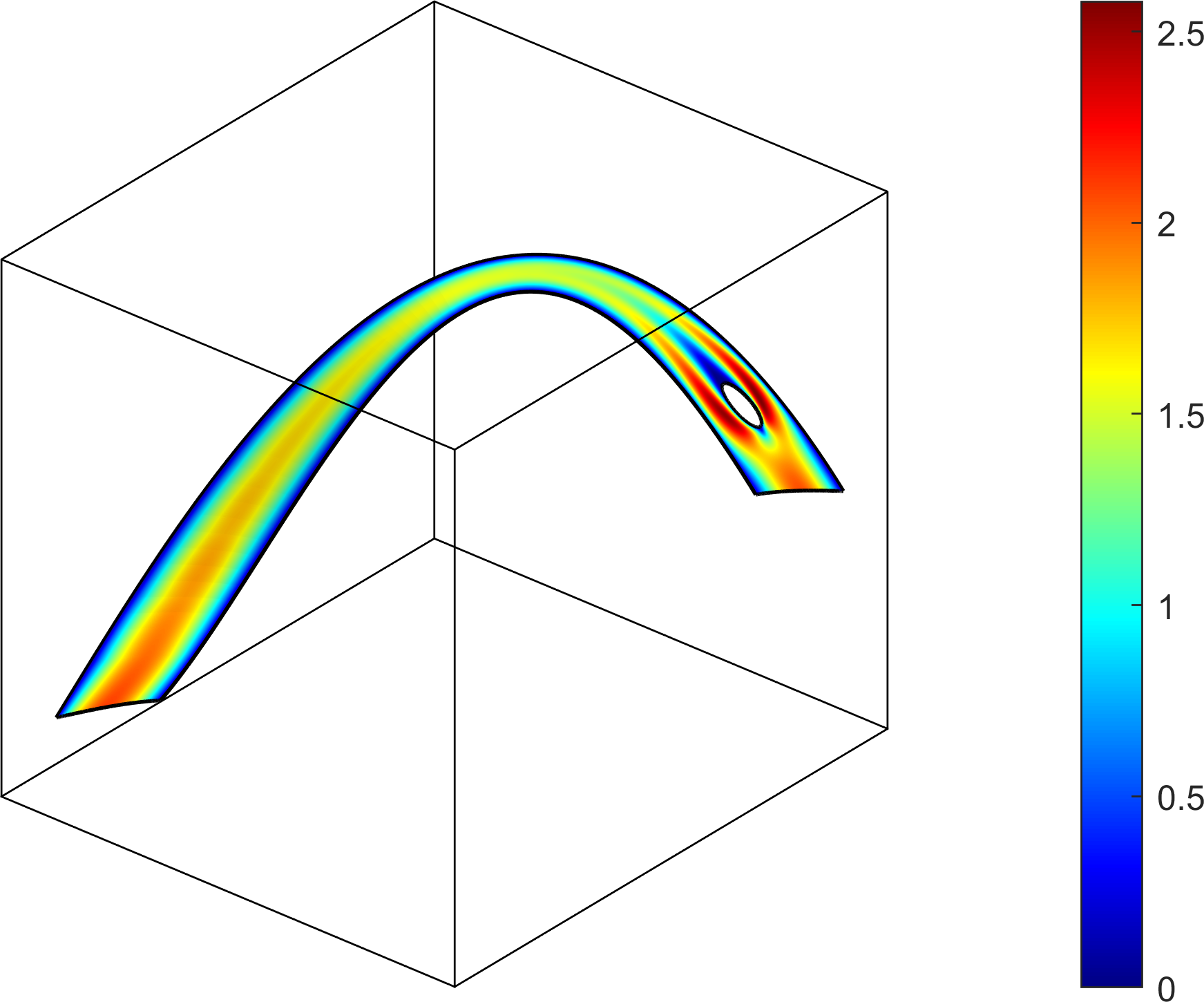}}
	\subfigure[Bulk Trace FEM - $p$]{\includegraphics[width=0.35\textwidth]{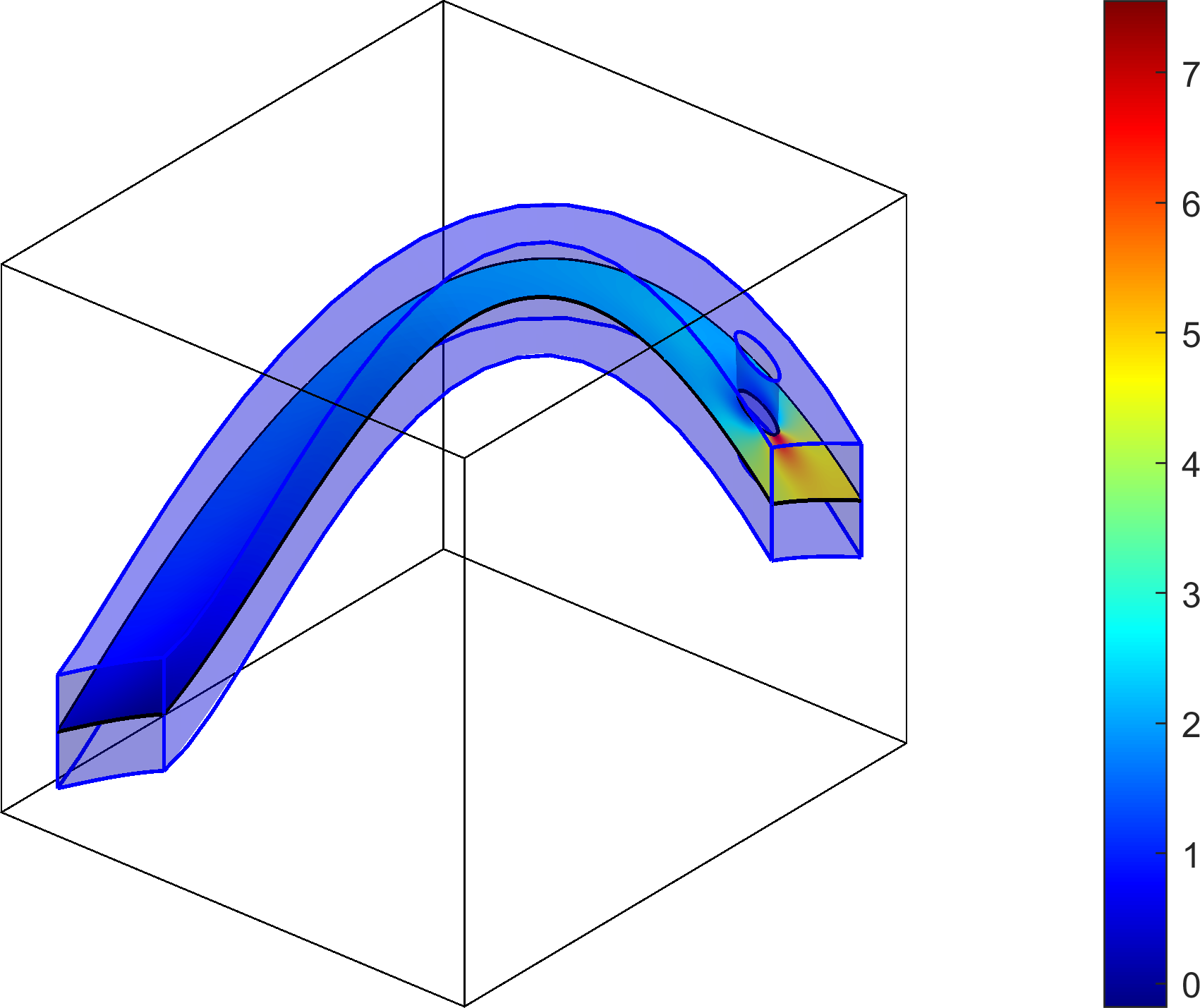}}\hspace{0.1\textwidth}
	\subfigure[Surface FEM - $p$]{\includegraphics[width=0.35\textwidth]{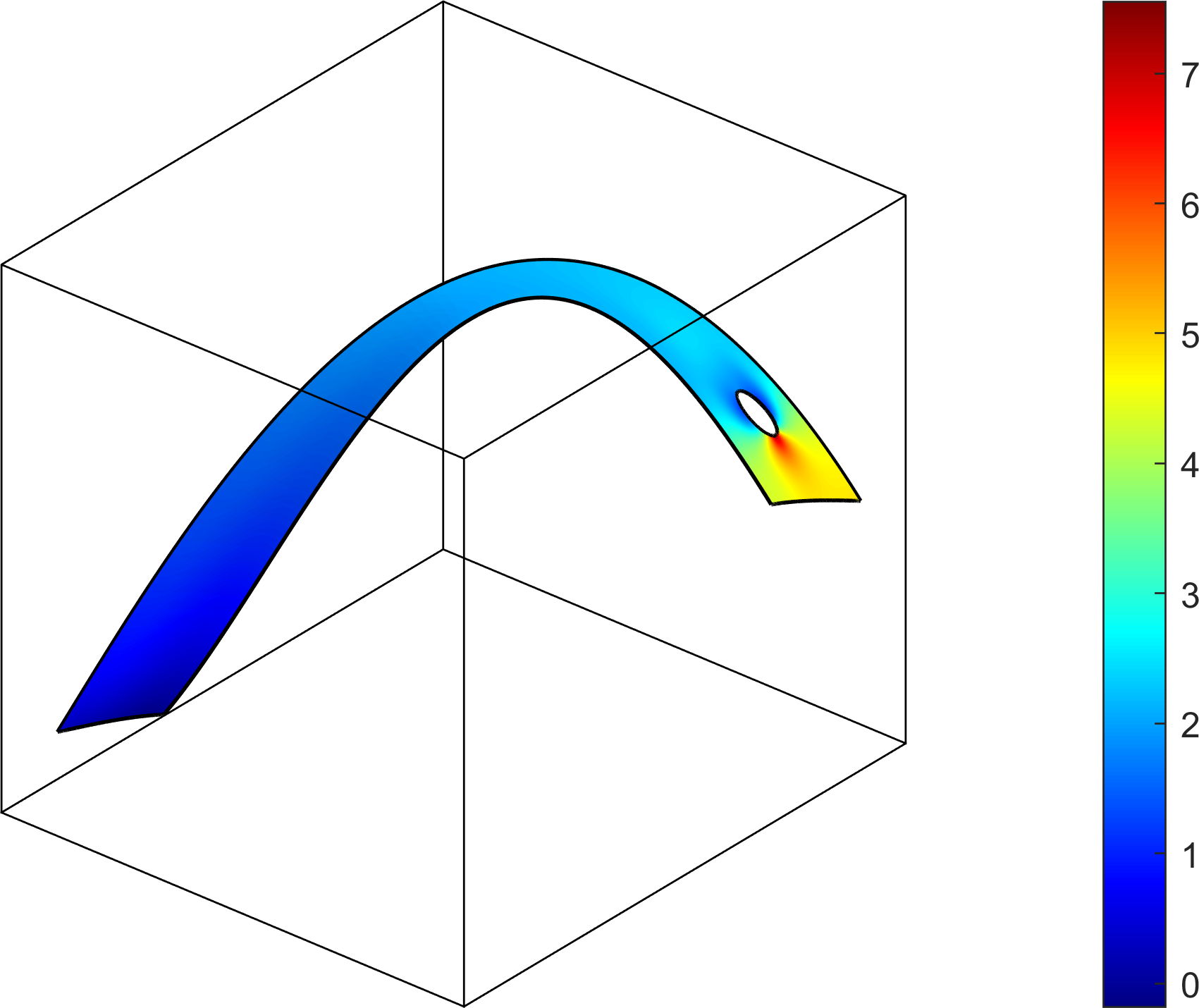}}
	
	\caption{\label{fig:TurekStatNSEQ-Res2} Results for mapping $\varphi_2$. The velocity magnitudes obtained with the Bulk Trace FEM on selected level sets are shown in (a) and (c), while in (d), a Surface FEM solution is shown on one level set. The pressure difference at two nodes on the cylindrical obstacle, obtained with the Bulk Trace FEM is shown in (b), the pressure field obtained with the Bulk Trace FEM in (e) and with the Surface FEM in (f).}
\end{figure}

\begin{figure}
	\centering
	
	\subfigure[Bulk Trace FEM - $\lVert\vek{u}\rVert$]{\includegraphics[width=0.35\textwidth]{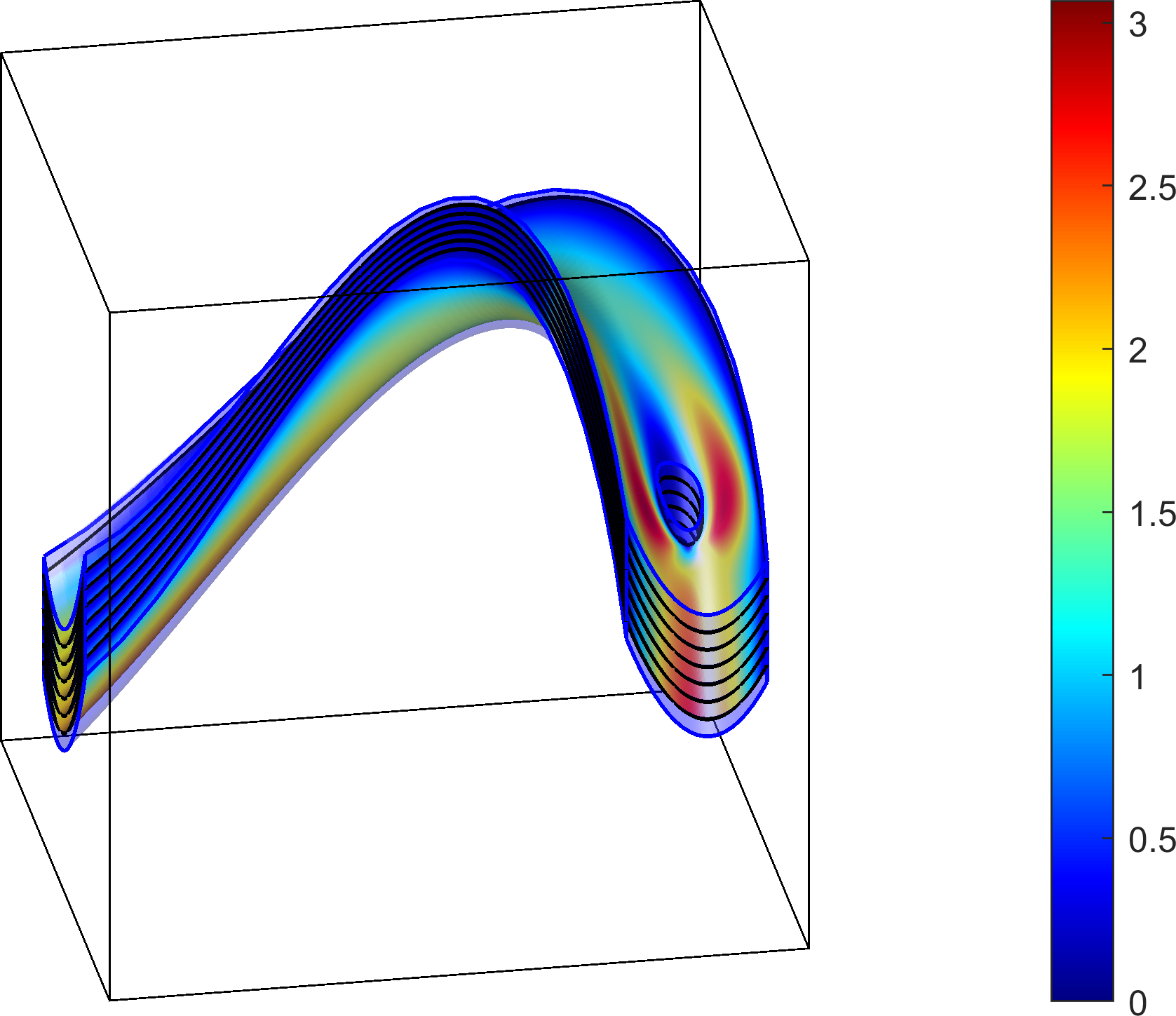}}\hspace{0.1\textwidth}
	\subfigure[$\Delta p$ over selected level sets $\phi$]{\includegraphics[width=0.35\textwidth,height=0.3\textwidth]{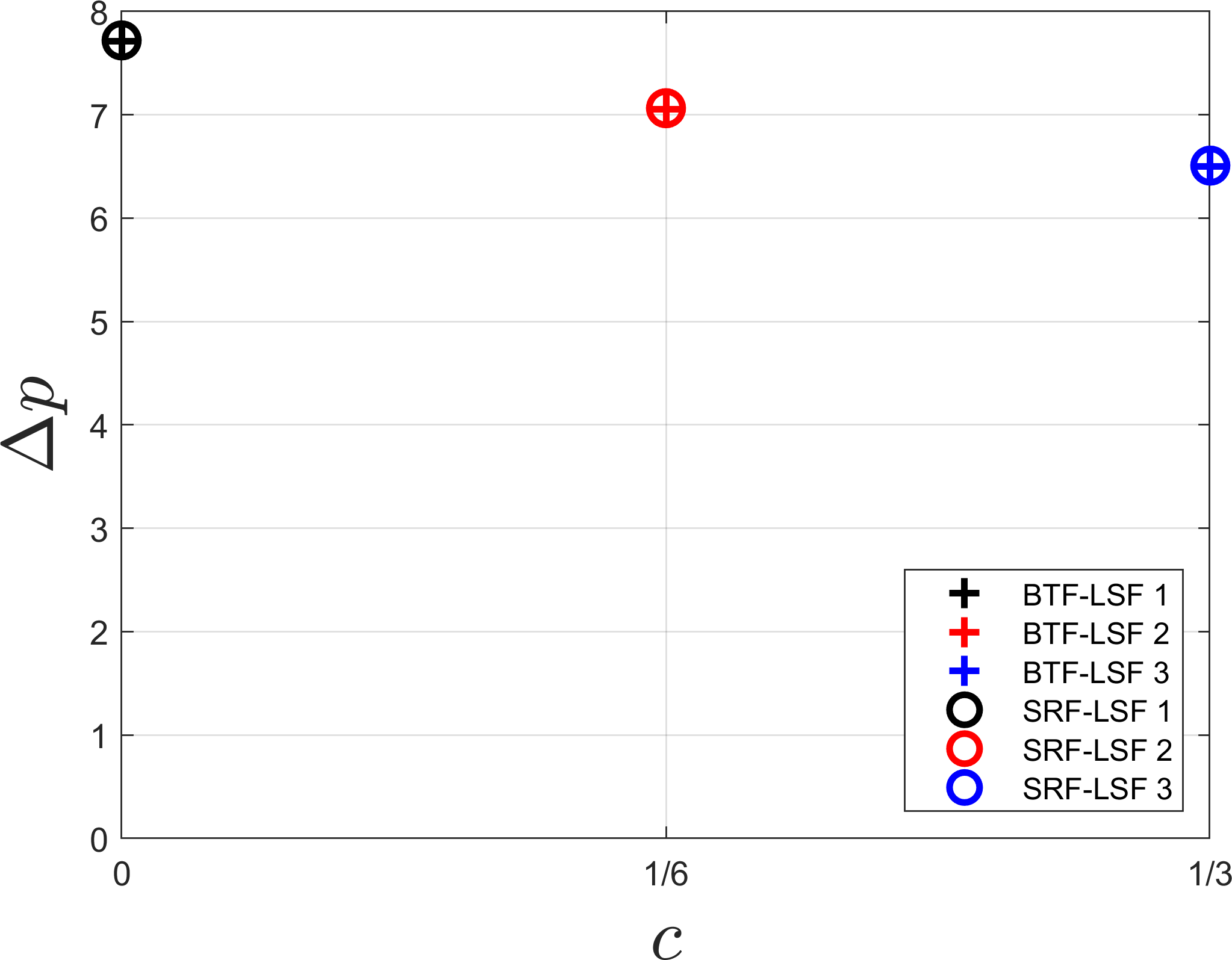}}
	\subfigure[Bulk Trace FEM - $\lVert\vek{u}\rVert$]{\includegraphics[width=0.35\textwidth]{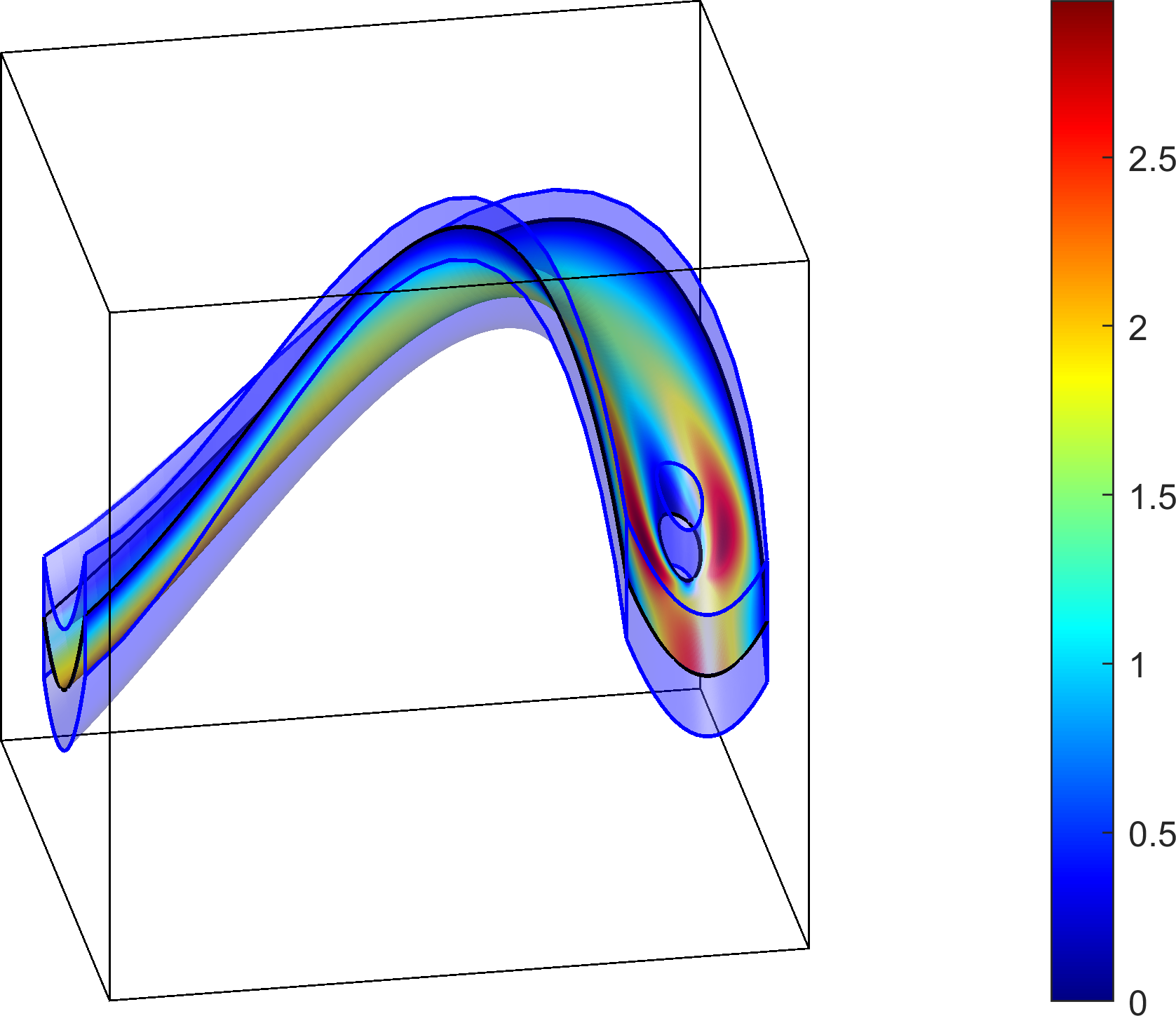}}\hspace{0.1\textwidth}
	\subfigure[Surface FEM - $\lVert\vek{u}\rVert$]{\includegraphics[width=0.35\textwidth]{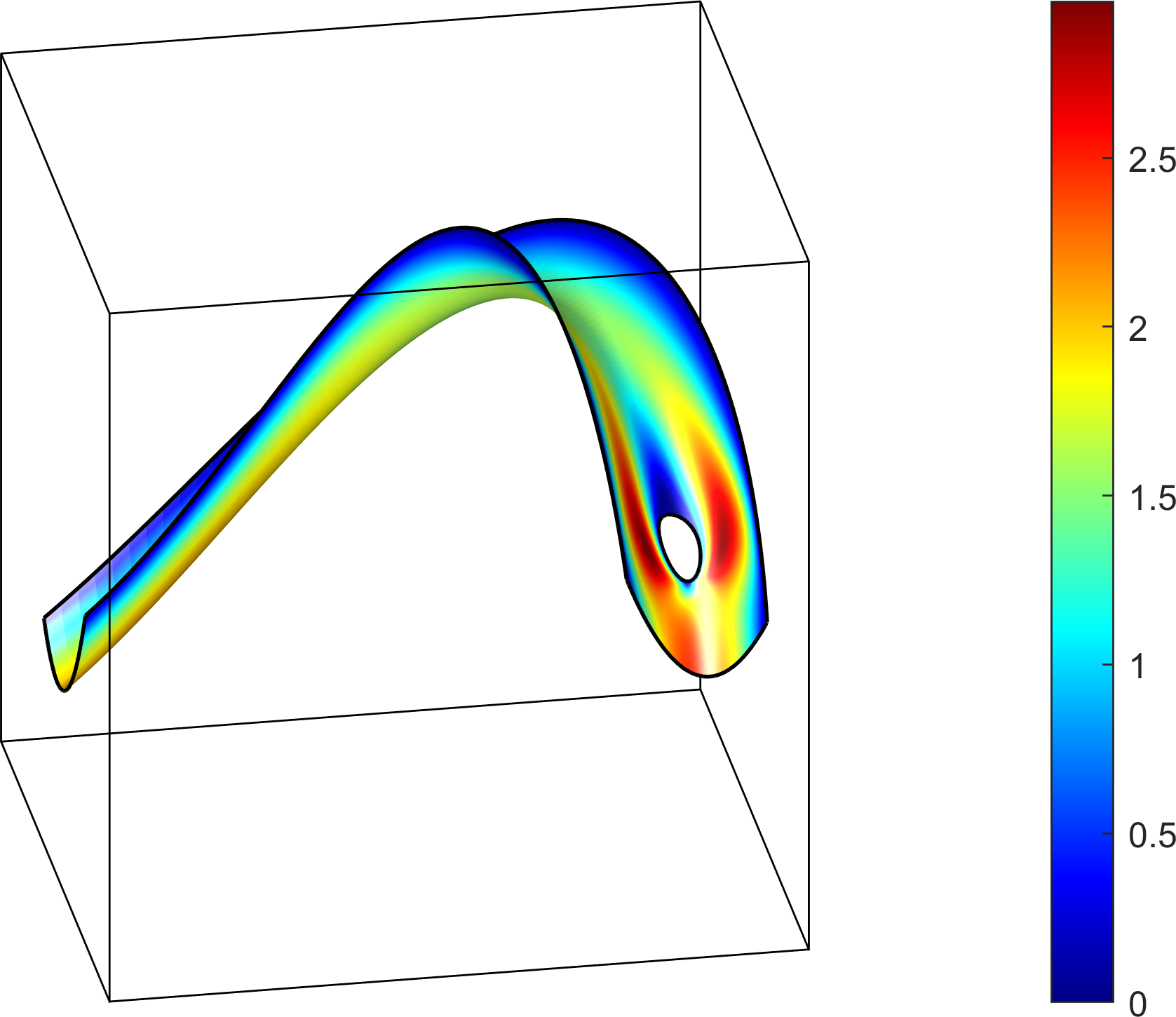}}
	\subfigure[Bulk Trace FEM - $p$]{\includegraphics[width=0.35\textwidth]{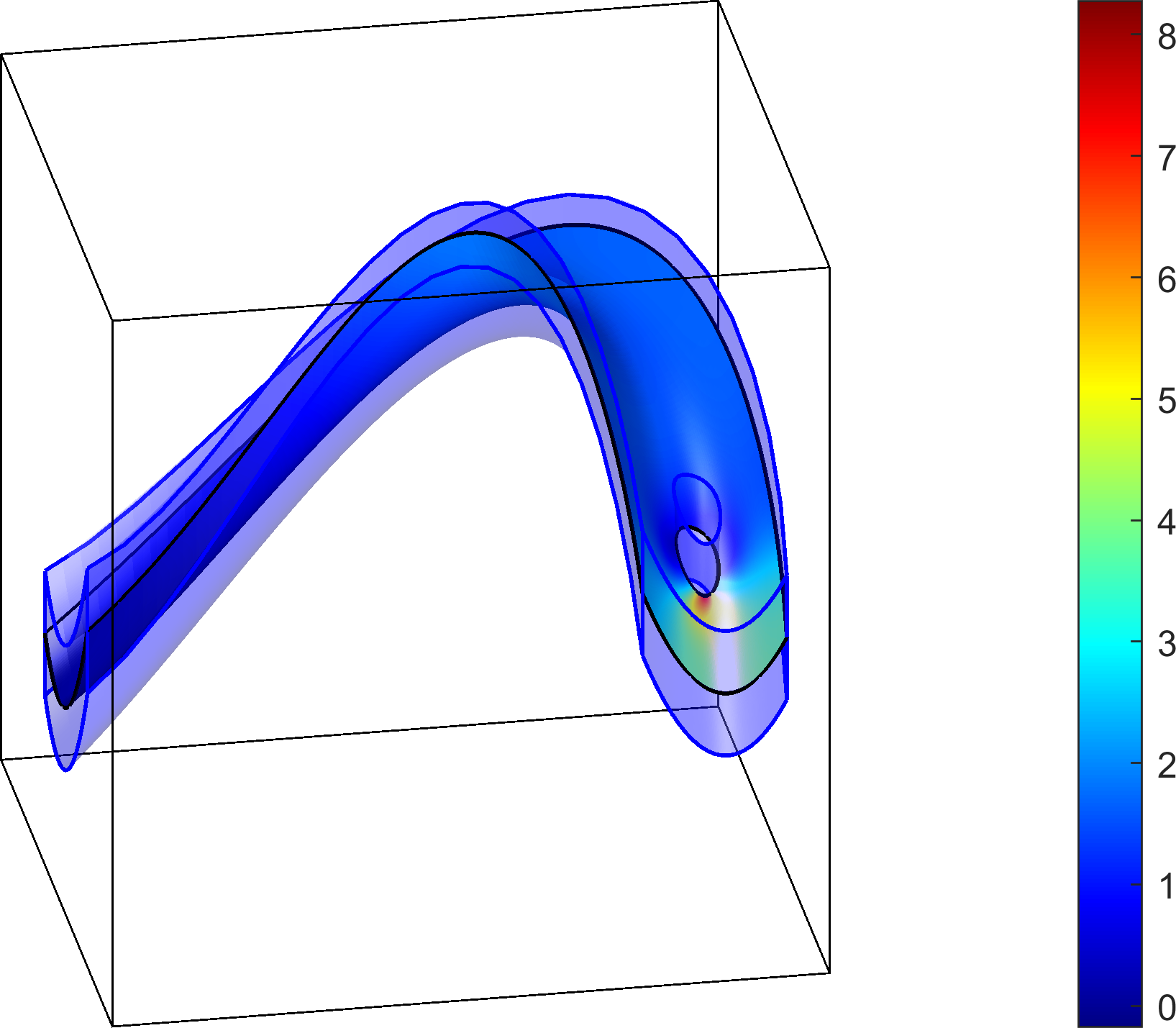}}\hspace{0.1\textwidth}
	\subfigure[Surface FEM - $p$]{\includegraphics[width=0.35\textwidth]{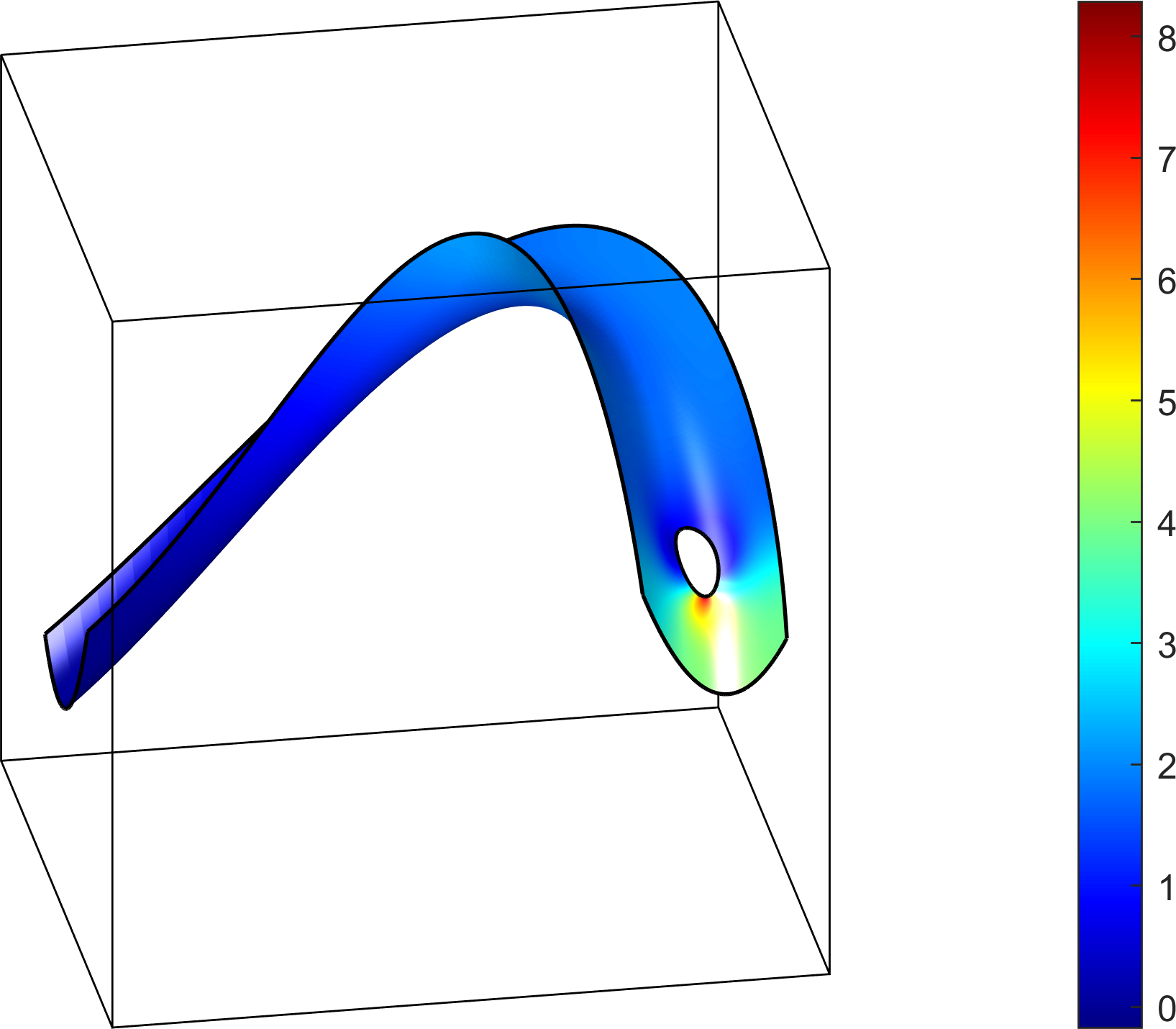}}
	
	\caption{\label{fig:TurekStatNSEQ-Res3} Results for mapping $\varphi_3$. The velocity magnitudes obtained with the Bulk Trace FEM on selected level sets are shown in (a) and (c), while in (d), a Surface FEM solution is shown on one level set. The pressure difference at two nodes on the cylindrical obstacle, obtained with the Bulk Trace FEM is shown in (b), the pressure field obtained with the Bulk Trace FEM in (e) and with the Surface FEM in (f).}
\end{figure}

The results of these test cases are shown in Figs.~\ref{fig:TurekStatNSEQ-Res1} to \ref{fig:TurekStatNSEQ-Res3}. To verify that the \emph{simultaneous} solution produces the same results for selected level sets as indepedent, successive simulations with the Surface FEM (SRF) \cite{Fries_2018a} on these level sets, the differences of the pressure at the front and back nodes of the cylindrical obstacle are compared, i.e., Figs.~\ref{fig:TurekStatNSEQ-Res1} to \ref{fig:TurekStatNSEQ-Res3} (b). The selected level sets $\Gamma_{\!c}$ with $c=\{0, 1/6, 1/3\}$ are used for the comparison. Furthermore, we visually compare the results for the level set with $c = 1/6$ for the velocities (Euclidean norm) and the pressure in Figs.~\ref{fig:TurekStatNSEQ-Res1} to \ref{fig:TurekStatNSEQ-Res3} (c) and (e) for the Bulk Trace FEM (BTF) and Figs.~\ref{fig:TurekStatNSEQ-Res1} to \ref{fig:TurekStatNSEQ-Res3} (d) and (f) for the Surface FEM (SRF). Excellent agreements are observed, confirming that the Bulk Trace FEM is successful in simultaneously solving for flow fields on all level sets.

\subsubsection{Driven cavity flow on manifolds}
\begin{figure}
	\centering
	
	\includegraphics[width=1\textwidth]{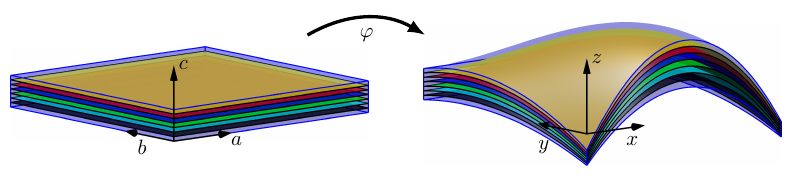}
	
	\caption{\label{fig:DrvCavMap} The map $\varphi$ for the geometry definition of the driven cavity test case. The bulk domain is shown in light blue and some arbitrarily selected level sets are shown in different colours.}
\end{figure}

\begin{figure}
	\centering
	
	\subfigure[]{\includegraphics[width=0.4\textwidth]{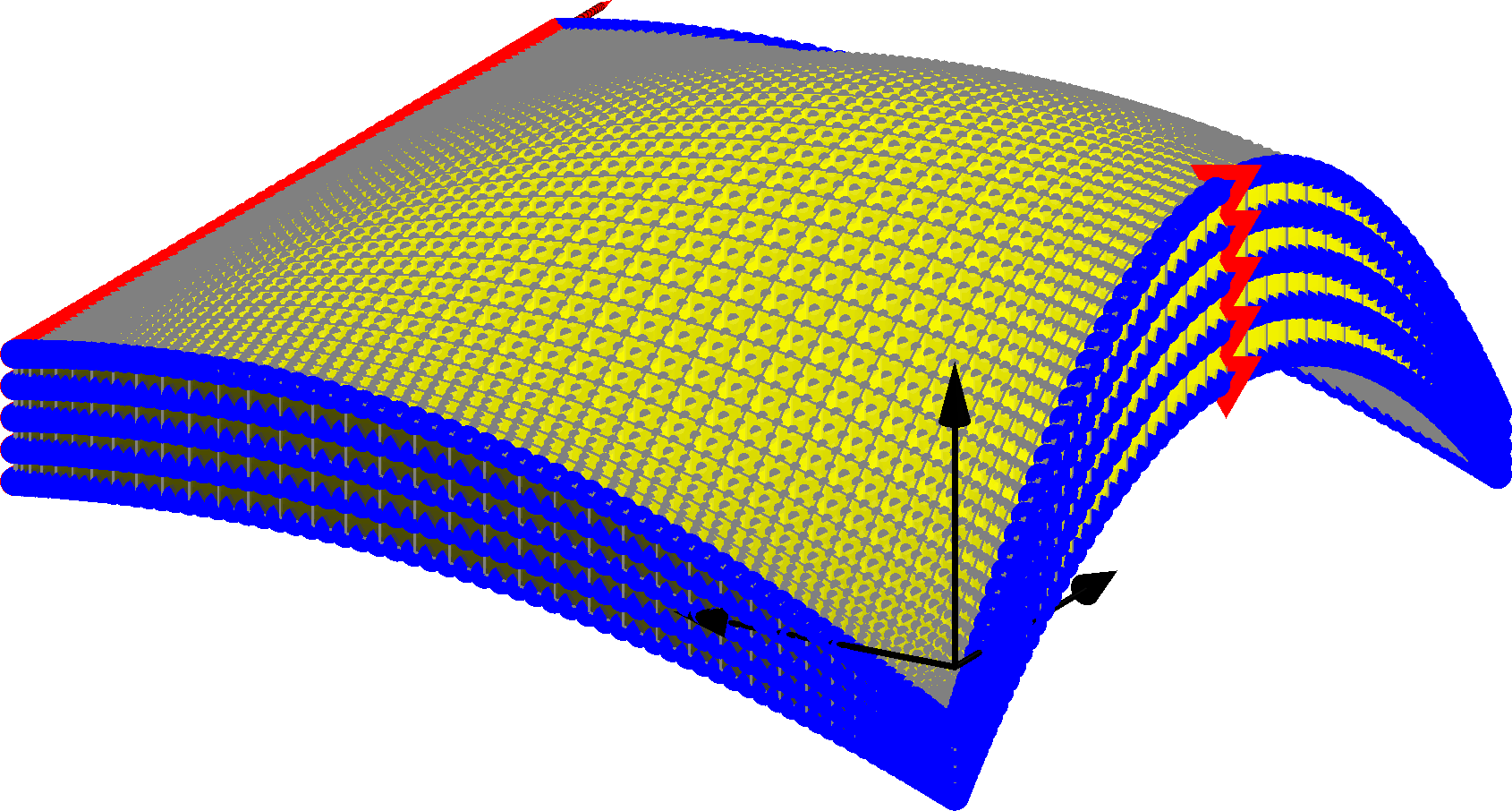}}
	\hspace{1.5cm}
	\subfigure[]{\includegraphics[width=0.4\textwidth]{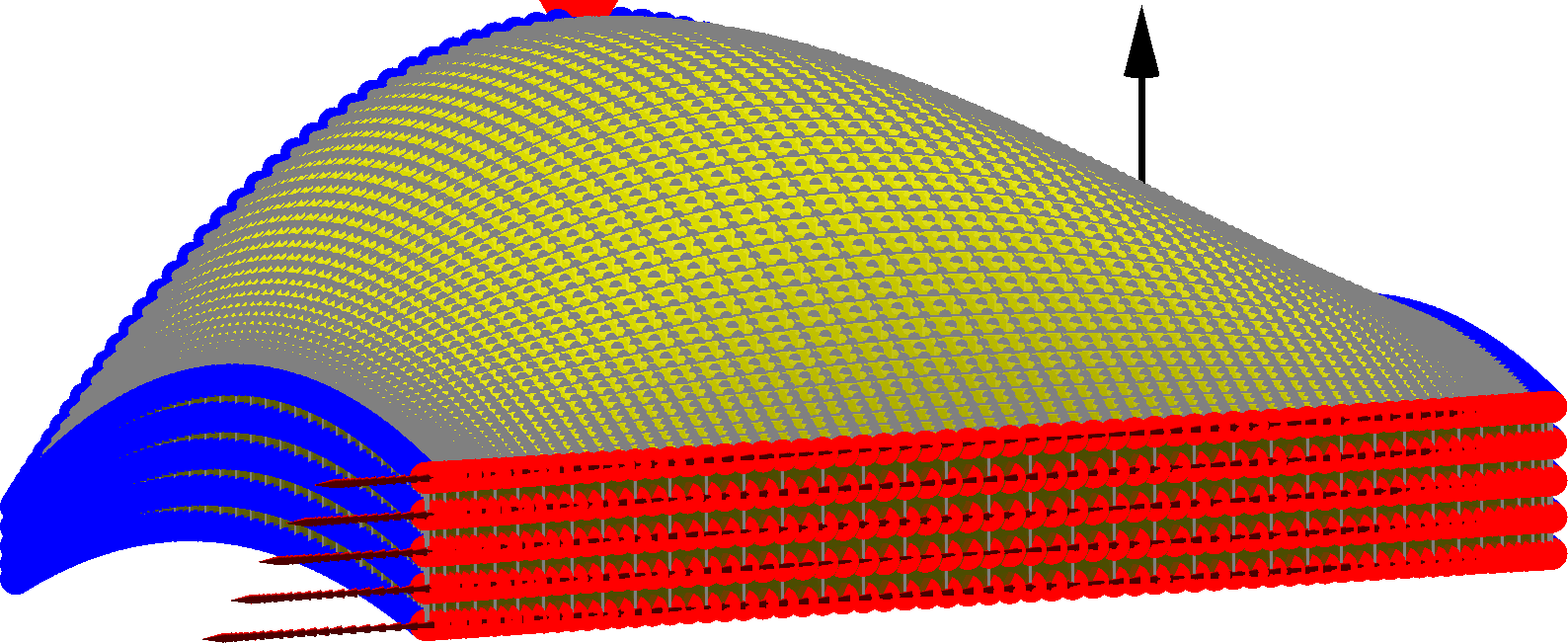}}
	
	\caption{\label{fig:DrvCavMeshBCs} The mesh with $40 \times 40$ elements in $x$-$y$-direction and $2$ elements in thickness direction from two different perspectives in (a) and (b), respectively. The Dirichlet boundary conditions are shown as red arrows for the prescribed non-zero velocity, the blue dots indicate the no-slip condition. The red triangles show the nodes at which the pressure boundary conditions are prescribed.}
\end{figure}

\begin{figure}
	\centering
	
	\subfigure[Bulk Trace FEM - $\lVert{\vek{u}}\rVert$]{\includegraphics[width=0.45\textwidth]{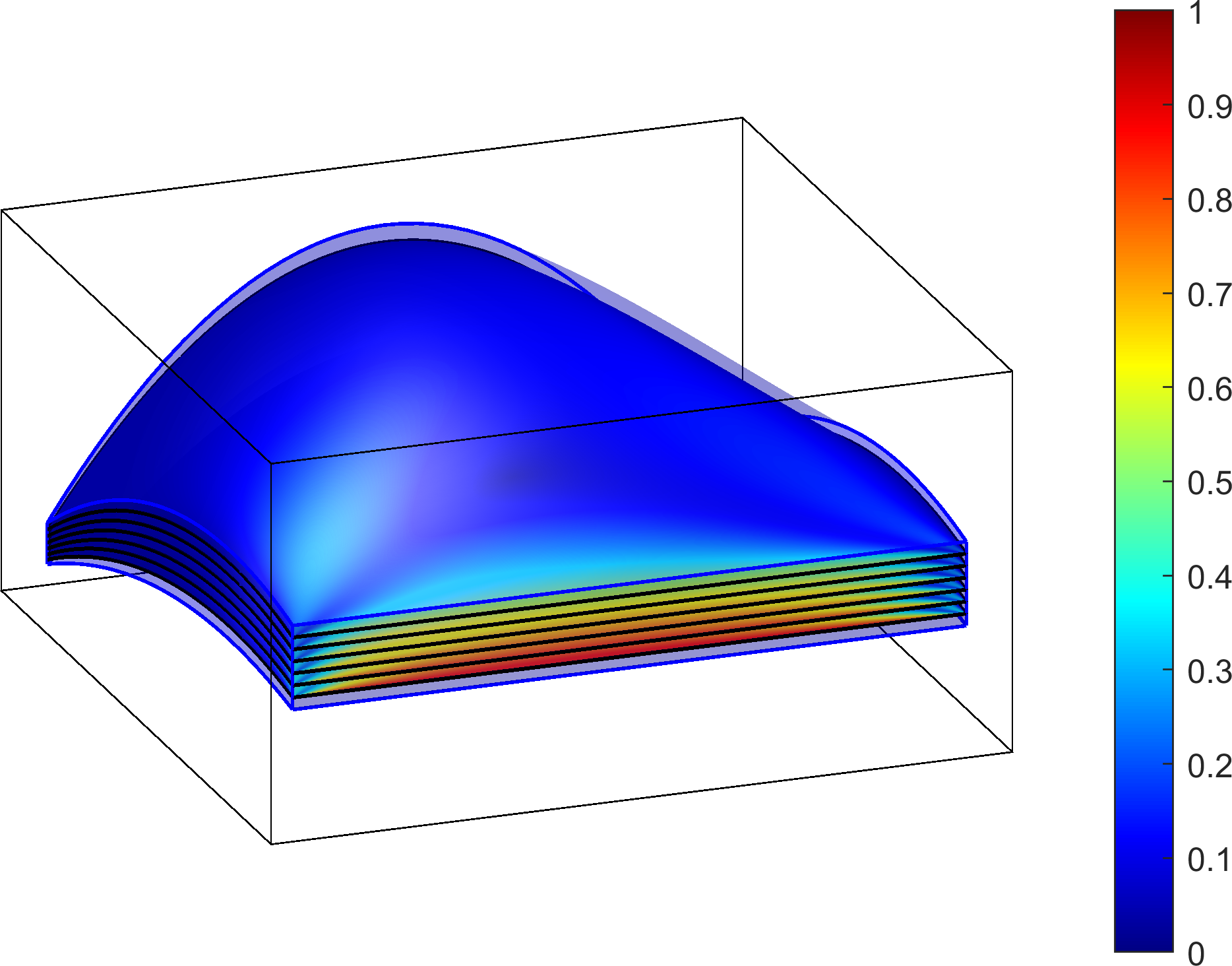}}\hspace{2cm}
	\subfigure[velocity profiles]{\includegraphics[width=0.37\textwidth]{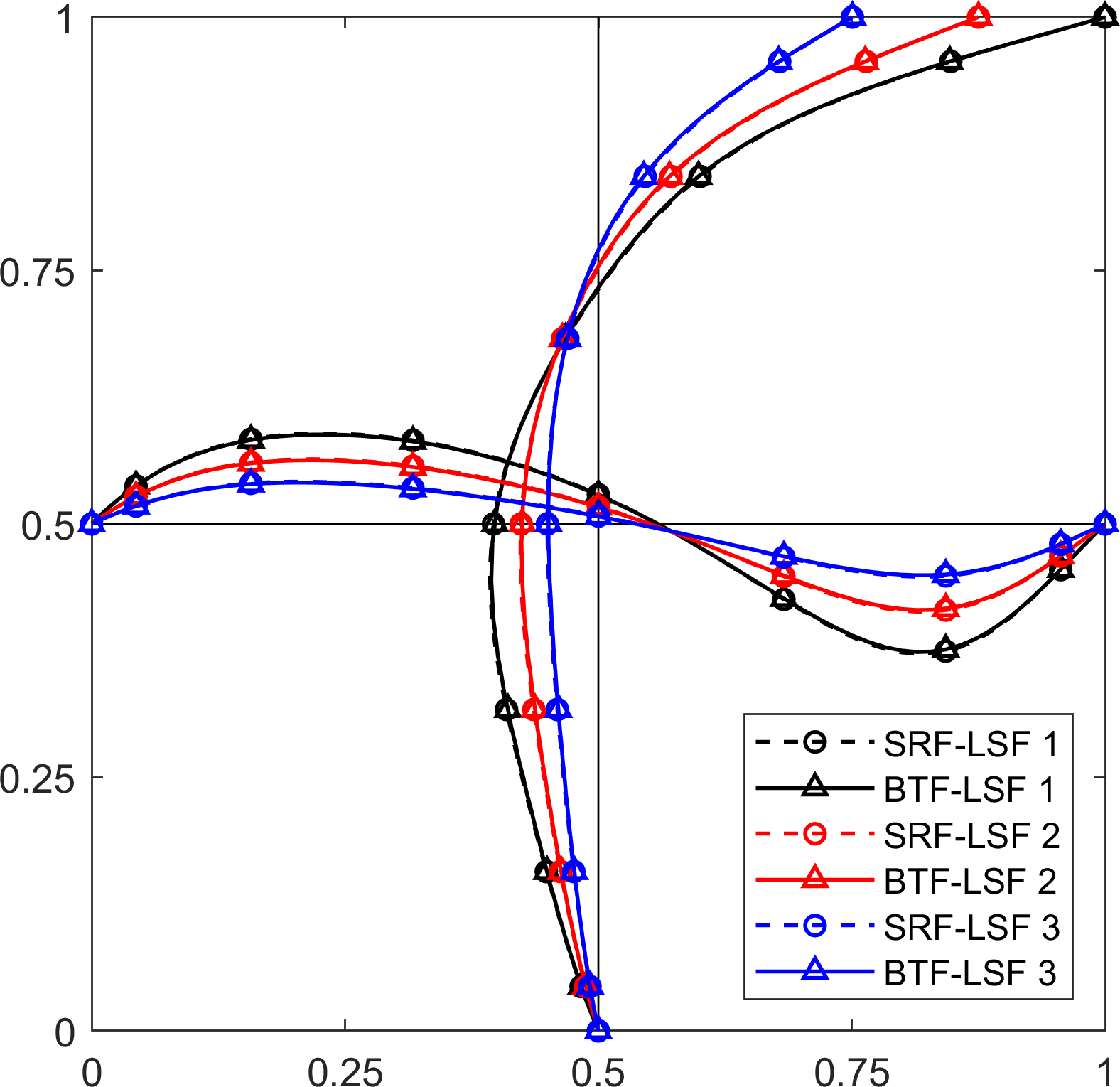}}
	\subfigure[Bulk Trace FEM - $\lVert{\vek{u}}\rVert$]{\includegraphics[width=0.45\textwidth]{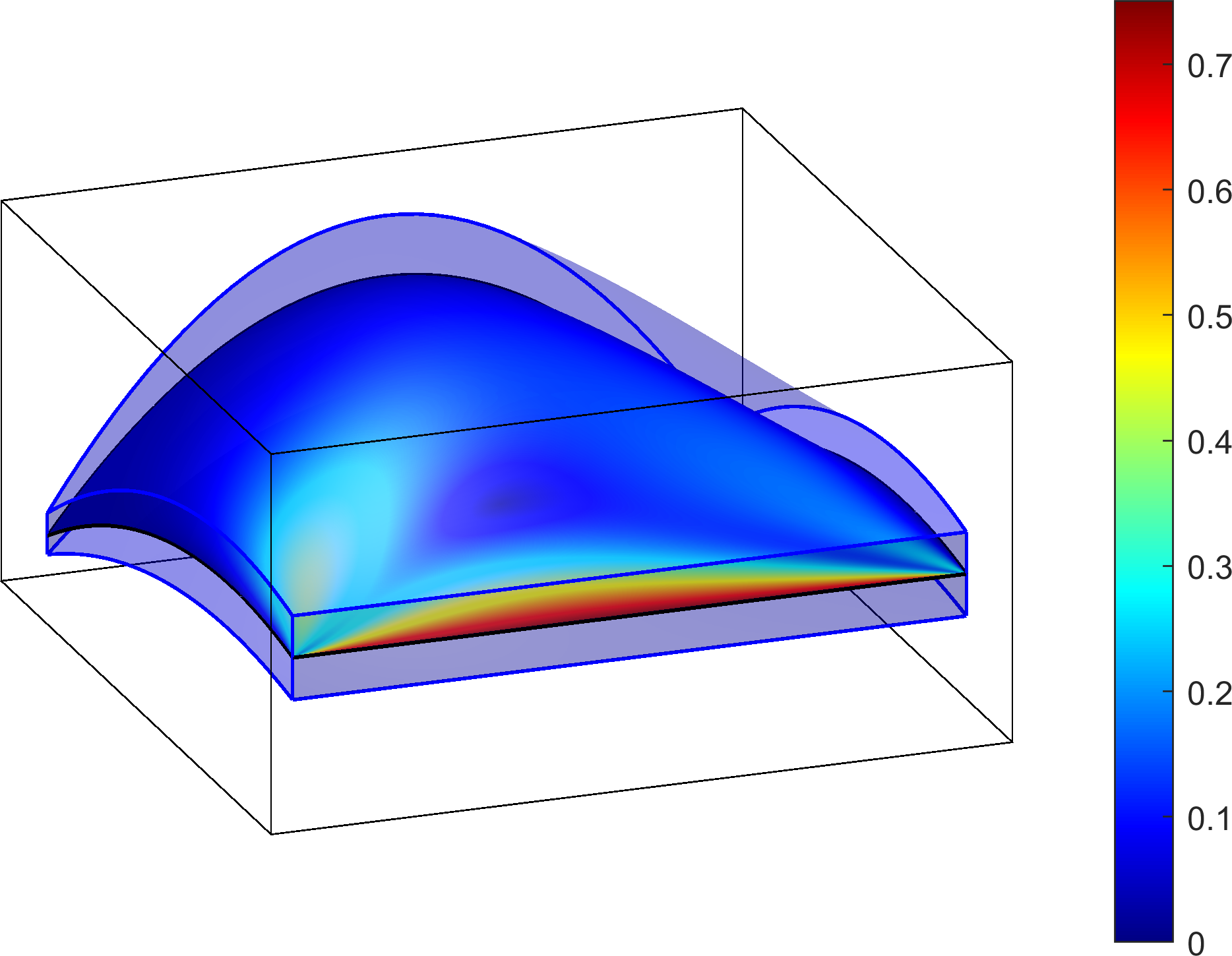}}\qquad
	\subfigure[Surface FEM -$\lVert{\vek{u}}\rVert$]{\includegraphics[width=0.45\textwidth]{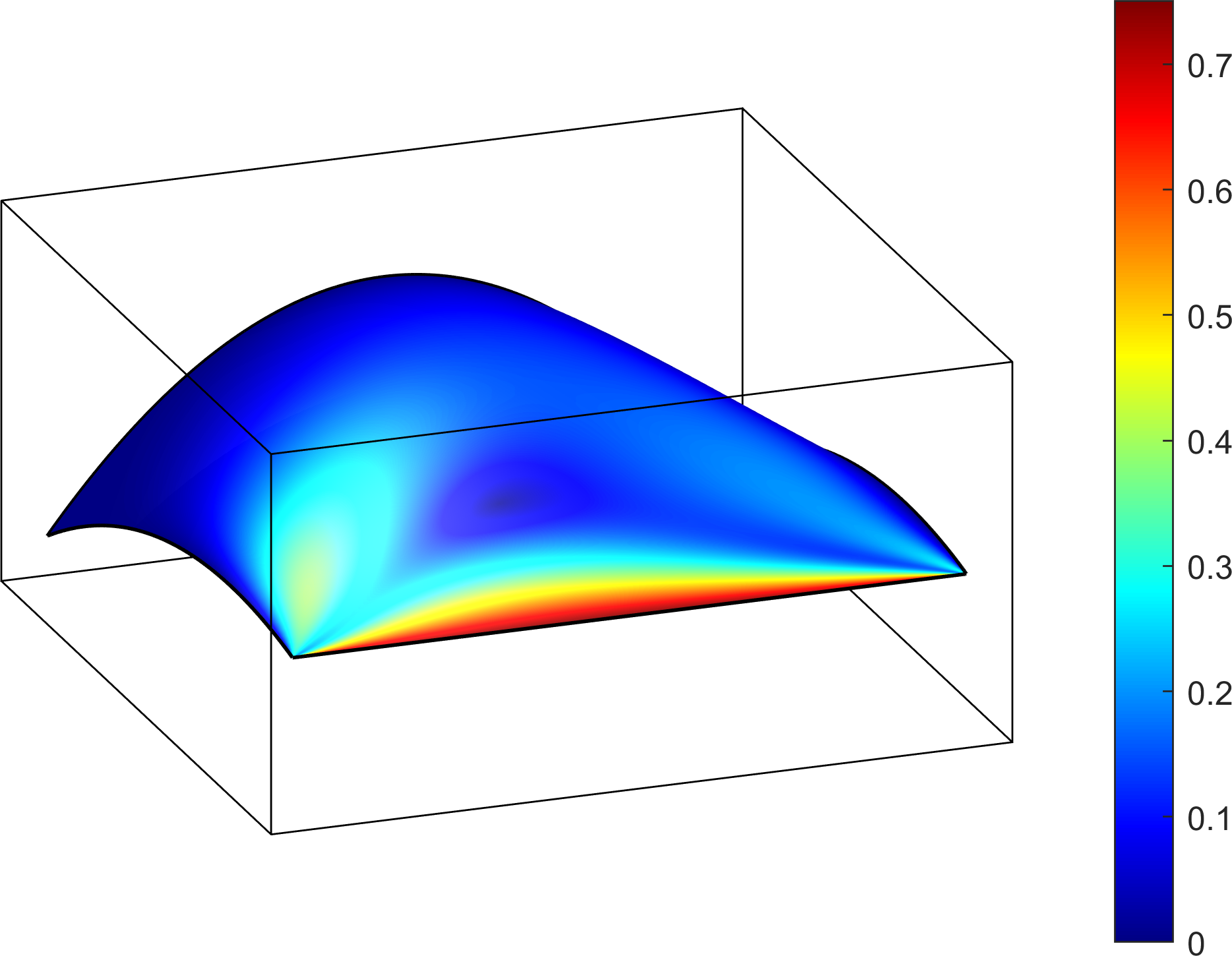}}
	
	\caption{\label{fig:DrvCavRes} Results for the simultaneous solution of the driven cavity test case. The velocity magnitudes obtained with the Bulk Trace FEM on selected level sets are shown in (a) and (c), while in (d), a Surface FEM solution is shown on one level set. Profiles of the velocity components $u$ along the vertical centre axis and $v$ along the horizontal centre axis are shown for three different surfaces (LSF) in (b).}
\end{figure}

This test case is inspired by the bench mark from Ghia et al. \cite{Ghia_1982a} and its application to stationary Navier--Stokes flow on one curved surface in \cite{Fries_2018a}. Herein, a flat bulk domain with embedded flat surfaces is mapped into a curved geometry. In the flat bulk domain, each point is defined by the coordinates $\left(a,b,c\right)$, i.e., $\mathcal{P}_{\mathrm{flat}} \left(\vek{a}\right) = \left[a,b,c\right]^{\mathrm{T}}$. These points are mapped with $\varphi$ into the Euclidean coordinate system $\left(x,y,z\right)$, i.e., $\mathcal{P}_{\mathrm{curved}} \left(\vek{x}\right)  = \left[x,y,z\right]^{\mathrm{T}} = \varphi\left(\mathcal{P}_{\mathrm{flat}}\right)$ with
\begin{equation*}
	\varphi = \vek{x}\left(\vek{a}\right) = \begin{Bmatrix} 
		x &=& a & \\ 
		y &=& b &\\ 
		z &=& \alpha & \cdot \left(0.1+c\right)^{1/2} \cdot \left(-1+8a + 2b - 8a^2\right) \cdot \left(1-b\right) + \sin(c) 
	\end{Bmatrix}.
\end{equation*}
The level-set function is defined as $\phi\left(\vek{a}\right)=c$. Fig.~\ref{fig:DrvCavMap} shows the applied mapping for this test case with $\alpha = 0.4$ for the bulk domain $\Omega$ with some embedded level sets $\Gamma_{\!c}$.
\\
\\
The shear viscosity is defined as $\mu = 0.01$. The Dirichlet boundary conditions along the straight part of the Dirichlet boundary $\partial \Omega_{\mathrm{D},\vek{u},\mathrm{s}} \in \left[x_{\mathrm{D}},y_{\mathrm{D}}=1,z_{\mathrm{D}}\right]$ are $u(z) = 1-4\cdot z_{\mathrm{D}}$ and $v=w=0$ to get more variability in the results for each embedded surface. The height along the straight Dirichlet boundary is defined as $z_{\mathrm{D}} \in [0,0.125]$ and $x_{\mathrm{D}} \in [0,1]$ is considered. On the other nodes of the Dirichlet boundary, i.e., $\partial \Omega_{\mathrm{D},\vek{u}} \backslash \partial \Omega_{\mathrm{D},\vek{u},\mathrm{s}}$, no-slip boundary conditions, i.e., $u=v=w=0$, are prescribed. At the nodes on $\partial \Omega_{\mathrm{N},p} = \left[0.5,0,z\right]^{\mathrm{T}}$, the pressure $p=0$ is prescribed. For the computations shown herein, a mesh of $40 \times 40 \times 2$ elements is used and shown together with the boundary conditions in Fig.~\ref{fig:DrvCavMeshBCs}.
\\
\\
The results of this test case are given in Fig.~\ref{fig:DrvCavRes}: (a) shows the velocity magnitudes on some selected level sets. Fig.~\ref{fig:DrvCavRes}(b) shows velocity profiles in analogy to \cite{Ghia_1982a} and in particular to Fig.~10 in \cite{Fries_2018a}. Along the horizontal centre line, the velocity profile for the velocity component $v$ and along the vertical centre line, the velocity profile for the velocity component $u$ are shown. The lines with the circular markers show the results for a selected surface for which the solution is obtained by a Surface FEM computation using Taylor--Hood elements. The triangular markers indicate the profiles for the Bulk Trace FEM solution obtained with equal-order elements for the velocities and pressure with order $q_{p}$ = $q_{\vek{u}} = 2$ and PSPG stabilization. The stabilization parameter is defined in analogy to SUPG/PSPG stabilization for Euclidean geometries \cite{Tezduyar_2003a} as
\begin{equation}
	\tau_{\mathrm{PSPG}} = \left[\left(\frac{2 \lVert \vek{u}_{\mathrm{el,node}}\rVert}{h_\mathrm{el}}\right)^2 + \left(\frac{4\mu}{h_\mathrm{el}^2}\right)^2\right]^{-1/2}. \label{eq:PSPGstabParamStat}
\end{equation}
Note that the norm of the velocities $\vek{u}_{\mathrm{el,node}}$ is evaluated within each element at each node instead of using an averaged velocity for the entire element because the velocities might be different on every surface which intersects with the element. The curves in Fig.~\ref{fig:DrvCavRes}(b) belong to the following level sets $\Gamma_{\!c}$: LSF 1 in the legend refers to $\Gamma_{\!c=0.0}$, LSF 2 in the legend refers to $\Gamma_{\!c=0.0625}$, and LSF 3 in the legend refers to $\Gamma_{\!c=0.125}$ where the index $c$ refers in this context to the height $c$ in the flat domain which is identical to the height $z$ of the straight Dirichlet boundary $\partial \Omega_{\mathrm{D},\vek{u},\mathrm{s}}$ of the mapped bulk domain. The curves show that the Surface FEM and Bulk Trace FEM solutions are almost identical on the selected level sets. Fig.~\ref{fig:DrvCavRes}(c) shows the velocity magnitudes on the level set $\Gamma_{\!c=0.0625}$ and Fig.~\ref{fig:DrvCavRes}(d) shows the same quantity obtained for this surface in a Surface FEM computation (with Taylor--Hood elements, no stabilization). Comparing these two figures, also verifies that visually the same solution is obtained by both numerical schemes and, therefore, that the Bulk Trace FEM leads to correct solutions for each embedded surface.

\section{Instationary Navier--Stokes flow}\label{sec:InstatNSEQ}

\subsection{Strong form for one level set}\label{subsec:InstNSEQ-sfEq1LS}
Instationary Navier--Stokes flow is characterized by the time-dependency of the physical fields, i.e., the velocity $\vek{u} = \vek{u}\left(\vek{x},t\right)$ and the pressure $p = p\left(\vek{x},t\right)$ are functions of space \emph{and} time $t$. Note that the domains of interest, i.e., the curved surfaces do not depend on time, i.e., they are fixed. The momentum equation for instationary Navier--Stokes flow is given as \cite{Fries_2018a,Jankuhn_2018a}
\begin{equation}
	\varrho\cdot\left(\dot{\vek{u}}_\ti\left(\vek{x},t\right)+\left(\vek{u}_\ti\cdot\nabla_{\Gamma}^{\mathrm{cov}}\right)\vek{u}_\ti-\vek{g}_\ti\left(\vek{x},t\right)\right)-\mat{P}\cdot\mathrm{div}_{\Gamma}\,\vek{\sigma}_\ti\left(\vek{x},t\right)=\vek0.\label{eq:MomentumEqtNSinstat}
\end{equation}
We consider the time interval $\tau = \left[0,T\right]$ and the space-time domain $\Gamma_{\!c} \times \tau$ in which Eqs.~(\ref{eq:ContinuityConstraint}) and (\ref{eq:MomentumEqtNSinstat}) are solved. Derivatives w.r.t.~time are denoted as $\dot{\square} = \nicefrac{\partial \square}{\partial t}$. Note that time is indicated by $t$ whereas $\square_\ti$ refers to some tangential/in-plane vector/tensor quantities.\\
\\
Additionally, the boundary conditions, i.e., Eqs.~(\ref{eq:DirBoundaryConditions}) and (\ref{eq:NeumBoundaryConditions}), are extended in time. There are prescribed velocities $\hat{\vek{u}}_\ti\left(\vek{x},t\right)$ along $\partial \Gamma_{\!c,\mathrm{D}} \times \tau$ and tractions $\hat{\vek{t}}_\ti\left(\vek{x},t\right)$ along $\partial \Gamma_{\!c,\mathrm{N}} \times \tau$ over time \cite{Fries_2018a,Fries_2024a}. Furthermore, the initial condition is defined as
\begin{equation}
	\vek{u}_\ti\left(\vek{x},t\right) = \vek{u}^0_\ti\left(\vek{x}\right), \,\text{with}\, \mathrm{div}_{\Gamma}\vek{u}^0_\ti = 0\,\,\text{and}\,\,\vek{u}^0_\ti \cdot \vek{n} = 0 \quad \forall \vek{x} \in \Gamma_{\!c} \,\text{at}\, t = 0. \label{eq:InitialCond}
\end{equation}

\subsection{Weak form for one level set}\label{subsec:InstNSEQ-wfEq1LS}

Analogously to Sec.~\ref{subsec:StStok-wfEq1LS} and \ref{subsec:stNSEQ-wfEq1LS}, the continuous weak form of the instationary Navier--Stokes problem is stated as follows \cite{Fries_2018a}: Given the fluid density $\varrho\in\mathbb{R}^{+}$, viscosity $\mu\in\mathbb{R}^{+}$, body force $\varrho\cdot\vek{g}_\ti\left(\vek{x},t\right)$ in $\Gamma_{\!c}\times\tau$, traction $\hat{\vek{t}}_\ti\left(\vek{x},t\right)$ on $\partial\Gamma_{\!c,\mathrm{N}}\times\tau$, and initial condition $u_{0}\left(\vek{x}\right)$ on $\Gamma_{\!c}$ at $t=0$ according to Eq.~(\ref{eq:InitialCond}), find the velocity field
$\vek{u}_\ti\left(\vek{x},t\right)\in \mathcal{S}^{\Gamma}_{\vek{u}} \times \tau$ and
pressure field $p\left(\vek{x},t\right)\in \mathcal{S}^{\Gamma}_{p} \times \tau$
such that for all test functions $\left(\vek{w}_{\vek{u},\ti},w_{p}\right)\in\mathcal{V}^{\Gamma}_{\vek{u}}\times\mathcal{V}^{\Gamma}_{p}$,
there holds in $\Gamma_{\!c}\times\tau$ 
\begin{align}
	\begin{split}
		\varrho\,\,\cdot\int_{\Gamma_{\!c}}\vek{w}_{\vek{u},\ti}\cdot\left(\dot{\vek{u}}_\ti+\left(\vek{u}_\ti\cdot\nabla_{\Gamma}^{\mathrm{cov}}\right)\vek{u}_\ti-\vek{g}_\ti\right)\mathrm{d}\Gamma \, &+ \int_{\Gamma_{\!c}}\nabla_{\Gamma}\vek{w}_{\vek{u},\ti}^{\mathrm{dir}}:\vek{\sigma}\left(\vek{u}_\ti,p\right)\mathrm{d}\Gamma\\ + \,\alpha\, \cdot \int_{\Gamma_{\!c}} \left(\vek{u} \cdot \vek{n} \right) \cdot \left(\vek{w}_{\boldsymbol{u}} \cdot \vek{n} \right) \mathrm{d}\Gamma &=  \int_{\partial\Gamma_{\!c,\mathrm{N}}}\!\!\!\vek{w}_{\vek{u},\ti}\cdot\hat{\vek{t}}_\ti\,\mathrm{d}\partial\Gamma,\label{eq:InstNSEQWeakFormMomentum1srf}
	\end{split} \\
	\int_{\Gamma_{\!c}}w_{p}\cdot\mathrm{div}_{\Gamma}\,\vek{u}_\ti\:\mathrm{d}\Gamma &=  0.\label{eq:InstNSEQWeakFormContinuity1srf}
\end{align}

\subsection{Weak form for all level sets in a bulk domain}\label{subsec:InstNSEQ-wfEqAllLS}
Integration over the level-set interval and application of the co-area formula for the domain and the boundary, respectively, leads to the weak form of the instationary Navier--Stokes equations for all level sets embedded in a bulk domain. The continuous weak form is omitted for brevity and we directly give the discrete weak form. Note that the space-time domain for \emph{one} considered surface was introduced as $\Gamma_{\!c} \times \tau$. For the case where \emph{all} surfaces $\Gamma_{\!c}$ over some bulk domain are considered, the space-time domain is defined as
\begin{equation}
	\Omega \times \tau = \bigg(\bigcup_{c \in \Phi} \Gamma_{\!c}\bigg) \times \tau
\end{equation}
with $\Phi = \left[\phi_{\min},\phi_{\max}\right]$. With that, the discrete weak form reads: Given the fluid density $\varrho\in\mathbb{R}^{+}$, viscosity $\mu\in\mathbb{R}^{+}$, body force $\varrho\cdot\vek{g}_\ti\left(\vek{x},t\right)$ in $\Omega\times\tau$, traction $\hat{\vek{t}}_\ti\left(\vek{x},t\right)$ on $\partial\Omega_{\mathrm{N}}\times\tau$, and initial condition $u_{0}\left(\vek{x}\right)$ at $t=0$ according to (\ref{eq:InitialCond}), find the velocity field
$\vek{u}_\ti\left(\vek{x},t\right)\in \mathcal{S}^{\Omega}_{\vek{u}} \times \tau$ and
pressure field $p\left(\vek{x},t\right)\in \mathcal{S}^{\Omega}_{p} \times \tau$
such that for all test functions $\left(\vek{w}_{\vek{u},\ti},w_{p}\right)\in\mathcal{V}^{\Omega}_{\vek{u}}\times\mathcal{V}^{\Omega}_{p}$,
there holds in $\Omega\times\tau$ 
\begin{align}
	\begin{split}
		\varrho\,\cdot\int_{\Omega^h}\vek{w}^h_{\vek{u},\ti}\cdot\left(\dot{\vek{u}}^h_\ti 	+\left(\vek{u}^h_\ti\cdot\nabla_{\Gamma}^{\mathrm{cov}}\right)\vek{u}^h_\ti-\vek{g}^h_\ti\right)\cdot \lVert \nabla \phi \rVert\;&\mathrm{d}\Omega \,+ \int_{\Omega^h}\big(\nabla^{\mathrm{dir}}_{\Gamma}\vek{w}^h_{\vek{u},\ti}:\vek{\sigma}_\ti\left(\vek{u}^h_\ti,p^h\right)\big)\cdot \lVert \nabla \phi \rVert\;\mathrm{d}\Omega \\ + \, \alpha \,\cdot \int_{\Omega^h} \left(\vek{u}^h \cdot \vek{n} \right) \cdot \left(\vek{w}^h_{\boldsymbol{u}} \cdot \vek{n} \right) \cdot \lVert \nabla \phi \rVert\;&\mathrm{d}\Omega =  \int_{\partial\Omega_{\mathrm{N}}^h}\!\!\!\vek{w}^h_{\vek{u},\ti}\cdot\hat{\vek{t}}^{h}_\ti \cdot \left(\vek{q} \cdot \vek{m}\right)\cdot \lVert \nabla \phi \rVert\;\mathrm{d}\partial\Omega,\label{eq:InstNSEQdiscWFmom}
	\end{split} \\
	\int_{\Omega^h}w^h_{p}\cdot\mathrm{div}_{\Gamma}\,\vek{u}^h_\ti\:\cdot \lVert \nabla \phi \rVert\; & \mathrm{d}\Omega =  0. \label{eq:InstNSEQdiscWFcont}
\end{align}
Note that we again omit the superscript $h$ at geometric quantities and differential operators for brevity. This discrete weak form leads to a system of non-linear semidiscrete equations in time $t \in \tau$
\begin{align}
	\mat{T}\cdot\underline{\dot{\vek{u}}}_\ti\left(t\right) + \left(\mat{D} + \mat{A}\left(\underline{\vek{u}}_\ti\right)\right) \cdot\underline{\vek{u}}_\ti\left(t\right) + \mat{C}^{\mathrm{T}} \cdot \vek{p}\left(t\right)+\mat{G} \cdot \underline{\vek{u}} &= \vek{f}\left(t\right), \\
	\mat{C}\cdot \underline{\vek{u}}_\ti\left(t\right) &= \vek{0},
\end{align}
with an initial condition $\underline{\vek{u}}_\ti\left(t=0\right)$. $\mat{T}$ is a mass matrix representing the time dependency, $\mat{D}$ contains the diffusion part, $\mat{A}$ is the advection matrix, $\mat{C}$ comes from the continuity equation, and $\mat{G}$ represents the penalty term. The vectors $\underline{\vek{u}}_\ti = [\tilde{\vek{u}}_\ti,\tilde{\vek{v}}_\ti,\tilde{\vek{w}}_\ti]^{\mathrm{T}}$ and $\vek{p}$ contain the sought values for the velocities and the pressure, respectively. Note that $\tilde{\vek{u}}_\ti = [u_{\ti,1},u_{\ti,2},\ldots,u_{\ti,n_q}]^{\mathrm{T}}$ is the vector of the nodal values of the velocity component $u_{\ti}$. This system of equations is advanced in time by the Crank--Nicolson method.
 
\subsection{Numerical results for instationary Navier--Stokes flow}\label{subsec:InstNSEQ-NumRes}
\subsubsection{Flow on a torus}
\begin{figure}
	\centering
	
	\subfigure[]{\includegraphics[width=0.28\textwidth]{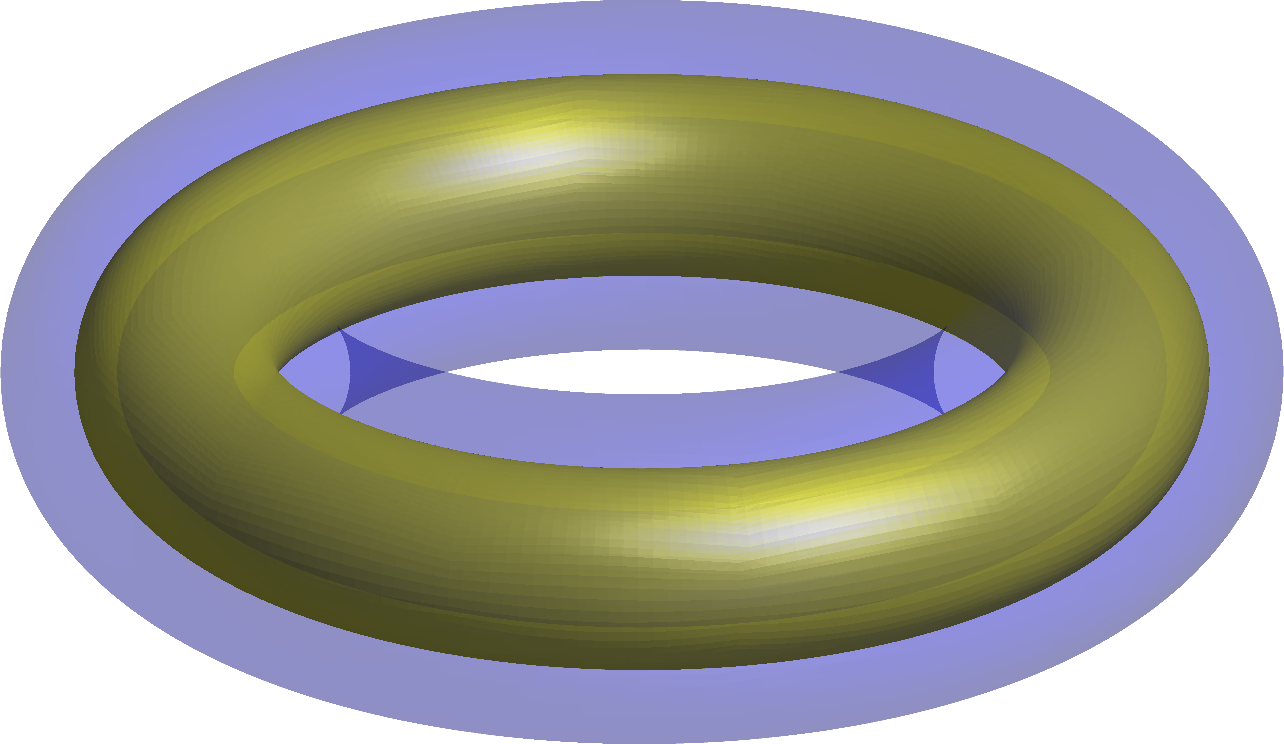}}
	\hspace{0.5cm}
	\subfigure[]{\includegraphics[width=0.28\textwidth]{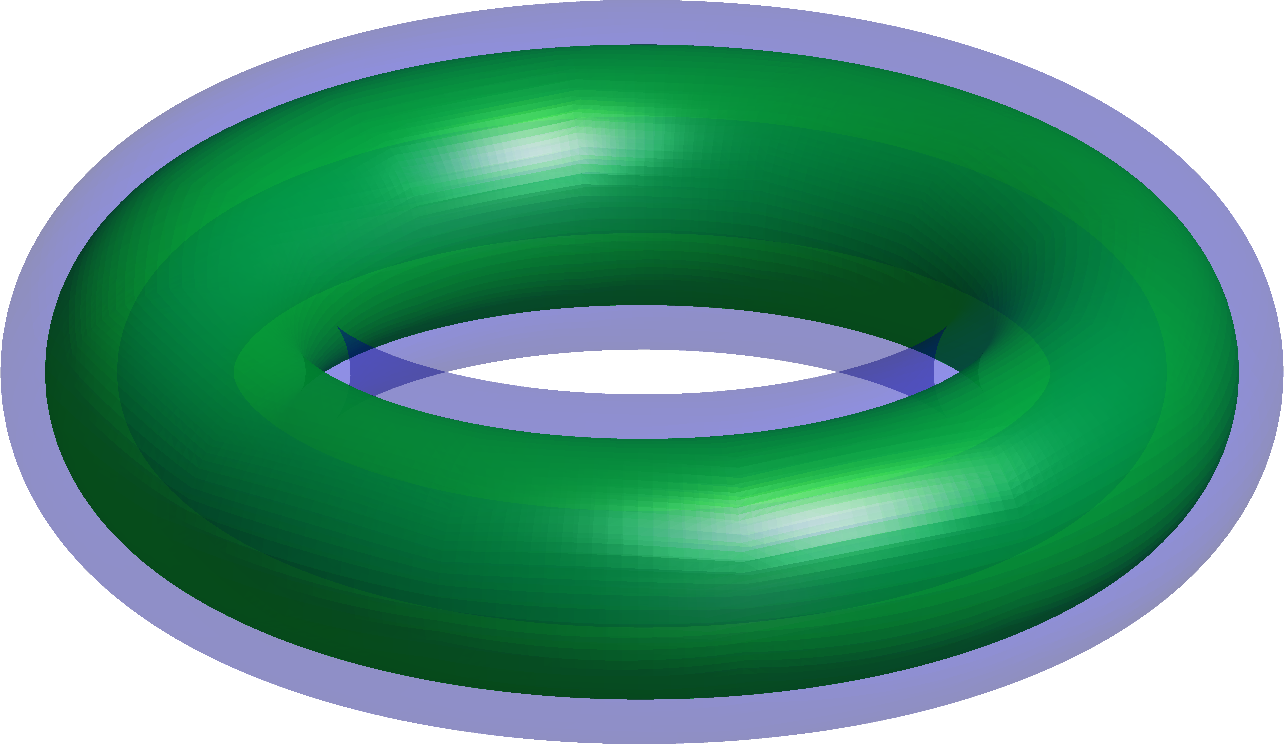}}
	\hspace{0.5cm}
	\subfigure[]{\includegraphics[width=0.28\textwidth]{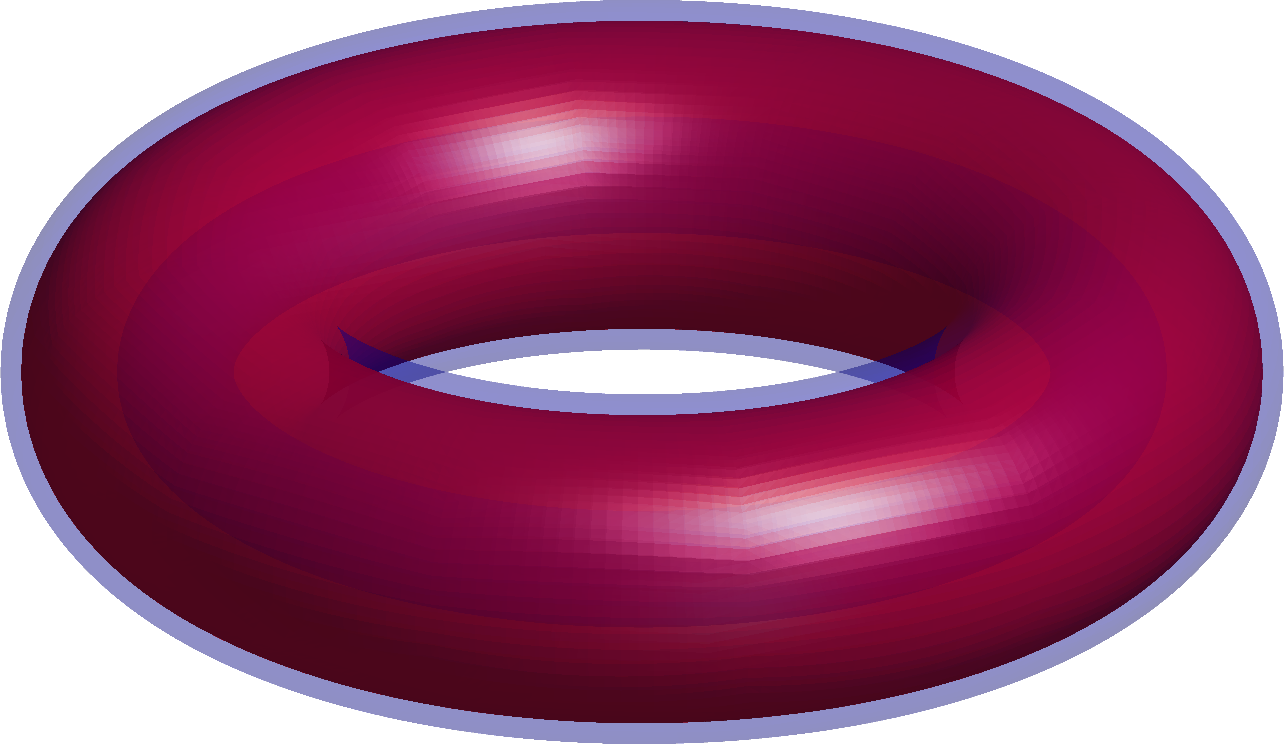}}
	
	\caption{\label{fig:Torus-setup} The bulk domain in light blue and three different torus surfaces in (a) yellow, (b) green, and (c) red.}
\end{figure}

\begin{figure}
	\centering
	
	\subfigure[$\Gamma_{\!c}\left(r=0.25\right),\,t=0$]{\includegraphics[width=0.35\textwidth]{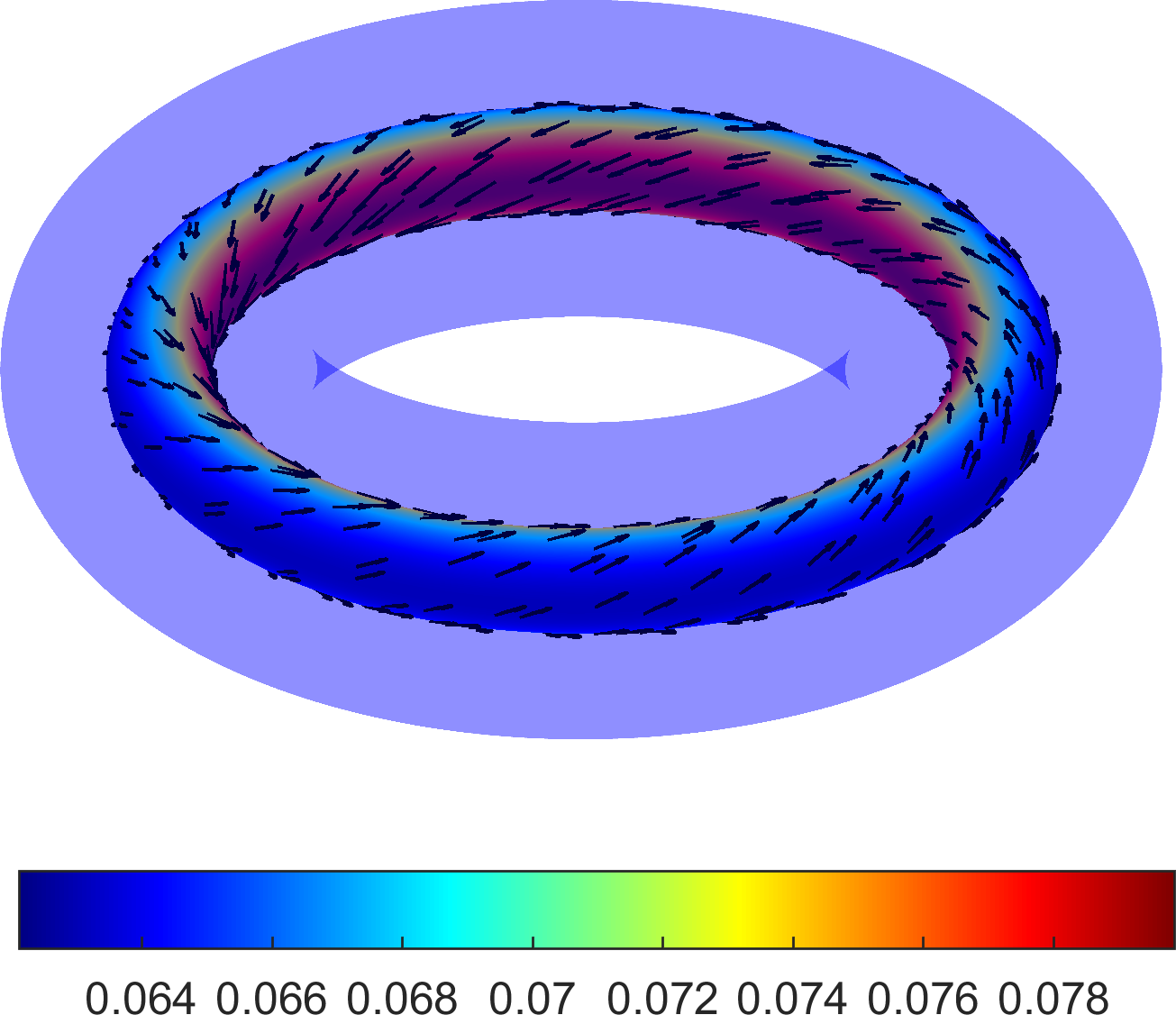}}\qquad
	\subfigure[$\Gamma_{\!c}\left(r=0.25\right),\,t=60$]{\includegraphics[width=0.35\textwidth]{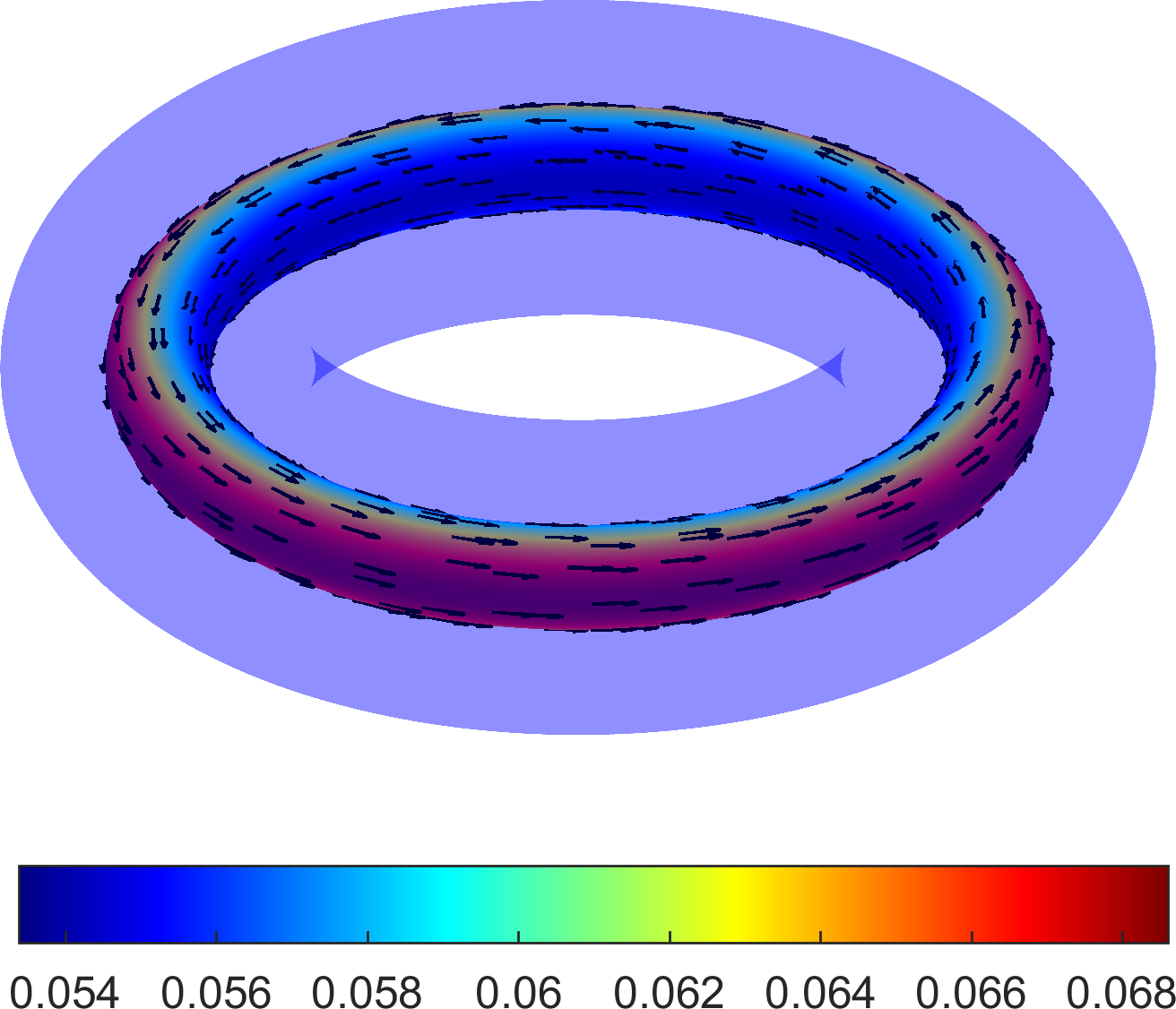}}\qquad
	\subfigure[$\Gamma_{\!c}\left(r=0.50\right),\,t=0$]{\includegraphics[width=0.35\textwidth]{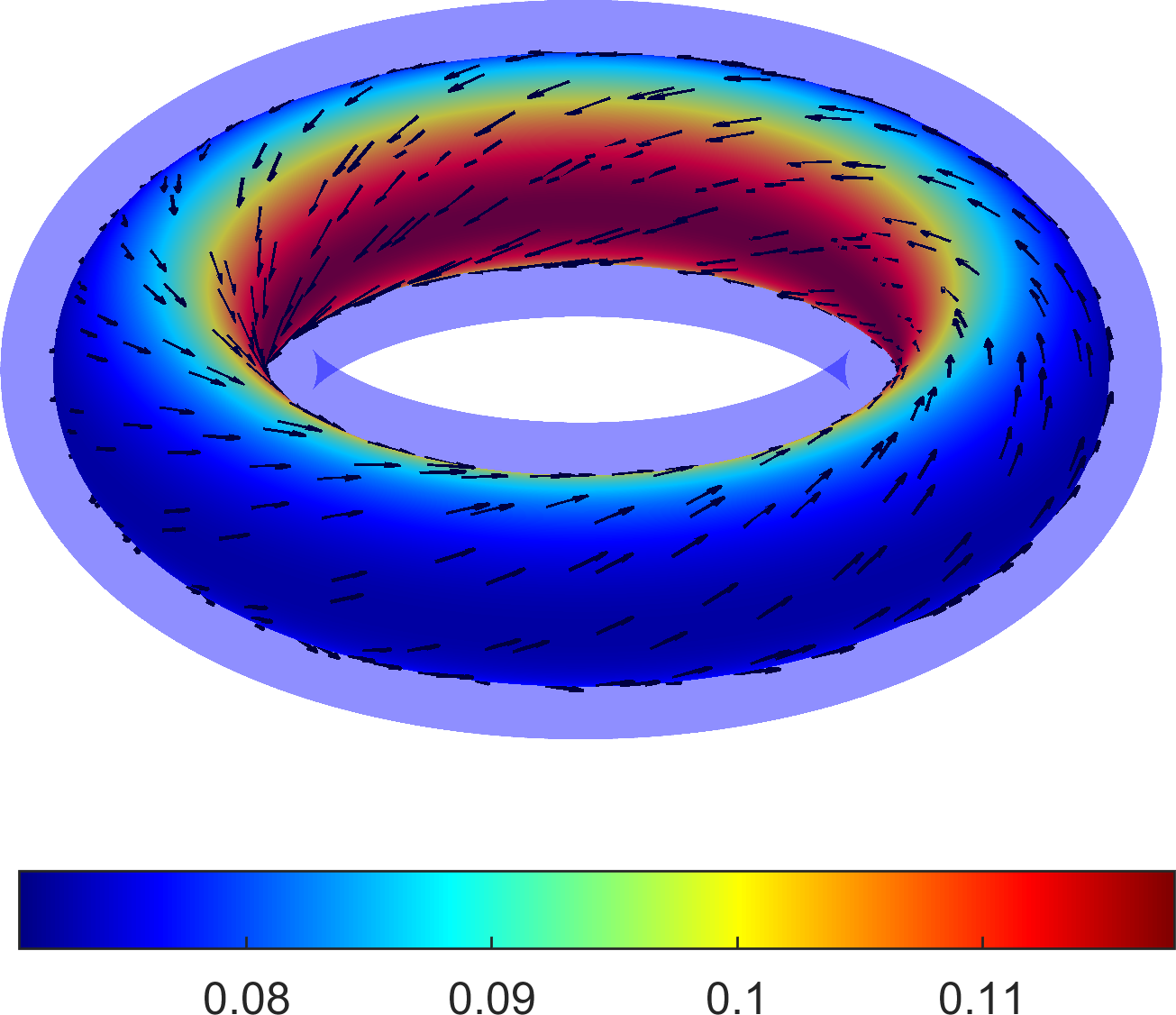}}\qquad
	\subfigure[$\Gamma_{\!c}\left(r=0.50\right),\,t=60$]{\includegraphics[width=0.35\textwidth]{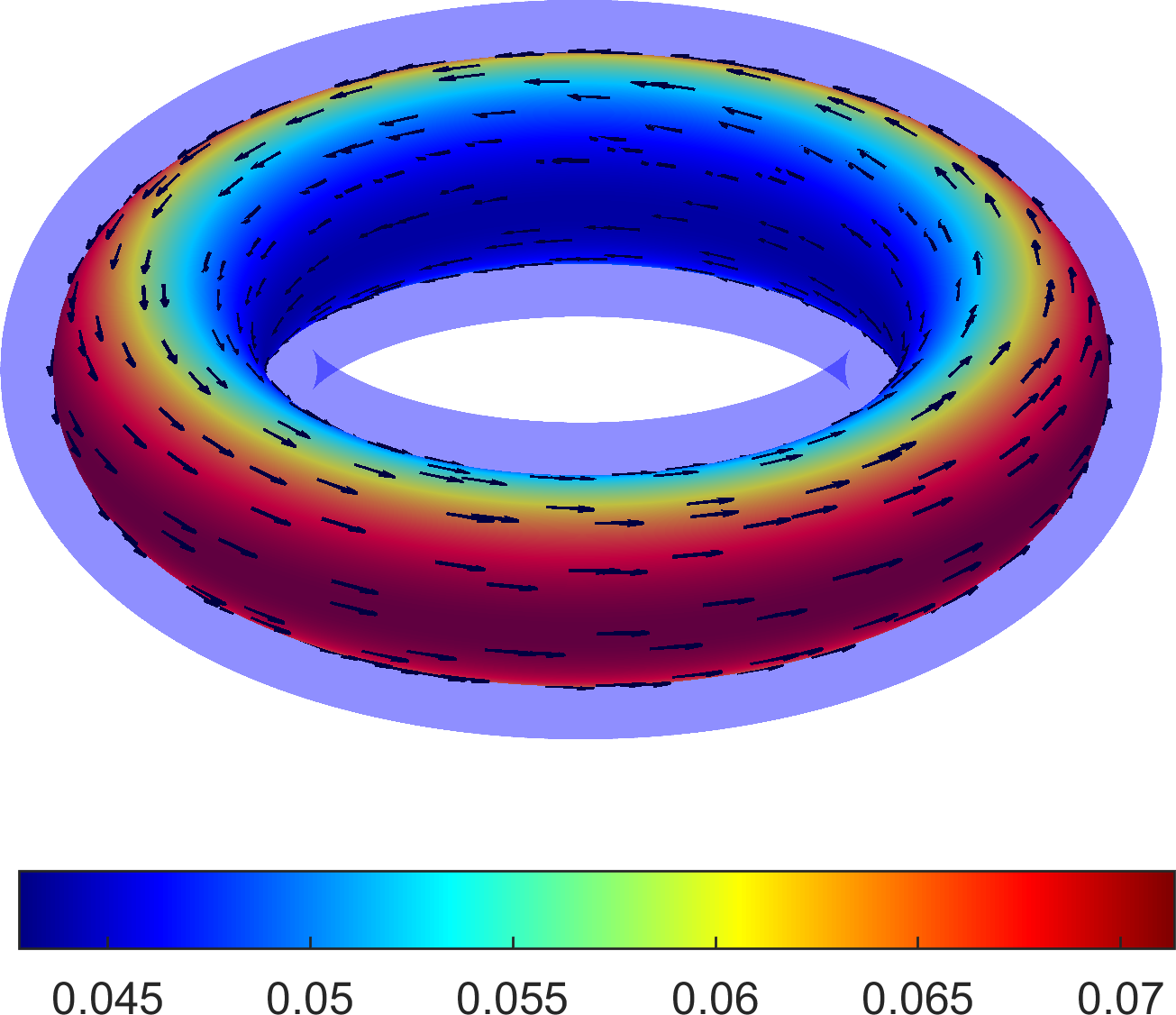}}\qquad
	\subfigure[$\Gamma_{\!c}\left(r=0.75\right),\,t=0$]{\includegraphics[width=0.35\textwidth]{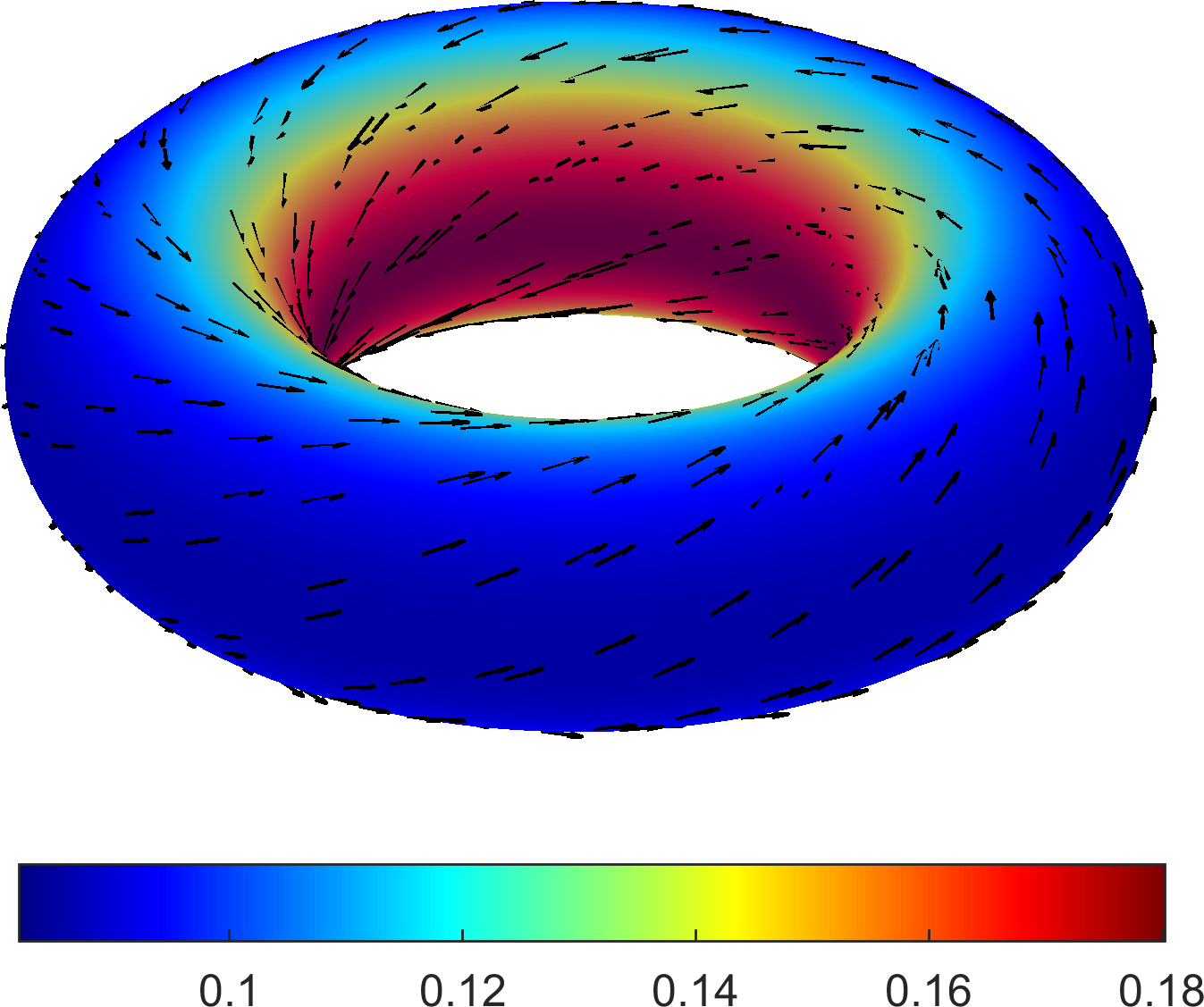}}\qquad
	\subfigure[$\Gamma_{\!c}\left(r=0.75\right),\,t=60$]{\includegraphics[width=0.35\textwidth]{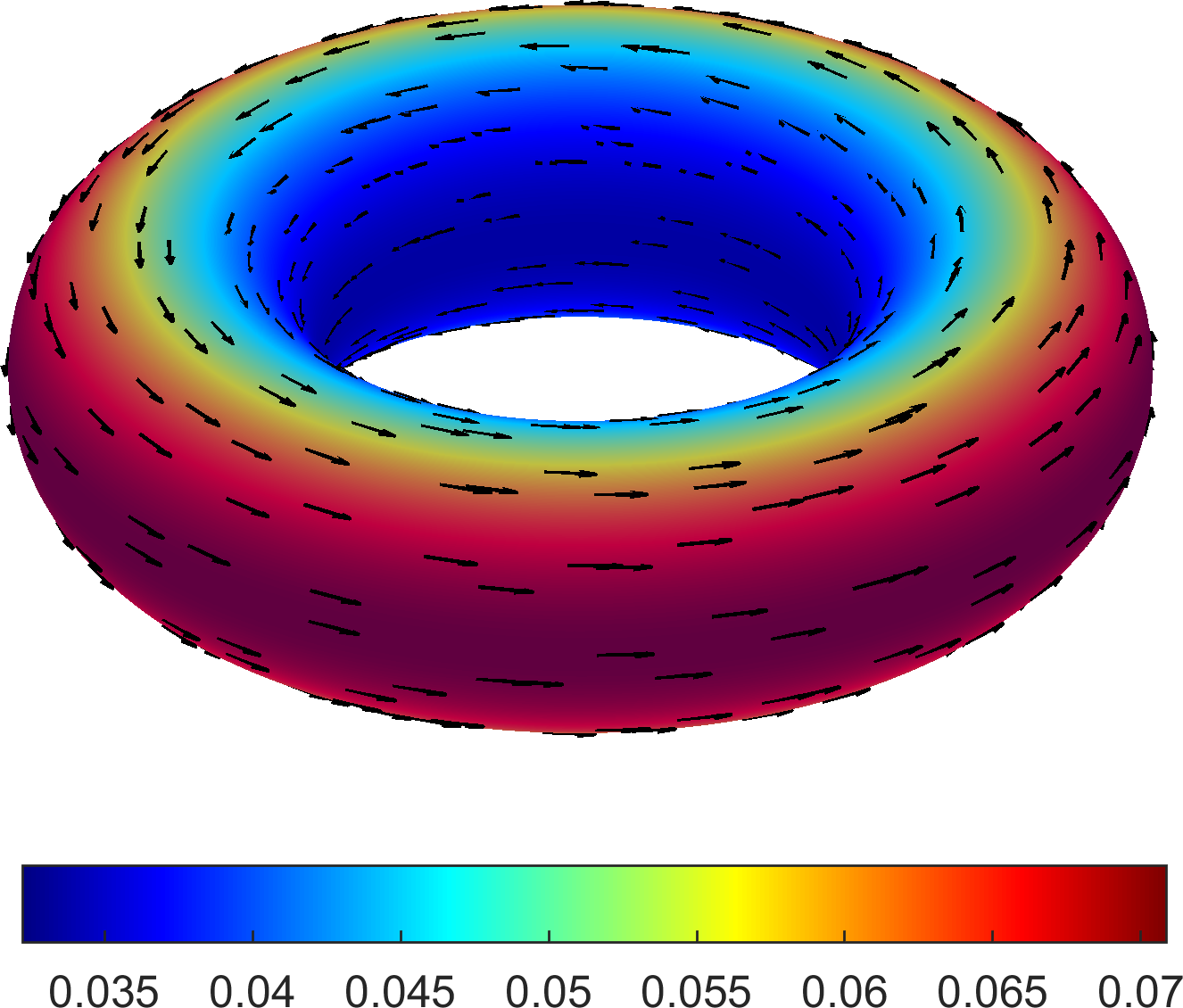}}\qquad
	
	\caption{\label{fig:TorusRes} Results for the simultaneous solution of flow on a torus with major radius $R=2$ and minor radius (a) and (b) $r = 0.25$, (c) and (d) $r = 0.5$ , and (e) and (f) $r = 0.75$ at different times. The light blue surface is the boundary surface of the bulk domain $\Omega$.}
\end{figure}

\begin{figure}
	\centering
	
	\includegraphics[width=0.75\textwidth]{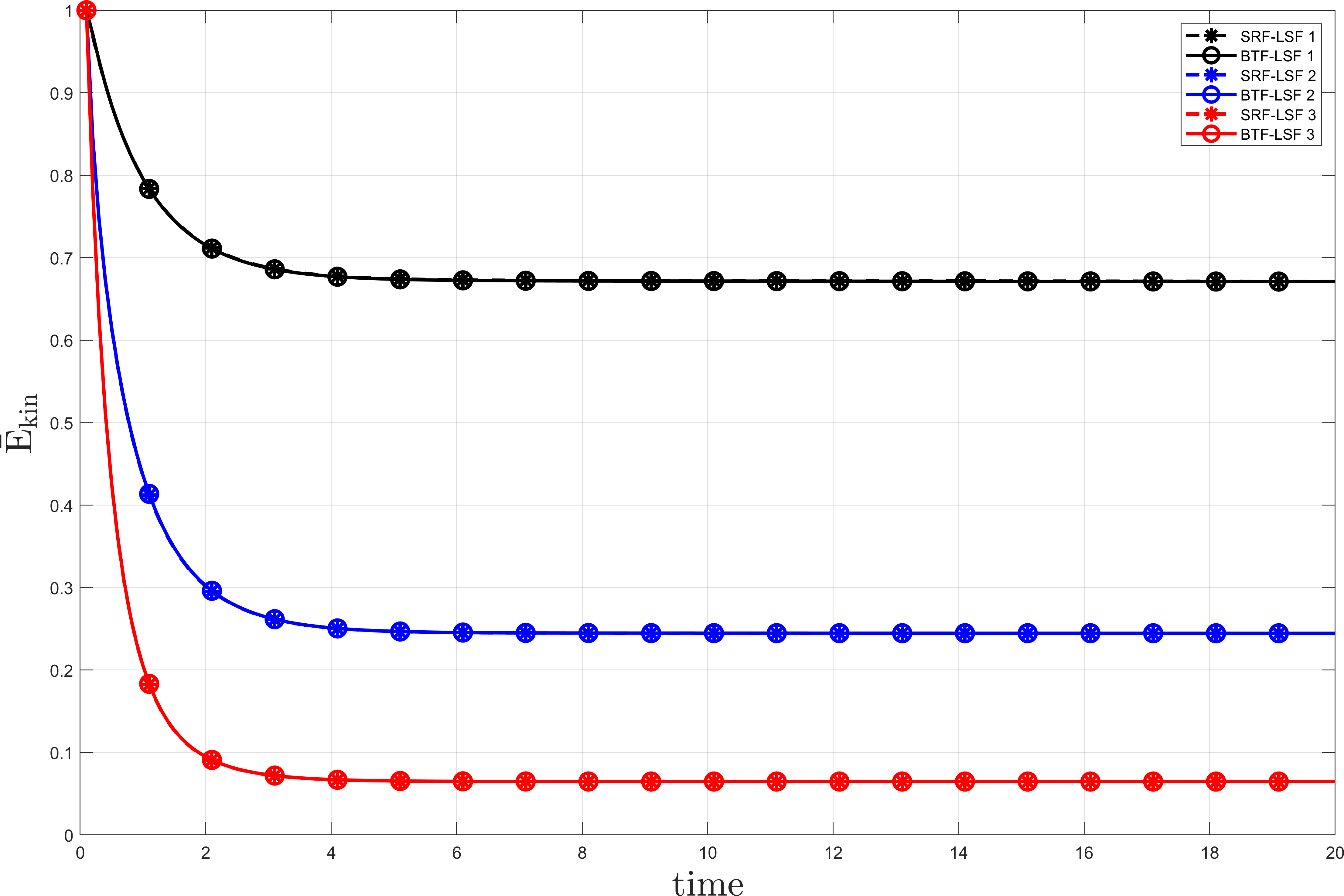}
	
	\caption{\label{fig:KinEngTorus} The normalized kinetic energy over the time for three selected torus surfaces (LSF 1 to 3), once from the simultaneous solution with the Bulk Trace FEM (BTF) and once obtained for each surface independently with the Surface FEM (SRF).}
\end{figure}

This test case is based on similar examples for a single surface which are presented in several publications to verify the approximation of the solution of surface Navier--Stokes equations, e.g., in \cite{Rank_2021a,Reuther_2018a} with finite elements, in \cite{Nitschke_2017a} with discrete exterior calculus, and in \cite{Yang_2020a} with finite differences. The considered surfaces $\Gamma_{\!c}$ are tori with a major radius $R = 2$ and minor radius $r \in \left[0.25,0.75\right]$. The tori are described by the level sets of the function $\phi = \left(\sqrt{x^2+y^2}-R\right)^2+z^2-r^2$. The bulk domain $\Omega$ is the toroidal ring bounded by  $\Gamma_{\!c}\left(r = 0.25\right)$ and $\Gamma_{\!c}\left(r = 0.75\right)$. Fig.~\ref{fig:Torus-setup} shows the bulk domain in light blue and three different embedded tori $\Gamma_{\!c}$.
On these compact manifolds, a flow takes place which is initiated by initial conditions of the velocity $\vek{u}_\ti = \left[u,v,w\right]^{\mathrm{T}}$ tangential to the toroidal surfaces and defined on the nodes with coordinates $\left[x,y,z\right]^{\mathrm{T}}$ of the bulk domain as
\begin{equation}
	\vek{u}_\ti = \begin{bmatrix}
		u \\ v \\ w
	\end{bmatrix} = \begin{bmatrix}
	-\frac{y + 2 x z}{8 \left(r_{xy}\right)} \\ 
	\frac{x - 2 y z}{8 \left(r_{xy}\right)}  \\ 
	\frac{\sqrt{r_{xy}}-2}{4 \sqrt{r_{xy}}}
	\end{bmatrix},
\end{equation}
with $r_{xy} = x^2+y^2$. This definition of the initial conditions is similar to \cite{Yang_2020a} where single surfaces are considered. The fluid's density is $\varrho = 1$ and the viscosity is set to $\eta = 1$. The time step size in the Crank-Nicoloson time stepping scheme is $\Delta t = 0.1$ for $\tau = [0,60]$. Taylor--Hood elements of order $q_{\vek{u}} = 2$, and $q_{p} = 1$ are used. For the geometry, we use again $q_{\mathrm{geom}} = 3$. Fig.~\ref{fig:TorusRes} shows the velocity on selected level sets at different times.
As can be seen in Figs.~\ref{fig:TorusRes} and \ref{fig:KinEngTorus}, the flow becomes stationary and the kinetic energy of each surface defined as $\mathrm{E}_{\mathrm{kin}} = \int_{\Gamma_{\!c}} \frac{1}{2} \varrho \lvert \vek{u} \rvert^2 \,\d\Gamma$ decreases to a constant value. Fig.~\ref{fig:KinEngTorus} shows the kinetic energy normalized by its value at the beginning $\bar{\mathrm{E}}_{\mathrm{kin}} = \mathrm{E}_{\mathrm{kin}} / \mathrm{E}_{\mathrm{kin}}(t=0)$ over the time for the three surfaces which are also shown in Fig.~\ref{fig:TorusRes}. This shows that the simultaneous solution with the Bulk Trace FEM leads to the same result when compared to solving each surface independently by the Surface FEM.

\subsubsection{Flow around an obstacle}
\begin{figure}
	\centering
	
	\subfigure[velocity]{\includegraphics[width=0.25\textwidth]{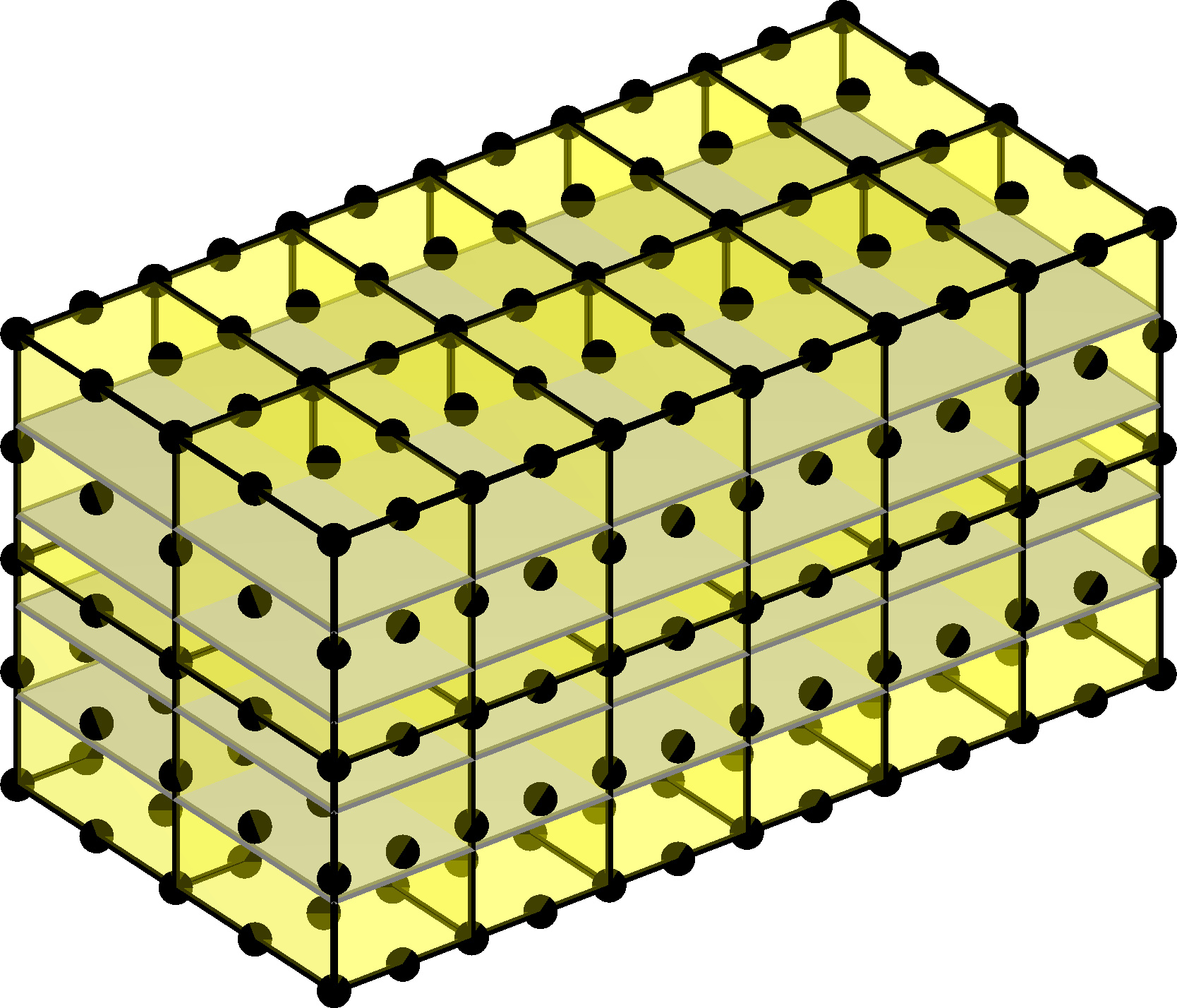}}
	\hspace{1cm}
	\subfigure[pressure - isotropic]{\includegraphics[width=0.25\textwidth]{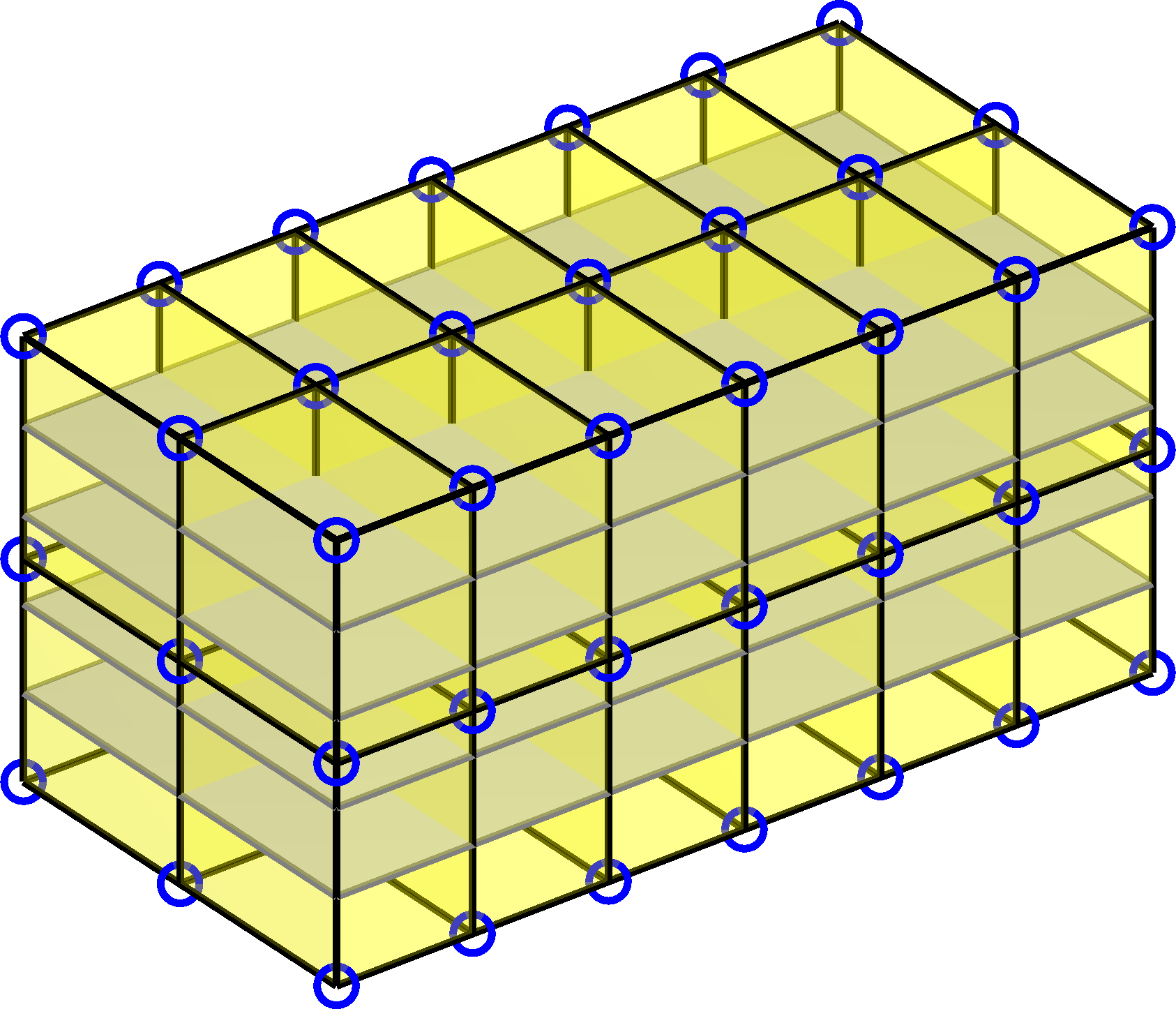}}
	\hspace{1cm}
	\subfigure[pressure - anisotropic]{\includegraphics[width=0.25\textwidth]{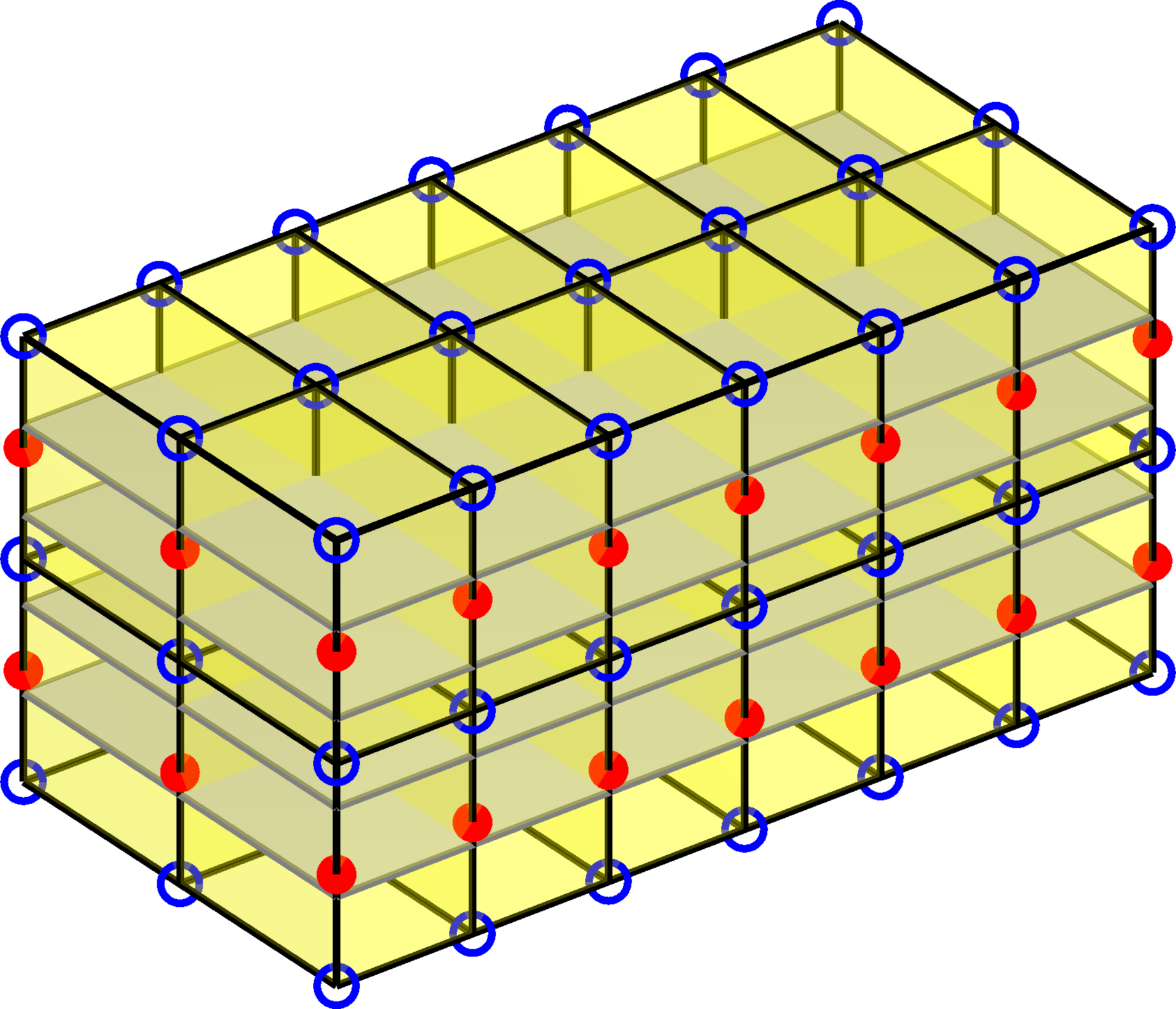}}
	
	\caption{\label{fig:AnisMeshGenEx} The anisotropic mesh concept for Taylor--Hood elements. (a) and (b): mesh pair with classical (isotropic) Taylor--Hood elements; (a) and (c): mesh pair with anisotropic Taylor--Hood elements. The red dots in (c) are the additional DOFs of the anisotropic case. Some level sets within the meshes are shown in grey.}
\end{figure}

\begin{figure}
	\centering
	
	\subfigure[$\varphi_1$ - isotropic THE]{\includegraphics[width=0.35\textwidth]{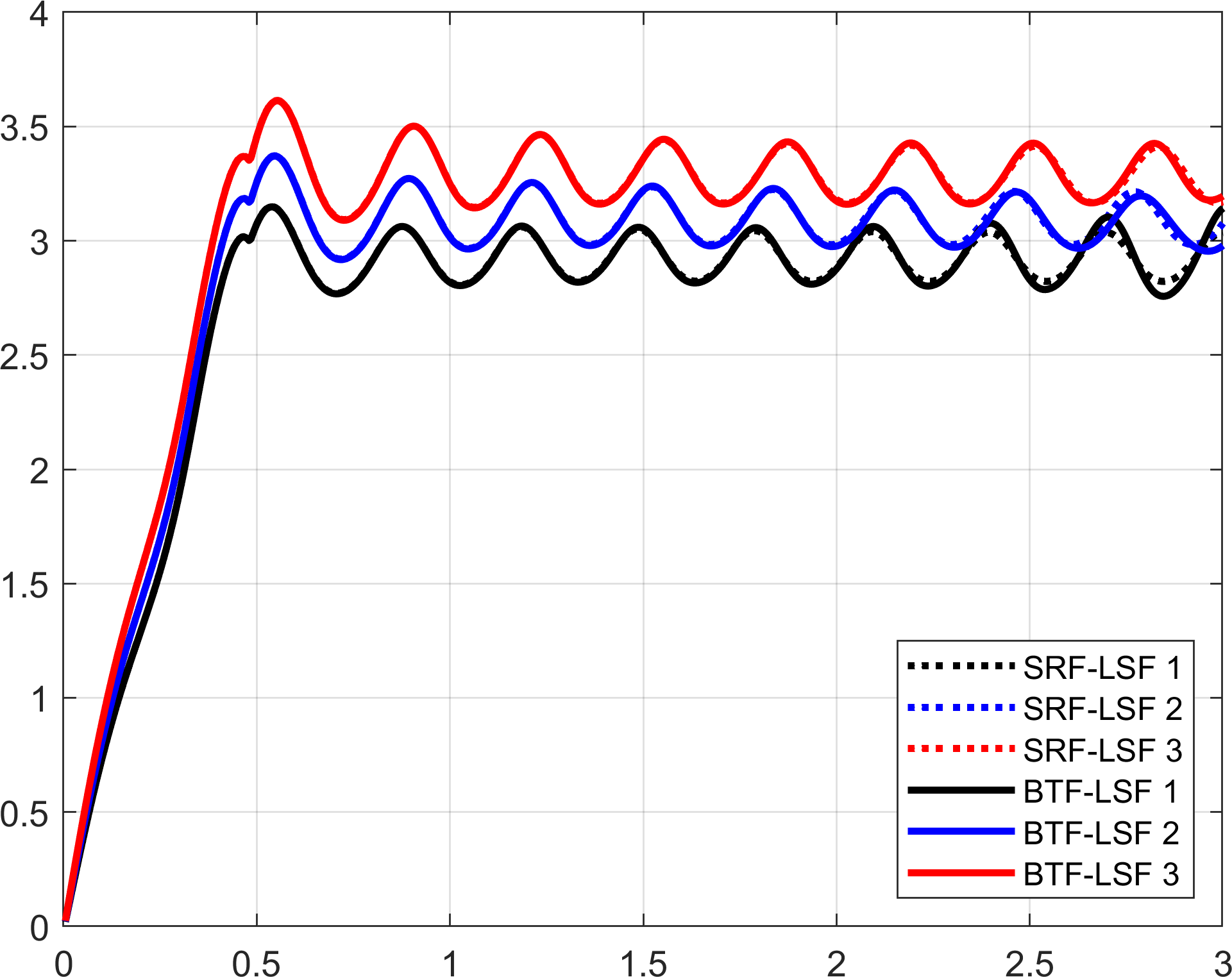}}\hspace{0.1\textwidth}
	\subfigure[$\varphi_1$ - anisotropic THE]{\includegraphics[width=0.35\textwidth]{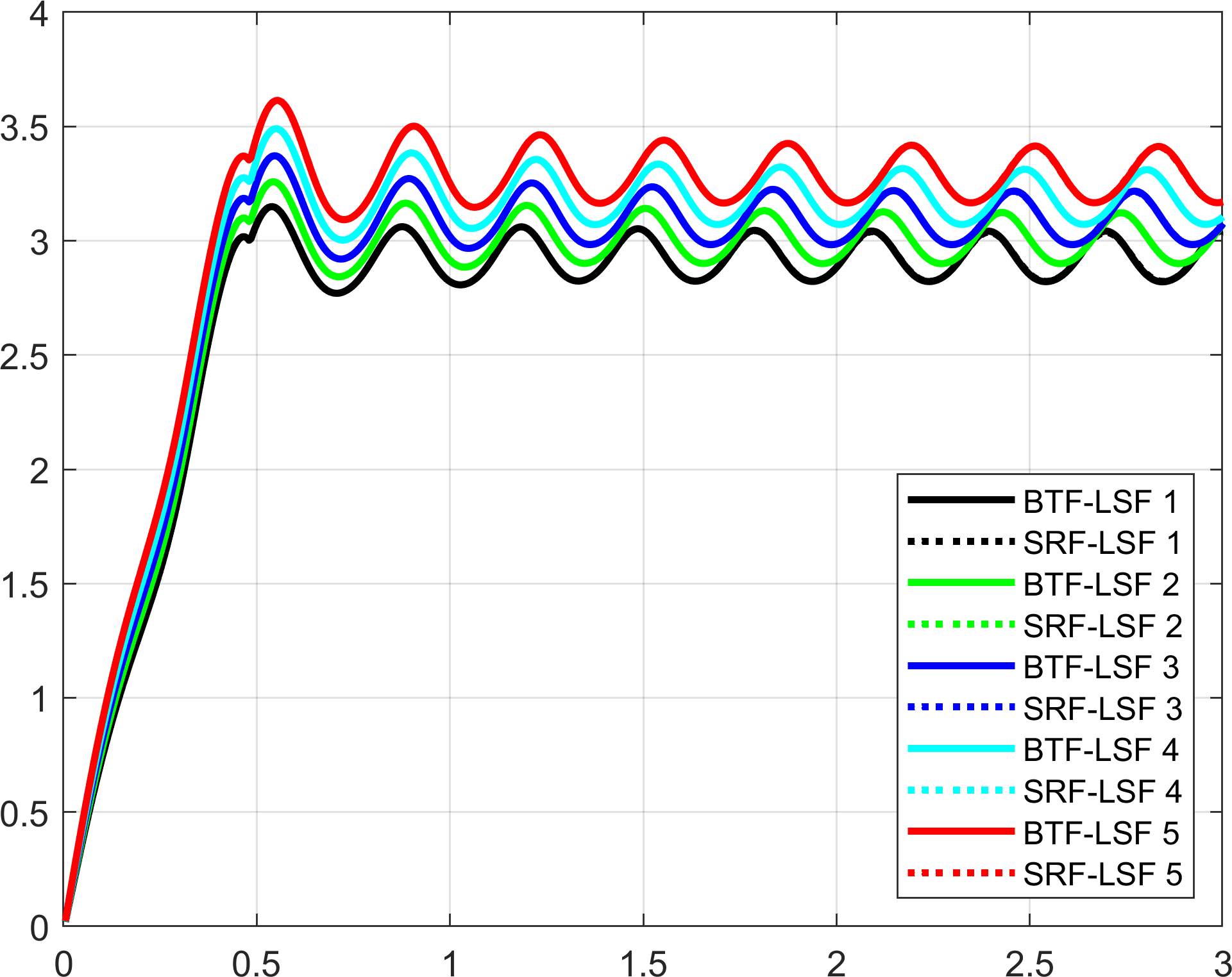}}
	\subfigure[$\varphi_2$ - isotropic THE]{\includegraphics[width=0.35\textwidth]{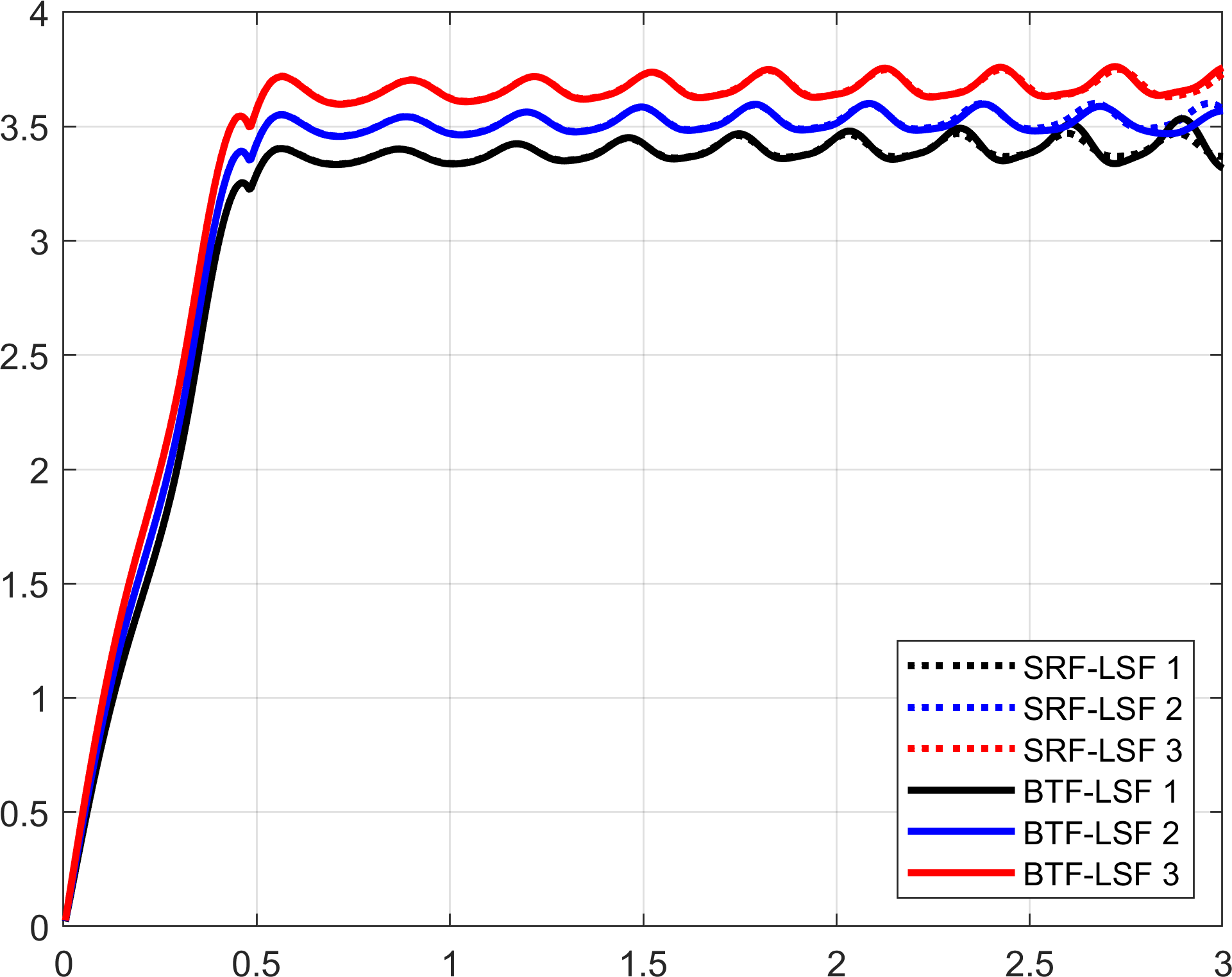}}\hspace{0.1\textwidth}
	\subfigure[$\varphi_2$ - anisotropic THE]{\includegraphics[width=0.35\textwidth]{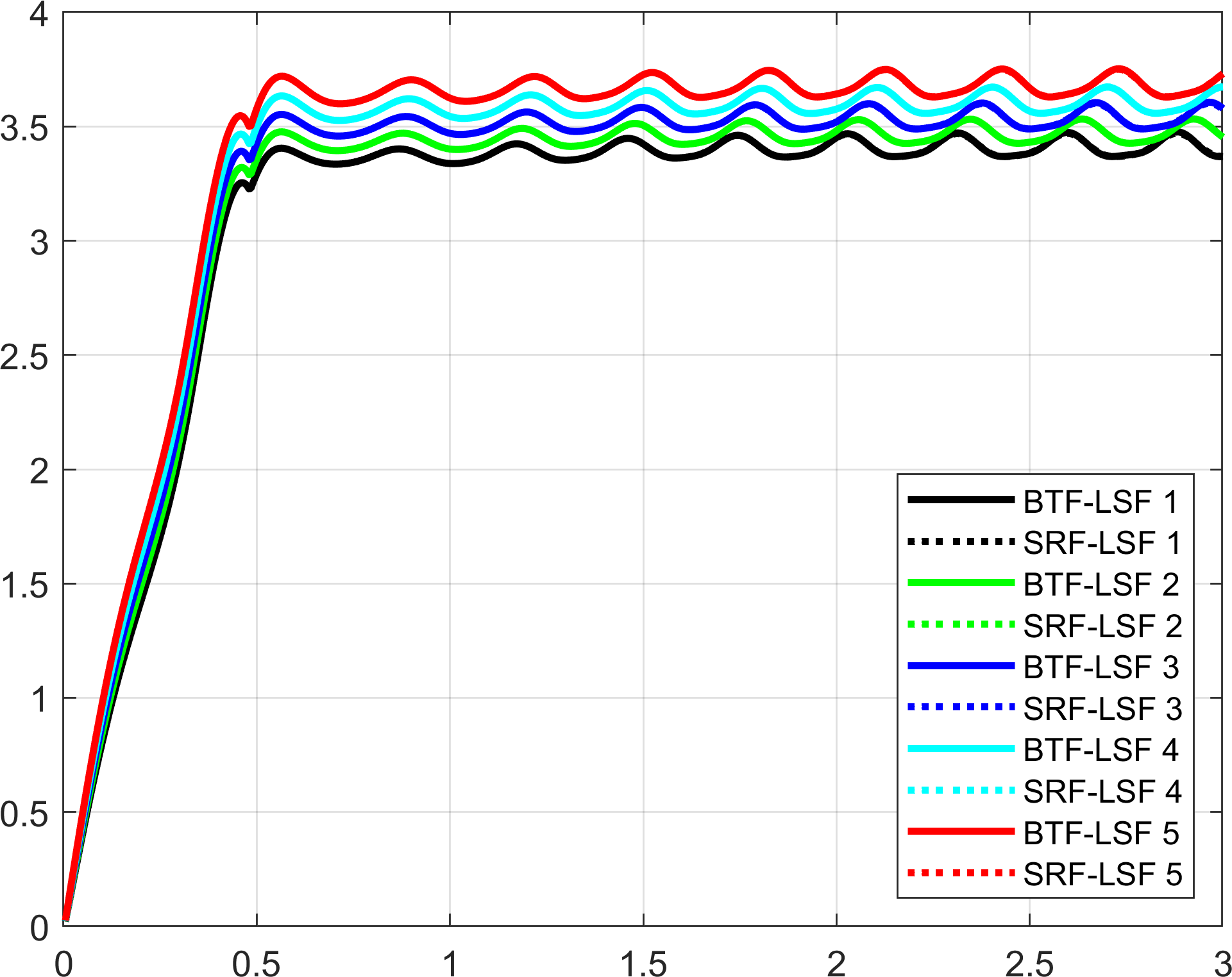}}
	\subfigure[$\varphi_3$ - isotropic THE]{\includegraphics[width=0.35\textwidth]{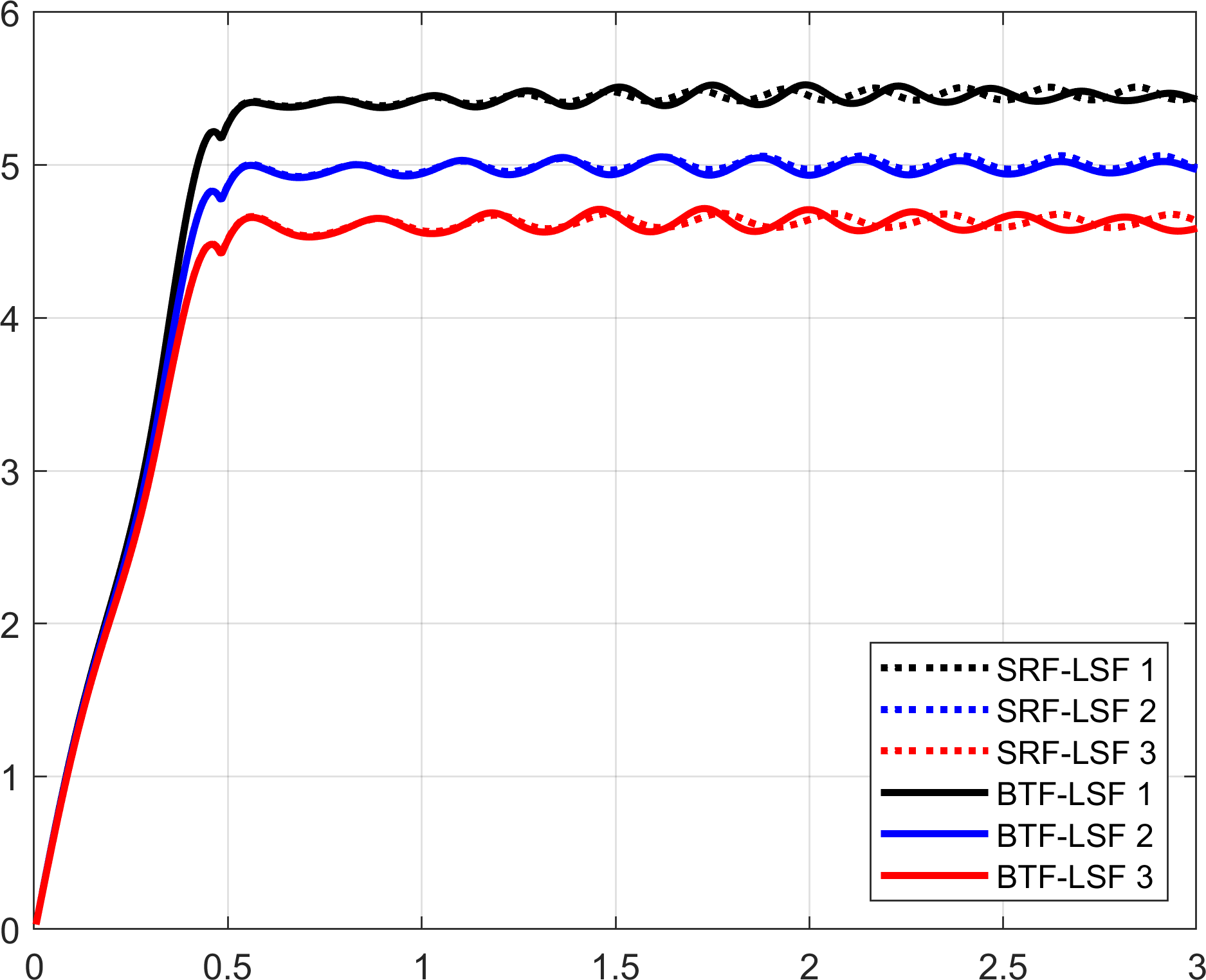}}\hspace{0.1\textwidth}
	\subfigure[$\varphi_3$ - anisotropic THE]{\includegraphics[width=0.35\textwidth]{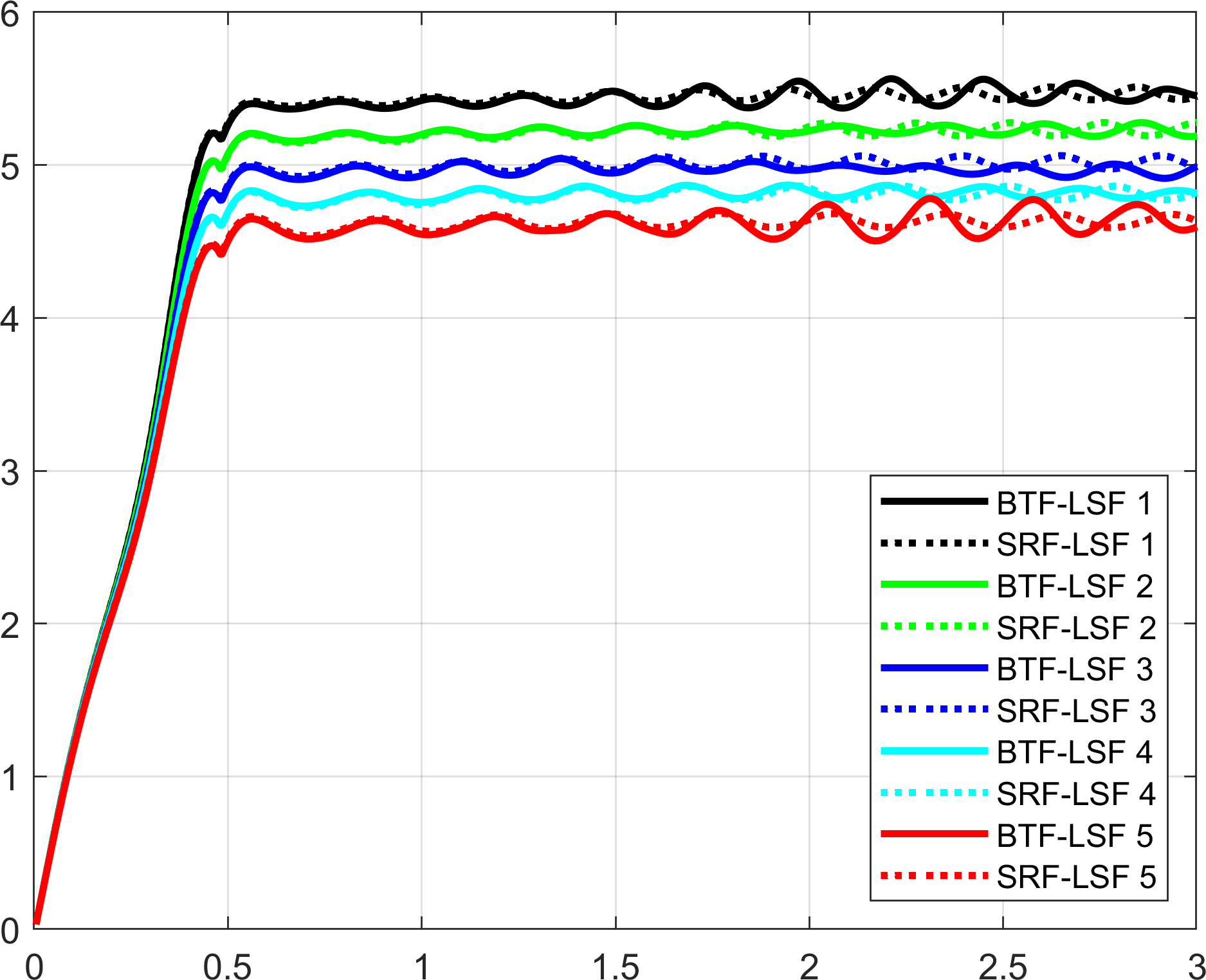}}
	
	\caption{\label{fig:TurekInstatNSEQ-ResPD} The pressure difference on the obstacle over time. (a), (c), and (e) computed with isotropic Taylor--Hood elements (THE) and (b), (d), and (f) computed with anisoptropic THE for mappings $\varphi_1$, $\varphi_2$, and $\varphi_3$, respectively.}
\end{figure}

\begin{figure}
	\centering
	
	\subfigure[BTF, $\varphi_1$, $\lVert \vek{u} \rVert$, $t = 1$]{\includegraphics[width=0.35\textwidth]{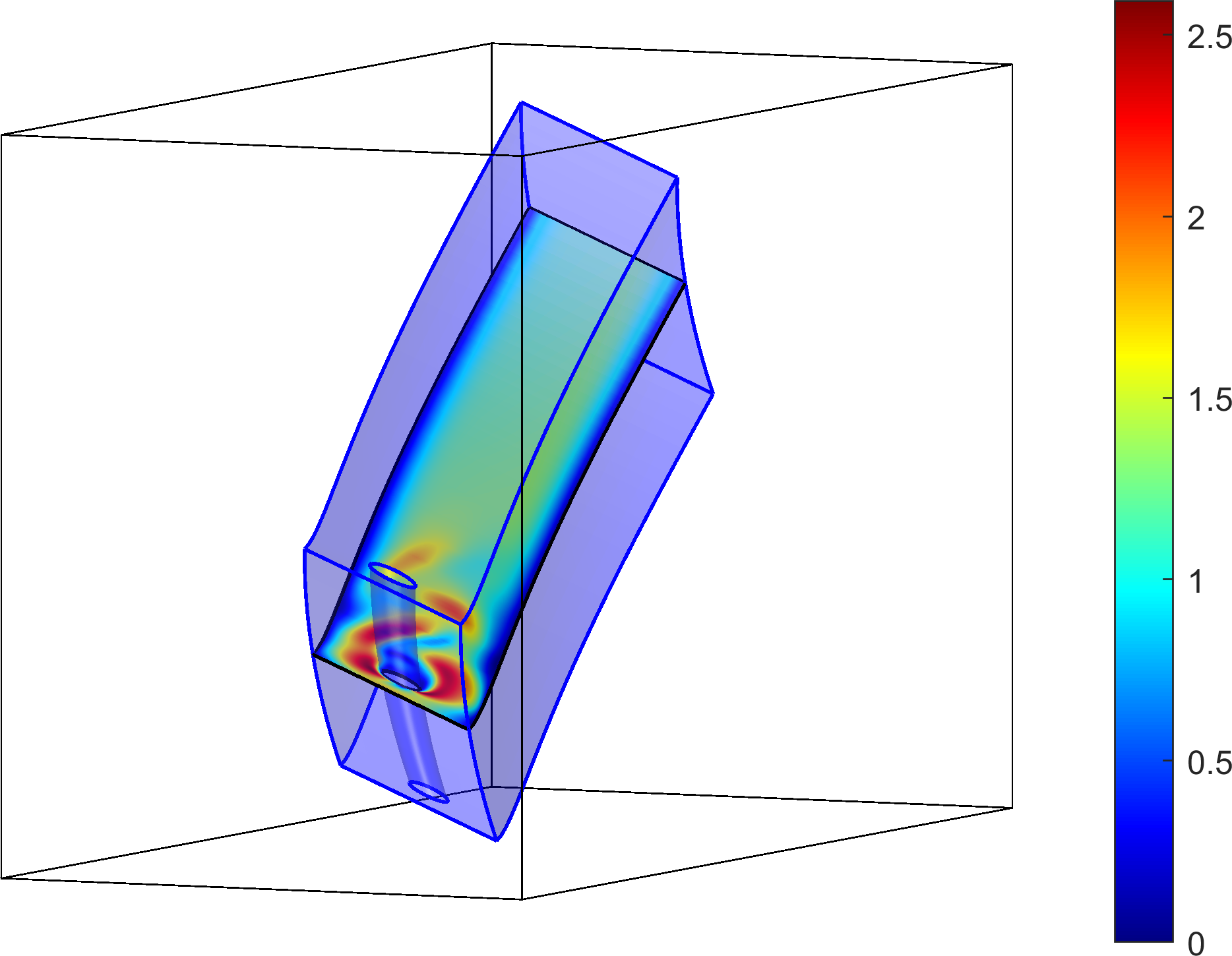}}\hspace{0.1\textwidth}
	\subfigure[SRF, $\varphi_1$, $\lVert \vek{u} \rVert$, $t = 1$]{\includegraphics[width=0.3\textwidth]{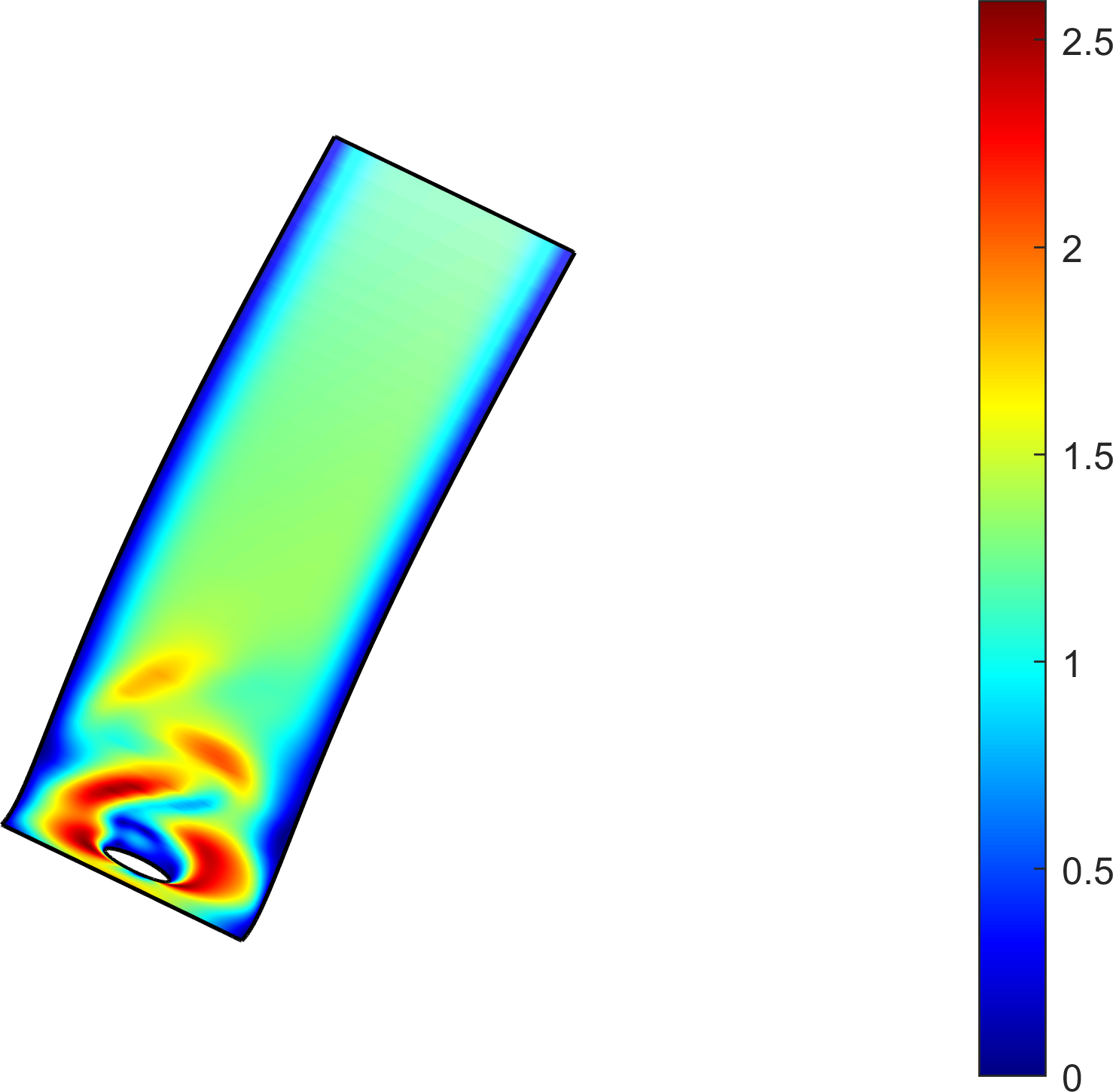}}
	\subfigure[BTF, $\varphi_1$, $\lVert \vek{u} \rVert$, $t = 2$]{\includegraphics[width=0.35\textwidth]{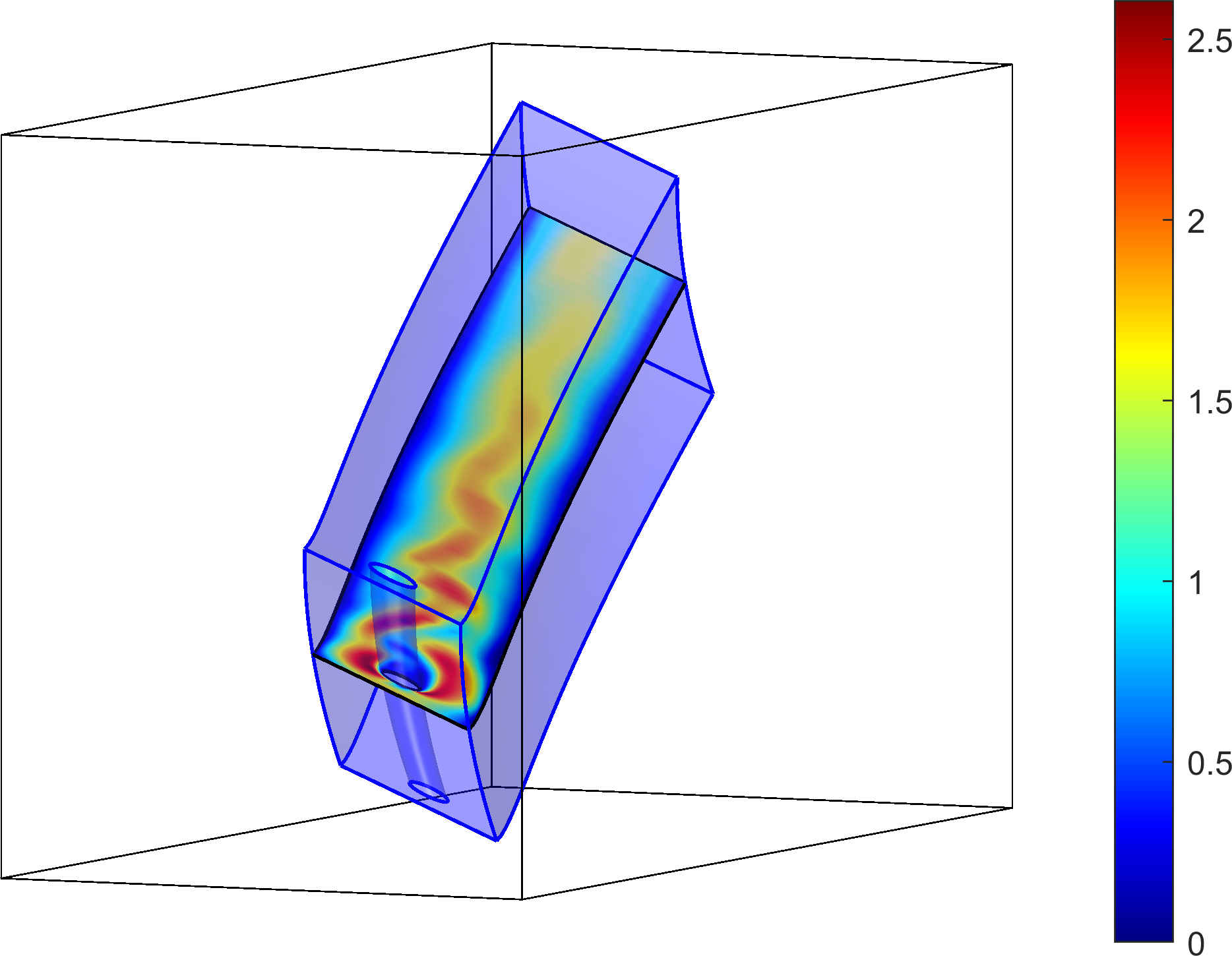}}\hspace{0.1\textwidth}
	\subfigure[SRF, $\varphi_1$, $\lVert \vek{u} \rVert$, $t = 2$]{\includegraphics[width=0.3\textwidth]{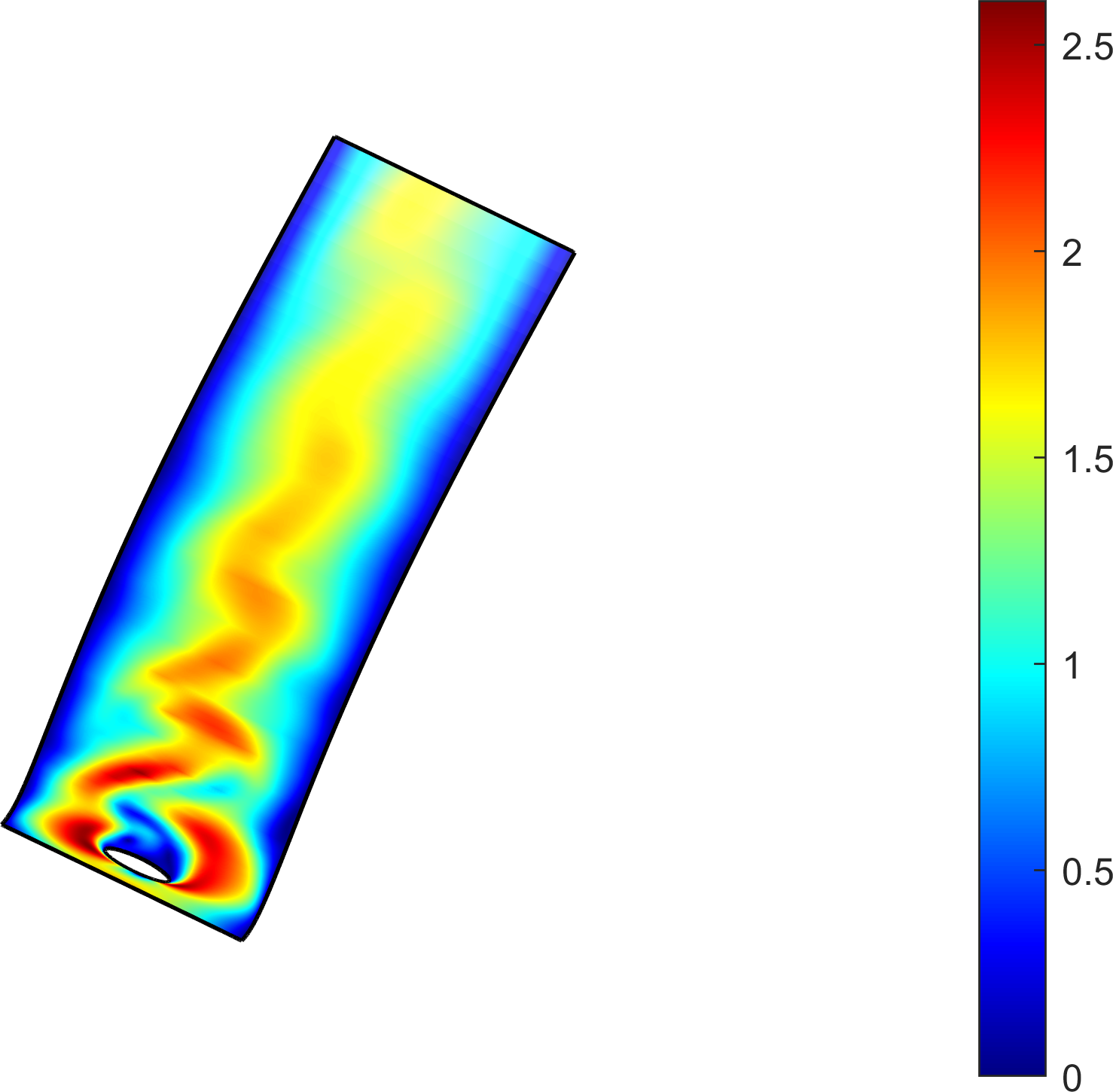}}
	\subfigure[BTF, $\varphi_1$, $\lVert \vek{u} \rVert$, $t = 3$]{\includegraphics[width=0.35\textwidth]{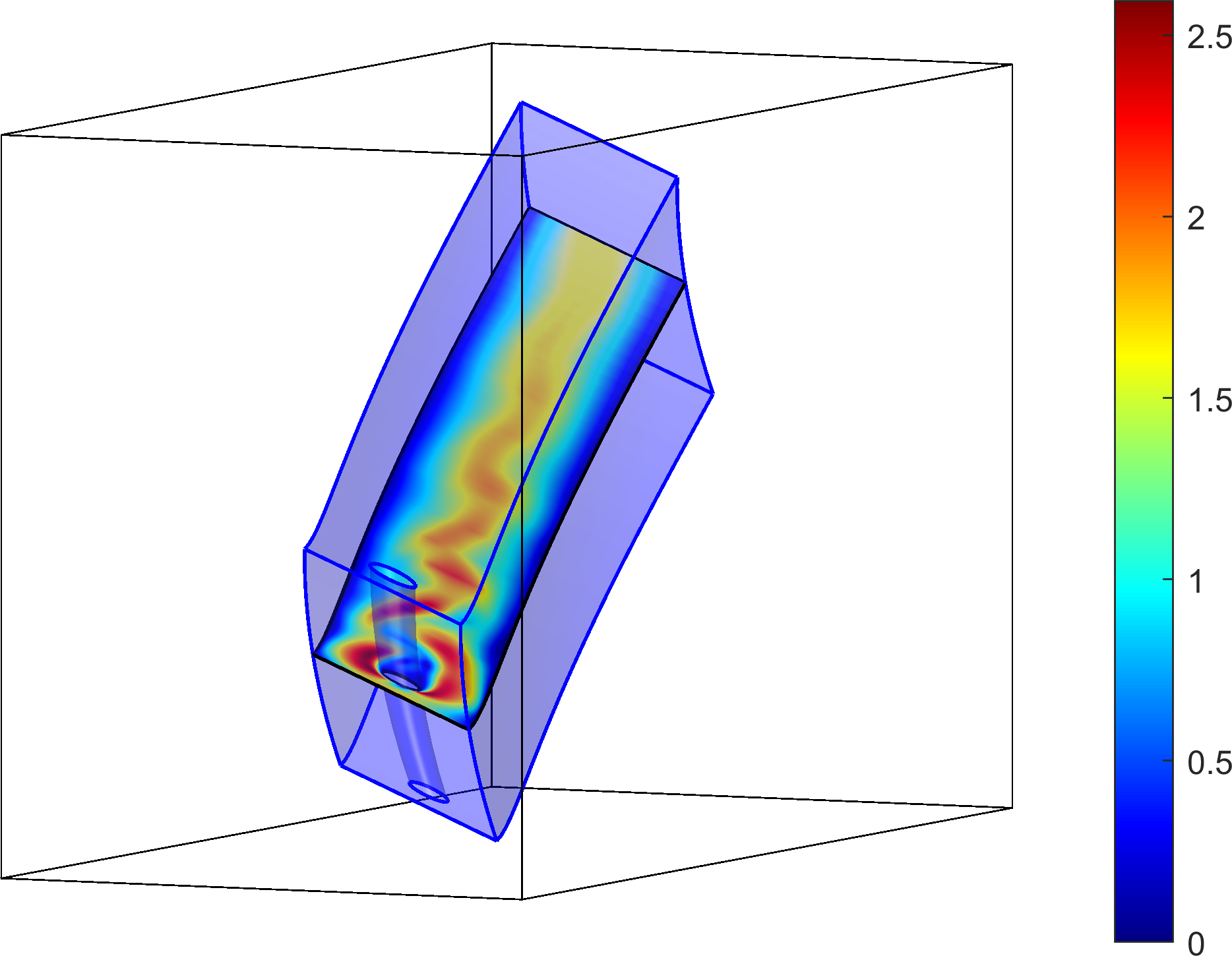}}\hspace{0.1\textwidth}
	\subfigure[SRF, $\varphi_1$, $\lVert \vek{u} \rVert$, $t = 3$]{\includegraphics[width=0.3\textwidth]{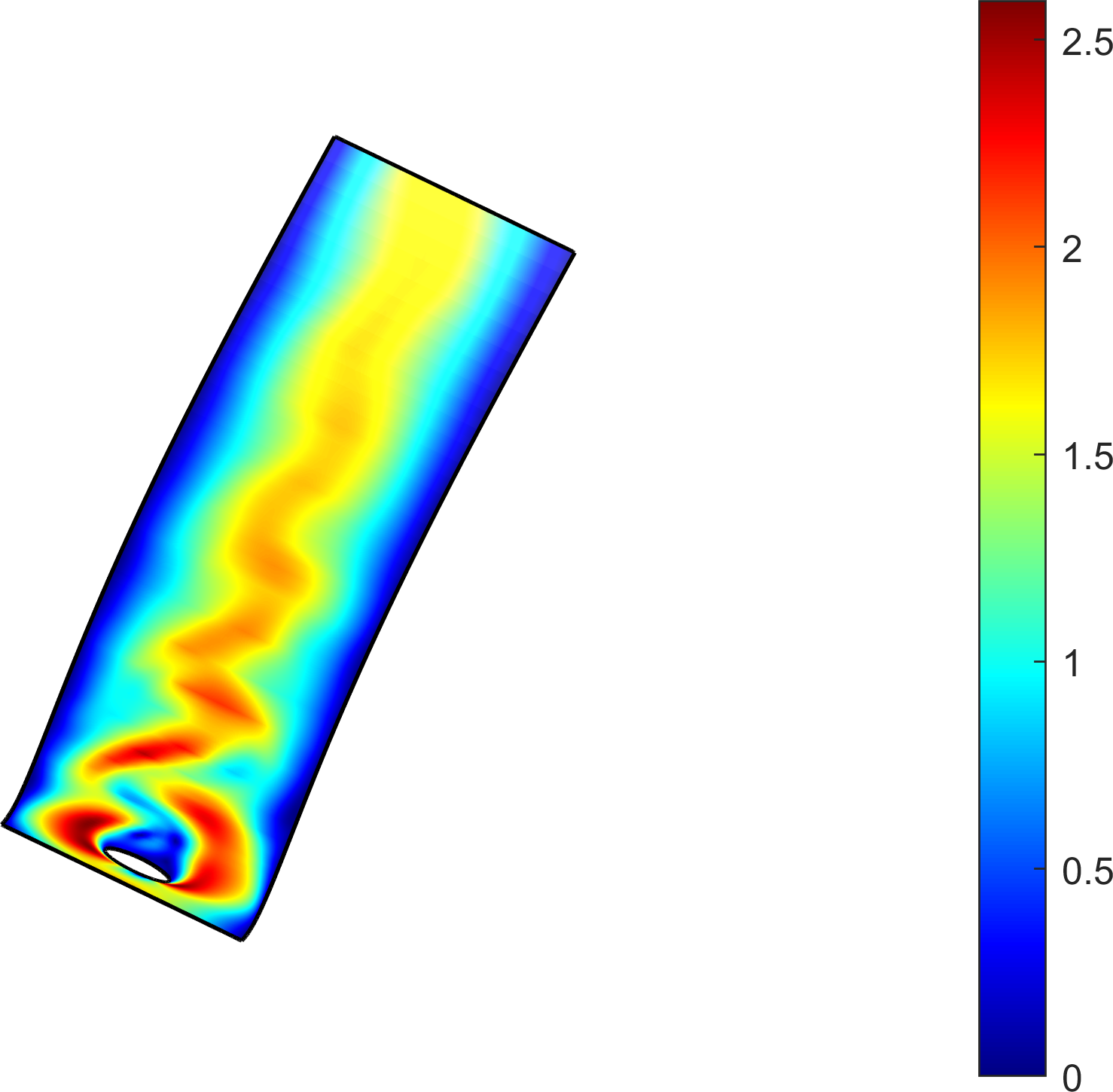}}
	
	\caption{\label{fig:TurekInstatNSEQ-Sol1} The velocity magnitudes with mapping $\varphi_1$ on $\Gamma_{\!c = 1/6}$. The left side shows Bulk Trace FEM results and the right side Surface FEM results.}
\end{figure}

\begin{figure}
	\centering
	
	\subfigure[BTF, $\varphi_1$, $p$, $t = 1$]{\includegraphics[width=0.35\textwidth]{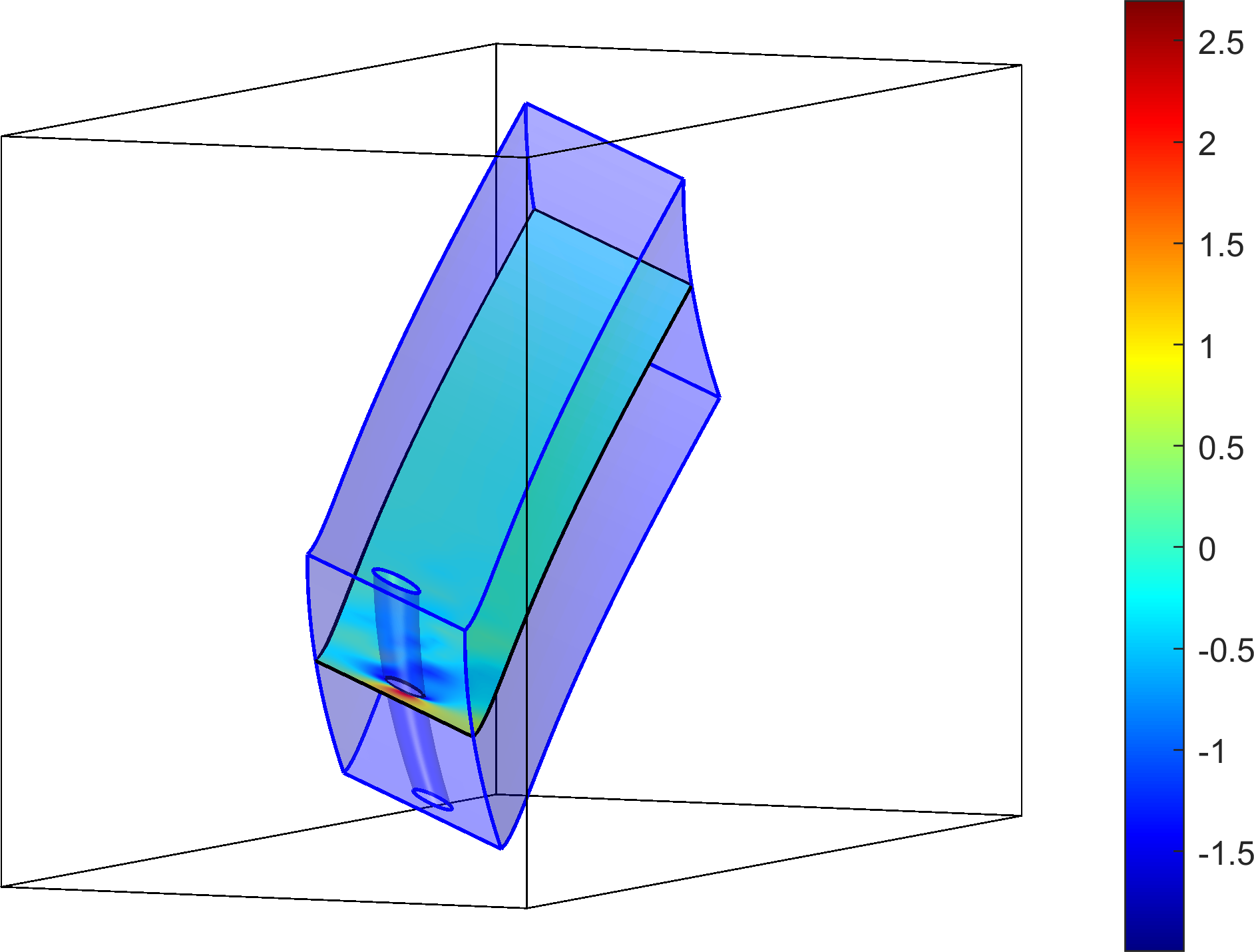}}\hspace{0.1\textwidth}
	\subfigure[SRF, $\varphi_1$, $p$, $t = 1$]{\includegraphics[width=0.3\textwidth]{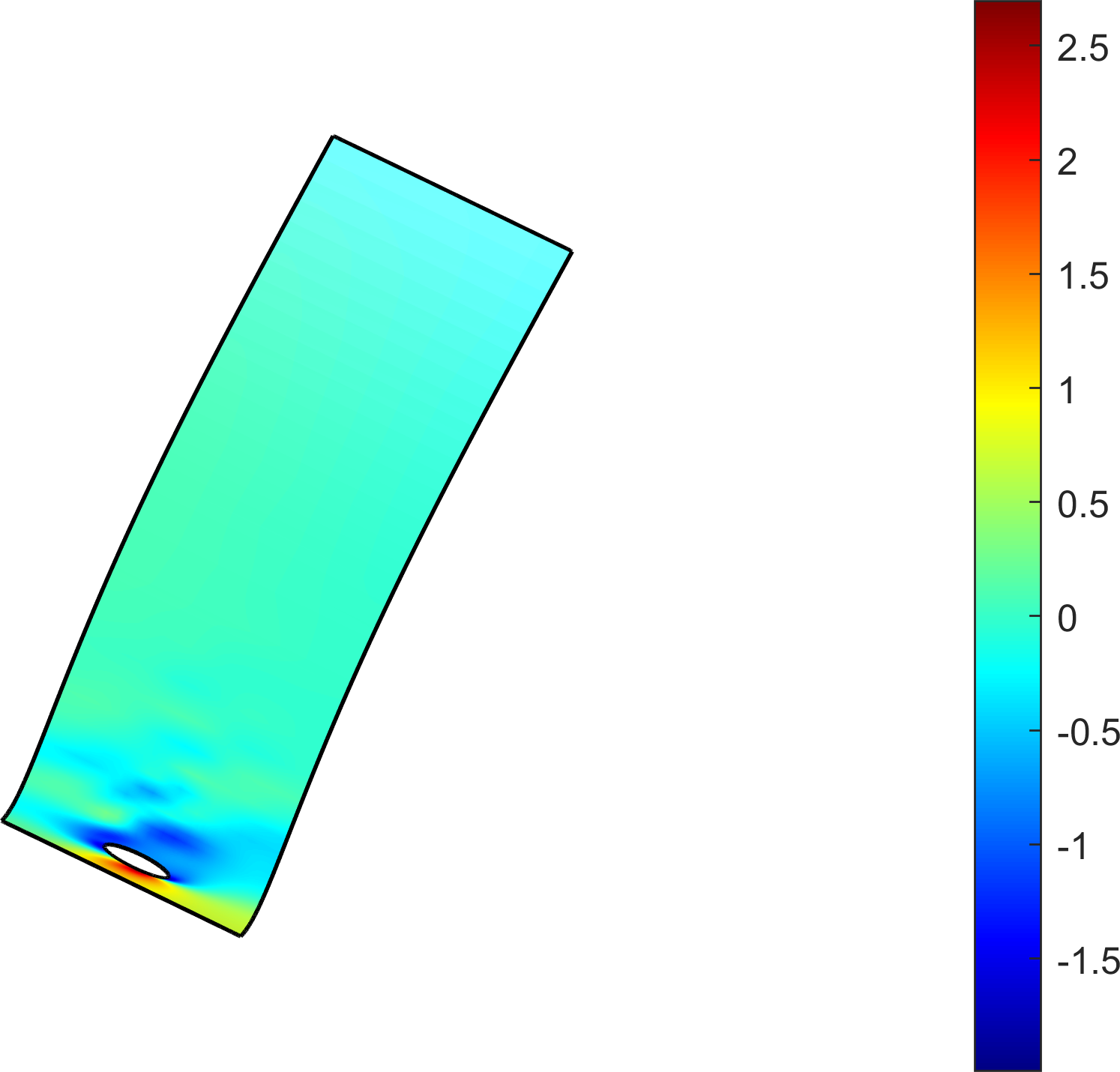}}
	\subfigure[BTF, $\varphi_1$, $p$, $t = 2$]{\includegraphics[width=0.35\textwidth]{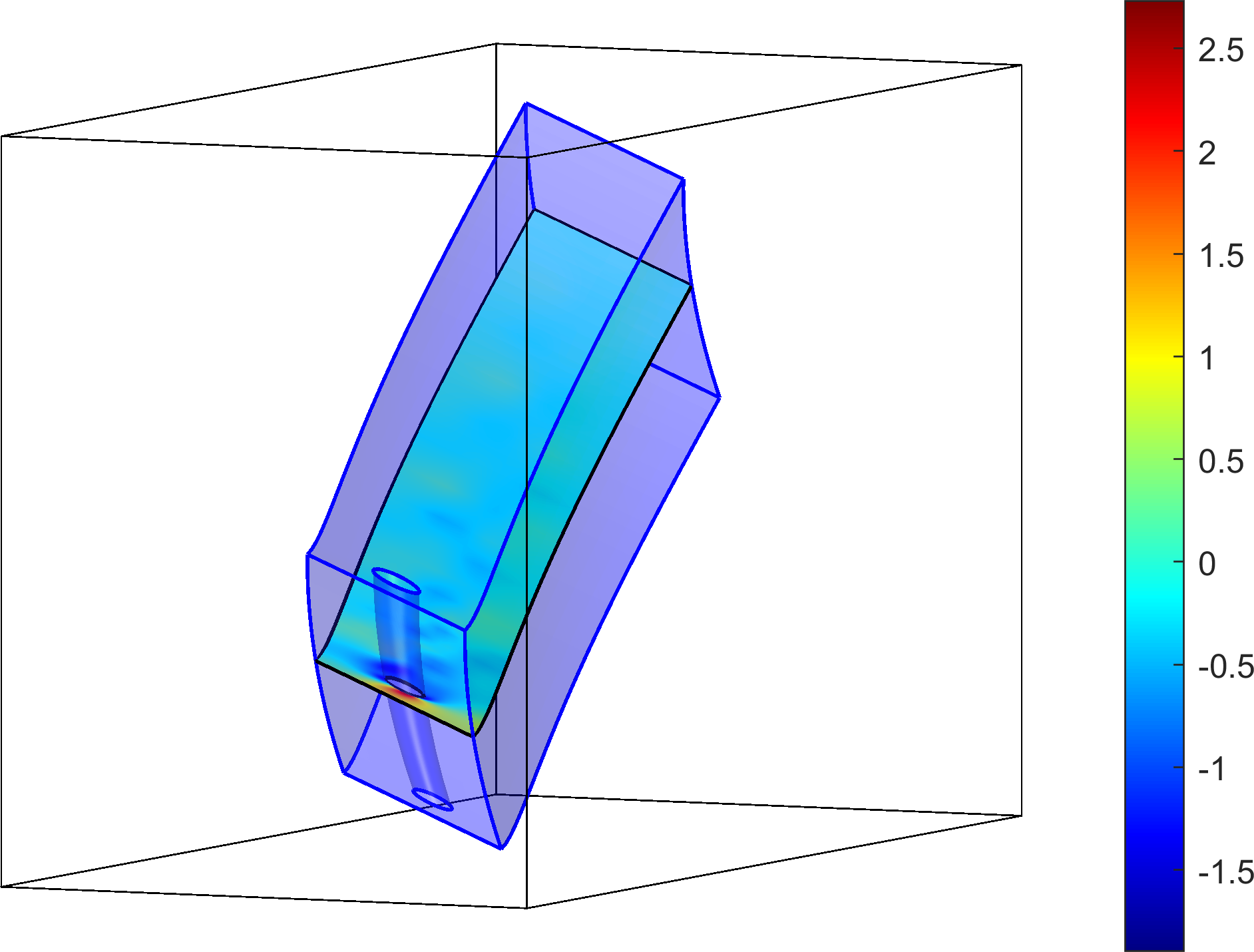}}\hspace{0.1\textwidth}
	\subfigure[SRF, $\varphi_1$, $p$, $t = 2$]{\includegraphics[width=0.3\textwidth]{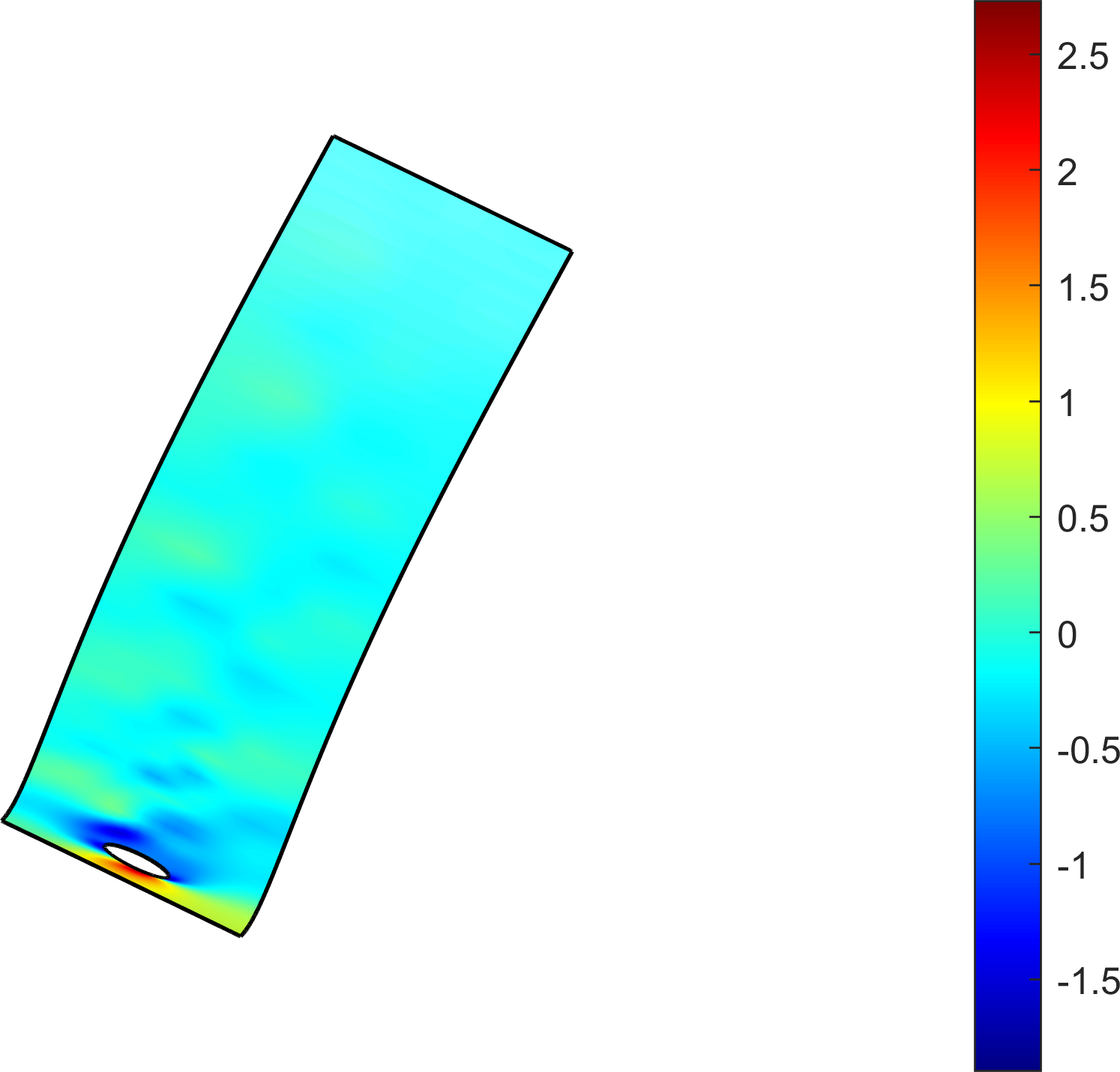}}
	\subfigure[BTF, $\varphi_1$, $p$, $t = 3$]{\includegraphics[width=0.35\textwidth]{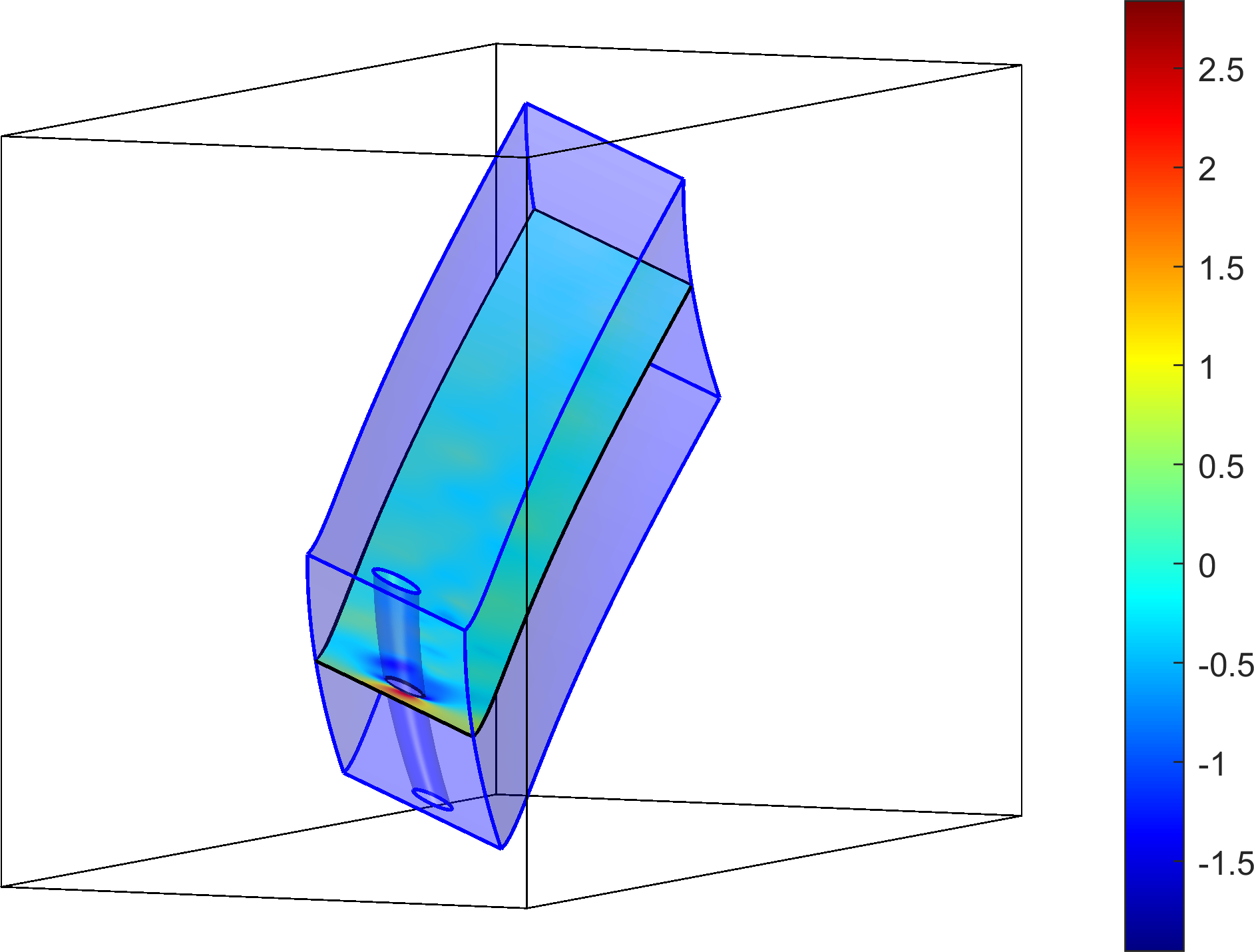}}\hspace{0.1\textwidth}
	\subfigure[SRF, $\varphi_1$, $p$, $t = 3$]{\includegraphics[width=0.3\textwidth]{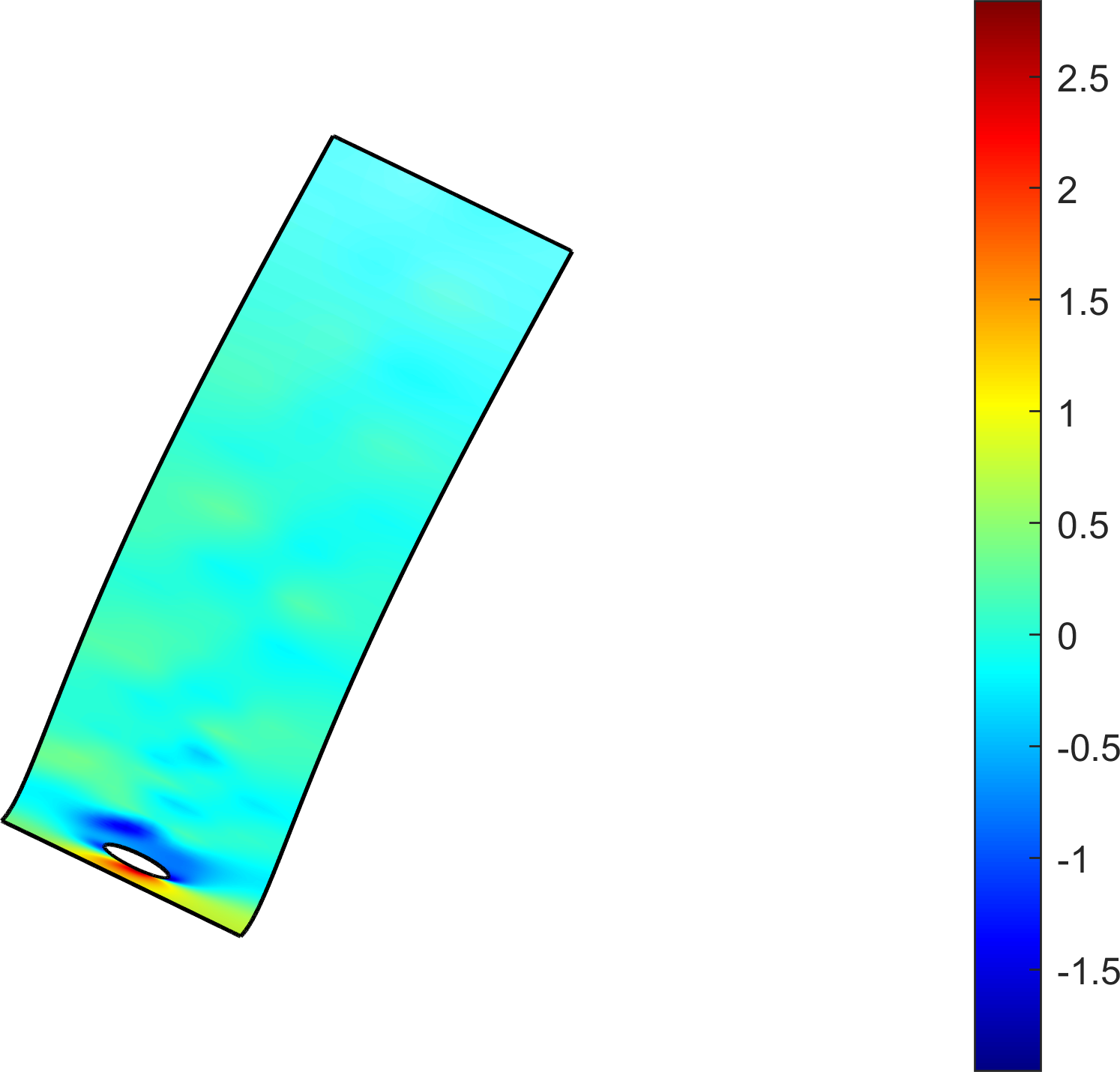}}
	
	\caption{\label{fig:TurekInstatNSEQ-Sol2} The pressure field of the flow with mapping $\varphi_1$ on $\Gamma_{\!c = 1/6}$. The left side shows Bulk Trace FEM results and the right side Surface FEM results.}
\end{figure}

\begin{figure}
	\centering
	
	\subfigure[BTF, $\omega^{\star}$, $t = 1$]{\includegraphics[width=0.35\textwidth]{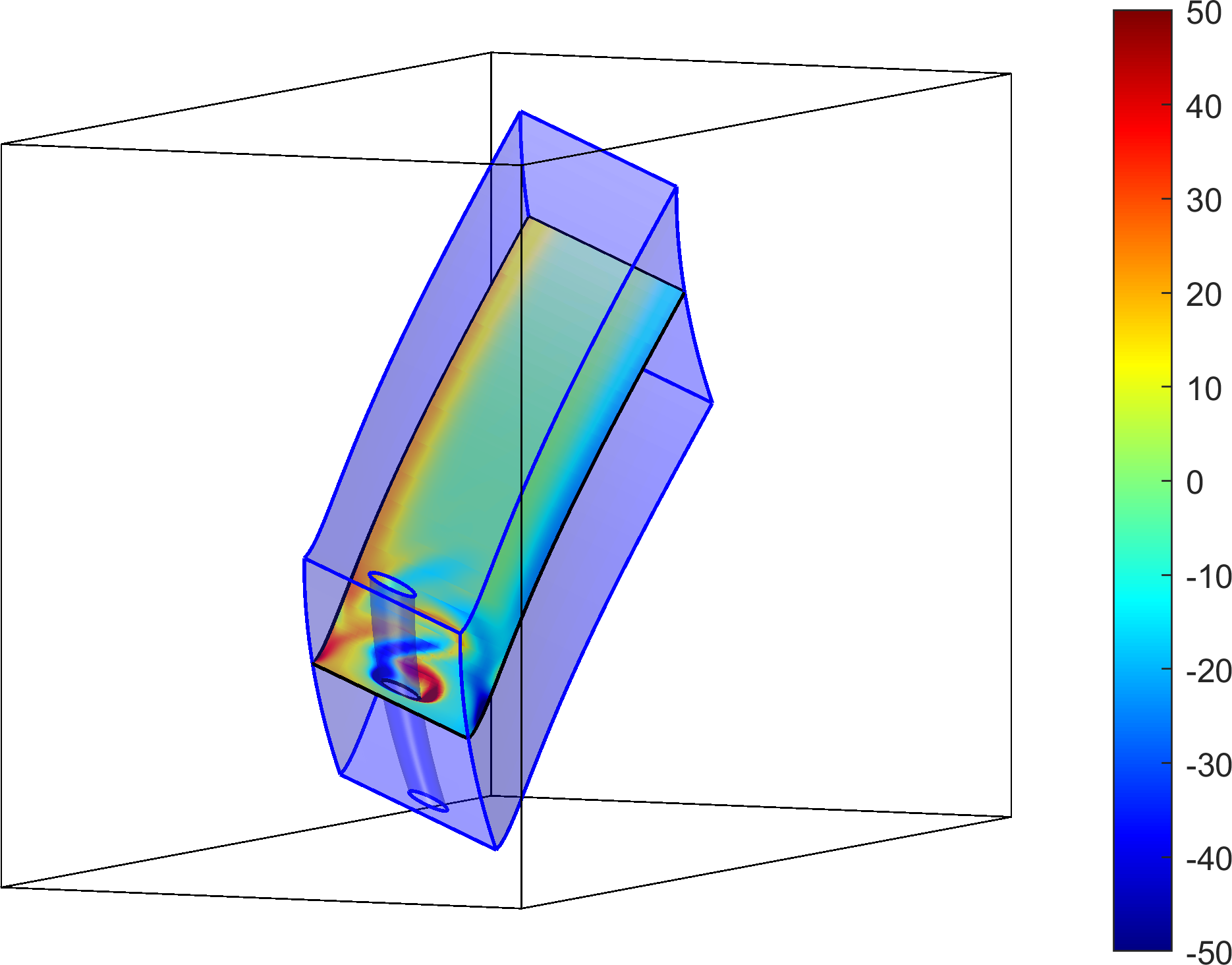}}\hspace{0.1\textwidth}
	\subfigure[SRF, $\omega^{\star}$, $t = 1$]{\includegraphics[width=0.3\textwidth]{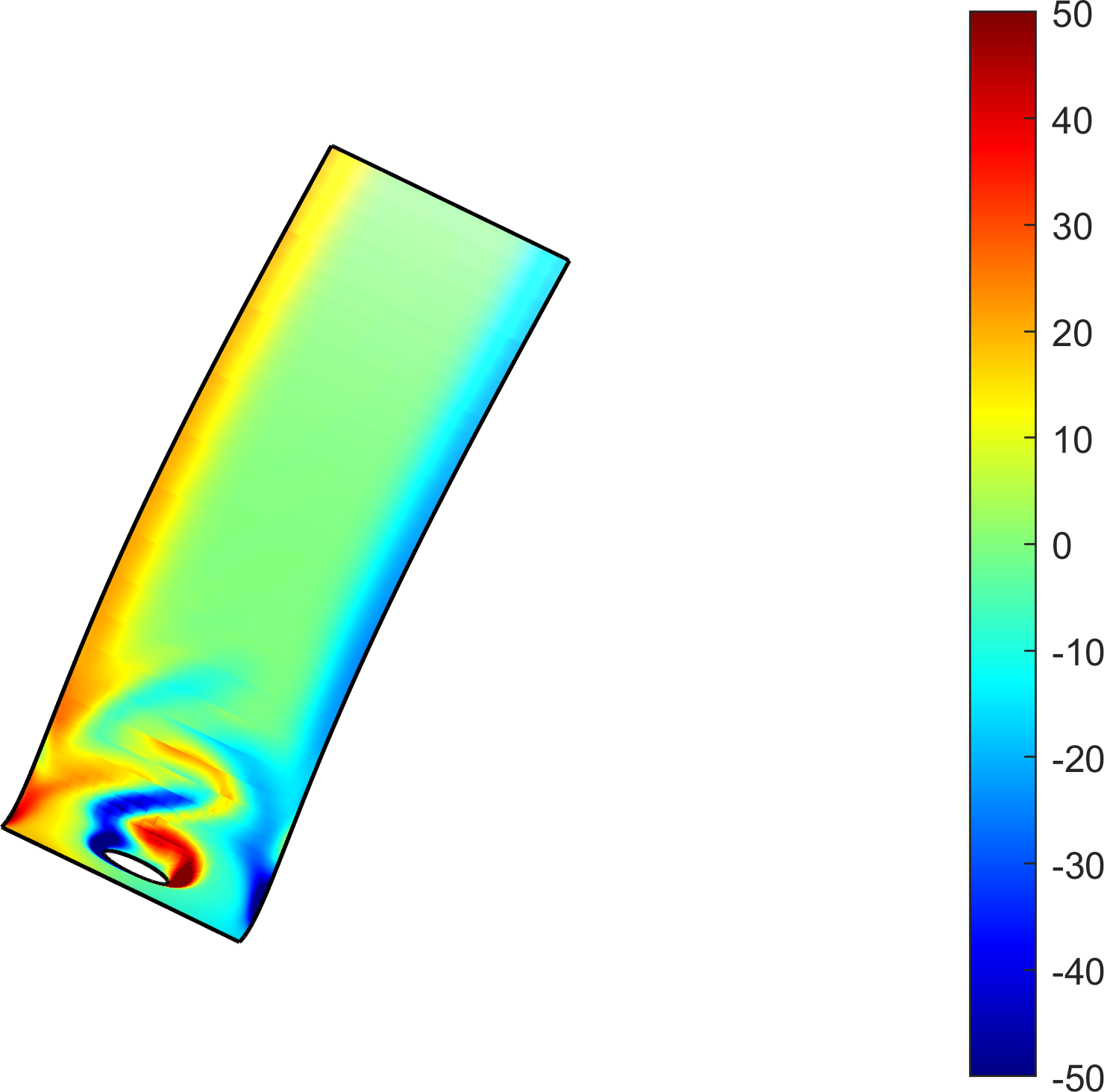}}
	\subfigure[BTF, $\omega^{\star}$, $t = 2$]{\includegraphics[width=0.35\textwidth]{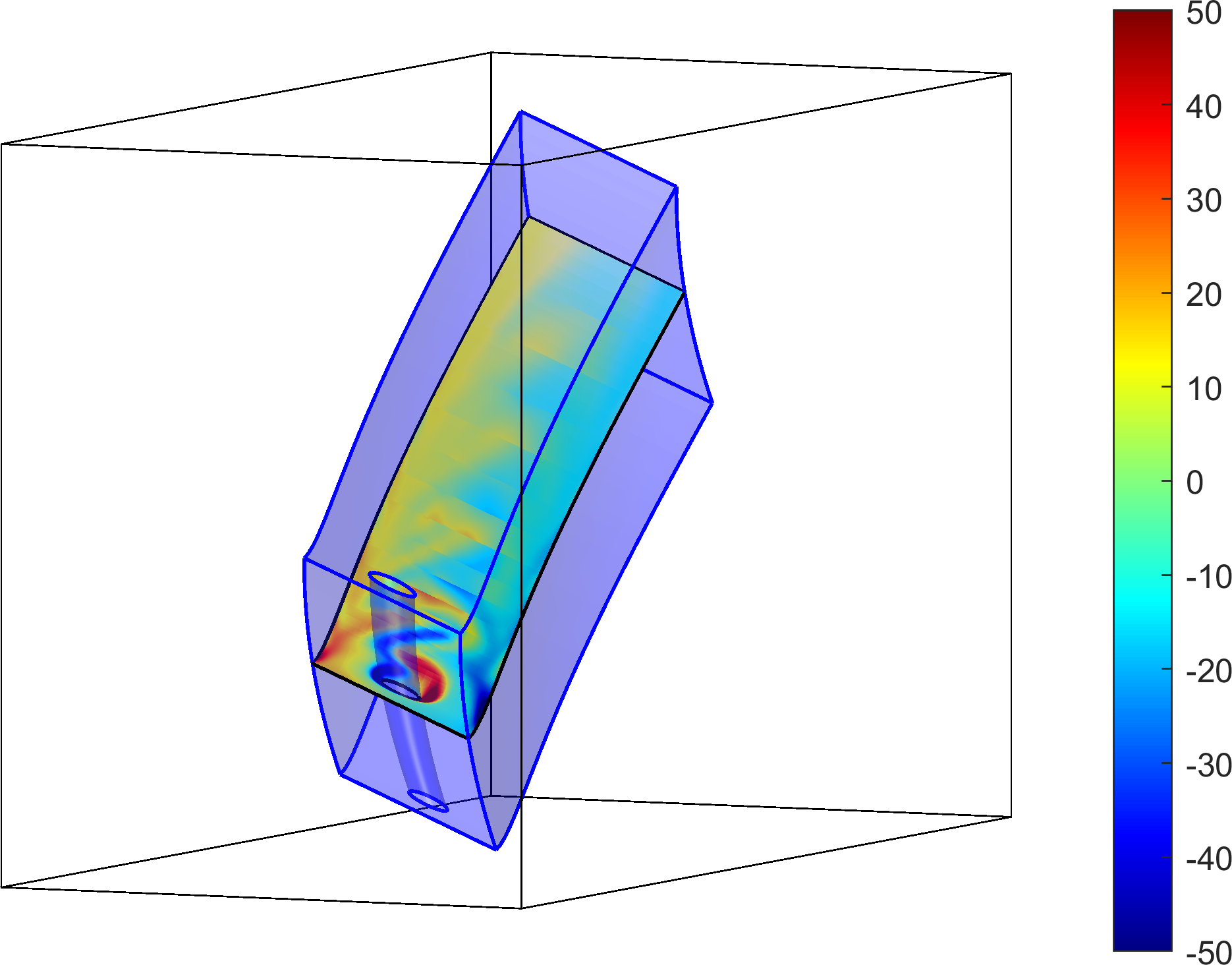}}\hspace{0.1\textwidth}
	\subfigure[SRF, $\omega^{\star}$, $t = 2$]{\includegraphics[width=0.3\textwidth]{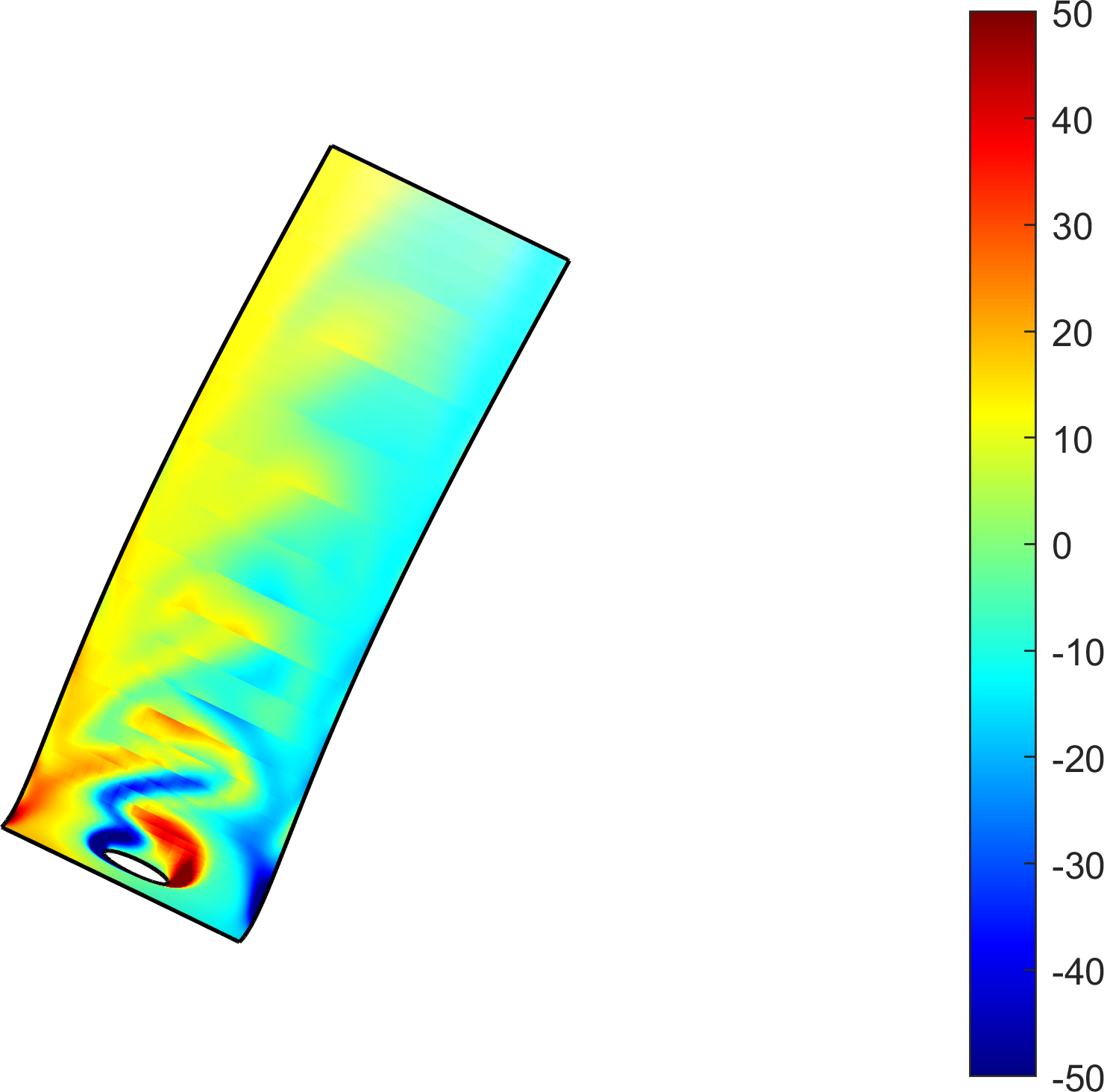}}
	\subfigure[BTF, $\omega^{\star}$, $t = 3$]{\includegraphics[width=0.35\textwidth]{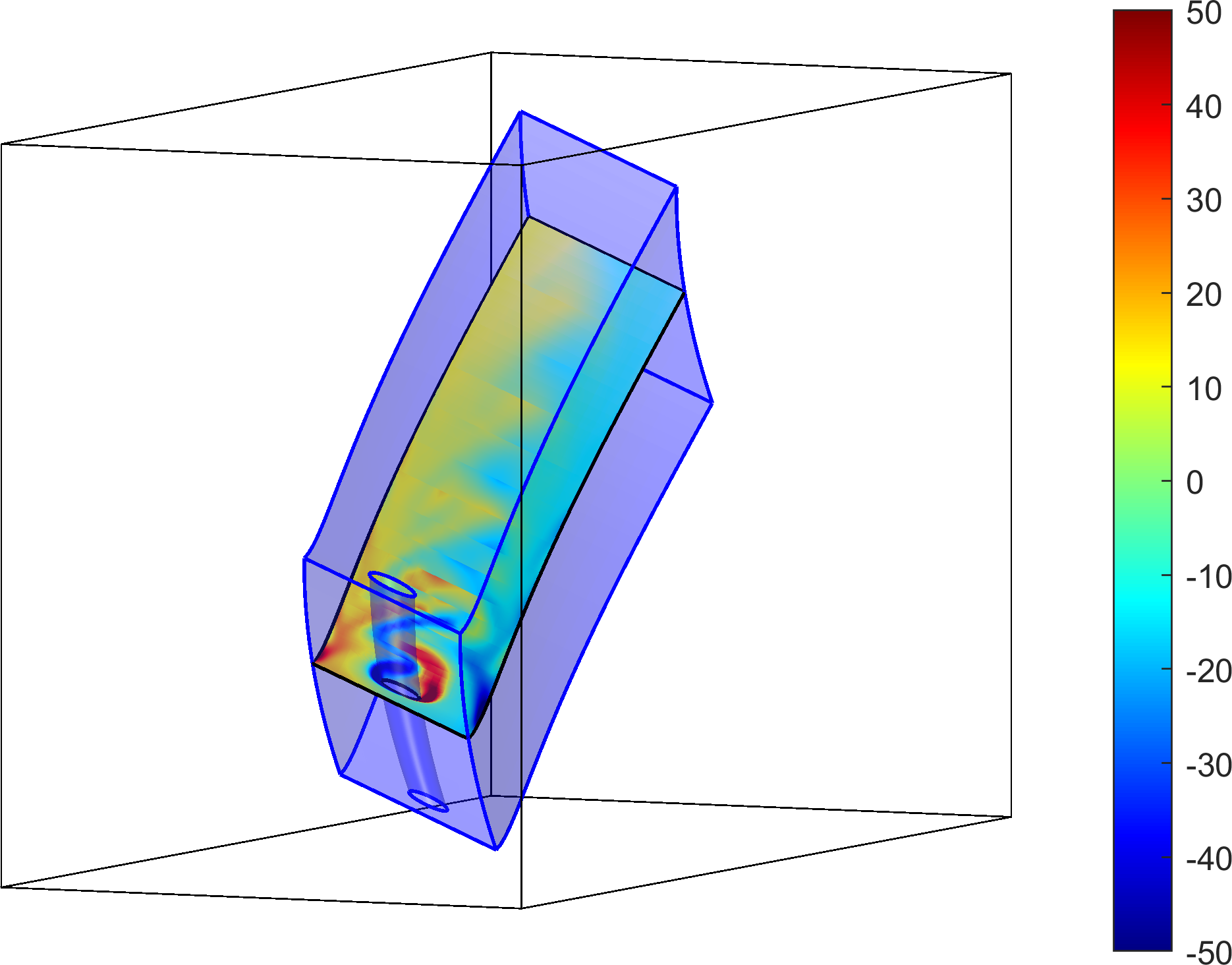}}\hspace{0.1\textwidth}
	\subfigure[SRF, $\omega^{\star}$, $t = 3$]{\includegraphics[width=0.3\textwidth]{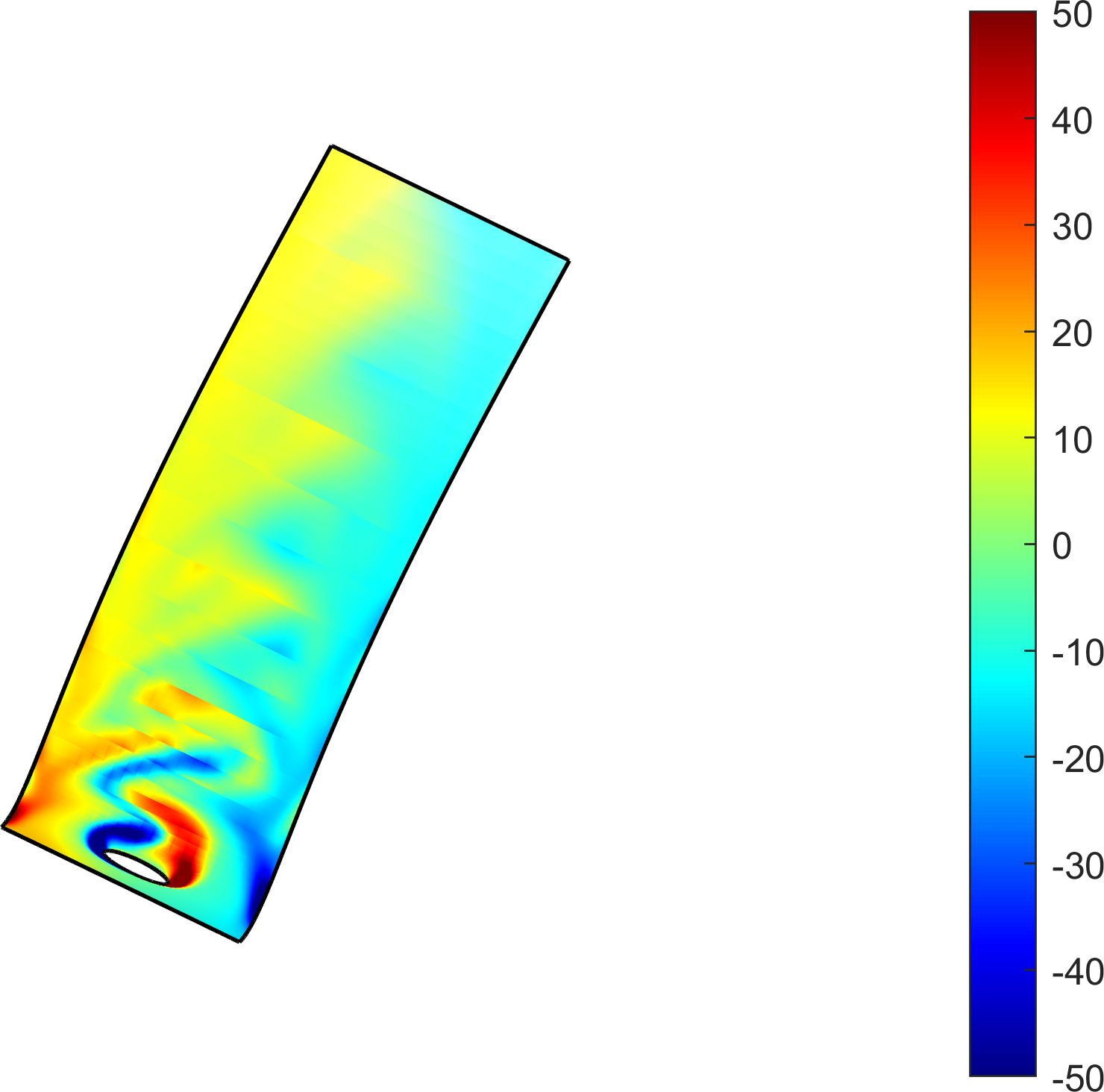}}
	
	\caption{\label{fig:TurekInstatNSEQ-Sol3} The scalar vorticity field $\omega^{\star}$ of the flow with mapping $\varphi_1$ on $\Gamma_{\!c = 1/6}$. The left side shows Bulk Trace FEM results and the right side Surface FEM results.}
\end{figure}

\begin{figure}
	\centering
	
	\subfigure[BTF, $\lVert \vek{u} \rVert$, $t = 1.5$]{\includegraphics[width=0.35\textwidth]{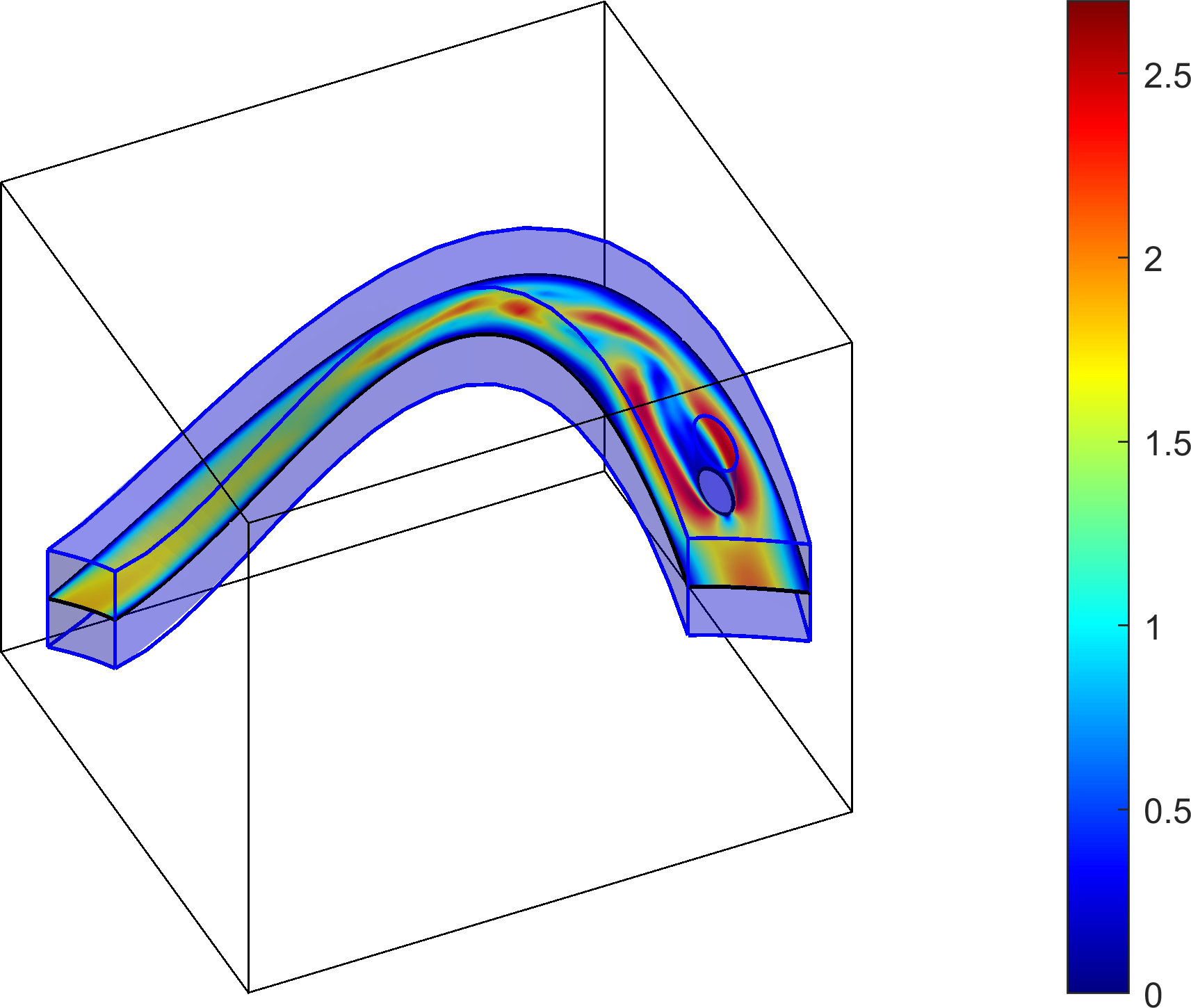}}\hspace{0.1\textwidth}
	\subfigure[SRF, $\lVert \vek{u} \rVert$, $t = 1.5$]{\includegraphics[width=0.35\textwidth]{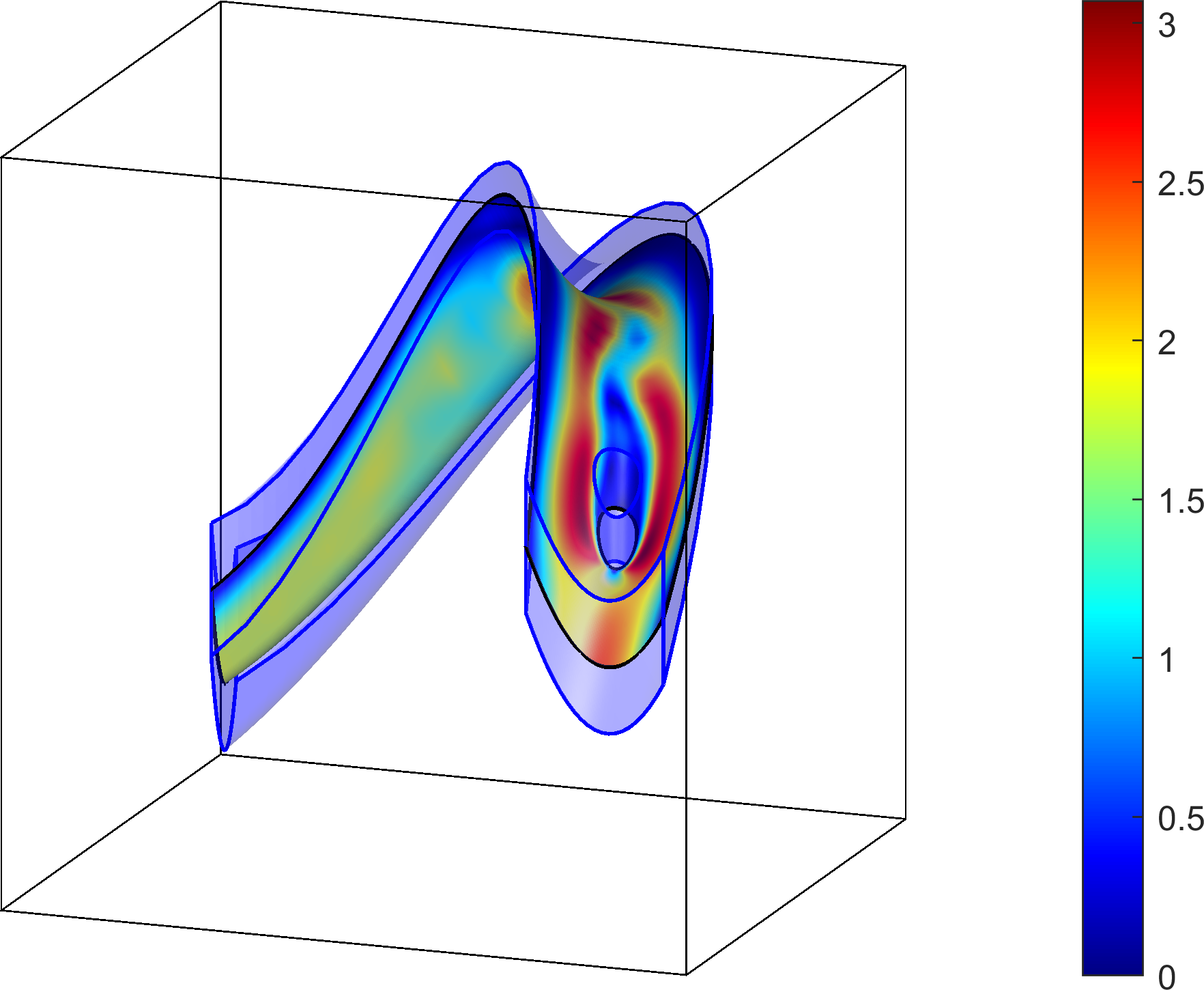}}
	\subfigure[BTF, $p$, $t = 1.5$]{\includegraphics[width=0.35\textwidth]{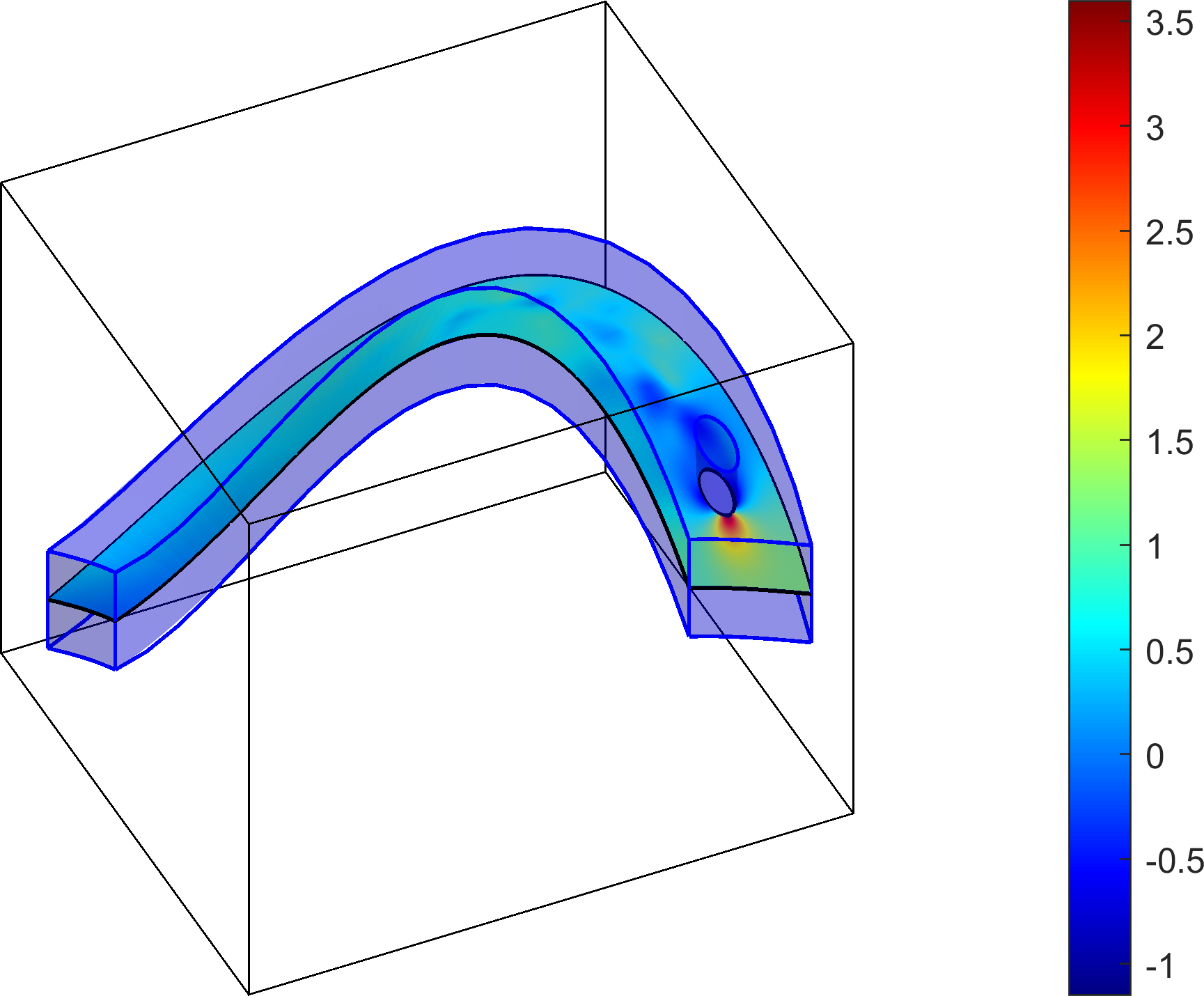}}\hspace{0.1\textwidth}
	\subfigure[SRF, $p$, $t = 1.5$]{\includegraphics[width=0.35\textwidth]{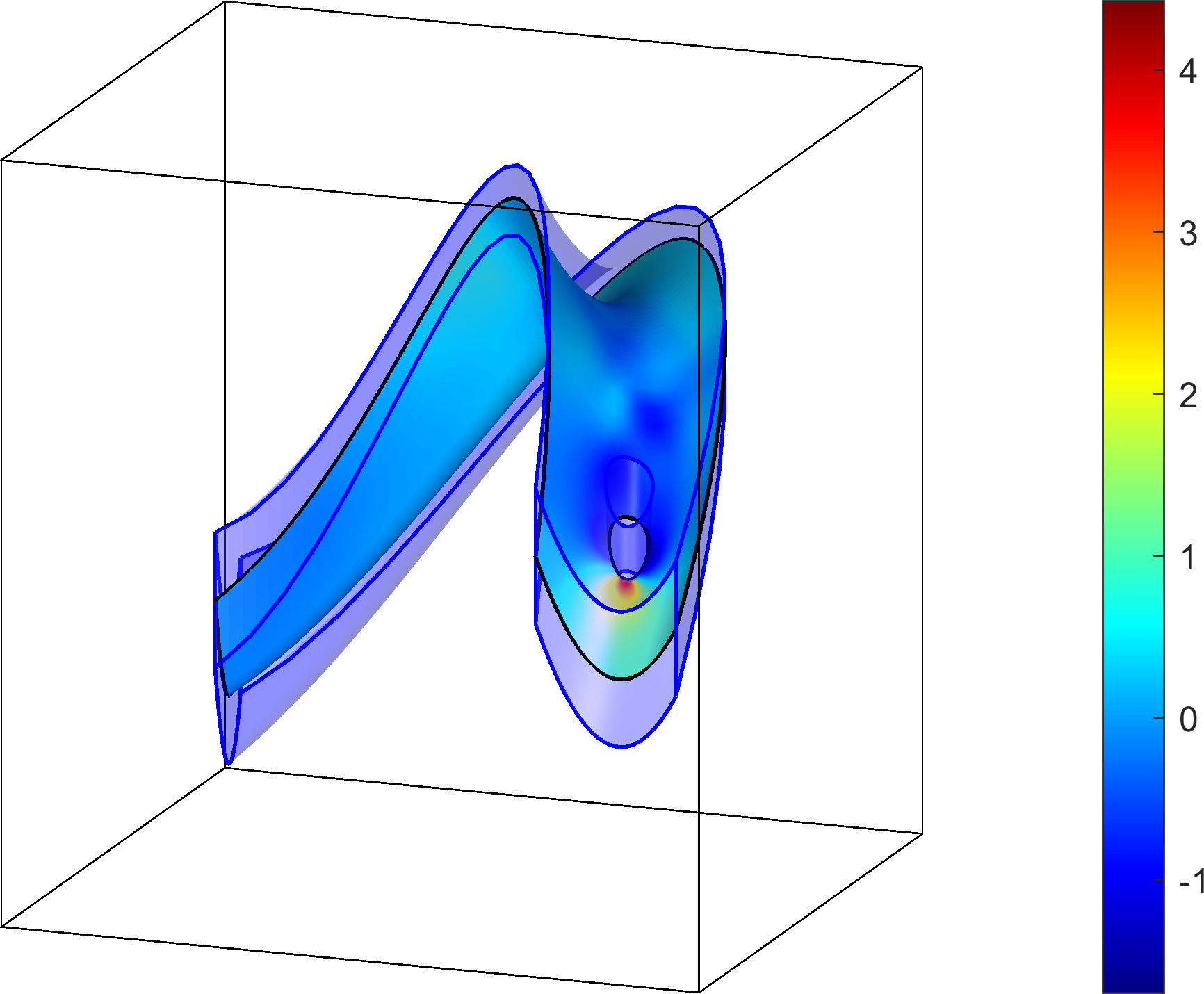}}
	\subfigure[BTF, $\omega^{\star}$, $t = 1.5$]{\includegraphics[width=0.35\textwidth]{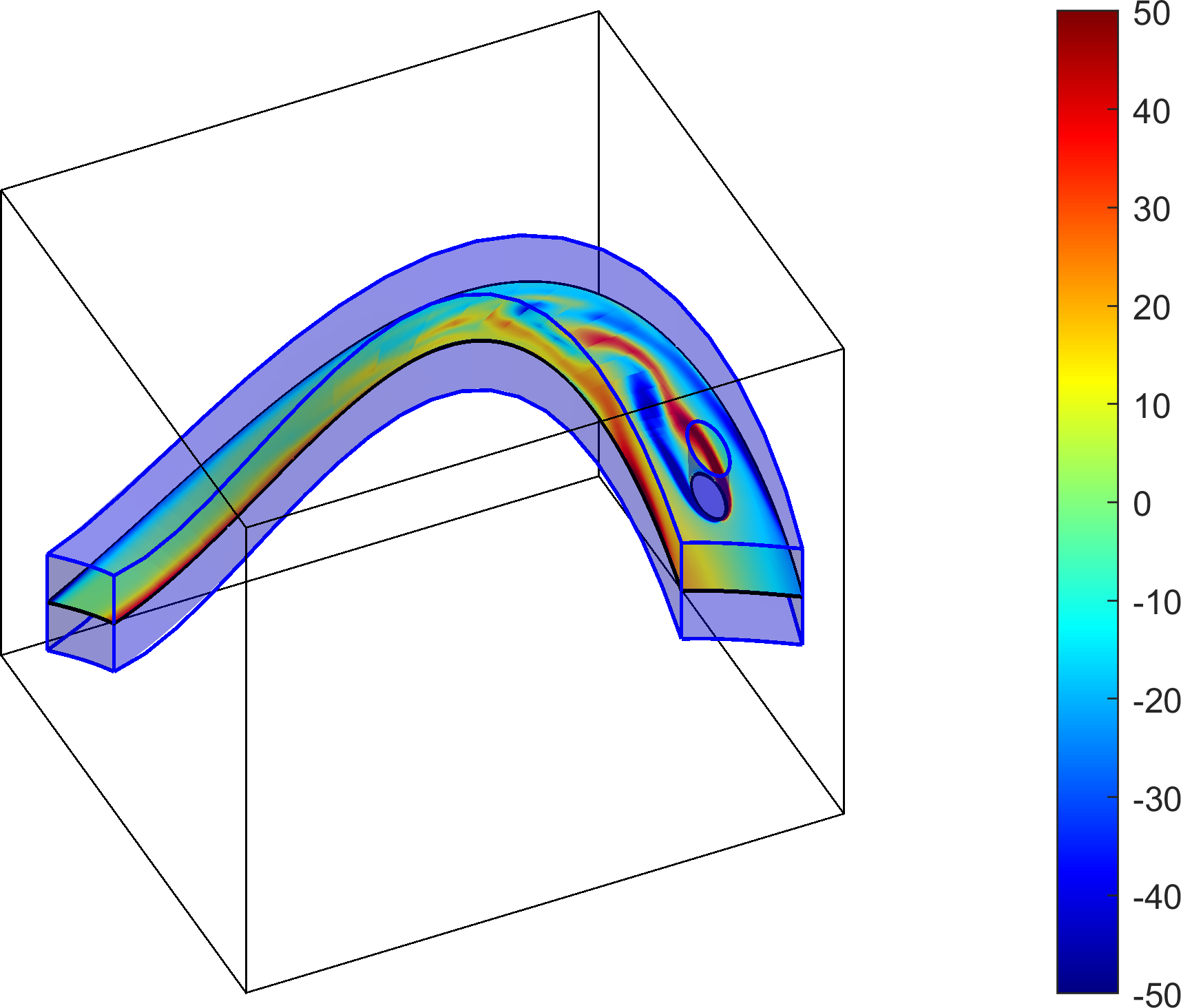}}\hspace{0.1\textwidth}
	\subfigure[SRF, $\omega^{\star}$, $t = 1.5$]{\includegraphics[width=0.35\textwidth]{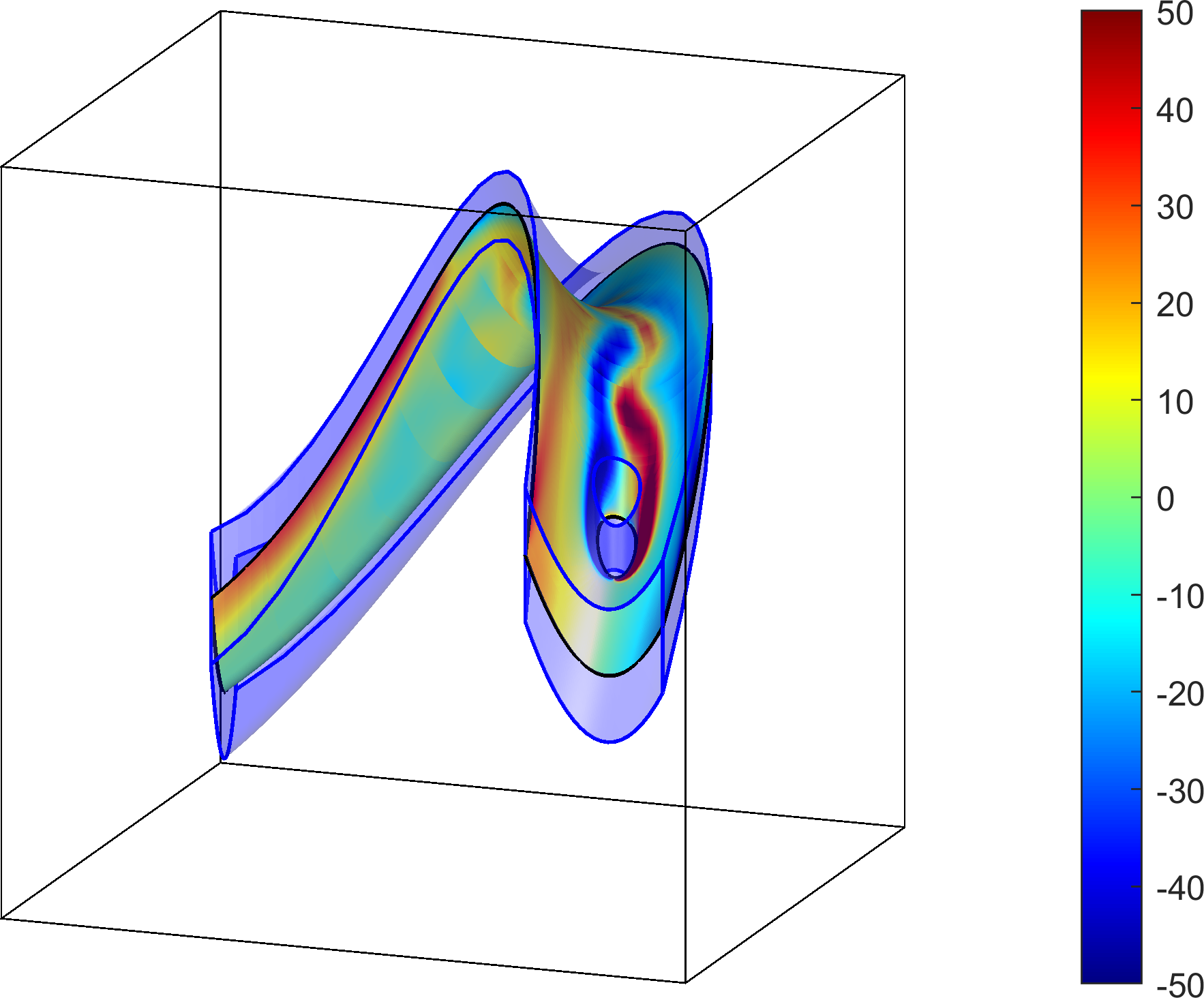}}
	
	\caption{\label{fig:TurekInstatNSEQ-Sol4} Results for mapping $\varphi_2$ on the left side and for mapping $\varphi_3$ on the right side.}
\end{figure}

In this section, we show the simultaneous solution of test cases on curved surfaces which were used in Sec.~\ref{subsec:CylFlowStat} in the context of stationary Navier--Stokes flow and are inspired by the Schäfer--Turek benchmark for a flat $2$-dimensional domain, c.f., \cite{Schaefer_1996a}. The definition of the geometry is the same as in Sec.~\ref{subsec:CylFlowStat}, i.e., mappings $\varphi_1$ to $\varphi_3$ as given in Eq.~(\ref{eq:TurekMaps}) and shown in Fig.~\ref{fig:TurekMaps}. Furthermore, the boundary conditions and the material parameters except for the viscosity are the same as used for the stationary Navier--Stokes flow above.\\
\\
The solutions for each surface are strongly nonlinear and, therefore, differ (significantly) from each other. Depending on the geometry and the fluid's parameters, the solution fields may be not smooth (enough) any more to use the classical (isotropic) Taylor--Hood element pairs for velocity and pressure. One possibility to overcome this problem is to introduce an anisotropic Taylor--Hood element pair. In the anisotropic case, the discretization of the bulk domain with respect to the level sets is as in the isotropic case, i.e., $q_p = q_{\vek{u}}-1$, while in the thickness direction (`normal' to the surfaces) the order is $q_p = q_{\vek{u}}$. Fig.~\ref{fig:AnisMeshGenEx} shows a comparison of some discretization using isotropic and anisotropic Taylor--Hood elements for a generic geometry. The black dots are degrees of freedom (DOFs) of the velocity, the blue circles are the DOFs for the pressure and the red dots are the additional DOFs of the pressure which result from the anisotropic order of the elements used in the mesh for the pressure discretization. Applying this concept in the Bulk Trace FEM for the instationary Navier--Stokes flow leads to satisfactory results. However, the disadvantage of the method is that the meshes must be somewhat aligned to the level sets.
Another strategy to overcome this problem and get sufficient results in the simultaneous solutions, is to use element pairs of equal-order for velocity and pressure together with a stabilization scheme, i.e., PSPG stabilization \cite{Hughes_1986e} or the Brezzi--Pitkäranta stabilization \cite{Brezzi_1984a} which is applied in the context of the Trace FEM for one single surface in \cite{Olshanskii_2018a}. For the PSPG stabilization, we add to the left hand side of Eq.~(\ref{eq:InstNSEQdiscWFmom}) the following term
\begin{equation}
	+ \sum_{i=1}^{n_{\mathrm{el}}} \int_{\Omega^{\mathrm{el},i}} \tau_{\mathrm{PSPG}} \cdot \frac{1}{\varrho} \cdot \left(\nabla_{\Gamma} w_p^h \right) \left[ \varrho\cdot\left(\dot{\vek{u}}^h_\ti+\left(\vek{u}_\ti^h\cdot\nabla_{\Gamma}^{\mathrm{cov}}\right)\vek{u}_\ti^h\right)-\mat{P}\cdot\mathrm{div}_{\Gamma}\,\vek{\sigma}_\ti\left(\vek{u}^h_\ti,p^h\right)-\vek{f}_\ti^h\right] \cdot \lVert \nabla \phi \rVert\,\mathrm{d}\Omega. \label{eq:PSPGinstatNSEQ}
\end{equation} 
which is to be evaluated as a sum over the element \emph{interiors} as usual in residual-based stabilization schemes. For the Brezzi--Pitkäranta stabilization \cite{Brezzi_1984a,Olshanskii_2018a}, we add
\begin{equation}
	+ \tau_p \int_{\Omega^h} \nabla_{\Gamma} w_p^h \cdot \nabla_{\Gamma} p^h \cdot \lVert \nabla \phi \rVert\,\mathrm{d}\Omega. \label{eq:BrezziPitk}
\end{equation}
Note that for Euclidean geometries, i.e., `classical' Navier--Stokes flows in $\mathbb{R}^d$, $d = \{2,3\}$ (not in a manifold context), the Brezzi--Pitkäranta stabilization is equivalent to the pressure term in the PSPG stabilization. This is not the case in the context of curved surfaces because the projector $\mat{P}$ is involved in the definition of the (Boussinesq--Scriven surface) stress tensor, see Eq.~\ref{eq:StressTens}. The stabilization parameter for the PSPG stabilization is defined as 
\begin{equation}
	\tau_{\mathrm{PSPG}} = \left[\left(\frac{2}{\Delta t}\right)^2 + \left(\frac{2 \lVert \vek{u}_{\mathrm{el,node}}\rVert}{h_\mathrm{el}}\right)^2 + \left(\frac{4\mu}{h_\mathrm{el}^2}\right)^2\right]^{-1/2}
\end{equation}
which is analogously to Eq.~(\ref{eq:PSPGstabParamStat}) and closely related to \cite{Tezduyar_2003a}. For the Brezzi--Pitkäranta stabilization, the parameter is chosen as $\tau_p = h_\mathrm{el}$, see \cite{Olshanskii_2018a}.\\
\\
For the following computations, the shear viscosity is set to $\mu = 0.0015$ for $\varphi_1$ and $\varphi_2$ and to $\mu = 0.002$ for $\varphi_3$. Fig.~\ref{fig:TurekInstatNSEQ-ResPD} shows the pressure difference between the front and the back node at the obstacle over time for different mappings and computations done with iso- and anisotropic Taylor--Hood elements. Note that for the case of anisotropic Taylor--Hood elements, two surfaces more are shown in the figures than for the isotropic Taylor--Hood elements because the anisotropic mesh includes more nodes due to the anisotropic pressure mesh. The bulk domain $\Omega$ is discretized with 3760 elements. The order for the velocities is $q_{\vek{u}} = 2$, and for the pressure $q_{p} = 1$. The geometry is again considered with elements of third order, $q_{\mathrm{geom}} = 3$. The continuous lines show results obtained with the Bulk Trace FEM and the dashed lines show the results obtained with the Surface FEM with Taylor--Hood elements in all subfigures. These plots show that good agreement is obtained. There is some offset between the results of the Bulk Trace FEM and the Surface FEM for the isotropic element pairs and for $\varphi_3$ also for the anisotropic case. However, this does not significantly change the frequency and the pressure values and, therefore, these figures are still an indication for good agreement between the simultaneous solution and the multiple solutions on individual surfaces.
Furthermore, we show plots of the velocity, pressure, and vorticity fields at different times for the three mappings in Figs.~\ref{fig:TurekInstatNSEQ-Sol1} to \ref{fig:TurekInstatNSEQ-Sol4}, respectively. The results of the simultaneous solutions shown in these figures are obtained with the anisotropic Taylor--Hood elements and are compared to solutions computed with Surface FEM for one selected surface defined by mapping $\varphi_1$. Fig.~\ref{fig:TurekInstatNSEQ-Sol4} shows one selected figure for each solution field for mappings $\varphi_2$ and $\varphi_3$. Further visualizations at other times and a comparison with the Surface FEM are omitted for brevity.
\\
\\
For lower viscosities, i.e., larger Reynolds numbers, isotropic Taylor--Hood elements may not lead to satisfactory results as seen in Fig.~\ref{fig:TurekInstatNSEQ-ResPD-001}. Herein, the viscosity for flows on surfaces defined by mapping $\varphi_1$ is set to $\eta = 0.001$. In Fig.~\ref{fig:TurekInstatNSEQ-ResPD-001} (a), isotropic Taylor--Hood elements are used and it can be clearly seen that these results are not usable due to the differences between the surfaces which result from the strong non-linearities in the solution fields. When anisotropic  Taylor--Hood elements or equal-order elements with $q_{\vek{u}} = q_{p} = 2$ and a stabilization technique are applied, the simultaneously computed results are clearly better. Although, there are offsets between the simultaneously and for each surface independently obtained solutions, the amplitudes and the frequencies are similar. Figs. \ref{fig:TurekInstatNSEQ-ResPD-001} (a), (b), (c), and (d) show the same flow situation computed with different methods, i.e., isotropic and anisotropic Taylor--Hood elements, Brezzi--Pitkäranta stabilization, and PSPG stabilization, respectively. For the other mappings the situation is similar for flows with lower viscosity, yet not shown for brevity.
\begin{figure}
	\centering
	
	\subfigure[$\varphi_1$ - isotropic THE]{\includegraphics[width=0.35\textwidth]{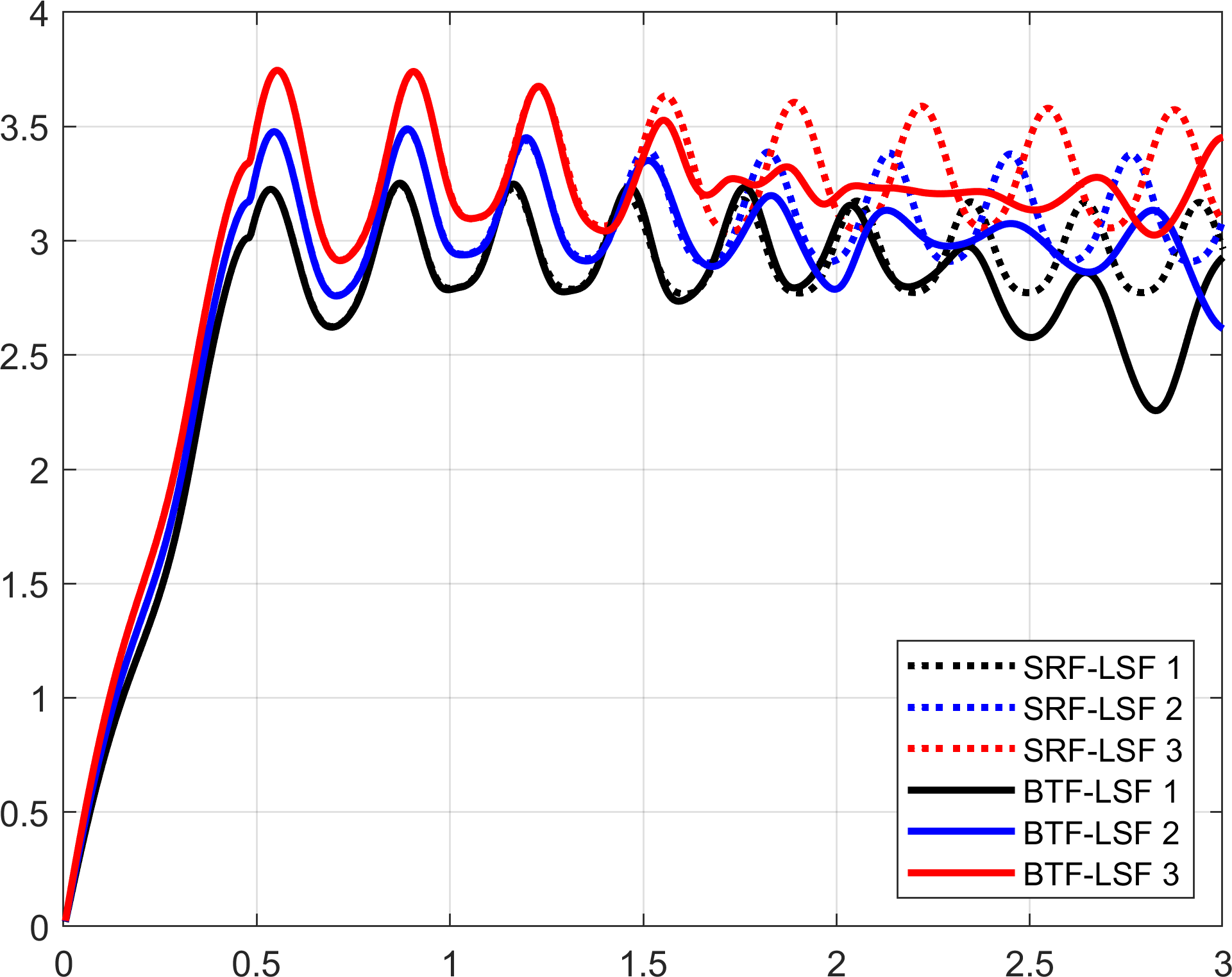}}\hspace{0.1\textwidth}
	\subfigure[$\varphi_1$ - anisotropic THE]{\includegraphics[width=0.35\textwidth]{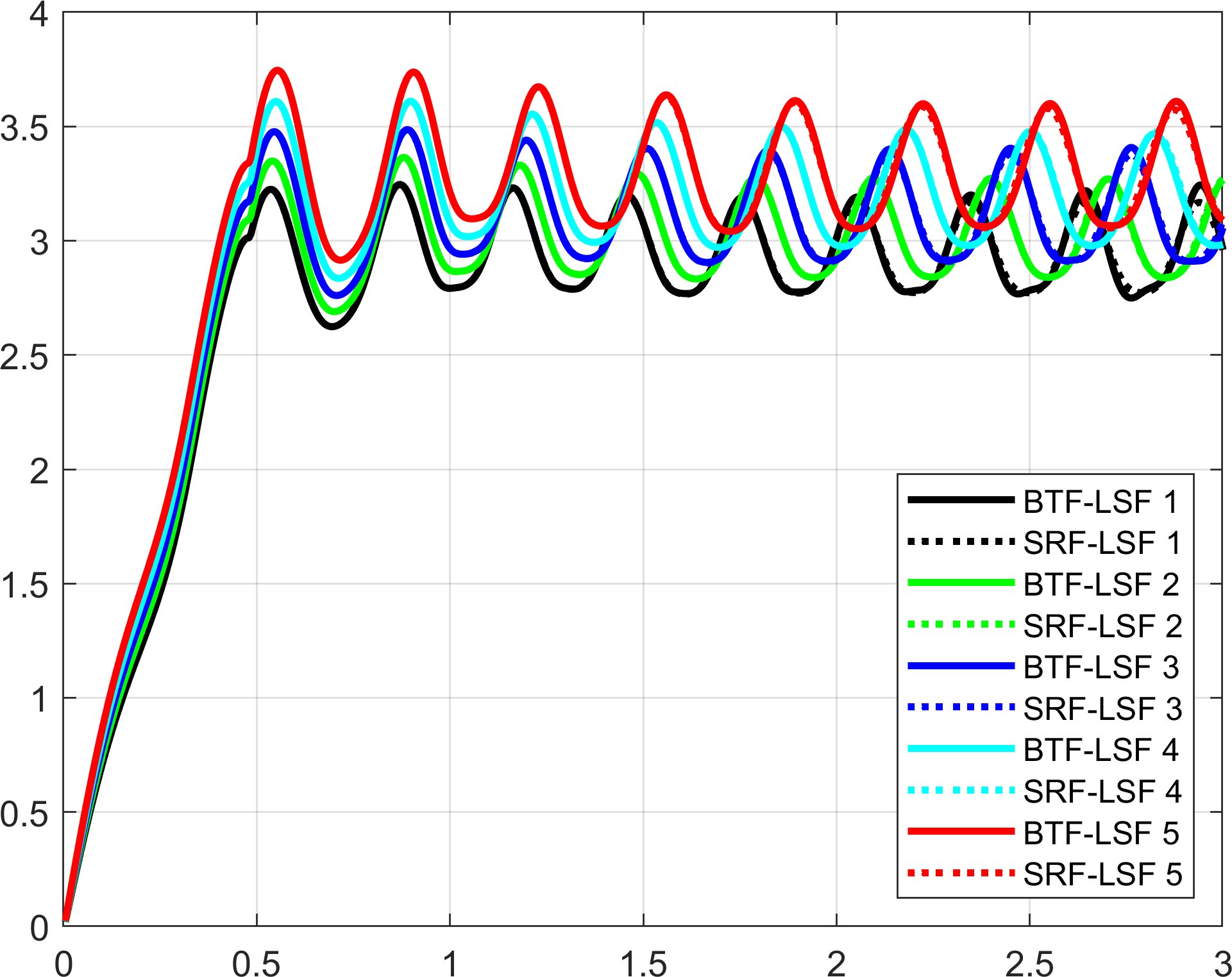}}
	\subfigure[$\varphi_1$ - BP - stabilization]{\includegraphics[width=0.35\textwidth]{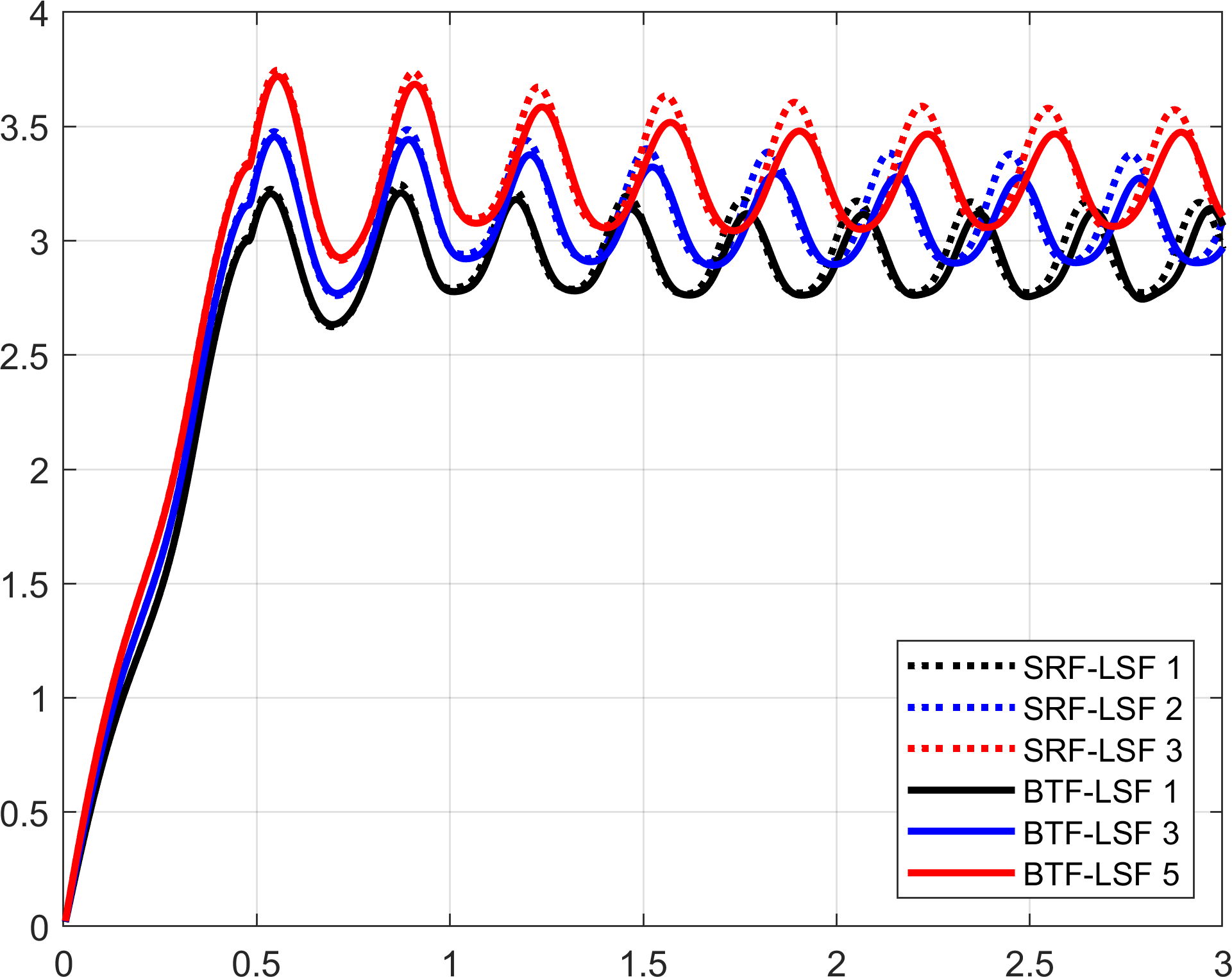}}\hspace{0.1\textwidth}
	\subfigure[$\varphi_1$ - PSPG stabilization]{\includegraphics[width=0.35\textwidth]{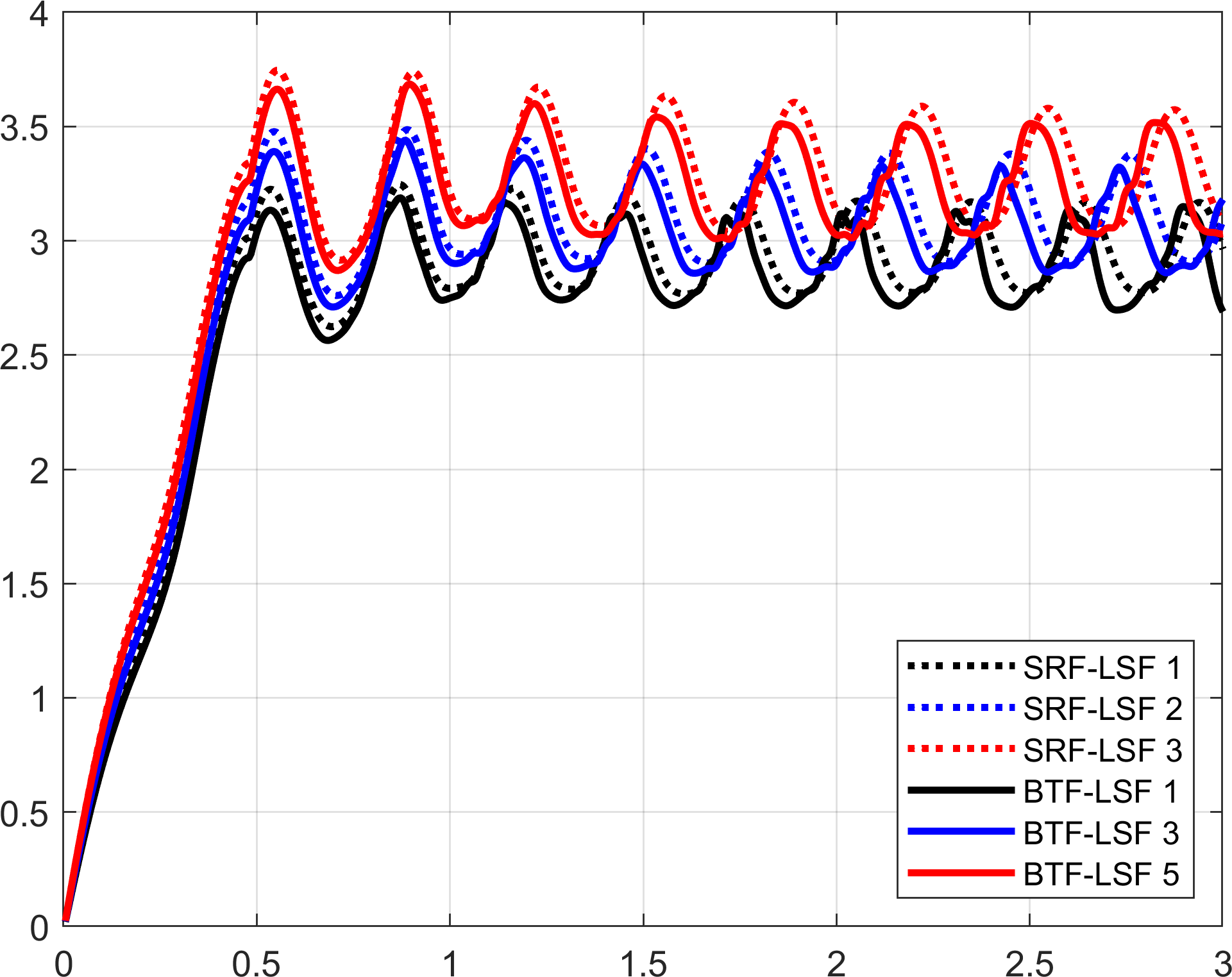}}
	\subfigure[$\varphi_2$ - anisotropic THE - $\lVert \vek{u} \rVert$, $t = 3$]{\includegraphics[width=0.35\textwidth]{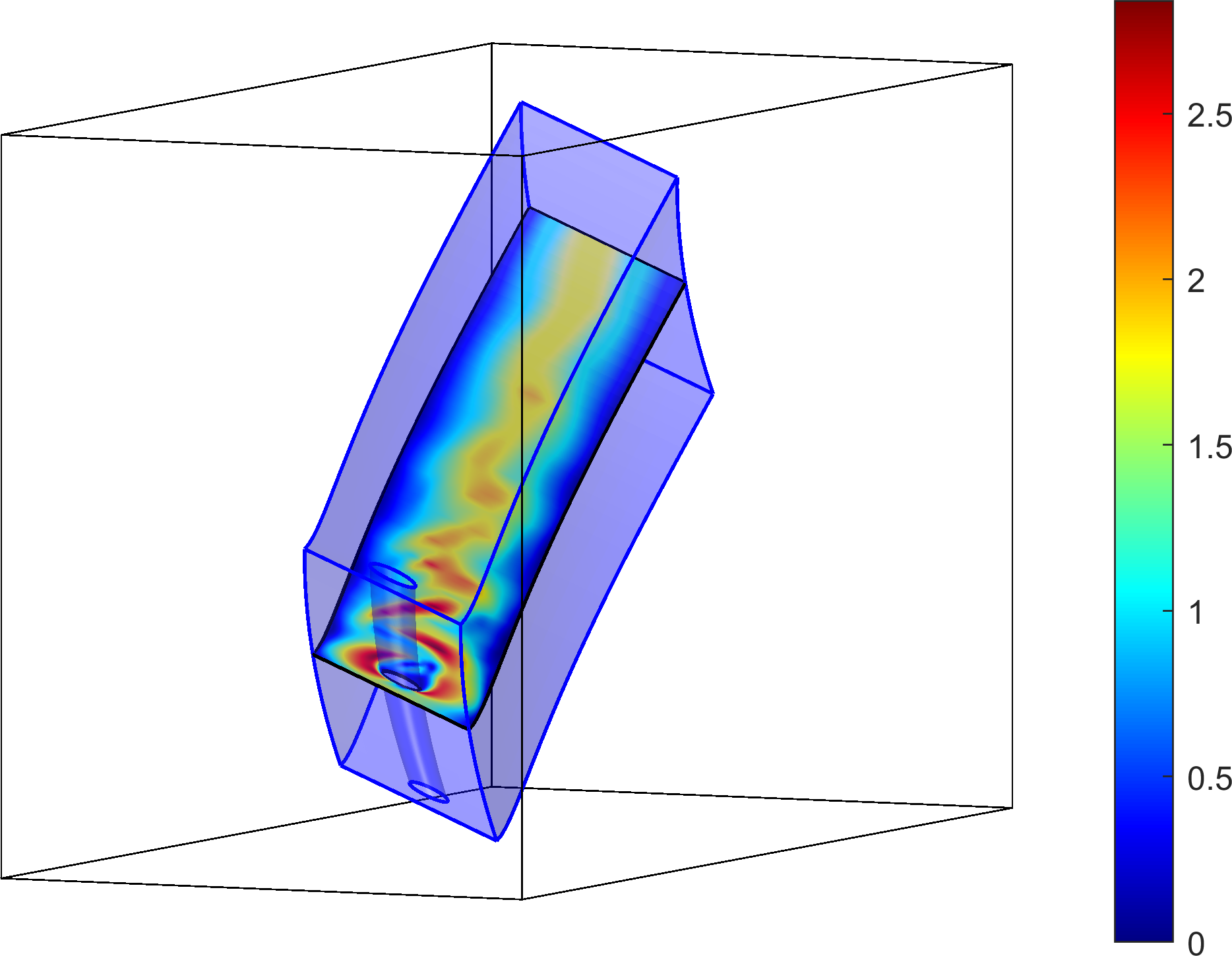}}\hspace{0.1\textwidth}
	\subfigure[$\varphi_3$ - anisotropic THE -  $\omega^{\star}$, $t = 3$]{\includegraphics[width=0.35\textwidth]{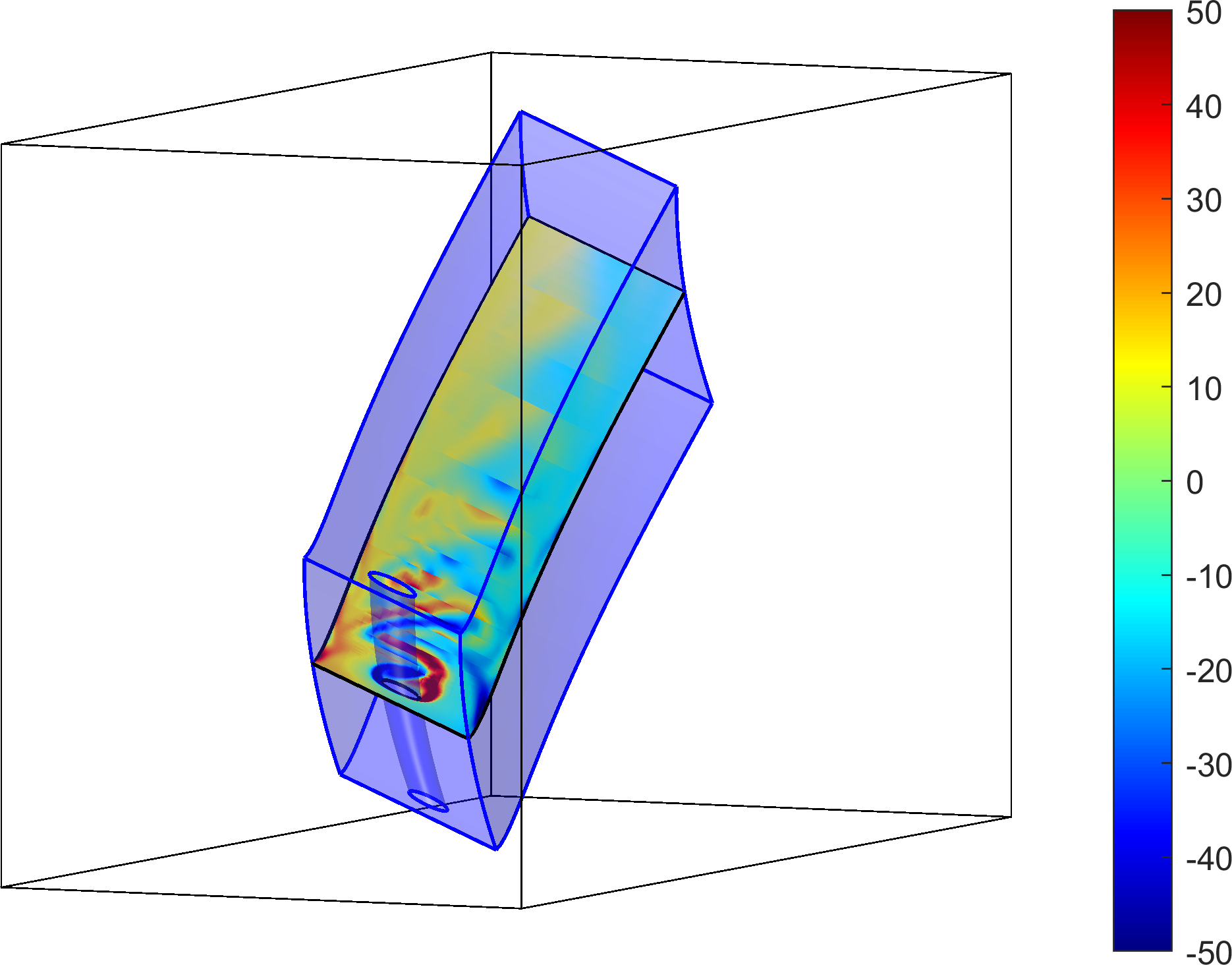}}
	
	\caption{\label{fig:TurekInstatNSEQ-ResPD-001} Results for Navier--Stokes flow with $\eta = 0.001$ on geometries implied by mapping $\varphi_1$. (a) to (d) show the pressure difference on the obstacle over time, where (a) is obtained using (isotropic) Taylor--Hood elements (THE), (b) results from anisotropic THE, in (c) the Brezzi--Pitkäranta (BP) stabilization and in (d) PSPG stabilisation is applied. In (e) the velocity magnitudes and in (f) the vorticity on one selected surface are shown.}
\end{figure}

\section{Conclusions and outlook}\label{sec:ConclOutlook}
The \emph{simultaneous} solution of Stokes (stationary) and Navier--Stokes flows (stationary and instationary) on multiple curved surfaces is proposed in this paper. Therefore, the governing equations for one \emph{single} surface in a coordinate-free formulation in their weak form are used as a starting point and combined  with co-area formulas to obtain a weak form which simultaneously applies for all level sets embedded in a three-dimensional bulk domain. The Bulk Trace FEM is employed to approximate the solution of these weak forms using arbitrary (higher-order) background meshes. These meshes are conforming to the boundary of the surfaces and, hence, drawbacks of the classical Trace FEM for single surface solutions, e.g., small cut-scenarios and, therefore, required stabilization techniques, do not occur in the Bulk Trace FEM. Numerical examples validate the method obtaining higher-order convergence rates for stationary Stokes flow. Good agreements to Surface FEM solutions on selected, individual surfaces for stationary and instationary Navier--Stokes flows are confirmed.\\
\\
The simultaneously generated results may be used in design processes to study different variants of geometries, e.g., how a changing curvature changes the flow patterns. Fields of applications could be in biomechanics, e.g., cell mechanics, physics, e.g., nuclear fusion where flows on a torus are important, and other engineering applications in design and optimization.\\
\\
Further research shall focus on improvements for instationary Navier--Stokes equations. Therefore, solution strategies should be obtained to better handle advection-induced instabilities in the flow. A detailed investigation of optimal stabilization parameters for Brezzi--Pitkäranta and PSPG stabilization techniques in the simultaneous solution is desirable. A combination of the work proposed herein with the simultaneous solution of structural shells or membranes to model biomechanical processes in cells is also an interesting topic for future research.



\printbibliography

\end{document}